\pdfoutput=1

\documentclass[11pt,twoside,a4paper,cmspaper,final,collab]{cms-tdr}
\def\svnVersion{361148}\def\cmsCernNoTag{CERN-PH-EP/2015-240}\def\cmsCernDate{\today}\def\cmsMessage{Published in the European Physical Journal C as \href{http://dx.doi.org/10.1140/epjc/s10052-016-4105-x}{\doi{10.1140/epjc/s10052-016-4105-x.}}}

\begin{document}\cmsNoteHeader{TOP-12-041}

\hyphenation{had-ron-i-za-tion}
\hyphenation{cal-or-i-me-ter}
\hyphenation{de-vices}
\RCS$Revision: 360507 $
\RCS$HeadURL: svn+ssh://svn.cern.ch/reps/tdr2/papers/TOP-12-041/trunk/TOP-12-041.tex $
\RCS$Id: TOP-12-041.tex 360507 2016-07-28 11:41:46Z cdiez $
\newlength\cmsFigWidth
\ifthenelse{\boolean{cms@external}}{\setlength\cmsFigWidth{0.85\columnwidth}}{\setlength\cmsFigWidth{0.4\textwidth}}
\ifthenelse{\boolean{cms@external}}{\providecommand{\cmsLeft}{top}}{\providecommand{\cmsLeft}{left}}
\ifthenelse{\boolean{cms@external}}{\providecommand{\cmsRight}{bottom}}{\providecommand{\cmsRight}{right}}
\providecommand{\CL}{CL\xspace}
\providecommand{\NA}{---\xspace}

\newcommand{\Madspin} {{\textsc{MadSpin}}\xspace}
\newcommand{\PowHel} {{\textsc{PowHel}}\xspace}
\newcommand{\Whizard}{\textsc{Whizard}\xspace}
\newcommand{\ttbb}{\ensuremath{\PQt\PAQt\PQb\PAQb}\xspace}
\newcommand{\ttb}{\ensuremath{\PQt\PAQt\PQb}\xspace}
\newcommand{\ttjj}{\ensuremath{\PQt\PAQt\mathrm{jj}}\xspace}
\newcommand{\ttH}{\ensuremath{\PQt\PAQt\PH}\xspace}
\newcommand{\ttHtobb}{\ensuremath{\PQt\PAQt\PH\,(\PQb\PAQb)}\xspace}
\newcommand{\ttcc}{\ensuremath{\PQt\PAQt\PQc\PAQc}\xspace}
\newcommand{\tttwob}{\ensuremath{\PQt\PAQt 2 \PQb}\xspace}
\newcommand{\abseta}{\ensuremath{|\eta|}\xspace}
\newcommand{\mbb}{\ensuremath{m_{\PQb\PQb}}\xspace}
\newcommand{\mjj}{\ensuremath{m_{\mathrm{jj}}}\xspace}
\newcommand{\Djj}{\ensuremath{\Delta R_{\mathrm{jj}}}\xspace}
\newcommand{\xsectheo}{\ensuremath{252.9\, \pm \,\xspace^{6.4}_{8.6} \text{(scale)} \pm 11.7 (\mathrm{PDF}+\alpha_s)\unit{pb}}\xspace}
\newcommand{\amcatnlo}{\textsc{MG5\_aMC@NLO}\xspace}
\newcommand{\hdamp}{\textsc{hdamp}\xspace}
\newcommand{\Vptmiss}{\ensuremath{{\vec p}_{\mathrm{T}}\hspace{-0.78em}/\kern0.45em}\xspace}

\newcolumntype{y}{D{,}{,}{-1}}
\newcolumntype{x}[1]{D{,}{\, \times\, }{#1}}

\cmsNoteHeader{TOP-12-041}
\title{Measurement of \texorpdfstring{\ttbar}{t-tbar} production with additional jet activity, including \texorpdfstring{\PQb}{b} quark jets, in the dilepton decay channel using pp collisions at \texorpdfstring{$\sqrt{s} = 8\TeV$}{sqrt(s) = 8 TeV}}

\titlerunning{\ttbar production with additional jet activity in pp collisions at 8\TeV}

\date{\today}

\abstract{
Jet multiplicity distributions in top quark pair (\ttbar) events are measured in pp collisions at a centre-of-mass energy of 8 TeV with the CMS detector at the LHC using a data set corresponding to an integrated luminosity of 19.7\fbinv. The measurement is performed in the dilepton decay channels ($\Pep \Pem$, $\Pgmp \Pgmm$, and $\Pe^{\pm}\Pgm^{\mp}$). The absolute and normalized differential cross sections for $\ttbar$ production are measured as a function of the jet multiplicity in the event for different jet transverse momentum thresholds and the kinematic properties of the leading additional jets. The differential $\ttb$ and $\ttbb$ cross sections are presented for the first time as a function of the kinematic properties of the leading additional $\PQb$ jets. Furthermore, the fraction of events without additional jets above a threshold is measured as a function of the transverse momenta of the leading additional jets and the scalar sum of the transverse momenta of all additional jets. The data are compared and found to be consistent with predictions from several perturbative quantum chromodynamics event generators and a next-to-leading order calculation.}

\hypersetup{%
pdfauthor={CMS Collaboration},%
pdftitle={Measurement of t-tbar production with additional jet activity, including b quark jets, in the dilepton channel using pp collisions at sqrt(s) = 8 TeV},%
pdfsubject={CMS},%
pdfkeywords={CMS, physics, top, cross section}
}

\maketitle
\section{Introduction}

Precise measurements of \ttbar production and decay properties~\cite{{Chatrchyan:2013faa},{Aad:2014kva},{bib:TOP-11-013_paper},{bib:2012hg}, {bib:TOP-12-018}, {bib:ATLAS_diffxsec_7TeV}, {bib:ATLAS_jetmulti_7TeV}, {bib:TOP-12-028},{Aad:2015pga}, {bib:ATLAS_jetmulti_7TeV}} provide crucial information for testing the expectations of the standard model (SM) and specifically of calculations in the framework of perturbative quantum chromodynamics (QCD) at high-energy scales. At the energies of the CERN LHC, about half of the \ttbar events contain jets with transverse momentum (\pt) larger than 30\GeV that do not come from the weak decay of the \ttbar system~\cite{bib:TOP-12-018}. In this paper, these jets will be referred to as ``additional jets'' and the events as ``\ttbar{}+jets''. The additional jets typically arise from initial-state QCD radiation, and their study provides an essential test of the validity and completeness of higher-order QCD calculations describing the processes leading to multijet events.

A correct description of these events is also relevant because \ttbar{}+jets processes constitute important backgrounds in the searches for new physics. These processes also constitute a challenging background in the attempt to observe the production of a Higgs boson in association with a \ttbar pair (\ttH), where the Higgs boson decays to a bottom (\PQb) quark pair (\bbbar), because of the much larger cross section compared to the \ttH signal. Such a process has an irreducible nonresonant background from \ttbar pair production in association with a \bbbar pair from gluon splitting. Therefore, measurements of \ttbar{}+jets and \ttbb production can give important information about the main background in the search for the \ttH process and provide a good test of next-to-leading-order (NLO) QCD calculations.

Here, we present a detailed study of the production of \ttbar events with additional jets and \PQb quark jets in the final state from pp~collisions at $\sqrt{s} = 8\TeV$ using the data recorded in 2012 with the CMS detector, corresponding to an integrated luminosity of 19.7 \fbinv. The \ttbar pairs are reconstructed in the dilepton decay channel with two oppositely charged isolated leptons (electrons or muons) and at least two jets. The analysis follows, to a large extent, the strategy used in the measurement of normalized \ttbar differential cross sections in the same decay channel described in Ref.~\cite{bib:TOP-12-028}.

The measurements of the absolute and normalized differential \ttbar cross sections are performed as a function of the jet multiplicity for different \pt thresholds for the jets, in order to probe the momentum dependence of the hard-gluon emission.
The results are presented in a visible phase space in which all selected final-state objects are produced within the detector acceptance and are thus measurable experimentally. The study extends the previous measurement at $\sqrt{s} = 7\TeV$~\cite{bib:TOP-12-018}, where only normalized differential cross sections were presented.

The absolute and normalized \ttbar{}+jets production cross sections are also measured as a function of the \pt and pseudorapidity ($\eta$)~\cite{bib:CMS} of the leading additional jets, ordered by \pt{}. The CMS experiment has previously published a measurement of the inclusive \ttbb production cross section~\cite{bib:ttbb_ratio:2014}. In the present analysis, the \ttbb and \ttb (referred to as ``\ttbb(\ttb)'' in the following) cross sections are measured for the first time differentially as a function of the properties of the additional jets associated with \PQb quarks, which will hereafter be called \PQb jets. The \ttbb process corresponds to events where two additional \PQb jets are generated in the visible phase space, while \ttb represents the same physical process, where only one additional \PQb jet is within the acceptance requirements. In cases with at least two additional jets or two \PQb jets, the cross section is also measured as a function of the angular distance between the two jets and their dijet invariant mass. The results are reported both in the visible phase space and extrapolated to the full phase space of the \ttbar system to facilitate the comparison with theoretical calculations.

Finally, the fraction of events that do not contain additional jets (gap fraction) is determined as a function of the threshold on the leading and subleading additional-jet \pt{}, and the scalar sum of all additional-jet \pt{}. This was first measured in Refs.~\cite{bib:TOP-12-018} and~\cite{bib:atlas2}.

The results are compared at particle level to theoretical predictions obtained with four different event generators: \MADGRAPH~\cite{bib:madgraph}, \MCATNLO~\cite{bib:mcatnlo}, \POWHEG~\cite{Frixione:2007vw}, and \amcatnlo~\cite{MG5_amcnlo}, interfaced with either \PYTHIA~\cite{Sjostrand:2006za} or \HERWIG~\cite{bib:herwig}, and in the case of \POWHEG with both. Additionally, the measurements as a function of the \PQb jet quantities are compared to the predictions from the event generator \PowHel~\cite{Garzelli:2014aba}.

This paper is structured as follows. A brief description of the CMS detector is provided in Section~\ref{sec:CMS}. Details of the event simulation generators and their theoretical predictions are given in Section~\ref{sec:theory}. The event selection and the method used to identify the additional radiation in the event for both \ttbar{}+jets and \ttbb(\ttb) studies are presented in Sections~\ref{sec:selection} and~\ref{sec:tt_addjets}. The cross section measurement and the systematic uncertainties are described in Sections~\ref{sec:syst} and~\ref{sec:diffxsec}. The results as a function of the jet multiplicity and the kinematic properties of the additional jets and \PQb jets are presented in Sections~\ref{sec:diffxsecNJets}--\ref{sec:diffxsecAddbJets}. The definition of the gap fraction and the results are described in Section~\ref{sec:gap}. Finally, a summary is given in Section~\ref{sec:summary}.

\section{The CMS detector}
\label{sec:CMS}
The central feature of the CMS apparatus is a superconducting solenoid of 6\unit{m} internal diameter, providing a magnetic field of 3.8\unit{T}. Within the solenoid volume are a silicon pixel and strip tracker, a lead tungstate crystal electromagnetic calorimeter (ECAL), and a brass and scintillator hadron calorimeter, each composed of a barrel and two endcap sections. Extensive forward calorimetry complements the coverage provided by the barrel and endcap detectors. Muons are measured in gas-ionization detectors embedded in the steel flux-return yoke outside the solenoid. A more detailed description of the CMS detector, together with a definition of the coordinate system used and the relevant kinematic variables, can be found in Ref.~\cite{bib:CMS}.

\section{Event simulation and theoretical predictions}
\label{sec:theory}

Experimental effects coming from event reconstruction, selection criteria, and detector resolution are modelled using Monte Carlo (MC) event generators interfaced with a detailed simulation of the CMS detector response using \GEANTfour (v.~9.4)~\cite{Agostinelli:2002hh}.

The \MADGRAPH (v.~5.1.5.11)~\cite{bib:madgraph} generator calculates the matrix elements at tree level up to a given order in $\alpha_s$. In particular, the simulated \ttbar sample used in this analysis is generated with up to three additional partons. The \Madspin~\cite{bib:madspin} package is used to incorporate spin correlations of the top quark decay products. The value of the top quark mass is chosen to be $m_{\PQt} = 172.5\GeV$, and the proton structure is described by the CTEQ6L1~\cite{bib:cteq} set of parton distribution functions (PDF). The generated events are subsequently processed with \PYTHIA (v.~6.426)~\cite{Sjostrand:2006za} for fragmentation and hadronization, using the MLM prescription for the matching of higher-multiplicity matrix element calculations with parton showers~\cite{Mangano:2006rw}. The \PYTHIA parameters for the underlying event, parton shower, and hadronization are set according to the Z2* tune, which is derived from the Z1 tune~\cite{Field:2010bc}. The Z1 tune uses the CTEQ5L PDFs, whereas Z2* adopts CTEQ6L.

In addition to the nominal \ttbar \MADGRAPH sample, dedicated samples are generated by varying the central value of the renormalization ($\mu_\mathrm{R}$) and factorization ($\mu_\mathrm{F}$) scales and the matrix element/parton showering matching scale (jet-parton matching scale). These samples are produced to determine the systematic uncertainties in the measurement owing to the theoretical assumptions on the modelling of \ttbar events, as well as for comparisons with the measured distributions. The nominal values of $\mu_\mathrm{R}$ and $\mu_\mathrm{F}$ are defined by the $Q^2$ scale in the event: $\mu_\mathrm{R}^2 =\mu_\mathrm{F}^2 = Q^2 = m_{\PQt}^2 + \sum{\pt^2(\text{jet})}$, where the sum runs over all the additional jets in the event not coming from the \ttbar decay. The samples with the varied scales use $\mu_\mathrm{R}^2 =\mu_\mathrm{F}^2 = 4Q^2$ and $Q^2/4$, respectively. For the nominal \MADGRAPH sample, a jet-parton matching scale of 40\GeV is chosen, while for the varied samples, values of 60 and 30\GeV are employed, respectively. These scales correspond to jet-parton matching thresholds of 20\GeV for the nominal sample, and 40 and 10\GeV for the varied ones.

The \POWHEG (v.~1.0 r1380) and \MCATNLO (v.~3.41) generators, along with the CT10~\cite{bib:CT10} and CTEQ6M~\cite{bib:cteq} PDFs, are used, respectively, for comparisons with the data. The \POWHEG generator simulates calculations of \ttbar production to full NLO accuracy, and is matched with two parton shower MC generators: the \PYTHIA (v.~6.426) Z2* tune (designated as \PYTHIA{6} in the following), and the \HERWIG~\cite{bib:herwig} (v.~6.520) AUET2 tune~\cite{bib:auet2tune} (referred to as \HERWIG{6} in the following). The parton showering in \PYTHIA\ is based on a transverse-momentum ordering of parton showers, whereas \HERWIG\ uses angular ordering. The \MCATNLO generator implements the hard matrix element to full NLO accuracy, matched with \HERWIG (v.~6.520) for the initial- and final-state parton showers using the default tune. These two generators, \POWHEG and \MCATNLO, are formally equivalent up to the NLO accuracy, but they differ in the techniques used to avoid double counting of radiative corrections that may arise from interfacing with the parton showering generators.

{\tolerance=1200
The cross section as a function of jet multiplicity and the gap fraction measurements are compared to the NLO predictions of the \POWHEG (v2)~\cite{Frixione:2007vw} and \amcatnlo~\cite{MG5_amcnlo} generators. The \POWHEG{} (v2) generator is matched to the \PYTHIA (v.~8.205) CUETP8M1 tune~\cite{Astalos:2015ivw} (referred to as \PYTHIA{8}), \HERWIG{6}, and \PYTHIA{6}. In these samples the \hdamp parameter of \textsc{powhegbox}, which controls the matrix element and parton shower matching and effectively regulates the high-\pt radiation, is set to $m_{\PQt}= 172.5\GeV$. The \amcatnlo generator simulates \ttbar events with up to two additional partons at NLO, and is matched to the \PYTHIA{8} parton shower simulation using the \textsc{FxFx} merging prescription~\cite{Frederix:2012ps}. The top quark mass value used in all these simulations is also 172.5\GeV and the PDF set is NNPDF3.0~\cite{Ball:2014uwa}. In addition, a \ttbar \MADGRAPH sample matched to \PYTHIA{8} for the parton showering and hadronization is used for comparisons with the data.
\par}

{\tolerance=1200
The \ttbb production cross sections are also compared with the predictions by the generator \PowHel~\cite{Garzelli:2014aba} ({\textsc HELAC-NLO}~\cite{Bevilacqua:2011xh} + \textsc{powhegbox}~\cite{Alioli:2010xd}), which implements the full \ttbb process at NLO QCD accuracy, with parton shower matching based on the \POWHEG NLO matching algorithm~\cite{Nason:2004rx, Frixione:2007vw}. The events are further hadronized by means of \PYTHIA (v.~6.428), using parameters of the Perugia 2011 C tune~\cite{Skands:2010ak}. In the generation of the events, the renormalization and factorization scales are fixed to $\mu_\mathrm{R} = \mu_\mathrm{F} = \HT/4$, where $\HT$ is the sum of the transverse energies of the final-state partons (\PQt, \PAQt, \PQb, \PAQb) from the underlying tree-level process, and the CT10 PDFs are used.
\par}

The SM background samples are simulated with \MADGRAPH, \POWHEG, or \PYTHIA, depending on the process. The \MADGRAPH generator is used to simulate $\PZ / \Pgg^*$ production (referred to as Drell--Yan, DY, in the following), \ttbar production in association with an additional boson (referred to as $\ttbar$+$\PZ$, $\ttbar$+$\PW$, and $\ttbar$+$\Pgg$), and \PW\ boson production with additional jets ($\PW$+jets in the following). Single top quark events ($\PQt\PW$ channel) are simulated using \POWHEG. Diboson ($\PW\PW$, $\PW\PZ$, and $\PZ\PZ$) and QCD multijet events are simulated using \PYTHIA. For the \ttb and \ttbb measurements, the expected contribution from SM \ttH processes, simulated with \PYTHIA, is also considered, although the final state has not yet been observed.

For comparison with the measured distributions, the events in the simulated samples are normalized to an integrated luminosity of $19.7\fbinv$ according to their predicted cross sections. These are taken from next-to-next-to-leading-order (NNLO) ($\PW$+jets~\cite{Melnikov:2006di} and DY~\cite{Melnikov:2006kv}), NLO + next-to-next-to-leading logarithmic (NNLL) (single top quark $\PQt\PW$ channel~\cite{bib:twchan}), NLO (diboson~\cite{bib:mcfm:diboson}, $\ttbar$+$\PZ$~\cite{bib:ttV}, $\ttbar$+$\PW$~\cite{bib:ttV}, and $\ttbar$+$\PH$~\cite{Heinemeyer:1559921}), and leading-order (LO) (QCD multijet~\cite{Sjostrand:2006za}) calculations. The contribution of QCD multijet events is found to be negligible.
The predicted cross section for the $\ttbar$+$\Pgg$ sample is obtained by scaling the LO cross section obtained with the \Whizard\ event generator~\cite{bib:whizard} by an NLO/LO $K$-factor correction~\cite{bib:ttgamma}. The \ttbar simulated sample is normalized to the total cross section $\sigma_{\ttbar} = \xsectheo$, calculated with the \textsc{Top++2.0} program to NNLO in perturbative QCD, including soft-gluon resummation to NNLL order~\cite{Czakon:2011xx}, and assuming $m_{\PQt} = 172.5\GeV$. The first uncertainty comes from the independent variation of the factorization and renormalization scales, $\mu_\mathrm{R}$ and $\mu_\mathrm{F}$, while the second one is associated with variations in the PDF and $\alpha_s$, following the PDF4LHC prescription with the MSTW2008 68\% confidence level (\CL) NNLO, CT10 NNLO, and NNPDF2.3 5f FFN PDF sets (see Refs.~\cite{{bib:PDF4LHC},{Alekhin:2011sk}} and references therein and Refs.~\cite{{Martin:2009bu},{Gao:2013xoa},{Ball:2012cx}}).

A number of additional pp simulated hadronic interactions (``pileup") are added to each simulated event to reproduce the multiple interactions in each bunch crossing from the luminosity conditions in the real data taking. Correction factors for detector effects (described in Sections~\ref{sec:selection} and~\ref{sec:syst}) are applied, when needed, to improve the description of the data by the simulation.

\section{Event reconstruction and selection}
\label{sec:selection}
The event selection is based on the decay topology of the \ttbar events, where each top quark decays into a \PW\ boson and a \PQb quark. Only the cases in which both \PW\ bosons decayed to a charged lepton and a neutrino are considered. These signatures imply the presence of isolated leptons, missing transverse momentum owing to the neutrinos from \PW\ boson decays, and highly energetic jets. The heavy-quark content of the jets is identified through \PQb tagging techniques. The same requirements are applied to select the events for the different measurements, with the exception of the requirements on the \PQb jets, which have been optimized independently for the \ttbar{}+jets and \ttbb(\ttb) cases. The description of the event reconstruction and selection is detailed in the following.

Events are reconstructed using a particle-flow (PF) algorithm, in which signals from all subdetectors are combined~\cite{bib:pf2010,CMS-PAS-PFT-09-001}. Charged particles are required to originate from the primary collision vertex~\cite{Chatrchyan:2014fea}, defined as the vertex with the highest sum of $\pt^2$ of all reconstructed tracks associated with it. Therefore, charged-hadron candidates from pileup events, \ie originating from additional pp interactions within the same bunch crossing, are removed before jet clustering on an event-by-event basis. Subsequently, the remaining neutral-particle component from pileup events is accounted for through jet energy corrections~\cite{Cacciari:2008gn}.

Muon candidates are reconstructed from tracks that can be linked between the silicon tracker and the muon system~\cite{Chatrchyan:2012xi}. The muons are required to have $\pt>20\GeV$, be within $|\eta|<2.4$, and have a relative isolation $I_{\text{rel}}<0.15$. The parameter $I_{\text{rel}}$ is defined as the sum of the \pt of all neutral and charged reconstructed PF candidates, except the muon itself, inside a cone of $\Delta R\equiv\sqrt{\smash[b]{(\Delta\eta)^2+(\Delta\phi)^2}} < 0.3$ around the muon direction, divided by the muon \pt, where $\Delta\eta$ and $\Delta\phi$ are the difference in pseudorapidity and azimuthal angle between the directions of the candidate and the muon, respectively. Electron candidates are identified by combining information from charged-track trajectories and energy deposition measurements in the ECAL~\cite{Khachatryan:2015hwa}, and are required to be within $|\eta|<2.4$, have a transverse energy of at least 20\GeV, and fulfill $I_{\text{rel}} < 0.15$ inside a cone of $\Delta R < 0.3$. Electrons from identified photon conversions are rejected. The lepton identification and isolation efficiencies are determined via a tag-and-probe method using \PZ boson events.

Jets are reconstructed by clustering the PF candidates, using the anti-$\kt$ clustering algorithm~\cite{Cacciari:2008gp,Cacciari:2011ma} with a distance parameter of $0.5$. The jet momentum is determined as the vectorial sum of all particle momenta in the jet, and is found in the simulation to be within 5 to 10\% of the true momentum over the entire \pt range and detector acceptance. Jet energy corrections are derived from the simulation, and are confirmed with in situ measurements with the energy balance of dijet and photon+jet events~\cite{Chatrchyan:2011ds}. The jet energy resolution amounts typically to 15\% at 10\GeV and 8\% at 100\GeV. Muons and electrons passing less stringent requirements compared to the ones mentioned above are identified and excluded from the clustering process. Jets are selected in the interval $|\eta|<2.4$ and with $\pt >20\GeV$. Additionally, the jets identified as part of the decay products of the \ttbar system (cf. Section~\ref{sec:tt_addjets}) must fulfill $\pt >30\GeV$. Jets originating from the hadronization of \PQb quarks are identified using a combined secondary vertex algorithm (CSV)~\cite{bib:btag004}, which provides a \PQb tagging discriminant by combining identified secondary vertices and track-based lifetime information.

The missing transverse energy (\ETslash) is defined as the magnitude of the projection on the plane perpendicular to the beams of the negative vector sum of the momenta of all reconstructed particles in an event~\cite{bib:MET}. To mitigate the effect of contributions from pileup on the \ETslash resolution, we use a multivariate correction where the measured momentum is separated into components that originate from the primary and the other collision vertices~\cite{Khachatryan:2014gga}. This correction improves the \ETslash resolution by ${\approx}5\%$.

Events are triggered by requiring combinations of two leptons ($\ell$ = e or $\mu$), where one fulfills a \pt threshold of 17\GeV and the other of 8\GeV, irrespective of the flavour of the leptons. The dilepton trigger efficiencies are measured using samples selected with triggers that require a minimum \ETslash or number of jets in the event, and are only weakly correlated to the dilepton triggers used in the analysis.

Events are selected if there are at least two isolated leptons of opposite charge. Events with a lepton pair invariant mass less than 20\GeV are removed to suppress events from heavy-flavour resonance decays, QCD multijet, and DY production. In the $\mu\mu$ and ee channels, the dilepton invariant mass is required to be outside a \PZ boson mass window of $91\pm15\GeV$, and \ETslash is required to be larger than 40\GeV.

For the \ttbar{}+jets selection, a minimum of two jets is required, of which at least one must be tagged as a \PQb jet. A loose CSV discriminator value is chosen such that the efficiency for tagging jets from \PQb (\PQc)  quarks is ${\approx}85\%$ (40\%), while the probability of tagging jets originating from light quarks ($\PQu$, $\PQd$, or $\PQs$) or gluons is around 10\%. Efficiency corrections, depending on jet \pt and $\eta$, are applied to account for differences in the performance of the \PQb tagging algorithm between data and simulation.

For the \ttbb(\ttb) selection, at least three \PQb-tagged jets are required (without further requirements on the minimum number of jets). In this case, a tighter discriminator value~\cite{bib:btag004} is chosen to increase the purity of the sample. The efficiency of this working point is approximately 70\% (20\%) for jets originating from a \PQb (\PQc) quark, while the misidentification rate for light-quark and gluon jets is around 1\%. The shape of the CSV discriminant distribution in simulation is corrected to better describe the efficiency observed in the data. This correction is derived separately for light-flavour and \PQb jets from a tag-and-probe approach using control samples enriched in events with a \PZ boson and exactly two jets, and \ttbar events in the $\Pe\mu$ channel with no additional jets~\cite{bib:HIG-13-029}.

\section{Identification of additional radiation in the event}
\label{sec:tt_addjets}
To study additional jet activity in the data, the identification of jets arising from the decay of the \ttbar system is crucial. In particular, we need to identify correctly the two \PQb jets from the top quark decays in events with more than two \PQb jets. This is achieved by following two independent but complementary approaches: a kinematic reconstruction~\cite{bib:Abbott:1997fv} and a multivariate analysis, optimized for the two cases under study, \ttbar{}+jets and \ttbb(\ttb), respectively. The purpose of the kinematic reconstruction is to completely reconstruct the \ttbar system based on \ETslash and the information on identified jets and leptons, taking into account detector resolution effects. This method is optimized for the case where the \PQb jets in the event only arise from the decay of the top quark pair.
The multivariate approach is optimized for events with more \PQb jets than just those from the \ttbar system. This method identifies the two jets that most likely originated from the top quark decays, and the additional \PQb jets, but does not perform a full reconstruction of the \ttbar system. Both methods are described in the following sections.

\subsection{Kinematic reconstruction in \texorpdfstring{$\ttbar$+jets}{t-tbar+jets} events}
\label{sec:ttjetsreco}
The kinematic reconstruction method was developed and used for the first time in the analysis from Ref.~\cite{bib:TOP-12-028}.
In this method the following constraints are imposed: \ETslash is assumed to originate solely from the two neutrinos; the \PW\ boson invariant mass is fixed to $80.4\GeV$~\cite{PDG2014}; and the top quark and antiquark masses are fixed to a value of $172.5\GeV$.
Each pair of jets and lepton-jet combination fulfilling the selection criteria is considered in the kinematic reconstruction. Effects of detector resolution are accounted for by randomly smearing the measured energies and directions of the reconstructed lepton and \PQb jet candidates by their resolutions. These are determined from the simulation of signal events by comparing the reconstructed \PQb jets and leptons matched to the generated \PQb quarks and leptons from top quark decays. For a given smearing, the solution of the equations for the neutrino momenta yielding the smallest invariant mass of the \ttbar system is chosen. For each solution, a weight is calculated based on the expected invariant mass spectrum of the lepton and \PQb jet from the top quark decays at the parton level. The weights are summed over 100 randomly smeared reconstruction attempts, and the kinematics of the top quark and antiquark are calculated as a weighted average.
Finally, the two jets and lepton-jet combinations that yield the maximum sum of weights are chosen for further analysis. Combinations with two \PQb-tagged jets are chosen over those with a single \PQb-tagged jet. The efficiency of the kinematic reconstruction, defined as the number of events with a solution divided by the total number of selected \ttbar{}+jets events, is approximately 94\%. The efficiency in simulation is similar to the one in data for all jet multiplicities. Events with no valid solution for the neutrino momenta are excluded from further analysis. In events with additional jets, the algorithm correctly identifies the two jets coming from the \ttbar decay in about 70\% of the cases.

After the full event selection is applied, the dominant background in the $\Pe\mu$ channel originates from other \ttbar decay channels and is estimated using simulation. This contribution corresponds mostly to leptonic $\tau$ decays, which are considered background in the \ttbar{}+jets measurements. In the $\Pe\Pe$ and $\Pgm\Pgm$ channels, the dominant background contribution arises from $\PZ / \Pgg^*$+jets production. The normalization of this background contribution is derived from data using the events rejected by the \PZ boson veto, scaled by the ratio of events failing and passing this selection, estimated from simulation~\cite{bib:TOP-11-002_paper}. The remaining backgrounds, including the single top quark $\PQt\PW$ channel, $\PW$+jets, diboson, and QCD multijet events, are estimated from simulation for all the channels.

\begin{figure*}[htbp!]
  \begin{center}
      \includegraphics[width=0.40 \textwidth]{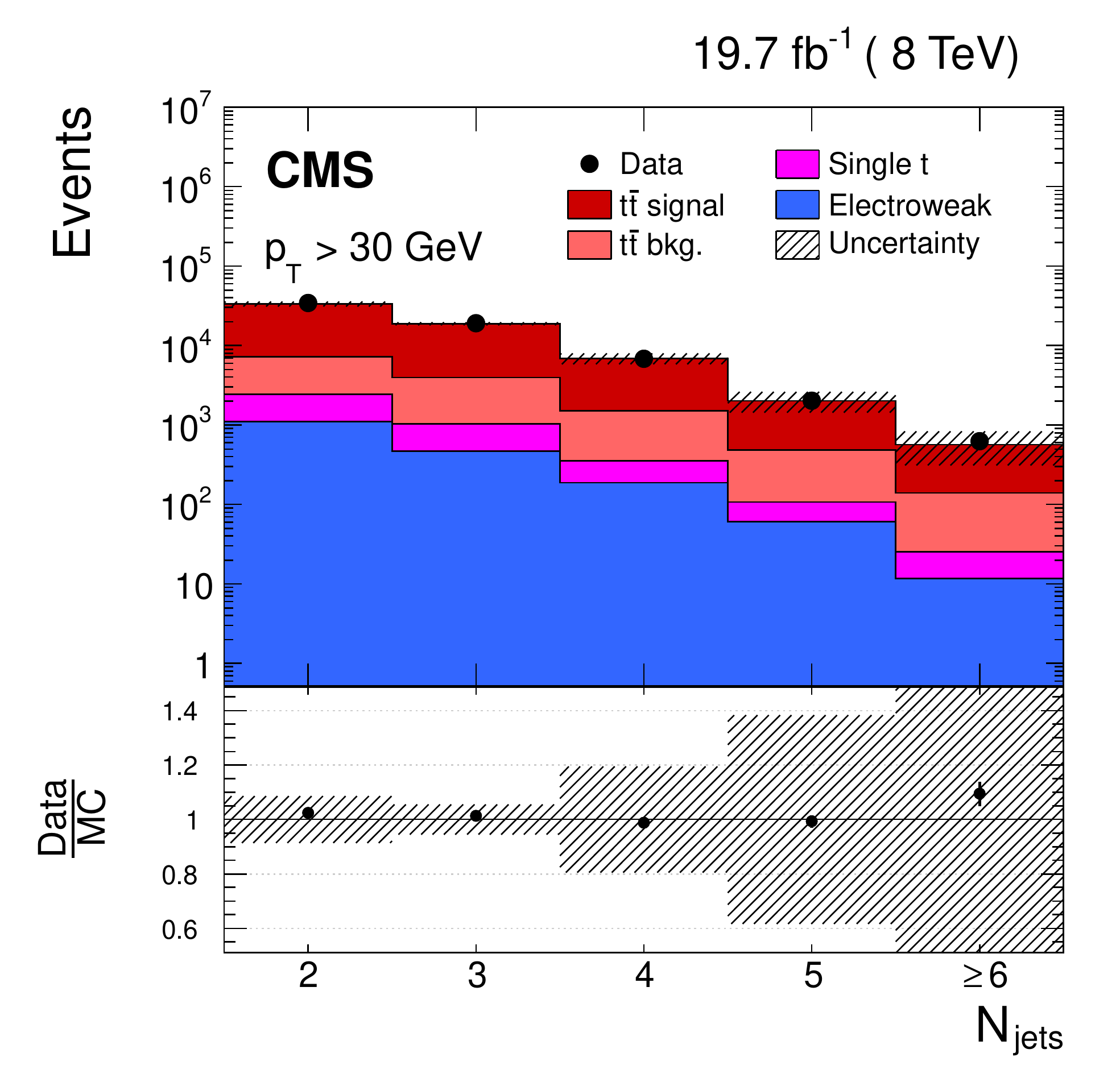}\\
      \includegraphics[width=0.40 \textwidth]{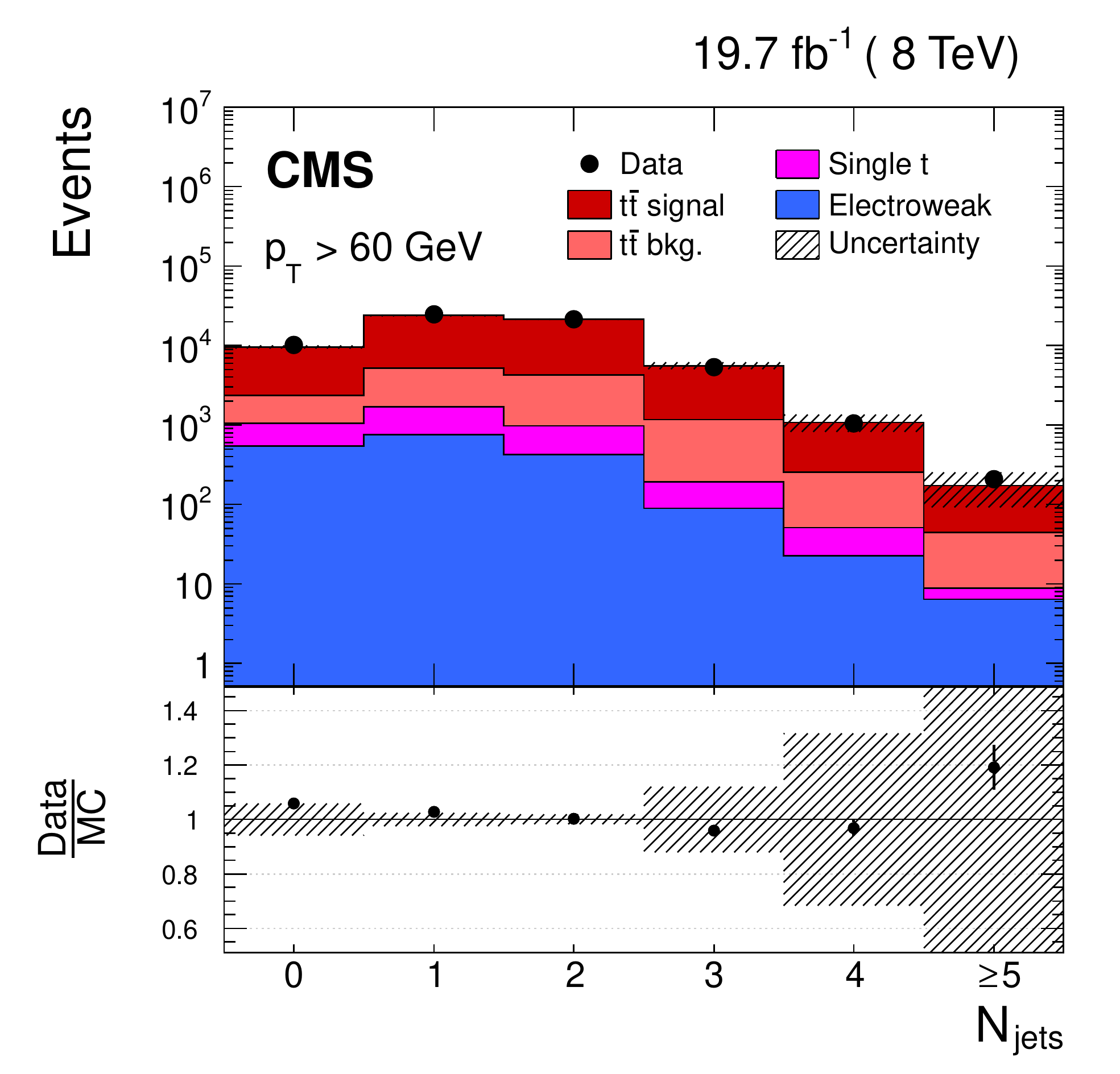}%
      \includegraphics[width=0.40 \textwidth]{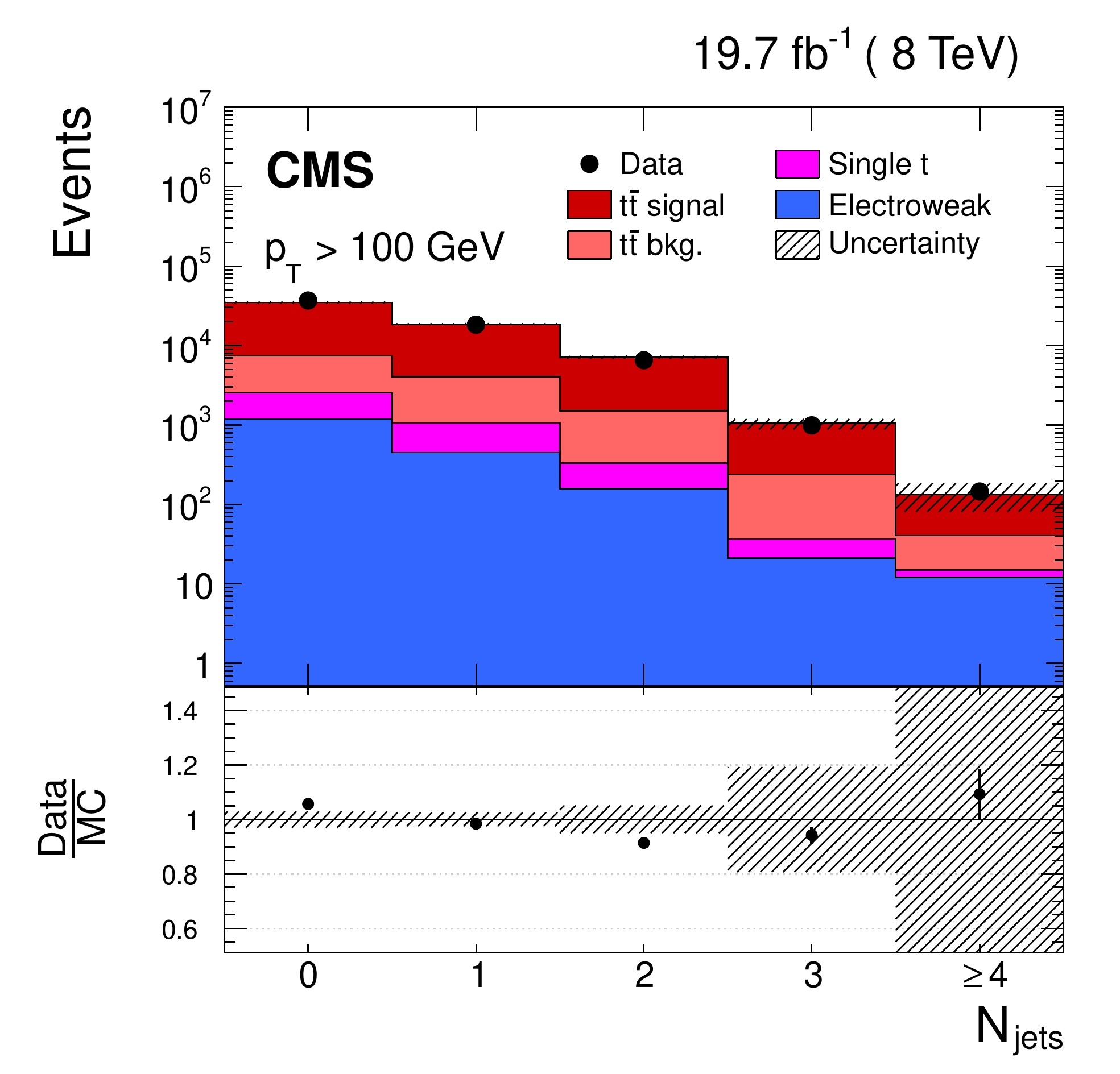}
\caption{Reconstructed jet multiplicity distribution after event selection in data (points) and from signal and background simulation (histograms) for all jets with \pt of at least 30\GeV (top), 60\GeV (bottom left), and 100\GeV (bottom right). The hatched regions correspond to the uncertainties affecting the shape of the distributions in the simulated signal \ttbar events and backgrounds (cf. Section~\ref{sec:syst}). The lower plots show the ratio of the data to the MC simulation prediction. Note that in all cases the event selection requires at least two jets with $\pt > 30\GeV$.}
\label{fig:ctrl:jetmult}
  \end{center}
\end{figure*}

\begin{figure*}[htbp!]
    \begin{center}
        \includegraphics[width=0.40 \textwidth]{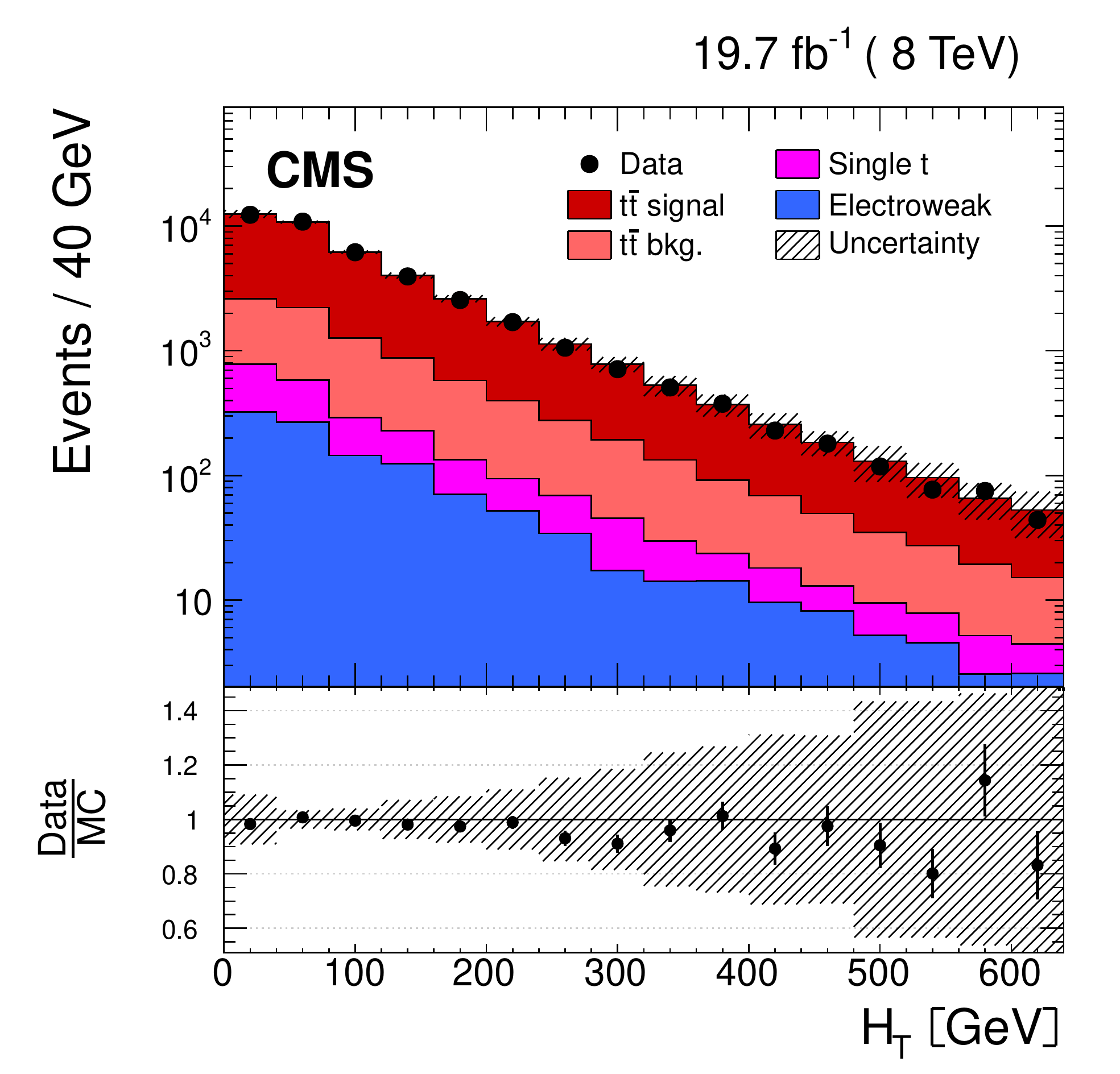}\\
        \includegraphics[width=0.40 \textwidth]{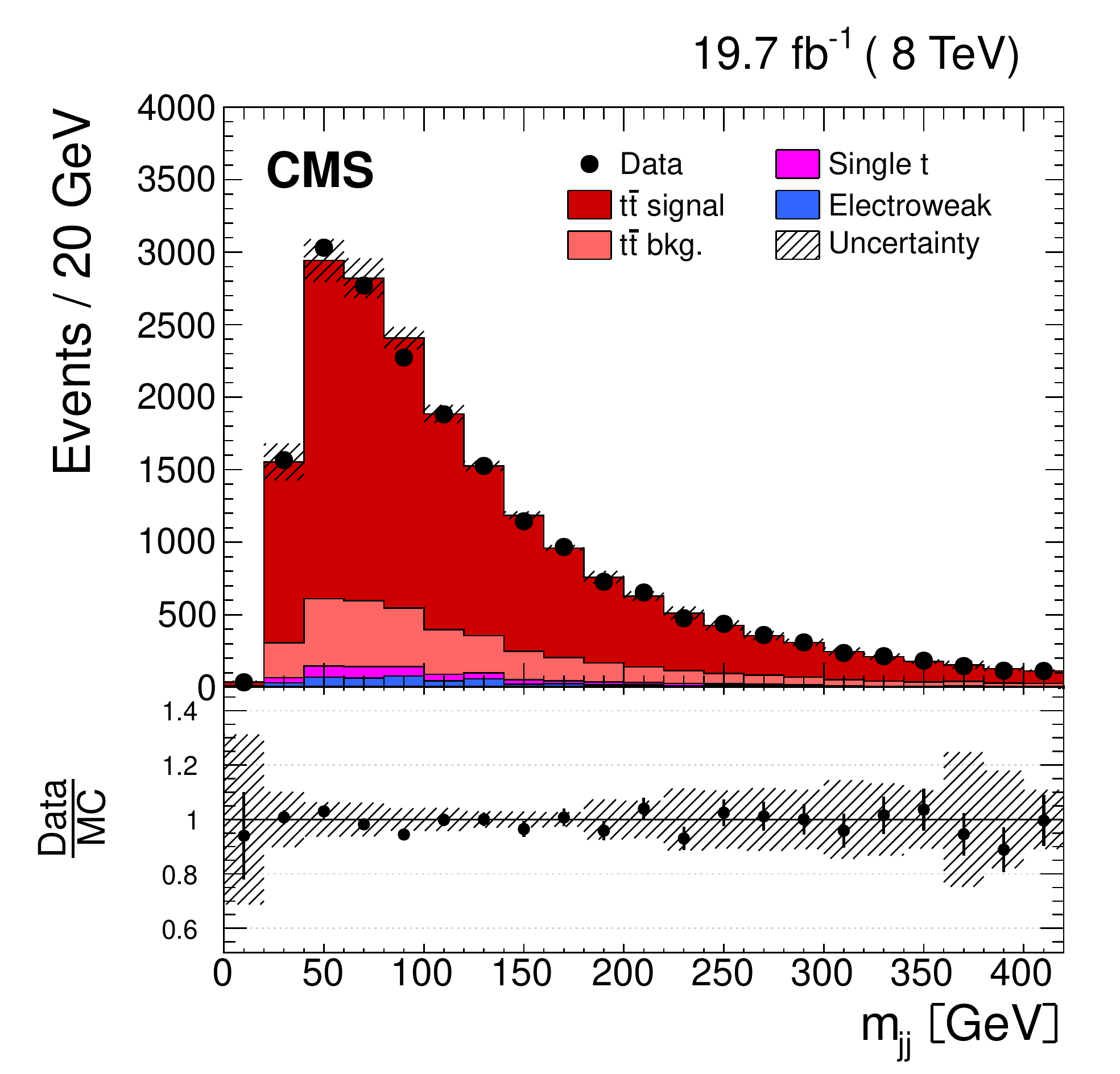}%
        \includegraphics[width=0.40 \textwidth]{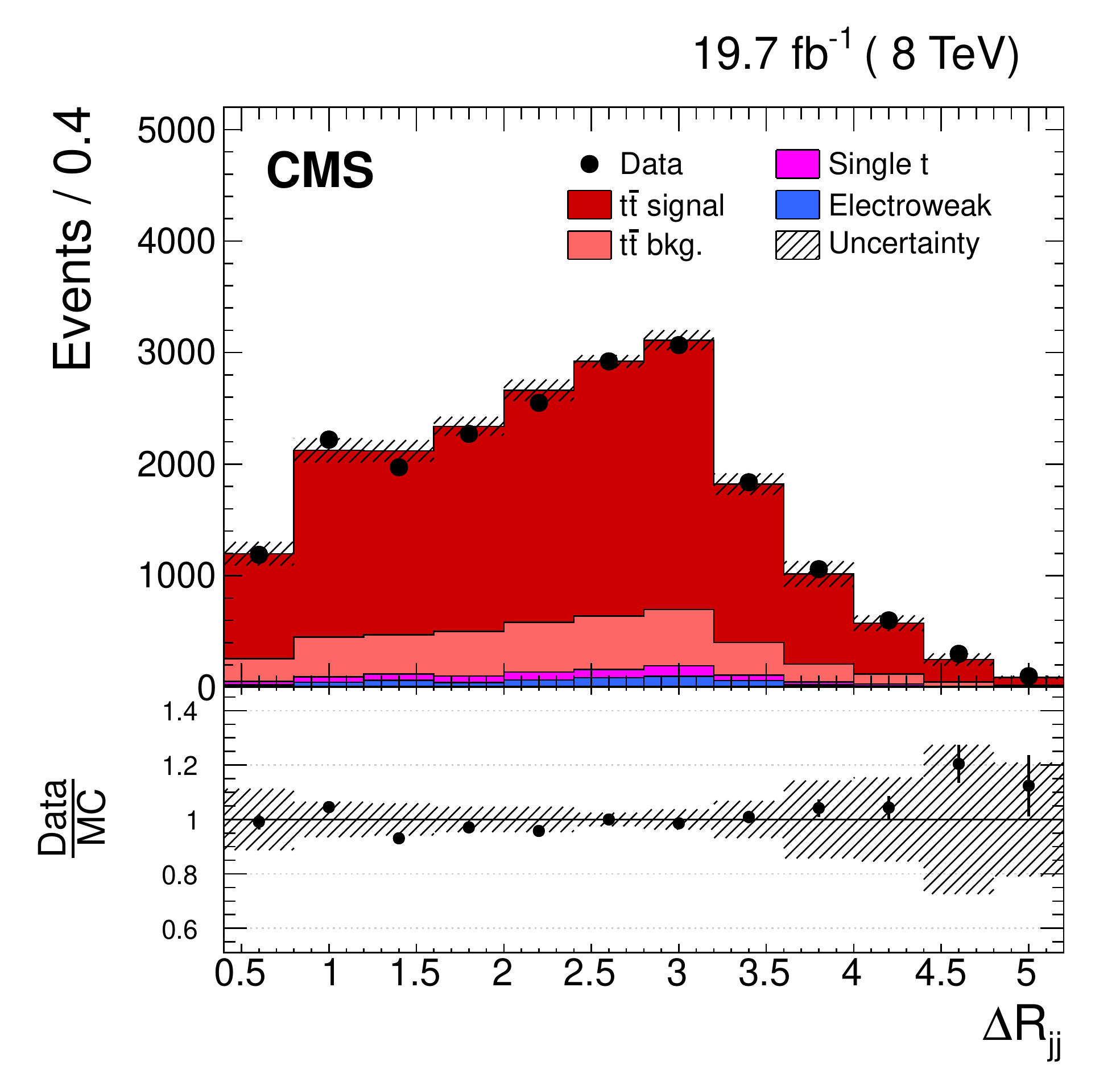}
  \caption{Distribution of the scalar sum of the \pt of all additional jets $\HT$ (top), the invariant mass of the leading and subleading additional jets \mjj (bottom left), and their angular distance \Djj (bottom right) in data (points) and from signal and background simulation (histograms). The hatched regions correspond to the uncertainties affecting the shape of the distributions in the simulated signal \ttbar events and backgrounds (cf. Section~\ref{sec:syst}). The lower plots show the ratio of the data to the MC simulation prediction.}
  \label{fig:leadjet12}
    \end{center}
   \end{figure*}

In Fig.\,\ref{fig:ctrl:jetmult}, the multiplicity distributions of the selected jets per event are shown for different jet \pt thresholds and compared to SM predictions. In this figure and the following ones, the \ttbar sample is simulated using \MADGRAPH{}+\PYTHIA{6}, where only \ttbar events with two leptons ($\Pe$ or $\mu$) from the \PW\ boson decay are considered as signal. All other \ttbar events, specifically those originating from decays via $\tau$ leptons, which are the dominant contribution, are considered as background. In the following figures, ``Electroweak'' corresponds to DY, $\PW$+jets, and diboson processes, and ``\ttbar bkg.'' includes the $\ttbar$+$\Pgg/\PW/\PZ$ events. The data are well described by the simulation, both for the low jet \pt threshold of 30\GeV and the higher thresholds of 60 and 100\GeV. The hatched regions in Figs.\,\ref{fig:ctrl:jetmult}--\,\ref{fig:leadjet12} correspond to the uncertainties affecting the shape of the simulated signal and background events (cf. Section~\ref{sec:syst}), and are dominated by modelling uncertainties in the former.

\begin{figure*}[htbp!]
    \begin{center}
        \includegraphics[width=0.40 \textwidth]{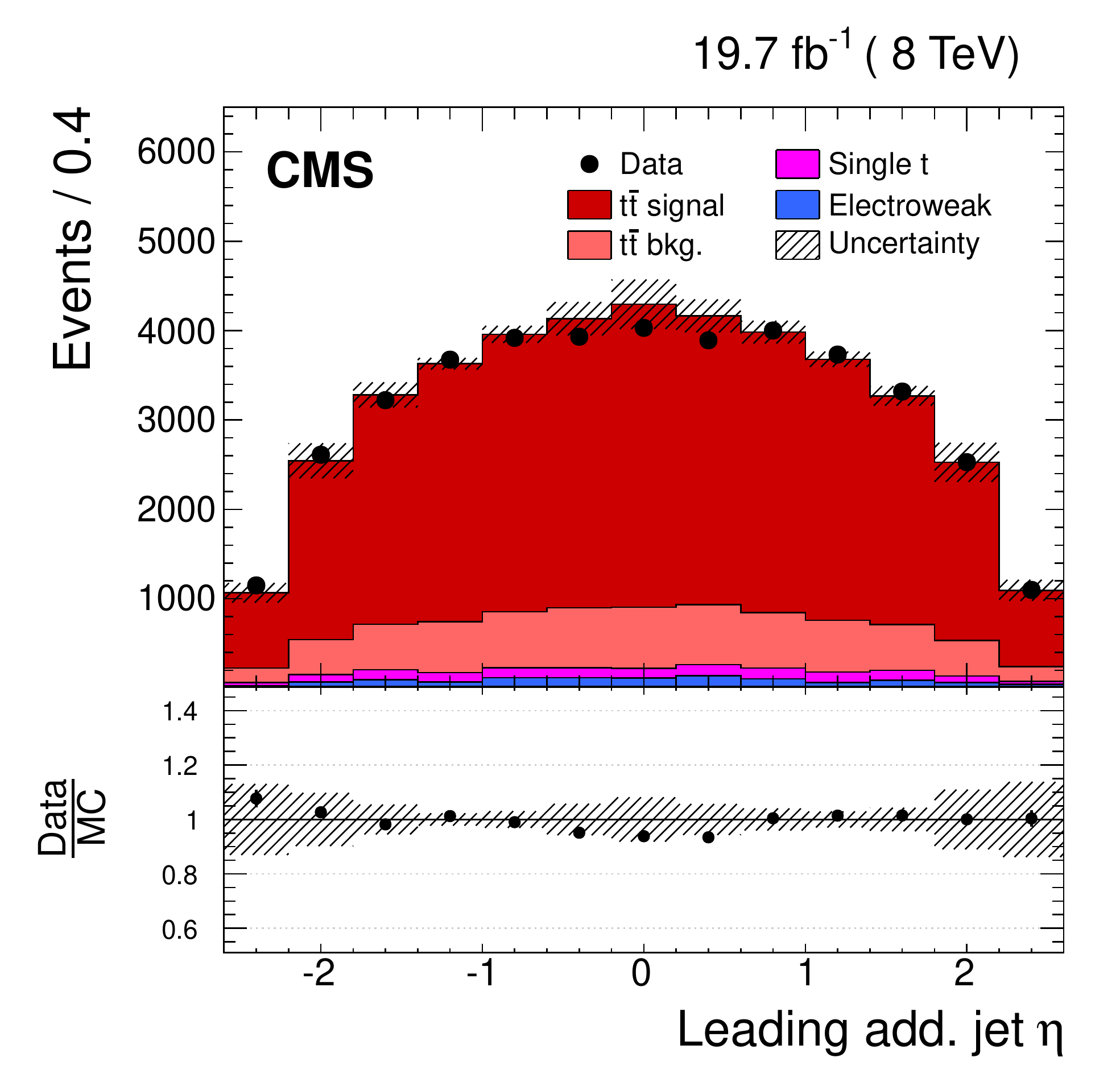}
        \includegraphics[width=0.40 \textwidth]{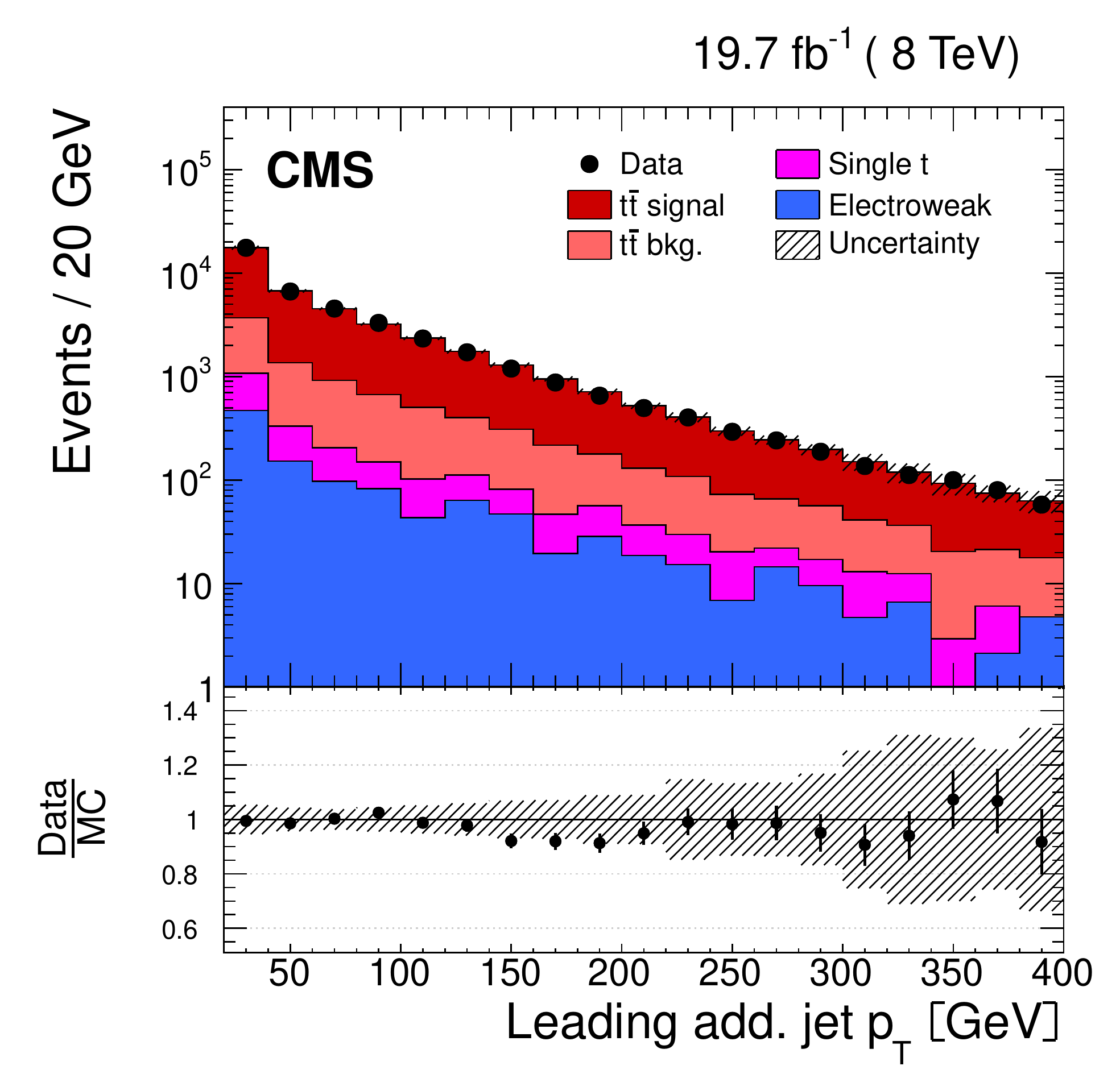}
        \includegraphics[width=0.40 \textwidth]{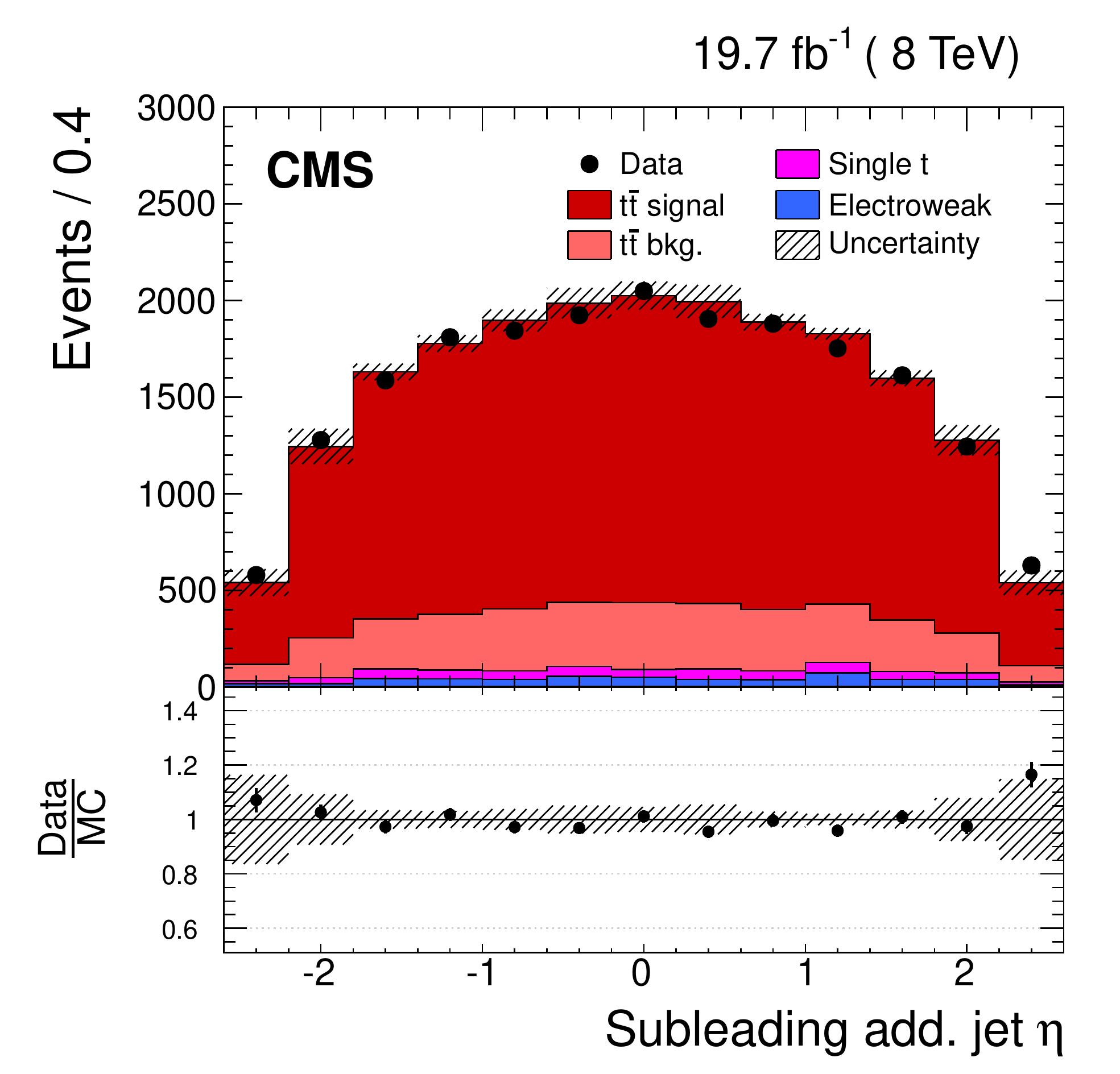}
        \includegraphics[width=0.40 \textwidth]{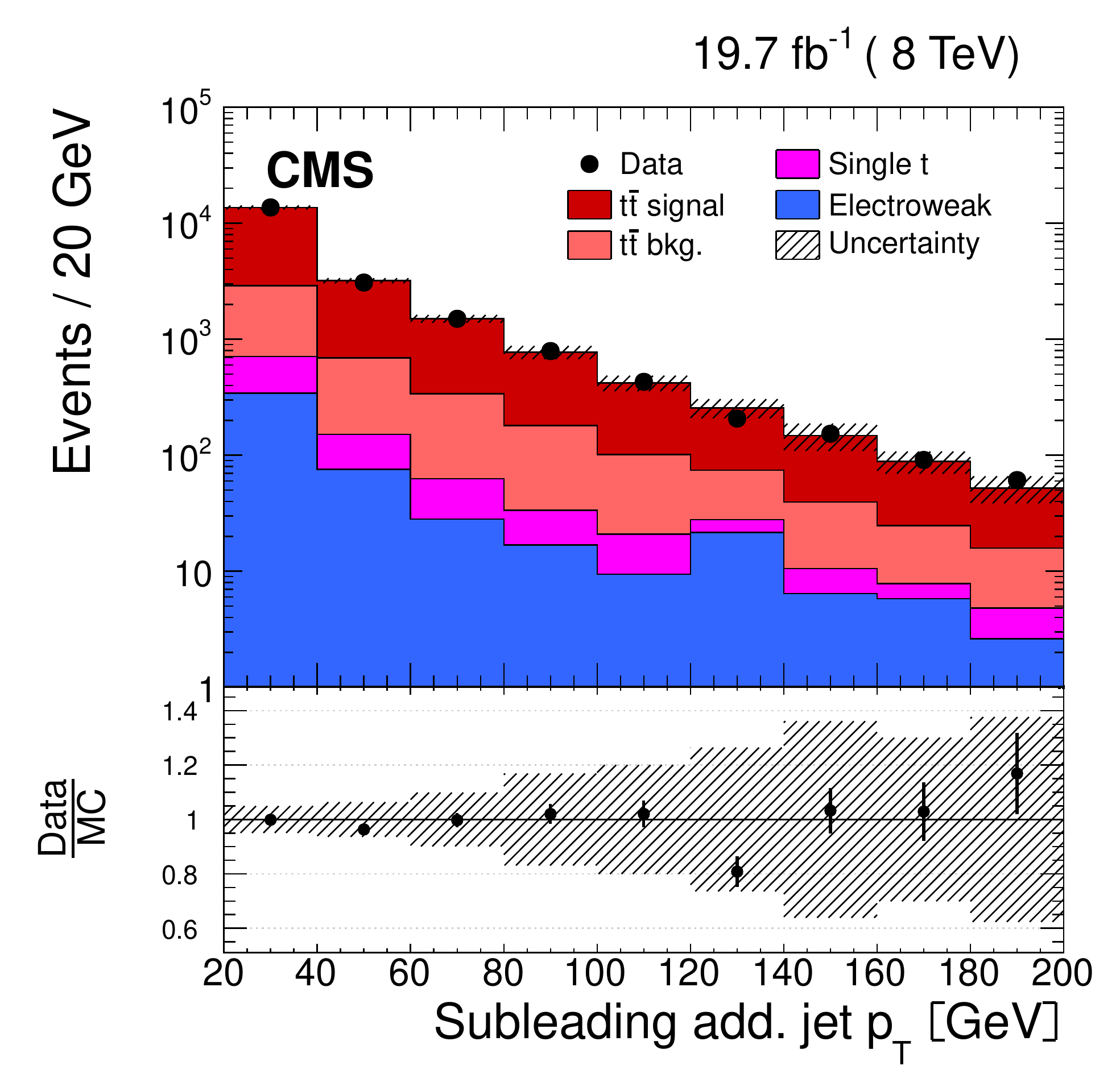}
  \caption{Distribution of the $\eta$ (left) and \pt (right) of the leading (top row) and subleading (bottom row) additional reconstructed jets in data (points) and from signal and background simulation (histograms). The hatched regions correspond to the uncertainties affecting the shape of the simulated distributions in the signal \ttbar events and backgrounds (cf. Section~\ref{sec:syst}). The lower plots show the ratio of the data to the MC simulation prediction.}
  \label{fig:leadjet}
    \end{center}
   \end{figure*}

Additional jets in the event are defined as those jets within the phase space described in the event selection (cf.~Section~\ref{sec:selection}) that are not identified by the kinematic reconstruction to be part of the \ttbar system. The $\eta$ and \pt distributions of the additional jets with the largest and second largest \pt in the event (referred to as the leading and subleading additional jets in the following) are shown in Fig.~\ref{fig:leadjet}. Three additional event variables are considered: the scalar sum of the \pt of all additional jets, $\HT$, the invariant mass of the leading and subleading additional jets, \mjj, and their angular separation, $\Djj=\sqrt{\smash[b]{(\Delta\eta)^2+(\Delta\phi)^2}}$, where $\Delta\eta$ and $\Delta\phi$ are the pseudorapidity and azimuthal differences between the directions of the two jets. These distributions are shown in Fig.~\ref{fig:leadjet12}. The predictions from the simulation, also shown in the figures, describe the data within the uncertainties.

\subsection{Identification of \texorpdfstring{\ttbar}{t-tbar} jets and additional jets in \texorpdfstring{\ttbb}{t-tbar-b-bbar} events}
\label{sec:ttbbreco}
The multivariate approach uses a boosted decision tree (BDT) to distinguish the \PQb jets stemming from the \ttbar system from those arising from additional radiation for final states with more than two \PQb jets. This method is optimized for \ttbb topologies in the dilepton final state of the \ttbar system.
The BDT is set up using the TMVA package~\cite{Hocker:2007ht}. To avoid any dependence on the kinematics of the additional jets, and especially on the invariant mass of the two additional jets, the method identifies the jets stemming from the \ttbar system by making use of properties of the \ttbar system that are expected to be mostly insensitive to the additional radiation. The variables combine information from the two final-state leptons, the jets, and \ETslash.
All possible pairs of reconstructed jets in an event are considered. For each pair, one jet is assigned to the \PQb jet and the other to the \PAQb\ jet. This assignment is needed to define the variables used in the BDT and is based on the measurement of the charge of each jet, which is calculated from the charge and the momenta of the PF constituents used in the jet clustering. The jet in the pair with the largest charge is assigned to the \PAQb, while the other jet is assigned to the \PQb. The efficiency of this jet charge pairing is defined as the fraction of events where the assigned \PQb and \PAQb\ are correctly matched to the corresponding generated b and \PAQb\ jets, and amounts to 68\%.

A total of twelve variables are included in the BDT. Some examples of the variables used are: the sum and difference of the invariant mass of the $\PQb\ell^+$ and $\PAQb\ell^-$ systems, $m^{\PQb\ell^+}\pm m^{\PAQb\ell^-}$; the absolute difference in the azimuthal angle between them, $ \lvert \Delta\phi^{ \PQb\ell^+,\PAQb\ell^- } \rvert $; the \pt of the $\PQb\ell^+$ and $\PAQb\ell^-$ systems, $\pt^{\PQb\ell^+}$ and $\pt^{\PAQb\ell^-}$; and the difference between the invariant mass of the two \PQb jets and two leptons and the invariant mass of the \bbbar pair, $m^{\PQb \PAQb \ell^+\ell^-}-m^{\bbbar}$. The complete list of variables can be found in Appendix~\ref{ap:mvaVariables}. The main challenge with this method is the large number of possible jet assignments, given four genuine \PQb jets and potential extra jets from additional radiation in each event. The basic methodology is to use the BDT discriminant value of each dijet combination as a  measure of the probability that the combination stems from the \ttbar system. The jets from the \ttbar system are then identified as the pair with the highest BDT discriminant. From the remaining jets, those \PQb-tagged jets with the highest \pt are selected as being the leading additional ones.

\begin{figure*}
\begin{center}
\includegraphics[width=0.40\textwidth]{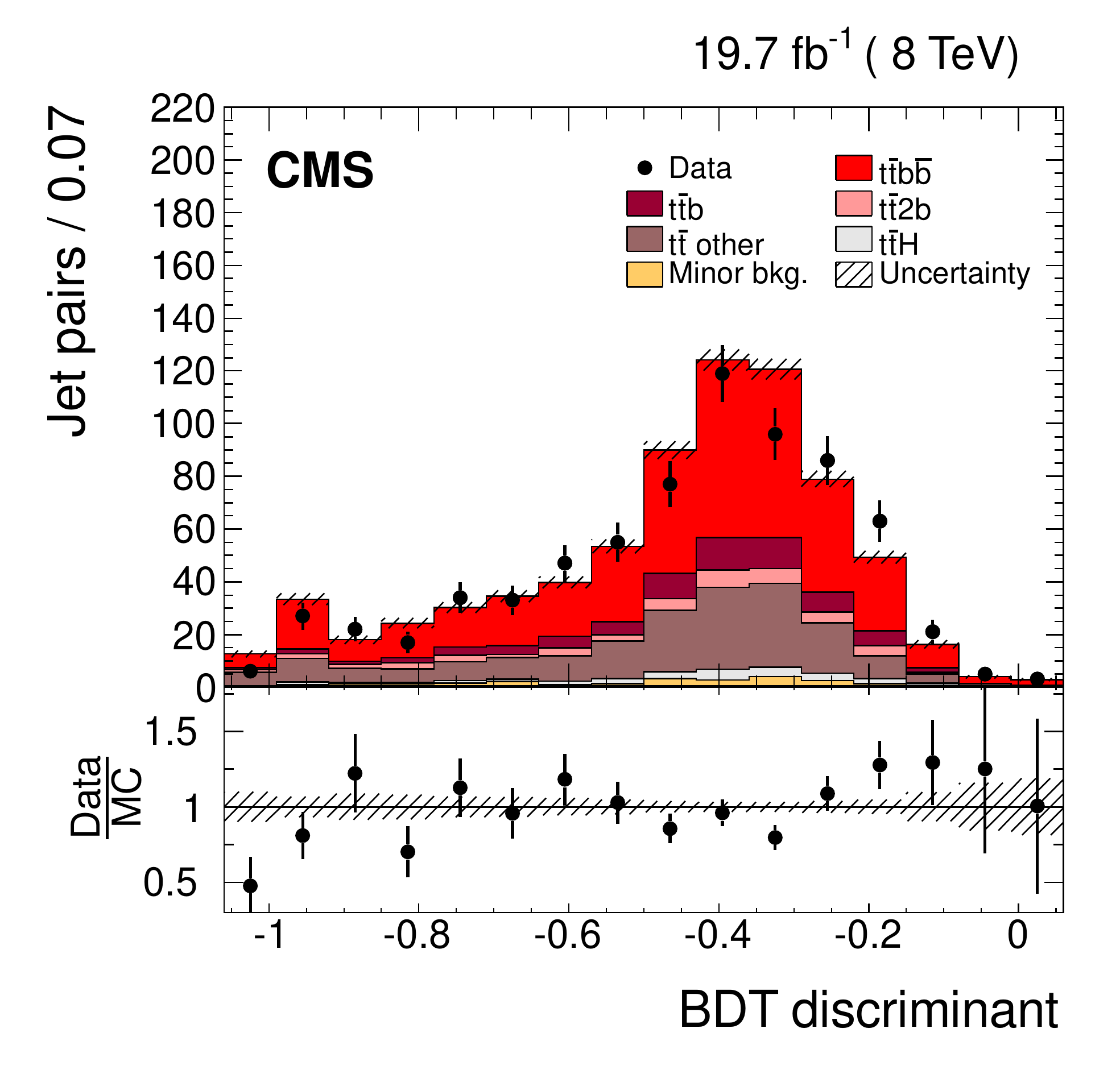}%
\includegraphics[width=0.40\textwidth]{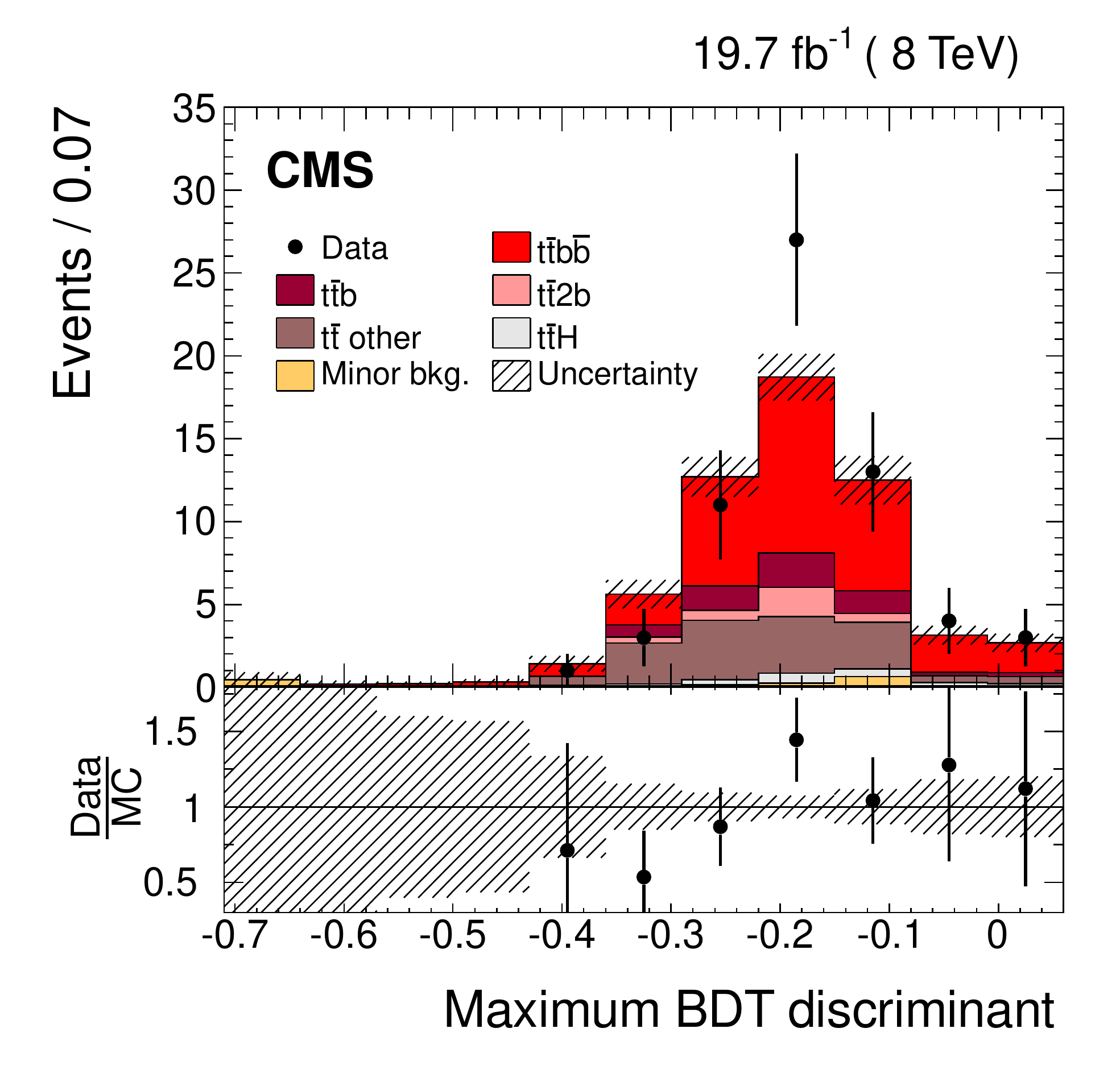}
\caption{The BDT discriminant of all dijet combinations in data (points) and from signal and background simulation (histograms) per event (left) and dijet combination with the highest discriminant per event (right) in events with at least four jets and exactly four \PQb-tagged jets. The distributions include the correction obtained with the template fit to the \PQb-tagged jet multiplicity (cf. Section~\ref{sec:ttbbreco}). The hatched area represents the statistical uncertainty in the simulated samples. ``Minor bkg." includes all non-\ttbar processes and $\ttbar$+$\PZ/\PW/\Pgg$. The lower plots show the ratio of the data to the MC simulation prediction. }
\label{fig:mvaTopJets:cpWeights}
\end{center}
\end{figure*}

The BDT training is performed on a large and statistically independent sample of simulated \ttH events with the Higgs boson mass varied over the range 110--140\GeV. The \ttbb events are not included in the training to avoid the risk of overtraining owing to the limited number of events in the available simulated samples. The simulated \ttHtobb sample is suited for this purpose since the four \PQb jets from the decay of the \ttbar system and the Higgs boson have similar kinematic distributions. Since it is significantly harder to identify the jets from the \ttbar system in \ttH events than in \ttbb events, where the additional \PQb jets arise from initial- or final-state radiation, a good BDT performance with \ttH events implies also a good identification in \ttbb events. The distributions of the BDT discriminant in data and simulation are shown in Fig.~\ref{fig:mvaTopJets:cpWeights} for all dijet combinations in an event, and for the combination with the highest weight that is assigned to the \ttbar system. The subset ``Minor bkg." includes all non-\ttbar processes and $\ttbar$+$\PZ/\PW/\Pgg$ events. There is good agreement between the data and simulation distributions within the statistical uncertainties.

The number of simulated events with correct assignments for the additional \PQb jets in \ttH events relative to the total number of events where those jets are selected and matched to the corresponding generator jets, is approximately 34\%. In \ttbb events, this fraction is about 40\%. This efficiency is high enough to allow the measurement of the \ttbar cross section as a function of the kinematic variables of the additional \PQb jets (the probability of selecting the correct assignments by choosing random combinations of jets is 17\% in events with four jets and 10\% in events with five jets). The relative increase in efficiency with respect to the use of the kinematic reconstruction for \ttbb is about 15\%. Additionally, the BDT approach improves the correlation between the generated and reconstructed variables, especially for the distribution of the invariant mass of the two leading additional \PQb jets \mbb and their angular separation $\Delta R_{\PQb\PQb} = \sqrt{\smash[b]{(\Delta\eta)^2+(\Delta\phi)^2}}$, where $\Delta\eta$ and $\Delta\phi$ are the pseudorapidity and azimuthal differences between the directions of the two \PQb jets.

\begin{figure*}[!htbp]
  \begin{center}
\includegraphics[width=0.40\textwidth]{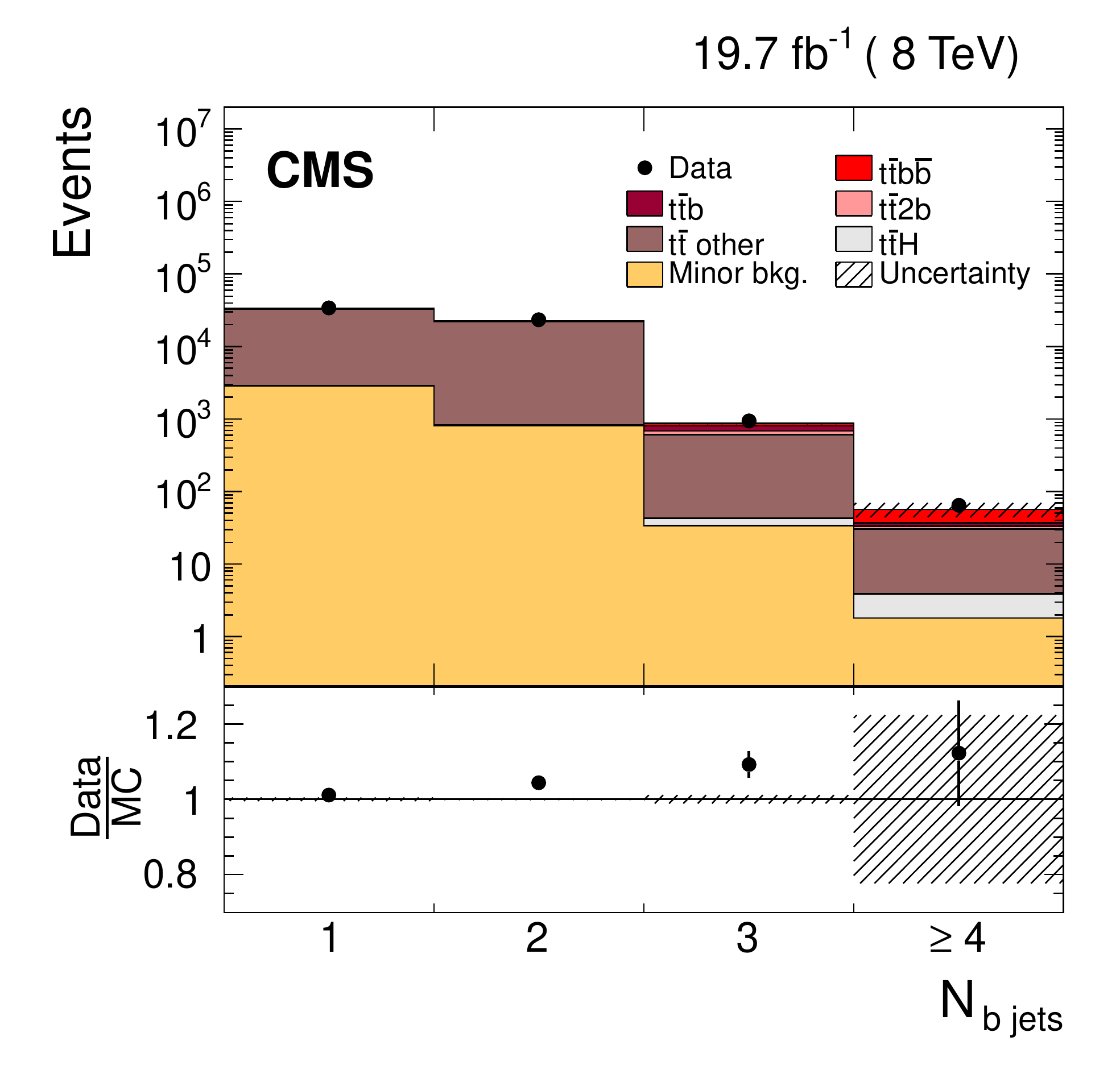}%
\includegraphics[width=0.40\textwidth]{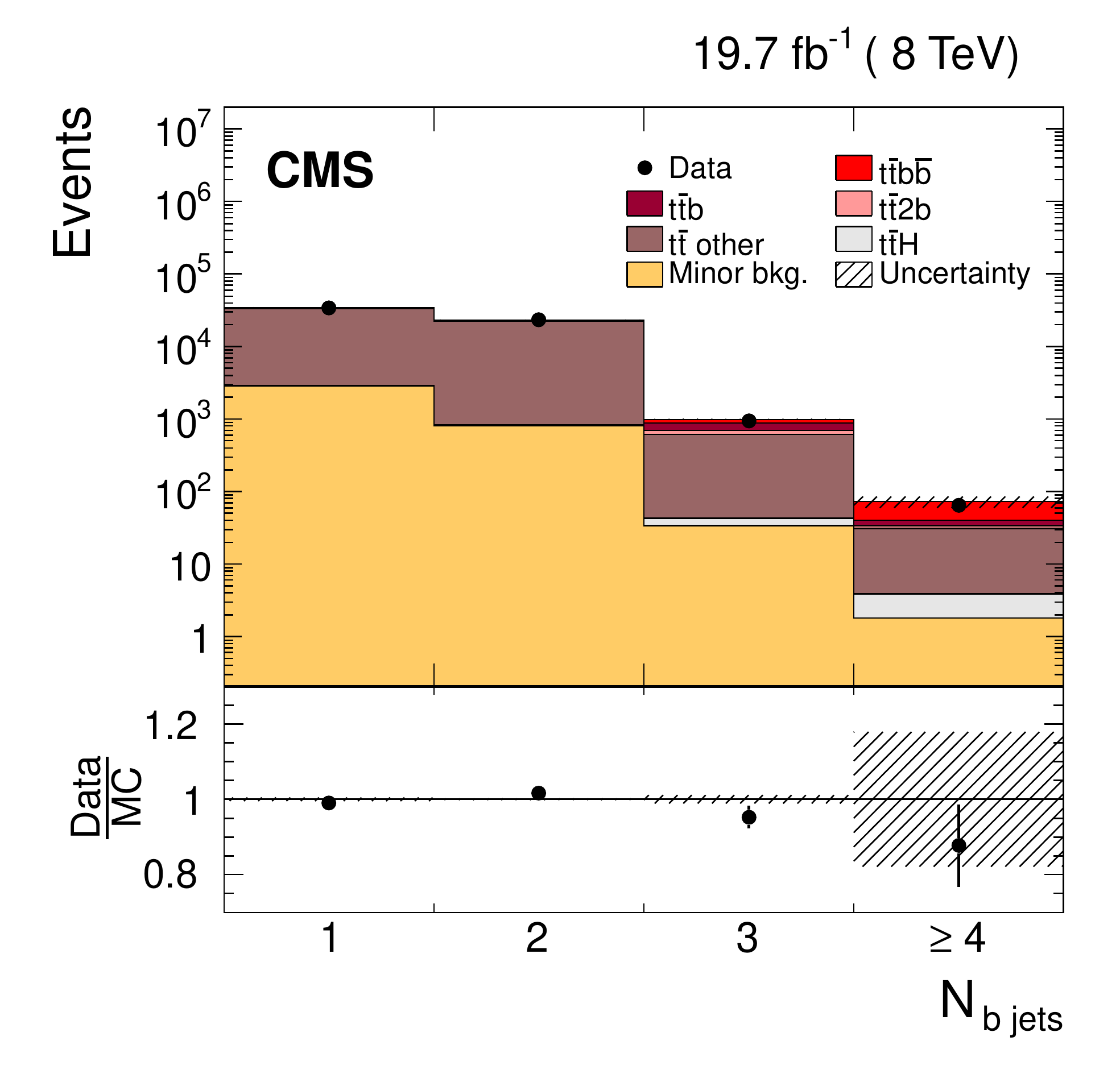}
\caption{The pre-fit distribution of the \PQb jet multiplicity in data (points) and from signal and background simulation (histograms) for events fulfilling the lepton selection criteria, having ${\ge} 2$ jets, ${\ge} 1$ \PQb-tagged jet (left), and the post-fit distribution (right). The hatched area represents the statistical uncertainty in the simulated samples. ``Minor bkg." includes all non-\ttbar processes and $\ttbar$+$\PZ/\PW/\Pgg$. The lower plots show the ratio of the data to the MC simulation prediction.}
\label{fig:control_btagMult}
  \end{center}
\end{figure*}

The expected fraction of events with additional \PQb jets is not properly modelled in the simulation, in agreement with the observation of a previous CMS measurement~\cite{bib:ttbb_ratio:2014}. This discrepancy between the \MADGRAPH{}+\PYTHIA simulation and data can be seen in the \PQb jet multiplicity distribution, as shown in Fig.~\ref{fig:control_btagMult}.

To improve the description of the data by the simulation, a template fit to the \PQb-tagged jet multiplicity distribution is performed using three different templates obtained from simulation. One template corresponds to the \ttb and \ttbb processes, defined at the generator level as the events where one or two additional \PQb jets are generated within the acceptance requirements, $\pt>20\GeV$ and $\abseta<2.4$, (referred to as ``\ttbar{}+HF"). The \ttbb and \ttb processes are combined into a single template because they only differ by the kinematic properties of the second additional \PQb jet. Details about the definition of the \PQb jets and the acceptance are given in Section~\ref{sec:diffxsec}. The second template includes the background contribution coming from \ttcc and \ttbar{}+light-jets events (referred to as ``\ttbar~other''), where \ttcc events are defined as those that have at least one \PQc jet within the acceptance and no additional \PQb jets. This contribution is not large enough to be constrained by data, therefore it is combined with the \ttbar{}+light-jets process in a single template. The third template contains the remaining background processes, including \tttwob, which corresponds to events with two additional \PQb hadrons that are close enough in direction to produce a single \PQb jet. This process, produced by collinear $\Pg \to \PQb \PAQb$ splitting, is treated separately owing to the large theoretical uncertainty in its cross section and insufficient statistical precision to constrain it with data. The normalizations of the first two templates are free parameters in the fit. The third is fixed to the corresponding cross section described in Section~\ref{sec:theory}, except for the cross section for the \tttwob process, which is corrected by a factor of $1.74{}_{-0.74}^{+0.69}$~\cite{bib:Zbb-xsec}. The normalization factors obtained for the template fit correspond to $1.66 \pm 0.43$ (\ttbar{}+HF) and $1.00 \pm 0.01$ (\ttbar~other). Details about the uncertainties in those factors are presented in Section~\ref{sec:syst_ttbb}. The improved description of the \PQb jet multiplicity can be seen in Fig.~\ref{fig:control_btagMult} (right).

\begin{figure*}[htbp!]
  \begin{center}
    \includegraphics[width=0.40\textwidth]{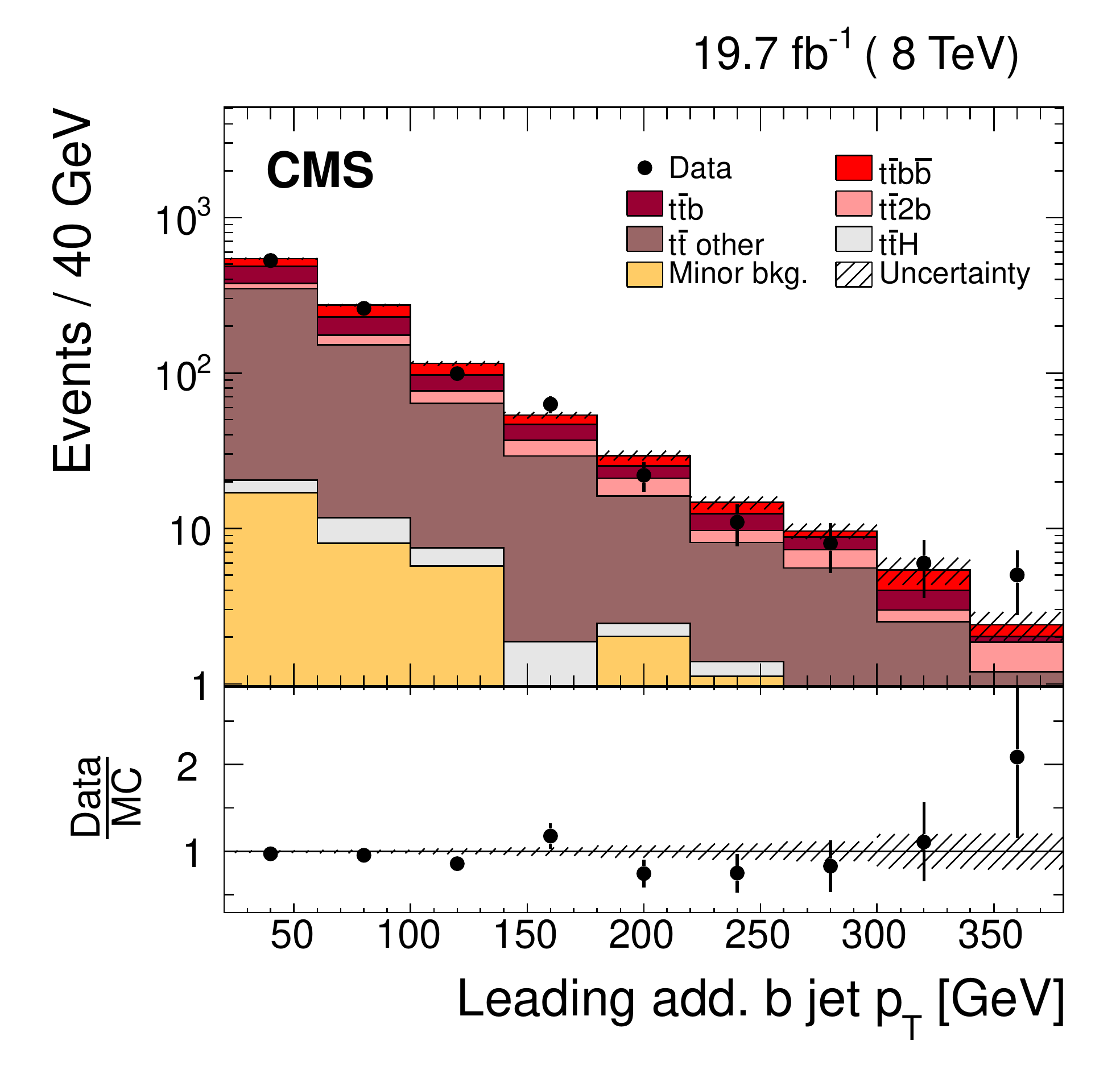}
    \includegraphics[width=0.40\textwidth]{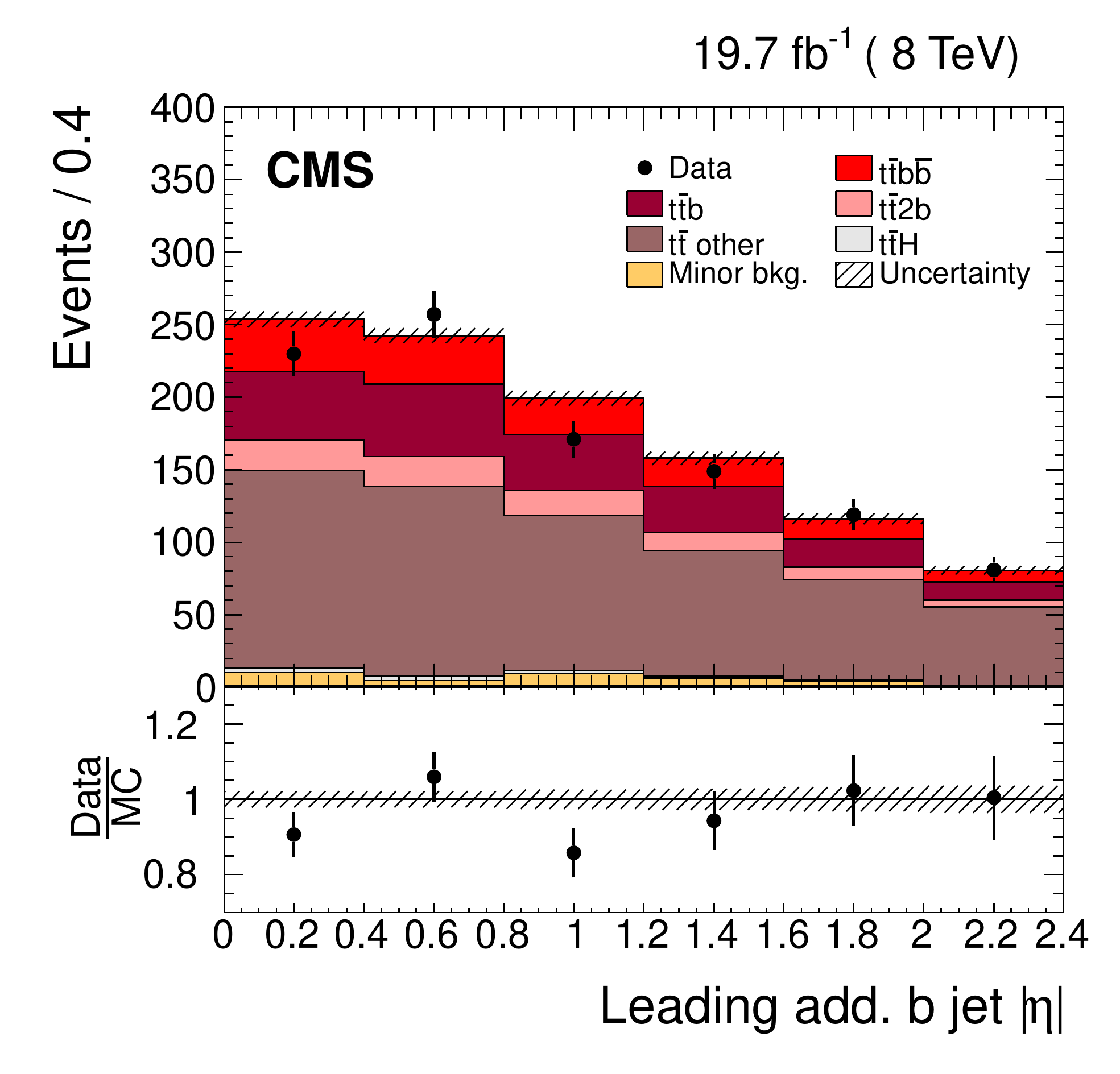}
    \includegraphics[width=0.40\textwidth]{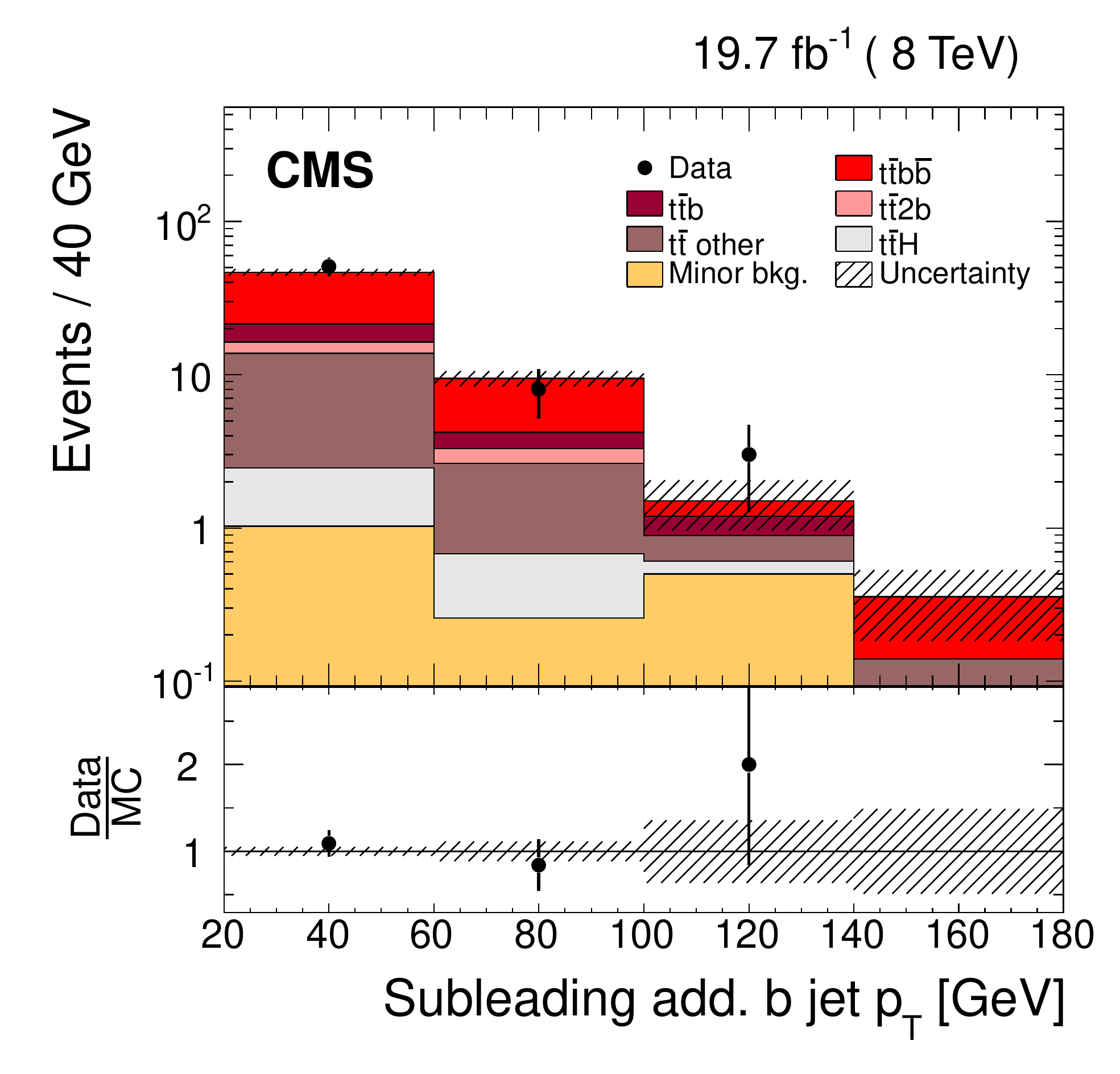}
    \includegraphics[width=0.40\textwidth]{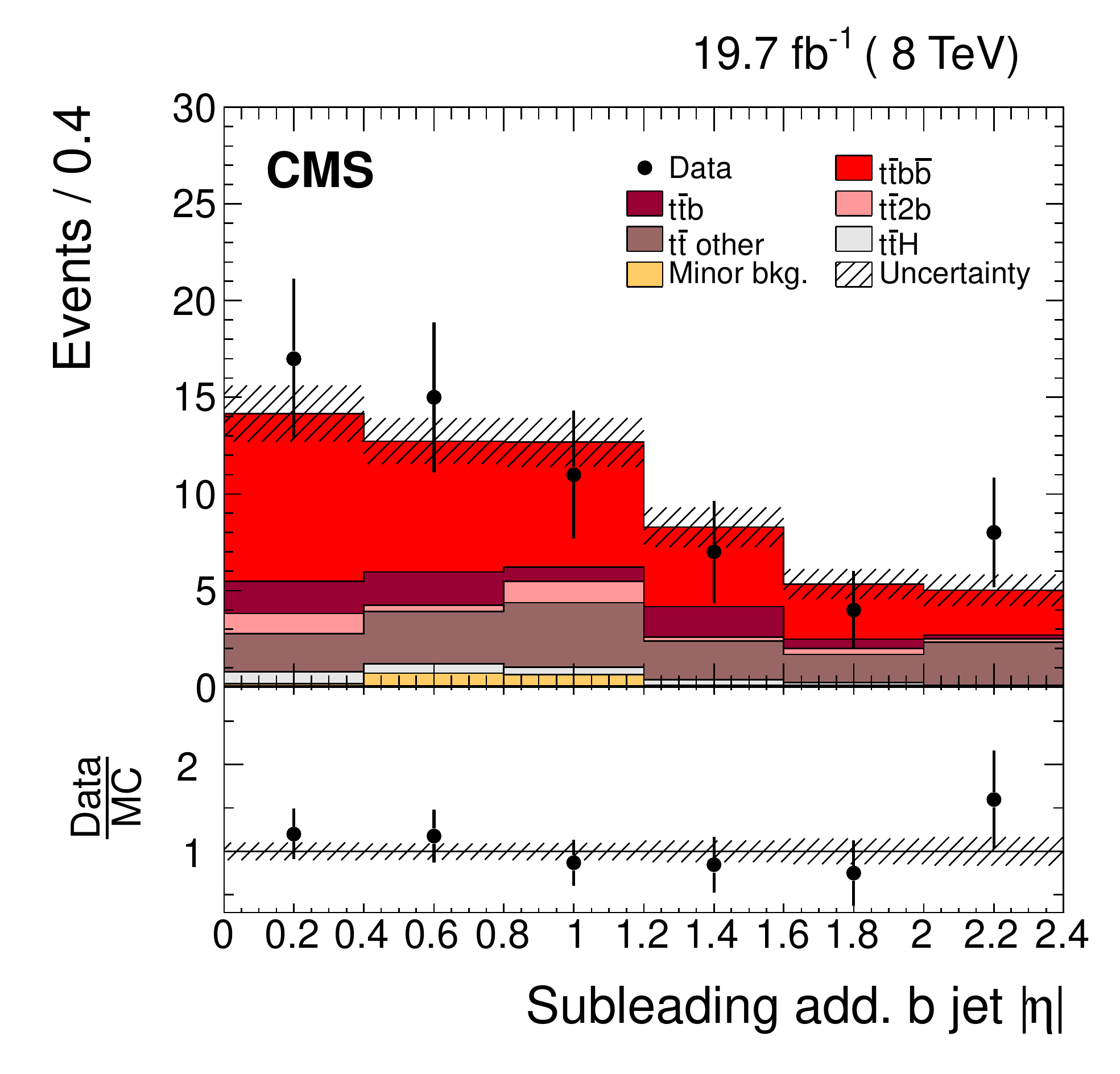}
    \includegraphics[width=0.40\textwidth]{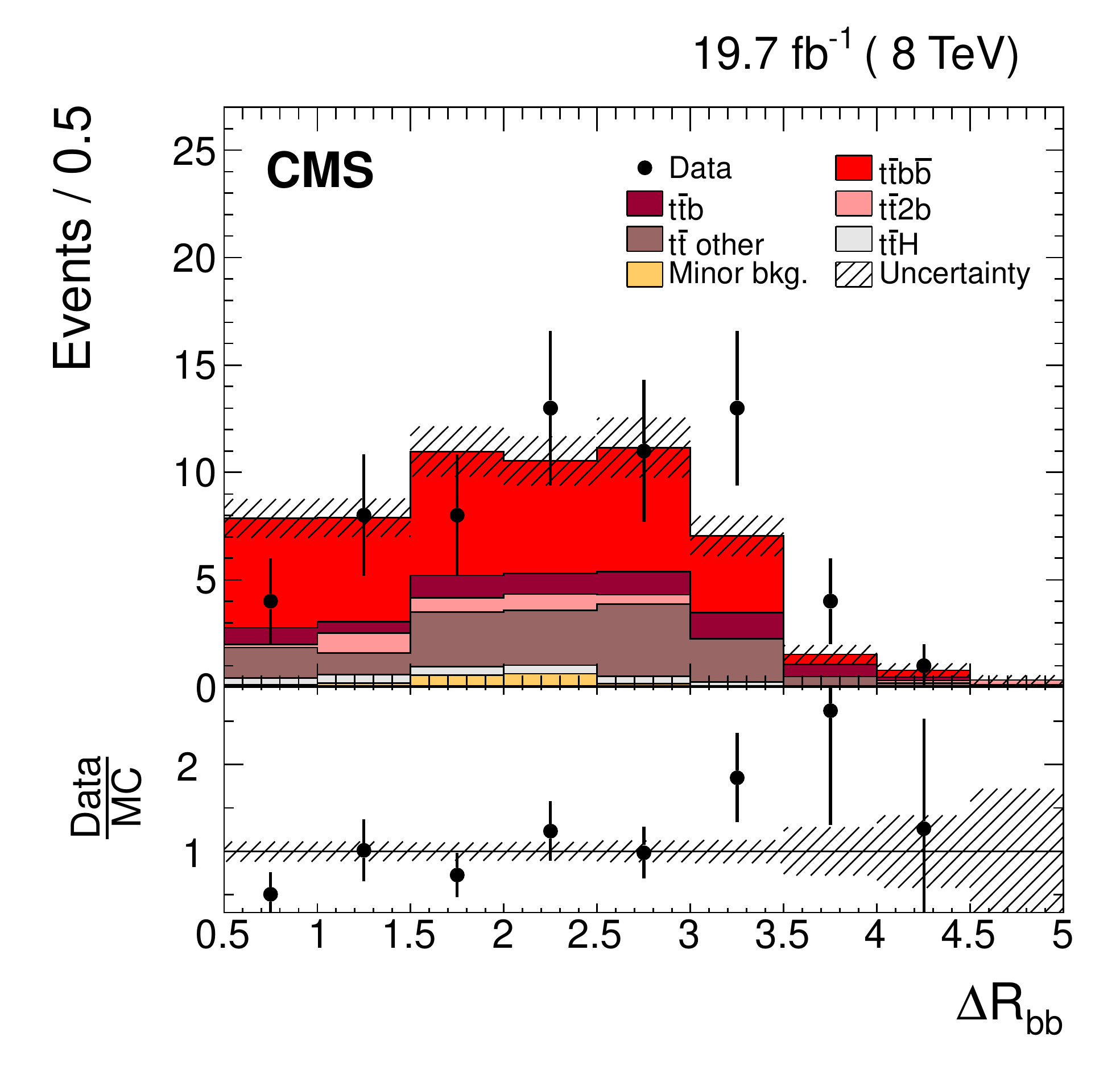}
    \includegraphics[width=0.40\textwidth]{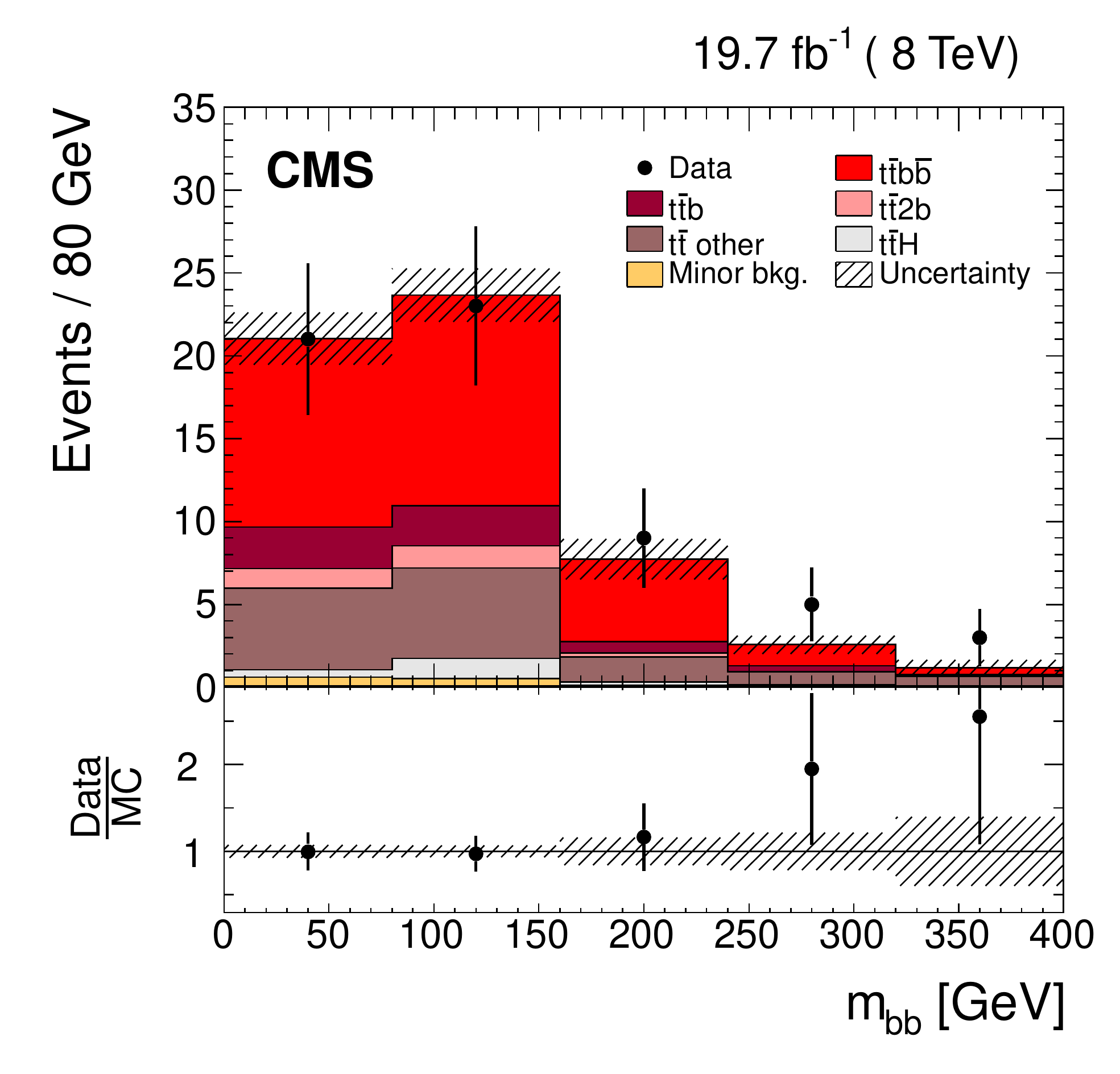}
    \caption{Distributions of the leading additional \PQb jet \pt (top left) and \abseta (top right), subleading additional \PQb jet \pt (middle left) and \abseta (middle right), $\Delta R_{\PQb\PQb}$ (bottom left), and \mbb (bottom right) from data (points) and from signal and background simulation (histograms). The hatched area represents the statistical uncertainty in the simulated samples. ``Minor bkg." includes all non-\ttbar processes and $\ttbar$+$\PZ/\PW/\Pgg$. The lower plots show the ratio of the data to the MC simulation prediction.}
    \label{fig:cp_bjets}
  \end{center}
\end{figure*}

Figure~\ref{fig:cp_bjets} (top) shows the \pt and \abseta distributions of the leading additional \PQb jet, measured in events with at least three \PQb-tagged jets (using the tighter discriminator value described in Section~\ref{sec:selection}), after the full selection and including all corrections. The distributions of the \pt and \abseta of the second additional \PQb jet in events with exactly four \PQb-tagged jets, $\Delta R_{\PQb\PQb}$, and \mbb are also presented. The dominant contribution arises from the \ttbb process. The \ttbar decays into $\tau$ leptons decaying leptonically are included as signal to increase the number of \ttb and \ttbb events both in data and simulation.
It has been checked that the distribution of the variables of relevance for this analysis do not differ between the leptons directly produced from $\PW$ boson decays and the leptons from $\tau$ decays within the statistical uncertainties in the selected \ttb and \ttbb events. In general, the variables presented are well described by the simulation, after correcting for the heavy-flavour content measured in data, although the simulation tends to predict smaller values of $\Delta R_{\PQb\PQb}$ than the data. After the full selection, the dominant background contribution arises from dilepton \ttbar events with additional light-quark, gluon, and \PQc jets, corresponding to about 50\% and 20\% of the total expected yields for the \ttb and \ttbb cases, respectively. Smaller background contributions come from single top quark production, \ttbar in association with \PW or \PZ bosons, and \ttbar events in the lepton+jets decay channels. The contribution from \ttHtobb is also small, amounting to 0.9\% and 3\% of the total expected events for the \ttb and \ttbb distributions. The contribution from background sources other than top quark production processes such as DY, diboson, or QCD multijet is negligible.

\section{Systematic uncertainties}
\label{sec:syst}
Different sources of systematic uncertainties are considered arising from detector effects, as well as theoretical uncertainties. Each systematic uncertainty is determined individually in each bin of the measurement by varying the corresponding efficiency, resolution, or model parameter within its uncertainty, in a similar way as in the CMS previous measurement of the \ttbar differential cross sections~\cite{bib:TOP-12-028}. For each variation, the measured differential cross section is recalculated and the difference with respect to the nominal result is taken as the systematic uncertainty. The overall uncertainty in the measurement is then derived by adding all contributions in quadrature, assuming the sources of systematic uncertainty to be fully uncorrelated.

\subsection{Experimental uncertainties}
The experimental sources of systematic uncertainty considered are the jet energy scale (JES), jet energy resolution (JER), background normalization, lepton trigger and identification efficiencies, \PQb tagging efficiency, integrated luminosity, pileup modelling, and kinematic reconstruction efficiency.

The experimental uncertainty from the JES is determined by varying the energy scale of the reconstructed jets as a function of their \pt and $\eta$ by its uncertainty~\cite{Chatrchyan:2011ds}. The uncertainty from the JER is estimated by varying the simulated JER by its $\eta$-dependent uncertainty~\cite{Chatrchyan:2011ds}.

The uncertainty from the normalization of the backgrounds that are taken from simulation is determined by varying the cross section used to normalize the sample, see Section~\ref{sec:theory}, by ${\pm}30\%$. This variation takes into account the uncertainty in the predicted cross section and all other sources of systematic uncertainty~\cite{bib:TOP-11-005_paper, bib:TOP-12-018, bib:TOP-12-028}. In the case of the tW background, the variation of ${\pm}30\%$ covers the theoretical uncertainty in the absolute rate, including uncertainties owing to the PDFs. The contribution from the DY process, as determined from data, is varied in the normalization by ${\pm}30\%$~\cite{{bib:TOP-11-002_paper},{Chatrchyan:2013faa}}.

The trigger and lepton identification efficiencies in simulation are corrected by lepton $\pt$ and $\eta$ multiplicative data-to-simulation scale factors. The systematic uncertainties are estimated by varying the factors by their uncertainties, which are in the range 1--2\%.

For the \ttbar{}+jets measurements, the \PQb tagging efficiency in simulation is also corrected by scale factors depending on the \pt and $\eta$ of the jet. The shape uncertainty in the \PQb tagging efficiency is then determined by taking the maximum change in the shape of the \pt and \abseta distributions of the \PQb jet, obtained by changing the scale factors. This is achieved by dividing the \PQb jet distributions in \pt and \abseta into two bins at the median of the respective distributions. The \PQb tagging scale factors for \PQb jets in the first bin are scaled up by half the uncertainties quoted in Ref.~\cite{bib:btag004}, while those in the second bin are scaled down, and vice versa, so that a maximum variation is assumed and the difference between the scale factors in the two bins reflects the full uncertainty. The changes are made separately in the \pt and \abseta distributions, and independently for heavy-flavour (b and c) and light-flavour (s, u, d, and gluon) jets, assuming that they are all uncorrelated. A normalization uncertainty is obtained by varying the scale factors up and down by half the uncertainties. The total uncertainty is obtained by summing in quadrature the independent variations.

{\tolerance=500
The uncertainty in the integrated luminosity is 2.6\%~\cite{bib:lumiPAS2013}. The effect of the uncertainty in the level of pileup is estimated by varying the inelastic pp cross section in simulation by ${\pm}5\%$.
\par}

The uncertainty coming from the kinematic reconstruction method is determined from the uncertainty in the correction factor applied to account for the small difference in efficiency between the simulation and data, defined as the ratio between the events with a solution and the total number of selected events.

\subsubsection{Specific systematic uncertainties associated with the \texorpdfstring{\ttbb(\ttb)}{t-tbar-b-bbar (t-tbar-b)} measurements}
\label{sec:syst_ttbb}
In the \ttbb(\ttb) measurements, an additional uncertainty associated with the template fit to the \PQb-tagged jet multiplicity distribution is considered. Since the input templates are known to finite precision, both the statistical and systematic uncertainties in the templates are taken into account. The considered systematic uncertainties that affect the shapes of the templates are those of the JES, the CSV discriminant scale factors following the method described in~\cite{bib:HIG-13-029}, the cross section of the \ttcc process, which is varied by ${\pm}50\%$~\cite{bib:HIG-13-029}, and the uncertainty in the \tttwob cross section. This is taken as the maximum between the largest uncertainty from the measurement described in Ref.~\cite{bib:Zbb-xsec} and the difference between the corrected cross section and the prediction by the nominal \MADGRAPH simulation used in this analysis. This results in a variation of the cross section of about ${\pm}40\%$. This uncertainty is included as a systematic uncertainty in the shape of the background template.

\subsection{Model uncertainties}
\label{sec:modelsyst}
The impact of theoretical assumptions on the measurement is determined by repeating the analysis, replacing the standard \MADGRAPH signal simulation by alternative simulation samples.
The uncertainty in the modelling of the hard-production process is assessed by varying the common renormalization and factorization scale in the \MADGRAPH signal samples up and down by a factor of two with respect to its nominal value of the $Q$ in the event (cf. Section~\ref{sec:theory}). Furthermore, the effect of additional jet production in \MADGRAPH is studied by varying up and down by a factor of two the threshold between jet production at the matrix element level and via parton showering.
The uncertainties from ambiguities in modelling colour reconnection (CR) effects are estimated by comparing simulations of an underlying-event (UE) tune including colour reconnection to a tune without it (Perugia 2011 and Perugia 2011 noCR tunes, described in Ref.~\cite{Skands:2010ak}). The modelling of the UE is evaluated by comparing two different Perugia 11 (P11) \PYTHIA tunes, mpiHi and TeV, to the standard P11 tune. The dependency of the measurement on the top quark mass is obtained using dedicated samples in which the mass is varied by ${\pm}1\GeV$ with respect to the default value used in the simulation. The uncertainty from parton shower modelling is determined by comparing two samples simulated with \POWHEG and \MCATNLO, using either \PYTHIA or \HERWIG for the simulation of the parton shower, underlying event, and hadronization.
The effect of the uncertainty in the PDFs on the measurement is assessed by reweighting the sample of simulated \ttbar signal events according to the 52 CT10 error PDF sets, at the 90\% \CL~\cite{bib:CT10}.

Since the total uncertainty in the \ttb and \ttbb production cross sections is largely dominated by the statistical uncertainty in the data, a simpler approach than for the \ttbar{}+jets measurements is chosen to conservatively estimate the systematic uncertainties: instead of repeating the measurement, the uncertainty from each source is taken as the difference between the nominal \MADGRAPH{}+\PYTHIA sample and the dedicated simulated sample at generator level. In the case of the uncertainty coming from the renormalization and factorization scales, the uncertainty estimated in the previous inclusive cross section measurement~\cite{bib:ttbb_ratio:2014} is assigned.

\subsection{Summary of the typical systematic uncertainties}
Typical values of the systematic uncertainties in the absolute differential cross sections are summarized in Table~\ref{tab:TypicalValSysUncertainties} for illustrative purposes. They are the median values of the distribution of uncertainties over all bins of the measured variables. Details on the impact of the different uncertainties in the results are given in Sections~\ref{sec:diffxsecNJets} to~\ref{sec:gap}.

In general, for the \ttbar{}+jets case, the dominant systematic uncertainties arise from the uncertainty in the JES, as well as from model uncertainties such as the renormalization, factorization, and jet-parton matching scales and the hadronization uncertainties. For the \ttb and \ttbb cross sections, the total uncertainty, including all systematic uncertainties, is only about 10\% larger than the statistical uncertainty. The experimental uncertainties with an impact on the normalization of the expected number of signal events, such as lepton and trigger efficiencies, have a negligible effect on the final cross section determination, since the normalization of the different processes is effectively constrained by the template fit.

\begin{table}[phtb]
  \centering
    \topcaption{Summary of the typical systematic uncertainties in the measurements of the \ttbar{}+jets and \ttbb(\ttb) absolute differential cross sections and their sources. The median of the distribution of uncertainties over all bins of each measured differential cross section is quoted.}
    \label{tab:TypicalValSysUncertainties}
    \begin{tabular}{l|c|c}
    \hline
     \multicolumn{3}{c}{Relative systematic uncertainty (\%)} \\
     \hline
     Source  & \ttbar{}+jets & \ttbb(\ttb) \\
      \hline
      \multicolumn{3}{c}{Experimental uncertainties} \\
      \hline
      Trigger efficiency               & 1.3 & 0.1 \\
      Lepton selection                 & 2.2 & 0.1 \\
      Jet energy scale                 & 6.8 & 11 \\
      Jet energy resolution            & 0.3 & 2.5 \\
      Background estimate              & 2.1 & 5.6 \\
      $\PQb$ tagging                        & 0.5 & 12 \\
      Kinematic reconstruction         & 0.3 &  \NA \\
      Pileup                           & 0.3 & 1.7 \\
      \hline
            \multicolumn{3}{c}{Model uncertainties} \\
      \hline
      Fact./renorm. scale              & 2.7 & 8.0 \\
      Jet-parton matching scale        & 1.3 & 3.0 \\
      Hadronization                    & 4.5 & 5.2 \\
      Top quark mass                   & 1.4 & 2.0 \\
      PDF choice                       & 0.3 & 0.9 \\
      Underlying event                 & 1.0 & 2.9 \\
      Colour reconnection              & 1.3 & 1.9 \\
      \hline
    \end{tabular}
\end{table}

\section{Differential \texorpdfstring{\ttbar}{t-tbar} cross section}
\label{sec:diffxsec}
The absolute differential \ttbar cross section is defined as:
\begin{equation}
\frac{\rd\sigma_{\ttbar}}{\rd x_i}=\frac{\sum_j A_{ij}^{-1} (N^j_{\text{data}}-N^j_{\text{bkg}})}{\Delta_x^i \mathcal{L}},
\label{eq:xsec}
\end{equation}
where $j$ represents the bin index of the reconstructed variable $x$, $i$ is the index of the corresponding generator-level bin, $N^j_{\text{data}}$ is the number of data events in bin $j$, $N^j_{\text{bkg}}$ is the number of estimated background events, $\mathcal{L}$ is the integrated luminosity, and $\Delta_x^i$ is the bin width. Effects from detector efficiency and resolution in each bin $i$ of the measurement are corrected by the use of a regularized inversion of the response matrix (symbolized by $A_{ij}^{-1}$) described in this section.

For the measurements of \ttbar{}+jets, the estimated number of background events from processes other than \ttbar production ($N_{\text{non \ttbar bkg}}$) is subtracted from the number of events in data ($N$). The contribution from other \ttbar decay modes is taken into account by correcting the difference $N$--$N_{\text{non \ttbar bkg}} $ by the signal fraction, defined as the ratio of the number of selected \ttbar signal events to the total number of selected \ttbar events, as determined from simulation. This avoids the dependence on the inclusive \ttbar cross section used for normalization.
For the \ttb and \ttbb production cross sections, where the different \ttbar contributions are fitted to the data, the expected contribution from all background sources is directly subtracted from the number of data events.

The normalized differential cross section is derived by dividing the absolute result, Eq.~(\ref{eq:xsec}), by the total cross section, obtained by integrating over all bins for each observable. Because of the normalization, the systematic uncertainties that are correlated across all bins of the measurement, \eg the uncertainty in the integrated luminosity, cancel out.

Effects from the trigger and reconstruction efficiencies and resolutions, leading to migrations of events across bin boundaries and statistical correlations among neighbouring bins, are corrected using a regularized unfolding method~\cite{bib:svd, bib:blobel, bib:TOP-12-028}. The response matrix $A_{ij}$ that corrects for migrations and efficiencies is calculated from simulated \ttbar events using \MADGRAPH. The generalized inverse of the response matrix is used to obtain the unfolded distribution from the measured distribution by applying a $\chi^2$ technique. To avoid nonphysical fluctuations, a smoothing prescription (regularization) is applied. The regularization level is determined individually for each distribution using the averaged global correlation method~\cite{bib:james}. To keep the bin-to-bin migrations small, the width of bins in the measurements are chosen according to their purity and stability.
The purity is the number of events generated and correctly reconstructed in a certain bin divided by the total number of reconstructed events in the same bin. The stability is the ratio of the number of events generated and reconstructed in a bin to the total number of events generated in that bin. The purity and stability of the bins are typically larger than 40--50\%, which ensures that the bin-to-bin migrations are small enough to perform the measurement. The performance of the unfolding procedure is tested for possible biases from the choice of the input model (the \ttbar \MADGRAPH simulation). It has been verified that by reweighting the \ttbar simulation the unfolding procedure based on the nominal response matrix reproduces the altered shapes within the statistical uncertainties. In addition, \ttbar samples simulated with \POWHEG and \MCATNLO are employed to obtain the response matrices used in the unfolding for the determination of systematic uncertainties of the model (Section~\ref{sec:modelsyst}). Therefore, possible effects from the unfolding procedure are already taken into account in the systematic uncertainties.

The differential cross section is reported at the particle level, where objects are defined as follows. Leptons from \PW\ boson decays are defined after final-state radiation, and jets are defined at the particle level by applying the anti-$\kt$ clustering algorithm with a distance parameter of 0.5~\cite{Cacciari:2008gp} to all stable particles, excluding the decay products from \PW\ boson decays into $\Pe\nu$, $\mu\nu$, and leptonic $\tau$ final states.
A jet is defined as a \PQb jet if it has at least one \PQb hadron associated with it. To perform the matching between \PQb hadrons and jets, the \PQb hadron momentum is scaled down to a negligible value and included in the jet clustering (so-called ghost matching~\citep{Cacciari:2008gn}). The \PQb jets from the \ttbar decay are identified by matching the \PQb hadrons to the corresponding original \PQb quarks.
The measurements are presented for two different phase-space regions, defined by the kinematic and geometric attributes of the \ttbar decay products and the additional jets. The visible phase space is defined by the following kinematic requirements:
\begin{itemize}
\item Leptons: $\pt>20\GeV$, $|\eta|<2.4$,
\item \PQb jets arising from top quarks: $\pt>30\GeV$, $|\eta|<2.4$,
\item Additional jets and \PQb jets: $\pt>20\GeV$, $|\eta|<2.4$.
\end{itemize}

The full phase space is defined by requiring only the additional jets or \PQb jets be within the above-mentioned kinematic range, without additional requirements on the decay products of the \ttbar system, and including the correction for the corresponding dileptonic branching fraction, calculated using the leptonic branching fraction of the \PW\ boson~\cite{PDG2014}.

In the following sections, the \ttbar differential cross section measured as a function of the jet multiplicity in the visible phase space and the results as a function of the kinematic variables of the additional jets in the event, measured in the visible and the full phase-space regions, are discussed. The absolute cross sections are presented as figures and compared to different predictions. The full results are given in tables in Appendix~\ref{sec:summarytables}, along with the normalized differential cross sections measurements.

\section{Differential \texorpdfstring{\ttbar}{t-tbar} cross sections as a function of jet multiplicity}
\sectionmark{\ttbar cross sections as a function of jet multiplicity}
\label{sec:diffxsecNJets}

\begin{figure*}[htbp!]
  \begin{center}
      \includegraphics[width=0.40 \textwidth]{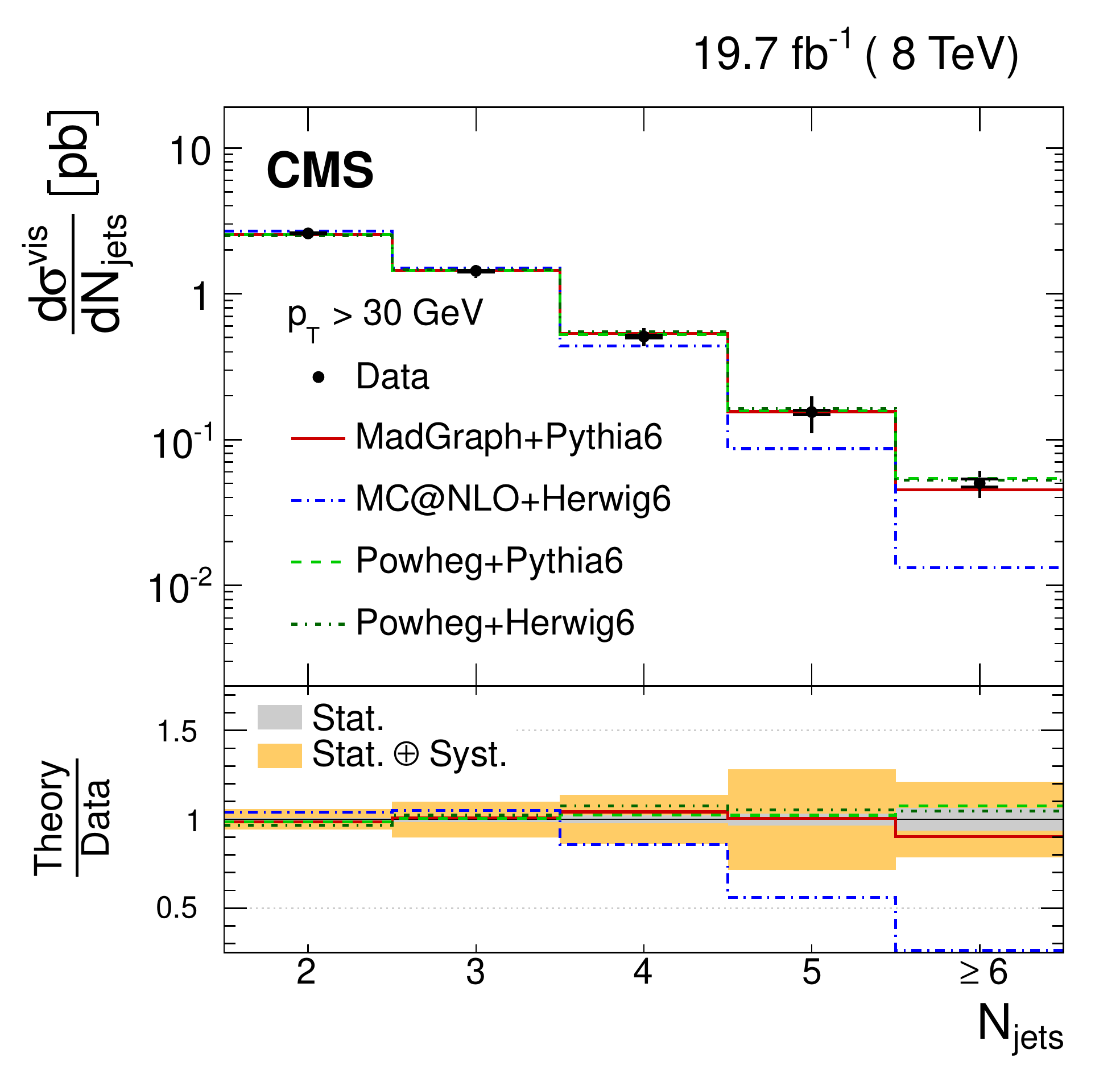}
      \includegraphics[width=0.40 \textwidth]{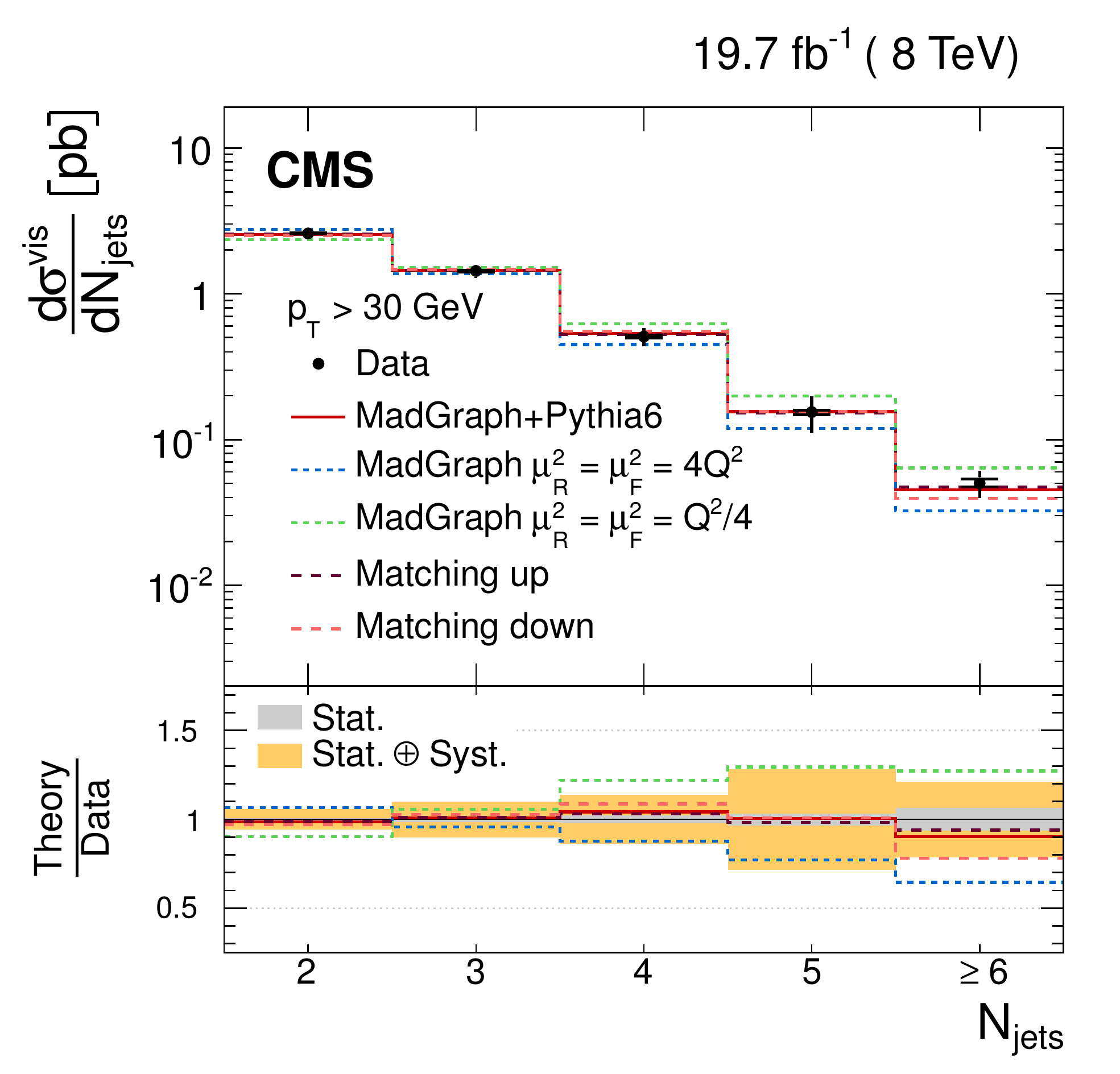}
      \includegraphics[width=0.40 \textwidth]{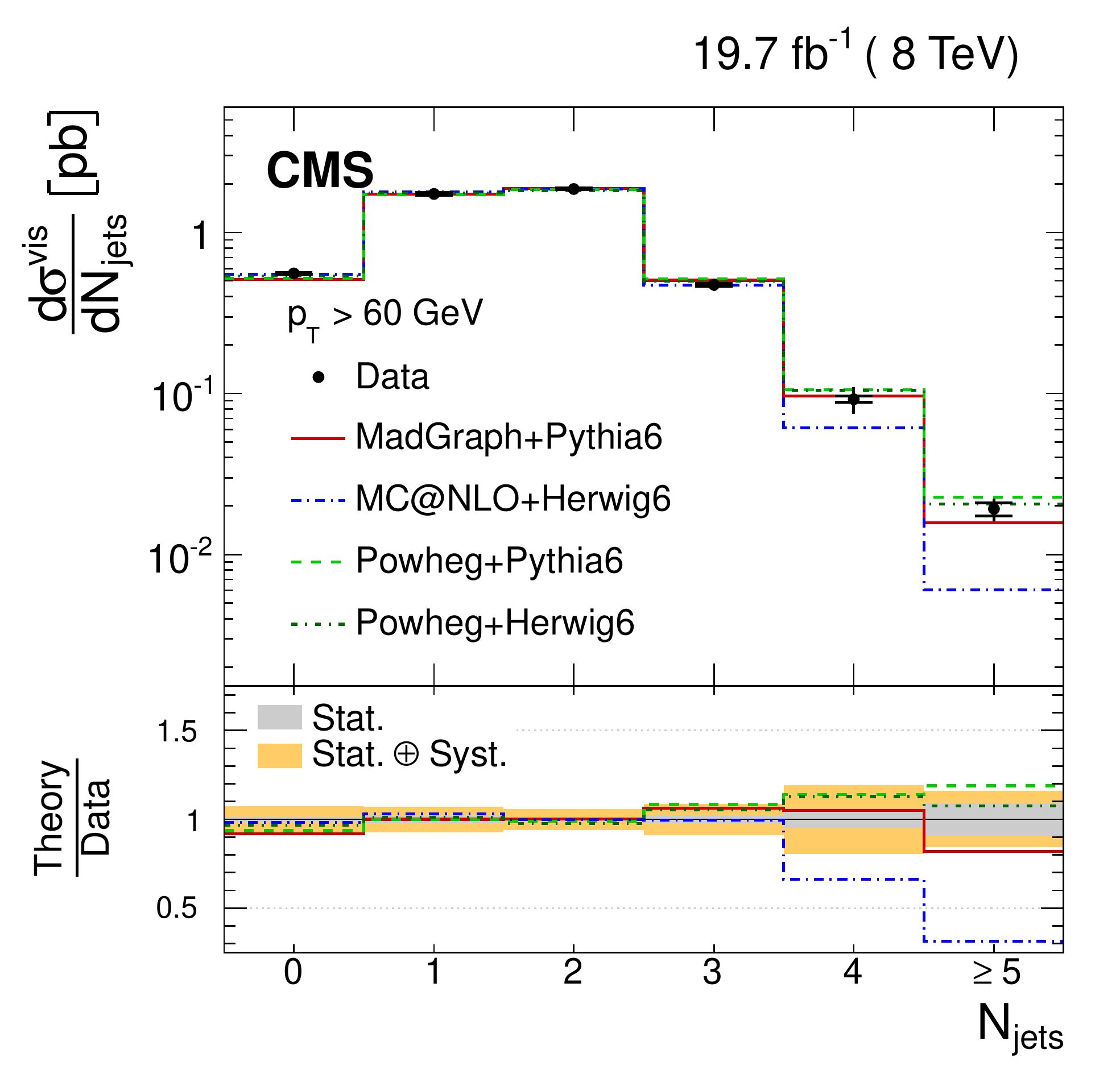}
      \includegraphics[width=0.40 \textwidth]{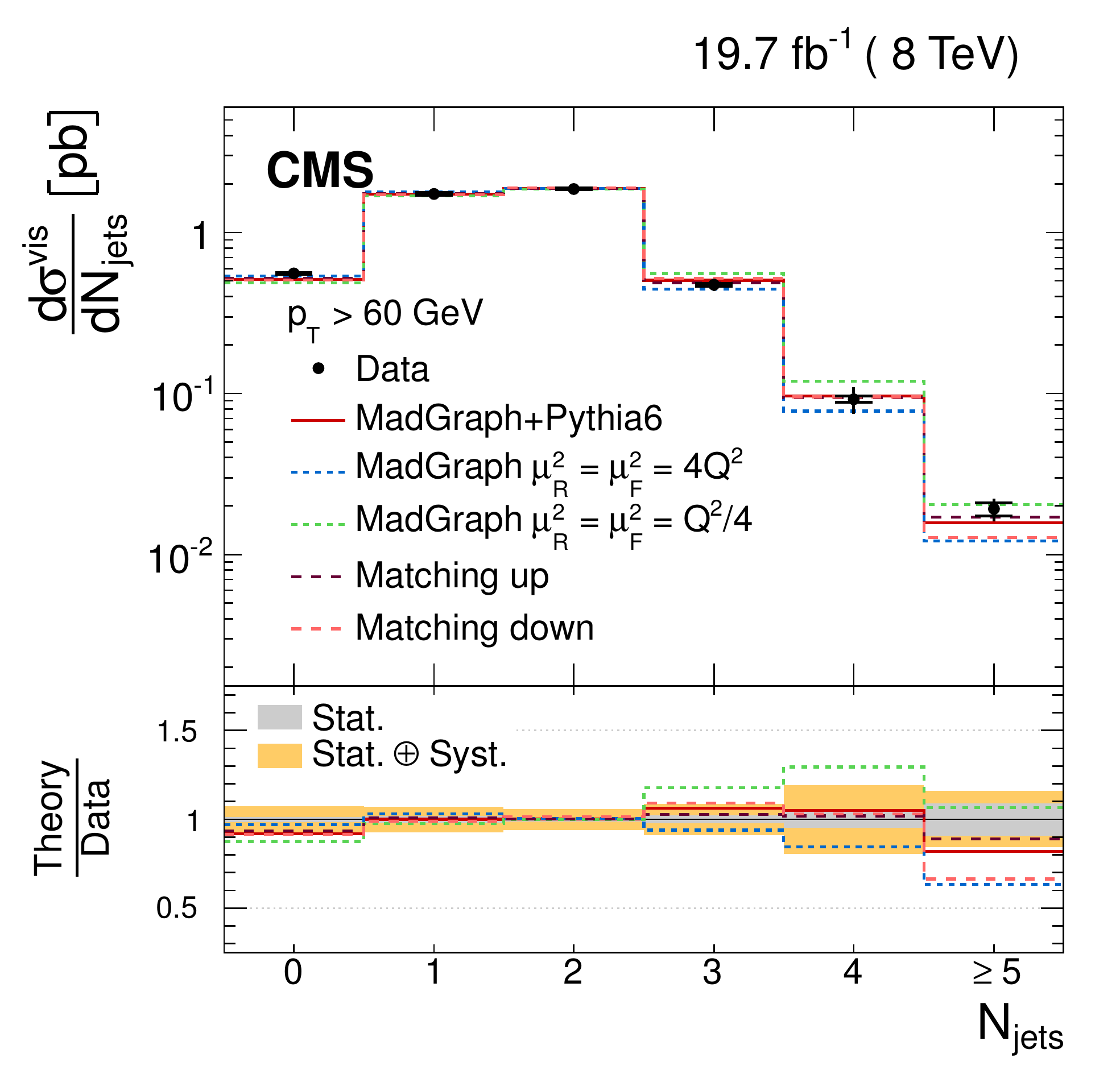}
      \includegraphics[width=0.40 \textwidth]{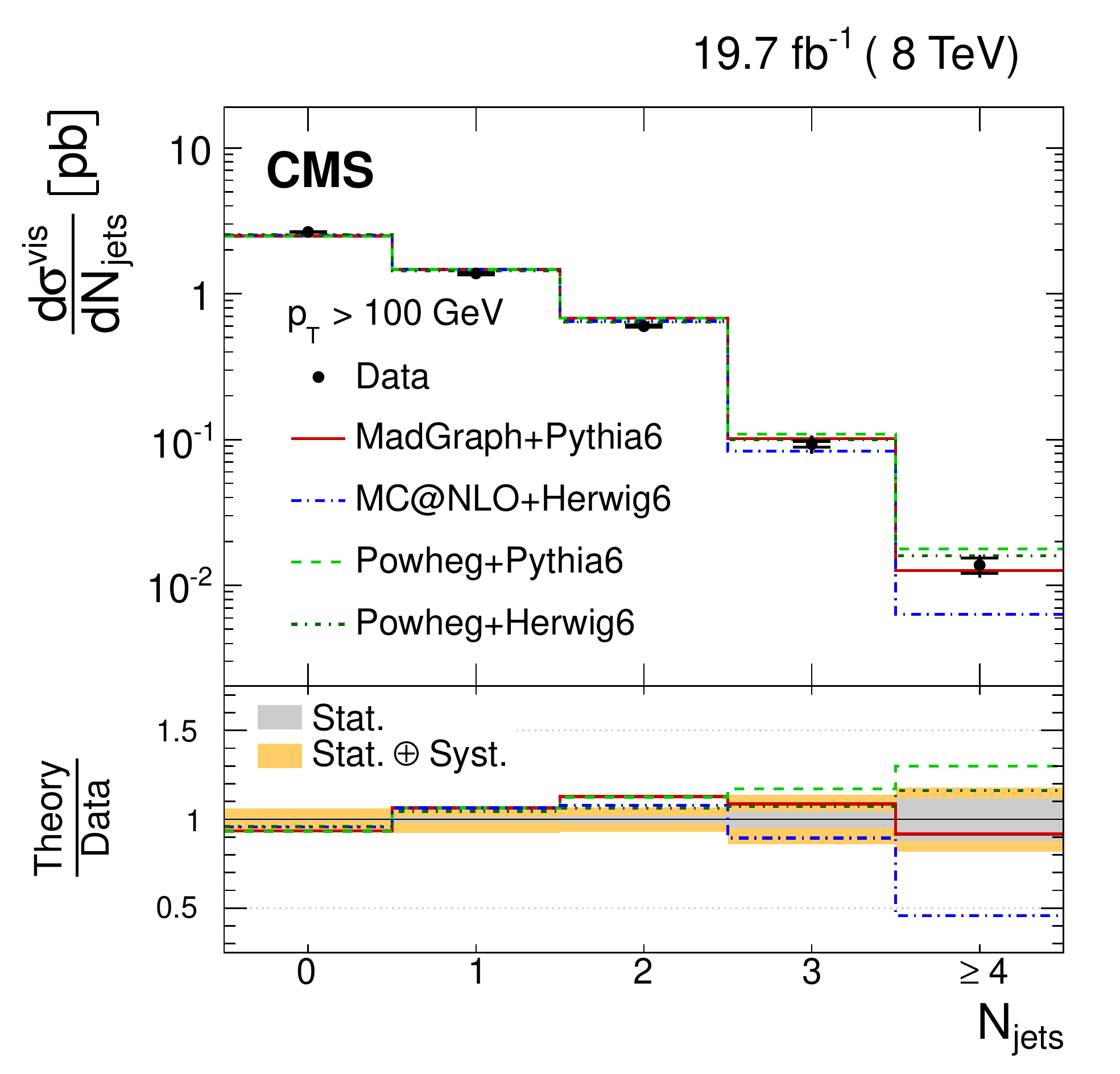}
      \includegraphics[width=0.40 \textwidth]{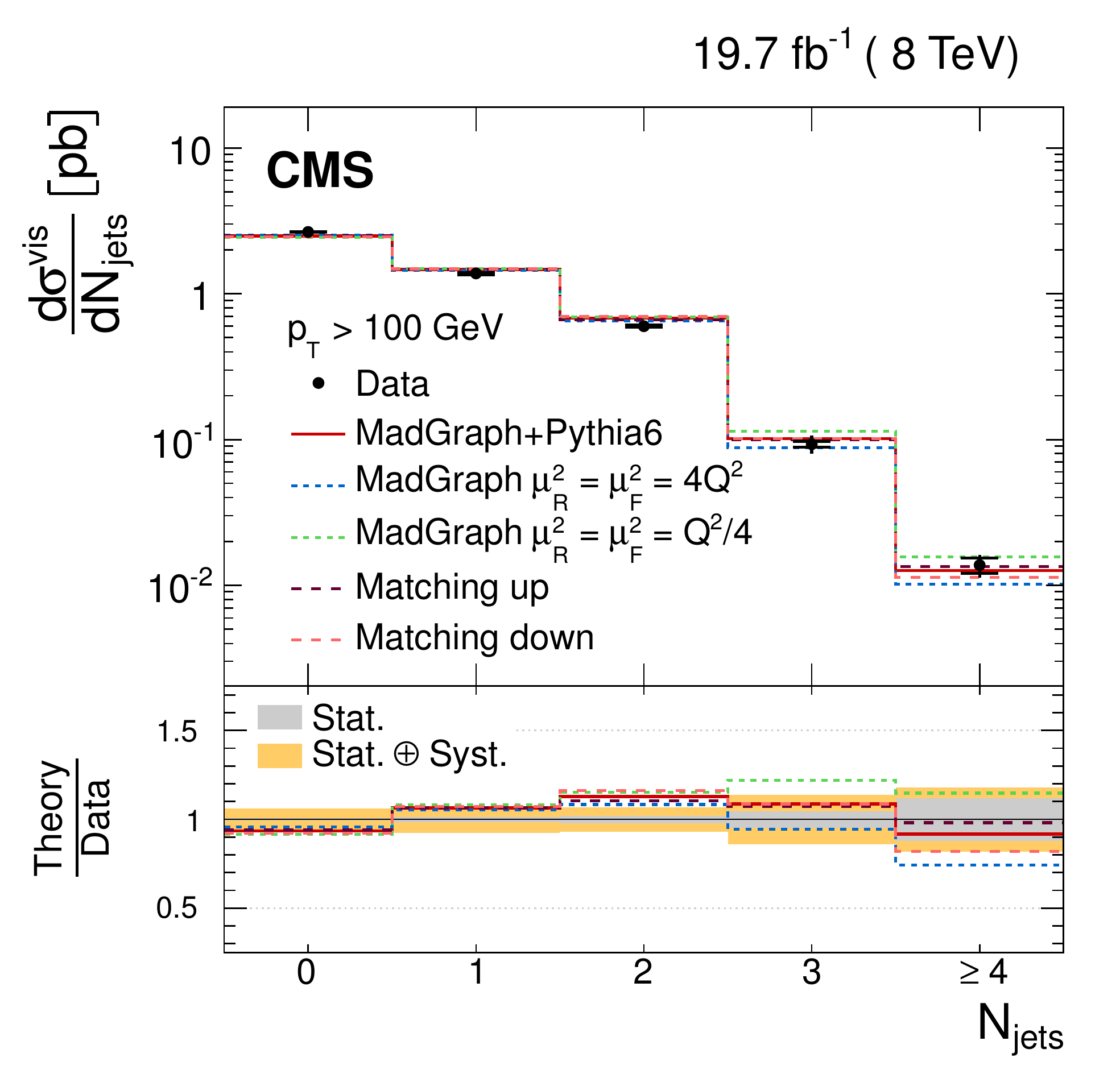}
\caption{Absolute differential \ttbar cross sections as a function of jet multiplicity for jets with $\pt>30\GeV$ (top row), 60\GeV (middle row), and 100\GeV (bottom row). In the figures on the left, the data are compared with predictions from \MADGRAPH interfaced with \PYTHIA{6}, \MCATNLO interfaced with \HERWIG{6}, and \POWHEG with \PYTHIA{6} and \HERWIG{6}. The figures on the right show the behaviour of the \MADGRAPH generator with varied renormalization, factorization, and jet-parton matching scales. The inner (outer) vertical bars indicate the statistical (total) uncertainties. The lower part of each plot shows the ratio of the predictions to the data.}
\label{fig:xsecjet}
  \end{center}
\end{figure*}

In Fig.~\ref{fig:xsecjet}, the absolute differential \ttbar cross section is shown for three different jet \pt thresholds: $\pt >30$, $60$, and $100\GeV$. The results are presented for a nominal top quark mass of $172.5\GeV$. The lower part of each figure shows the ratio of the predictions from simulation to the data. The light and dark bands in the ratio indicate the statistical and total uncertainties in the data for each bin, which reflect the uncertainties for a ratio of 1.0. All predictions are normalized to the measured cross section in the range shown in the histogram, which is evaluated by integrating over all bins for each observable. The results are summarized in Table~\ref{tab:dilepton:SummaryResultsJetMult}, together with the normalized cross sections. In general, the \MADGRAPH generator interfaced with \PYTHIA{6}, and \POWHEG interfaced both with \HERWIG{6} and \PYTHIA{6}, provide reasonable descriptions of the data. The \MCATNLO generator interfaced with \HERWIG{6} does not generate sufficiently large jet multiplicities, especially for the lowest jet \pt threshold. The sensitivity of \MADGRAPH to scale variations is investigated through the comparison of different renormalization, factorization, and jet-parton matching scales with respect to the nominal \MADGRAPH simulation. Variations in the jet-parton matching threshold do not yield large effects in the cross section, while the shape and normalization are more affected by the variations in the renormalization and factorization scales, which lead to a slightly worse description of the data up to high jet multiplicities, compared to their nominal values.

\begin{figure*}[htbp!]
  \begin{center}
      \includegraphics[width=0.40 \textwidth]{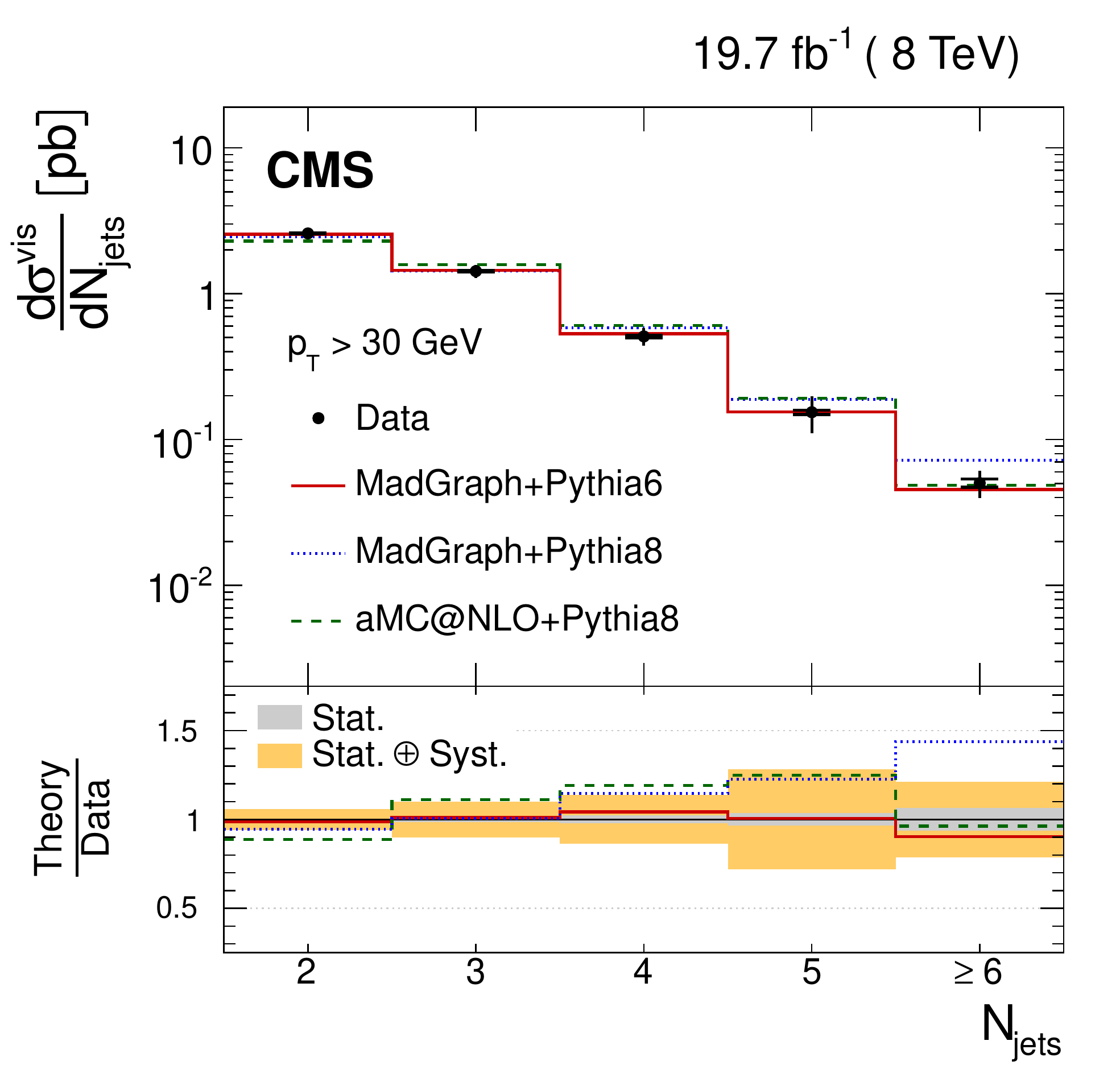}
      \includegraphics[width=0.40 \textwidth]{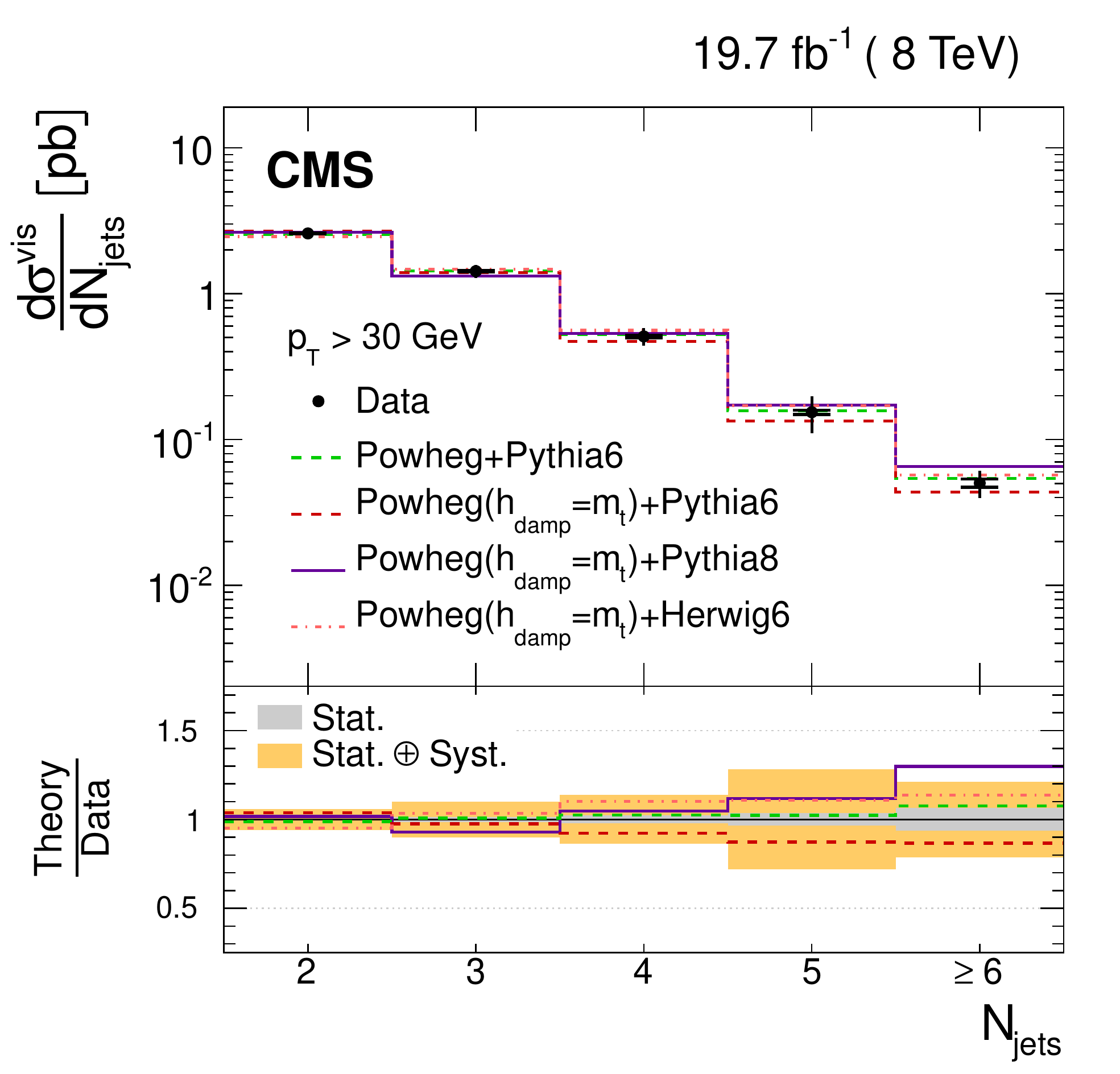}
      \includegraphics[width=0.40 \textwidth]{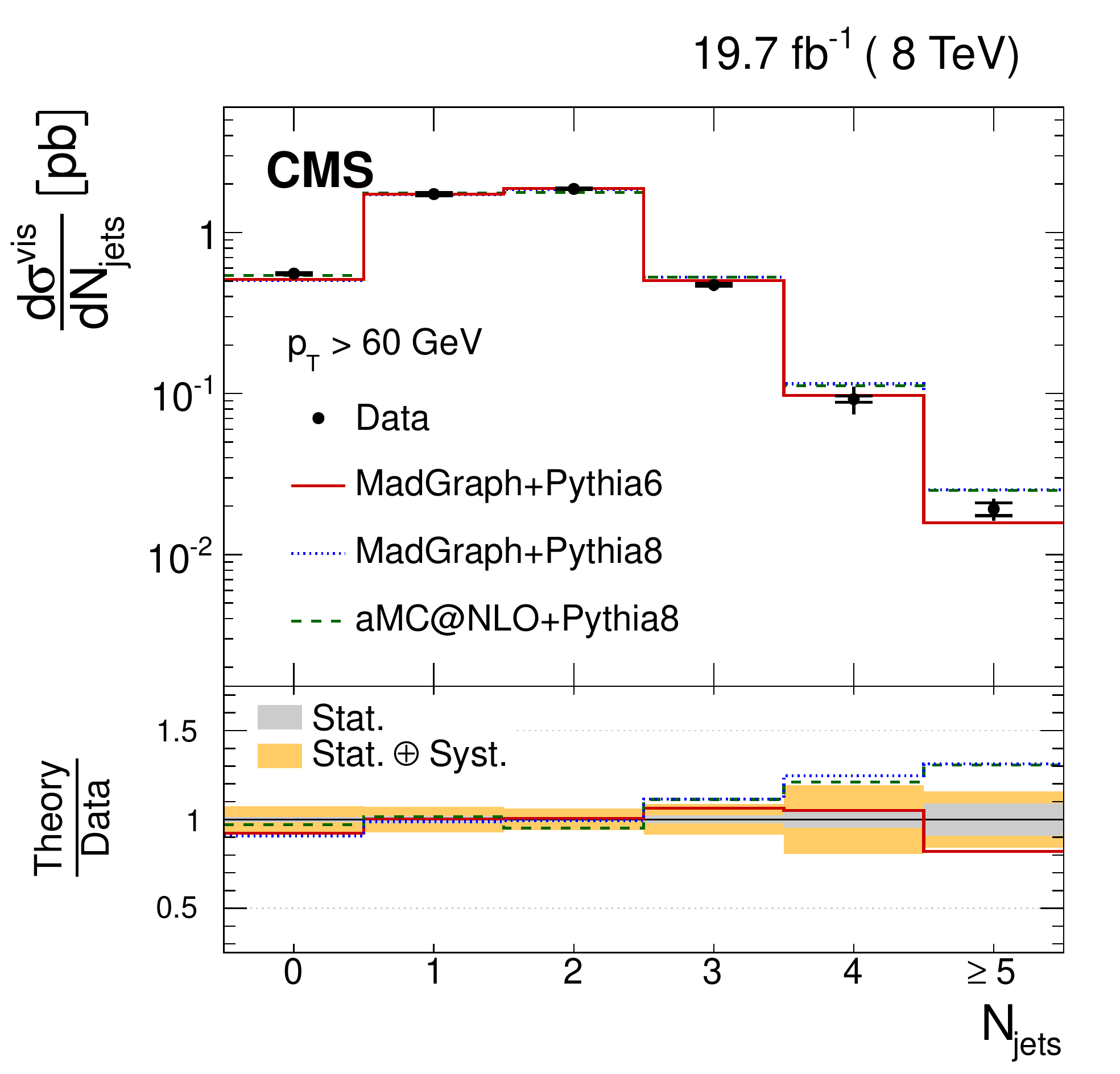}
      \includegraphics[width=0.40 \textwidth]{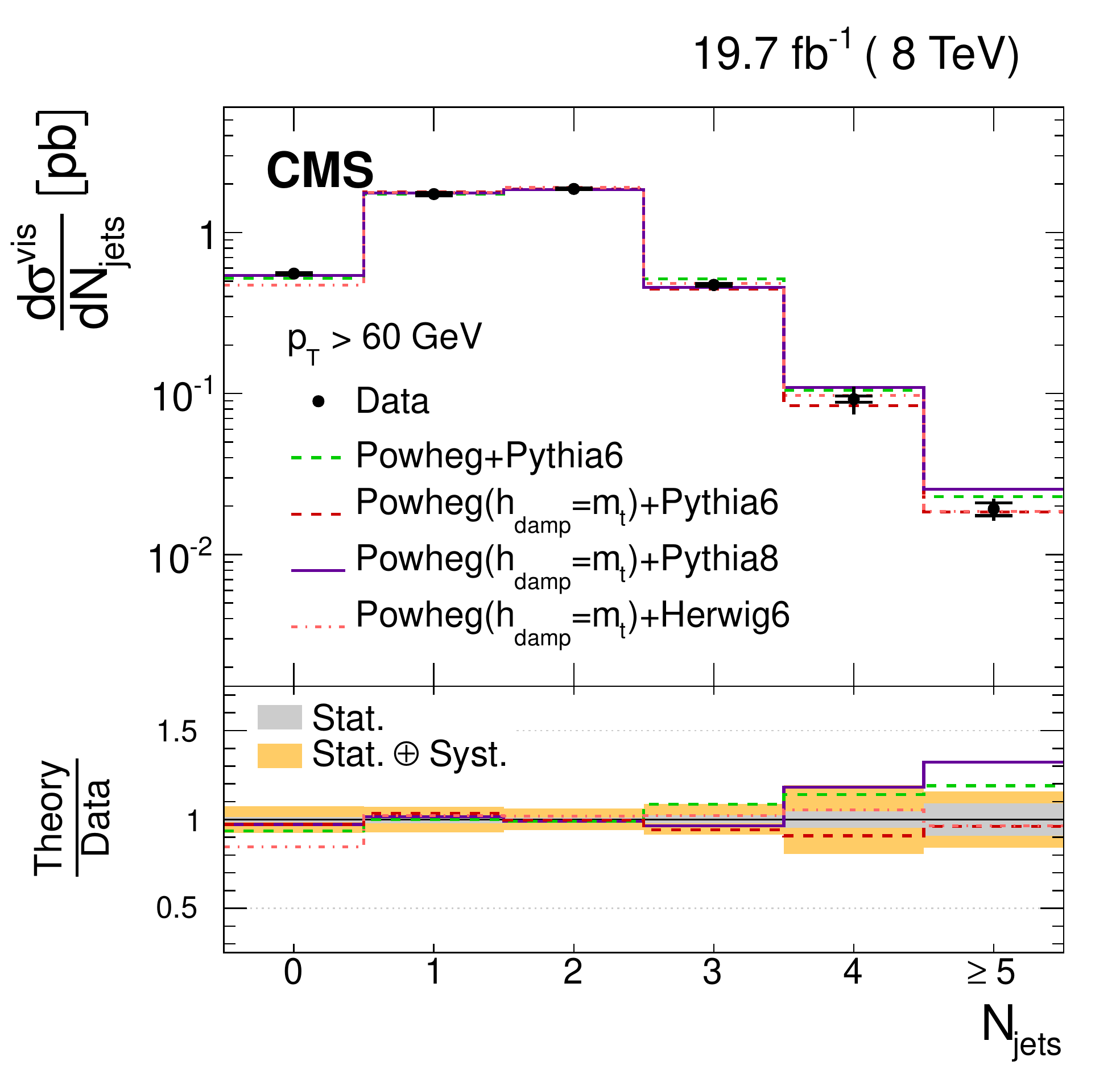}
      \includegraphics[width=0.40 \textwidth]{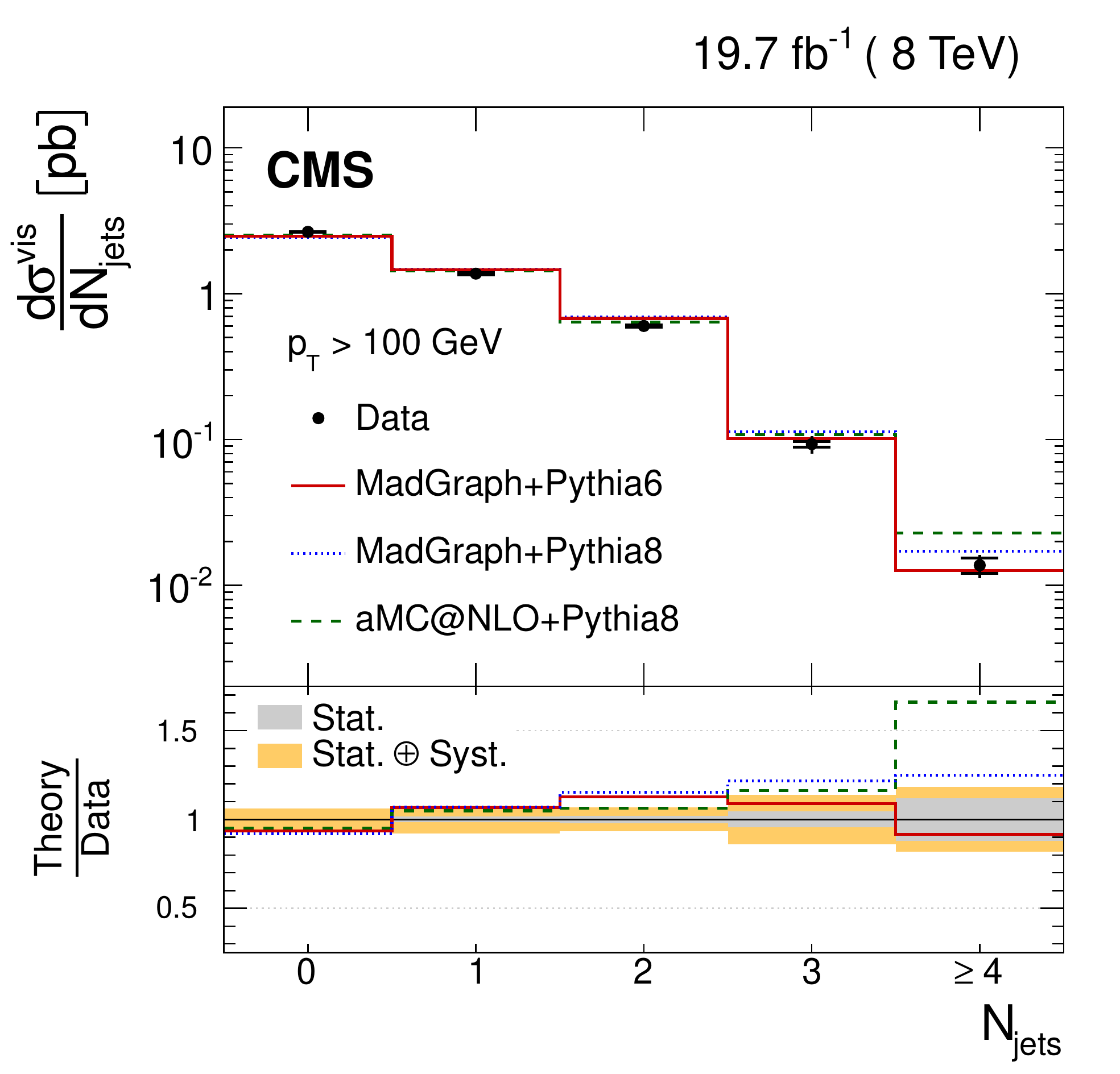}
      \includegraphics[width=0.40 \textwidth]{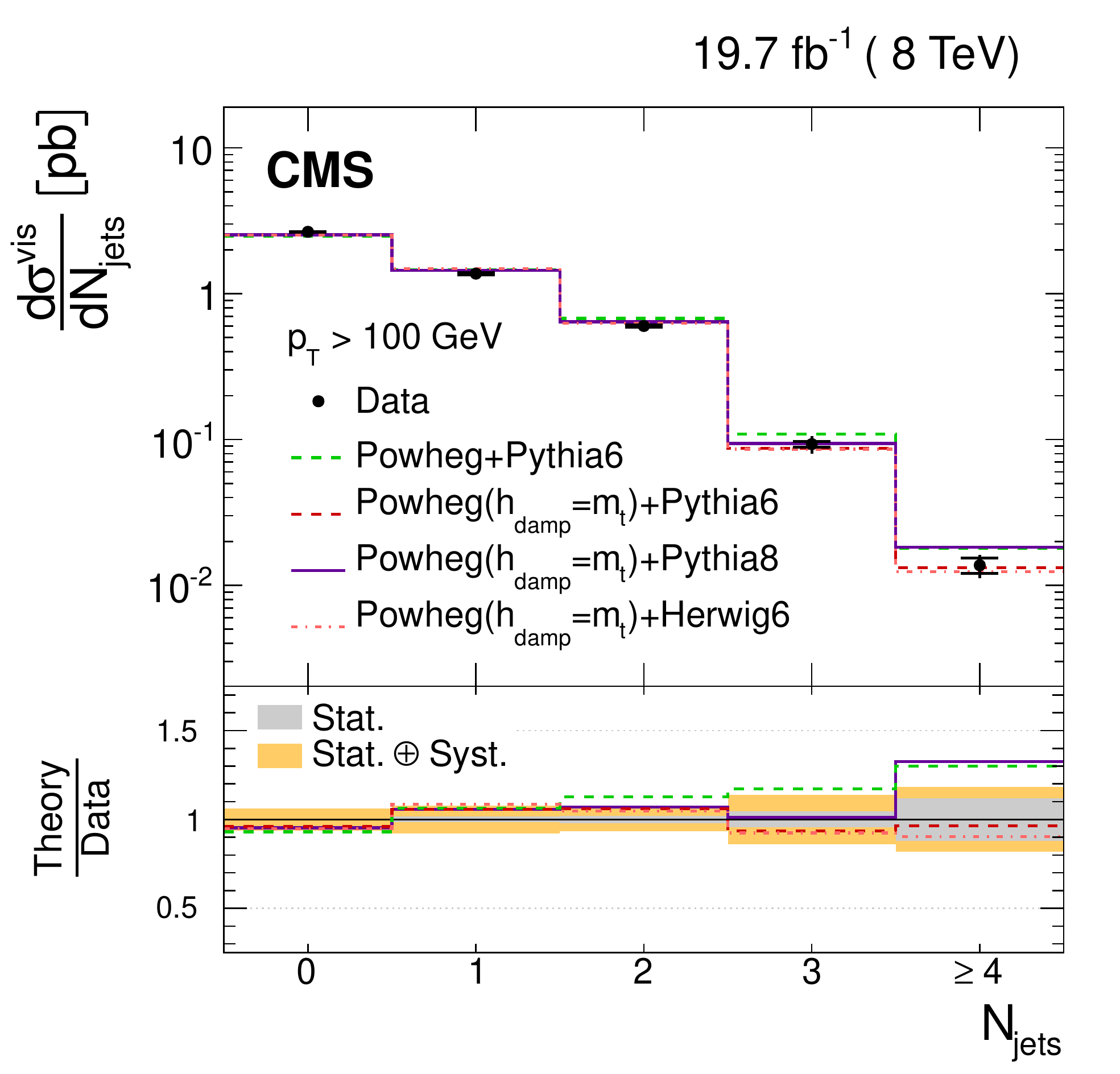}
\caption{Absolute differential \ttbar cross sections as a function of jet multiplicity for jets with $\pt>30\GeV$ (top row), 60\GeV (middle row), and 100\GeV (bottom row). In the figures on the left, the data are compared with predictions from \MADGRAPH interfaced with \PYTHIA{6} and \PYTHIA{8}, and \amcatnlo interfaced with \PYTHIA{8}. The figures on the right show the behaviour of the \POWHEG generator without and with \hdamp set to $m_{\PQt}$, matched with different versions and tunes of \PYTHIA and \HERWIG{6}. The inner (outer) vertical bars indicate the statistical (total) uncertainties. The lower part of each plot shows the ratio of the predictions to the data.}
\label{fig:xsecjetNewMC}
  \end{center}
\end{figure*}

{\tolerance=400
In Fig.~\ref{fig:xsecjetNewMC}, the results are compared to the predictions from \MADGRAPH and \amcatnlo interfaced with \PYTHIA{8}, and the \POWHEG generator with the \hdamp parameter set to $m_{\PQt}=172.5\GeV$ (labelled \POWHEG (h$_{\text{damp}}=m_{\PQt}$) in the legend), interfaced with \PYTHIA{6}, \PYTHIA{8}, and \HERWIG{6}. The \MADGRAPH and \amcatnlo simulations interfaced with \PYTHIA{8} predict larger jet multiplicities than measured in the data for all the considered \pt thresholds. In general, no large deviations between data and the different \POWHEG predictions are observed.
\par}

The total systematic uncertainty in the absolute differential cross section ranges between 6 to 30\%, while for the normalized cross section it varies from 2\% up to 20\% for the bins corresponding to the highest number of jets. In both cases, the dominant experimental systematic uncertainty arises from the JES, having a maximum value of 16\% for the absolute cross section bin with at least six jets and $\pt> 30\GeV$. Typical systematic uncertainty values range between 0.5 and 8\%, while the uncertainty in the normalized cross section is 0.5--4\%.
Regarding the modelling uncertainties, the most relevant ones are the uncertainty in the renormalization and factorization scales and the parton shower modelling, up to 6\% and 10\%, respectively. The uncertainties from the assumed top quark mass used in the simulation and the jet-parton matching threshold amount to 1--2\%. Other modelling uncertainties such as PDF, CR, and UE have slightly smaller impact. These uncertainties cancel to a large extent in the normalized results, with typical contributions below 0.5\%.
The total contribution from the integrated luminosity, lepton identification, and trigger efficiency, which only affect the normalization, is 3.5\%. This contribution is below 0.1\% for every bin in the normalized results. The uncertainty from the estimate of the background contribution is around 2\% for the absolute cross sections and typically below 0.5\% for the normalized results.

\section{Differential \texorpdfstring{\ttbar}{t-tbar} cross sections as a function of the kinematic variables of the additional jets}
\sectionmark{\ttbar cross sections as a function of the kinematics of additional jets}
\label{sec:diffxsecJets}
The absolute and normalized differential cross sections are measured as a function of the kinematic variables of the additional jets in the visible phase space defined in Section~\ref{sec:diffxsec}. The results are compared to predictions from four different generators: \POWHEG interfaced with \PYTHIA{6} and \HERWIG{6}, \MCATNLO{}+\HERWIG{6}, and \MADGRAPH{}+\PYTHIA{6} with varied renormalization, factorization, and jet-parton matching scales. All predictions are normalized to the measured cross section over the range of the observable shown in the histogram in the corresponding figures.

\begin{figure*}[htbp!]
  \begin{center}
      \includegraphics[width=0.40 \textwidth]{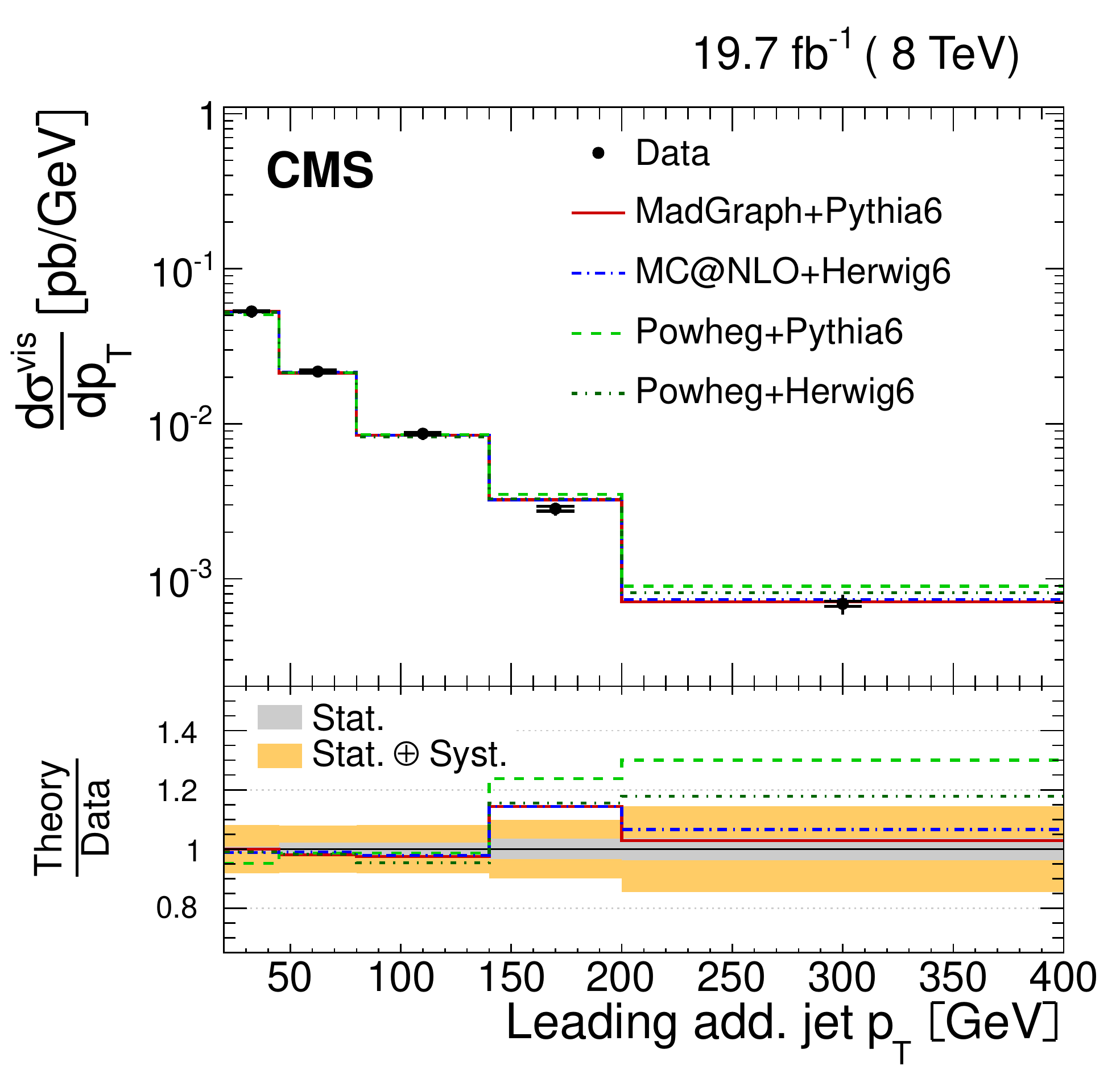}%
      \includegraphics[width=0.40 \textwidth]{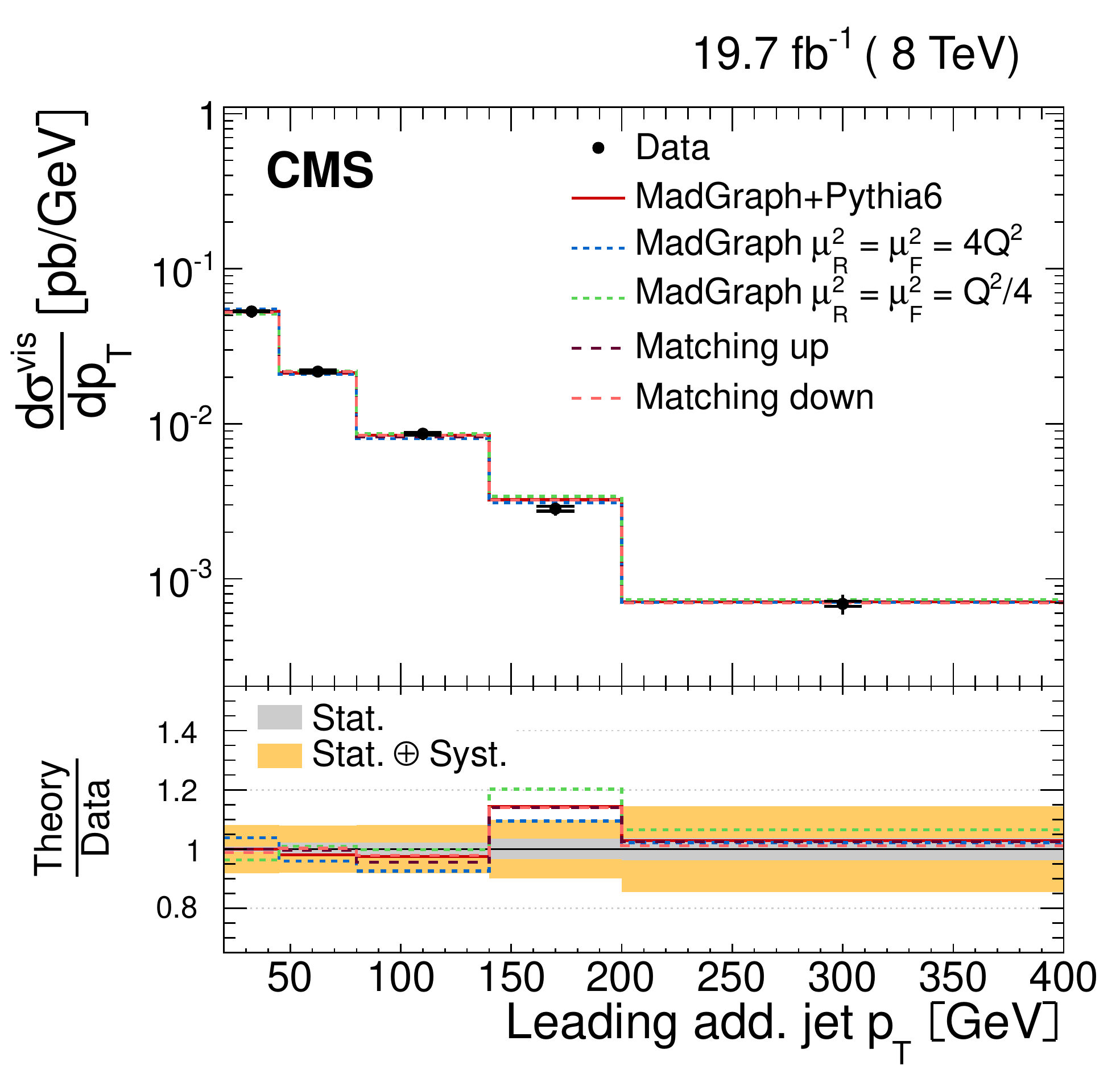}\\
      \includegraphics[width=0.40 \textwidth]{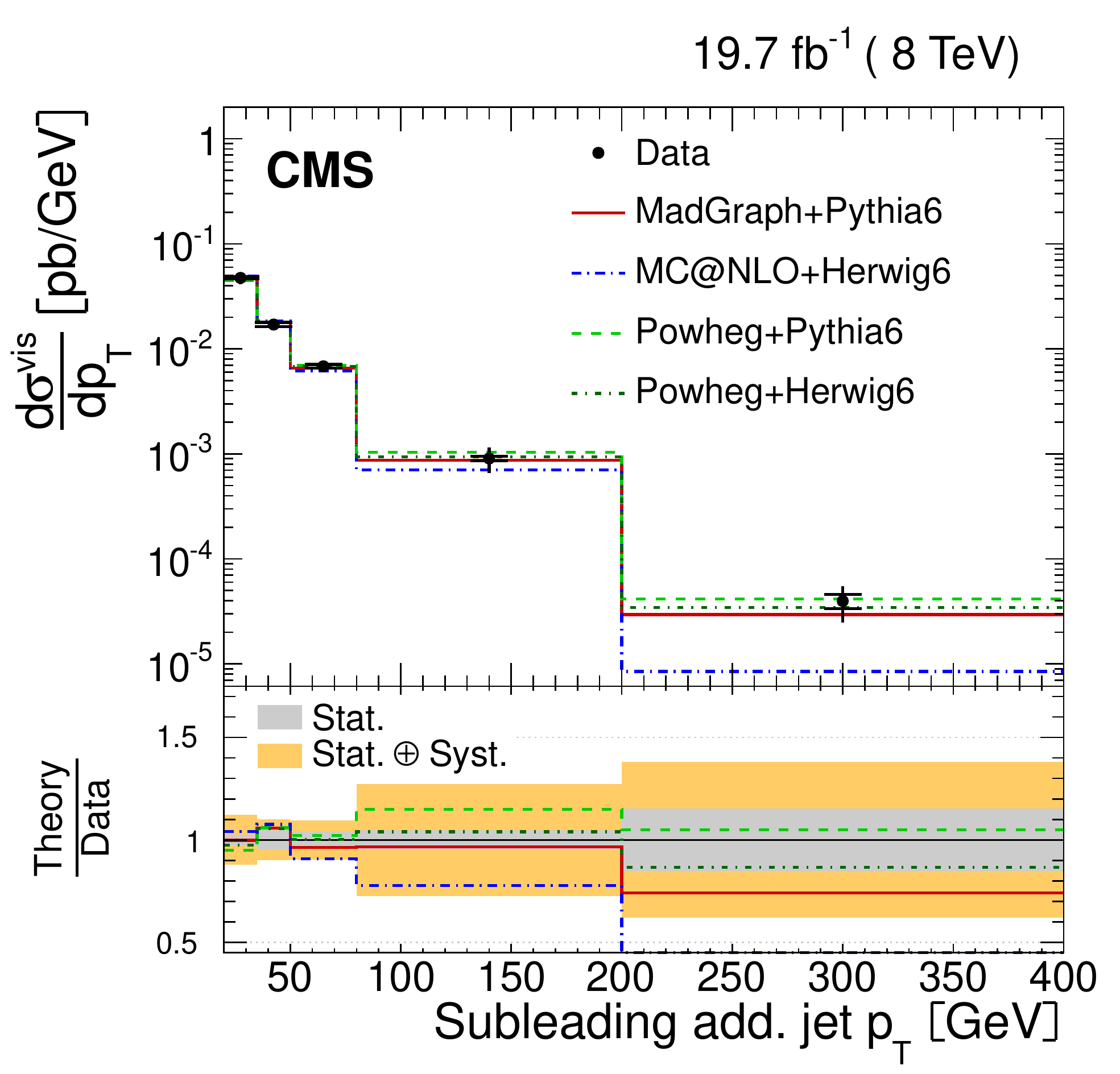}%
      \includegraphics[width=0.40 \textwidth]{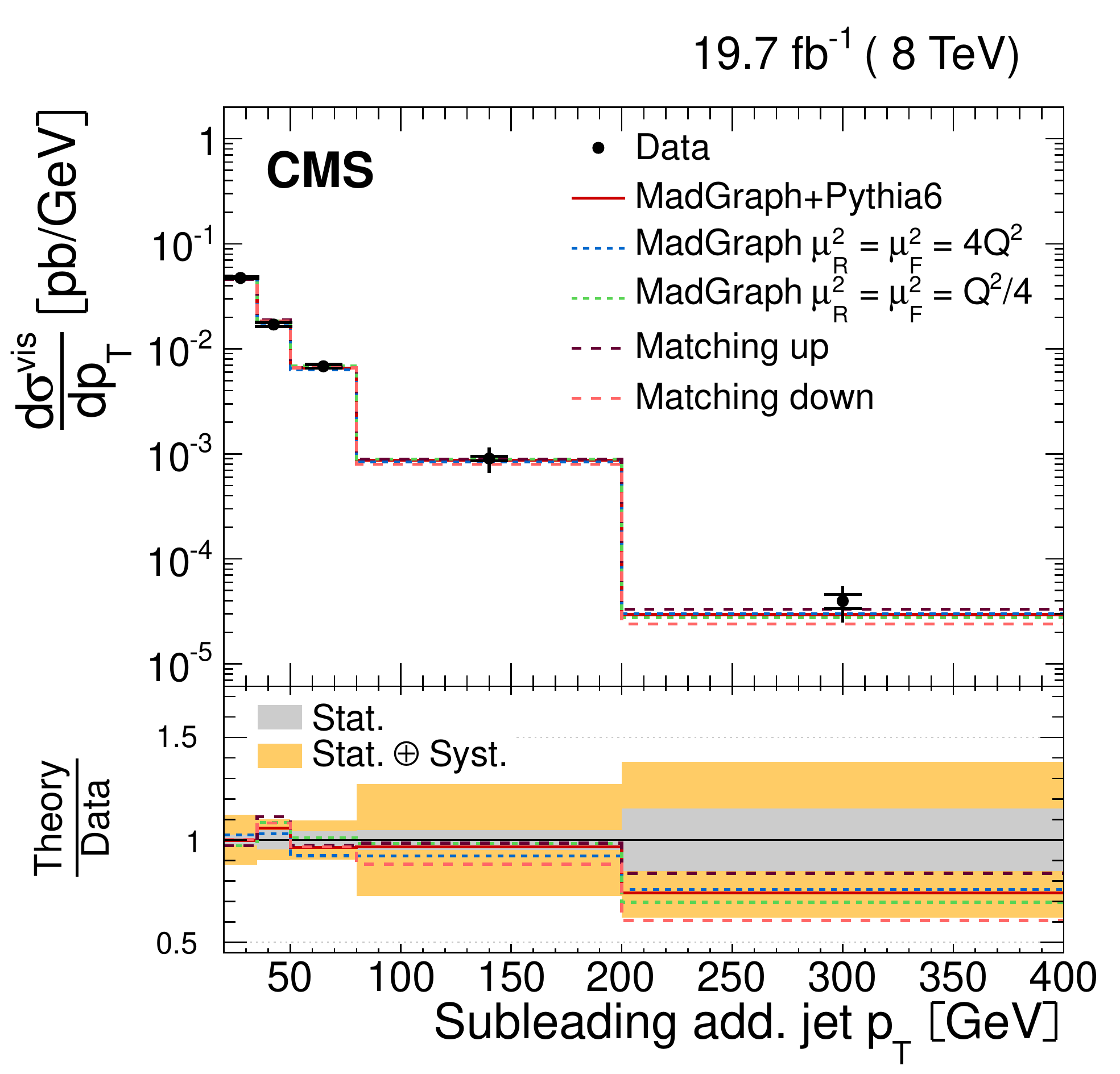}
      \includegraphics[width=0.40 \textwidth]{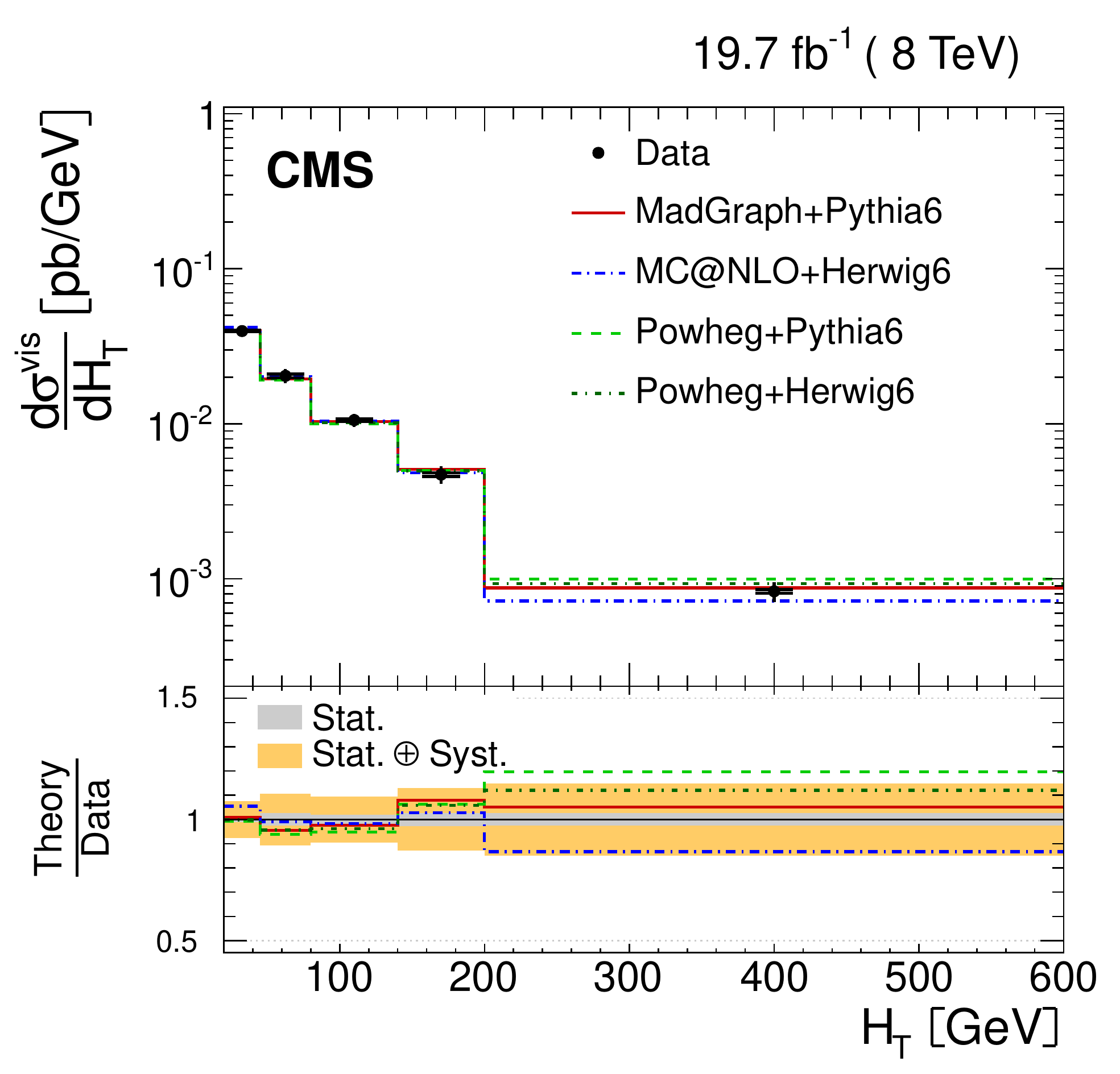}%
      \includegraphics[width=0.40 \textwidth]{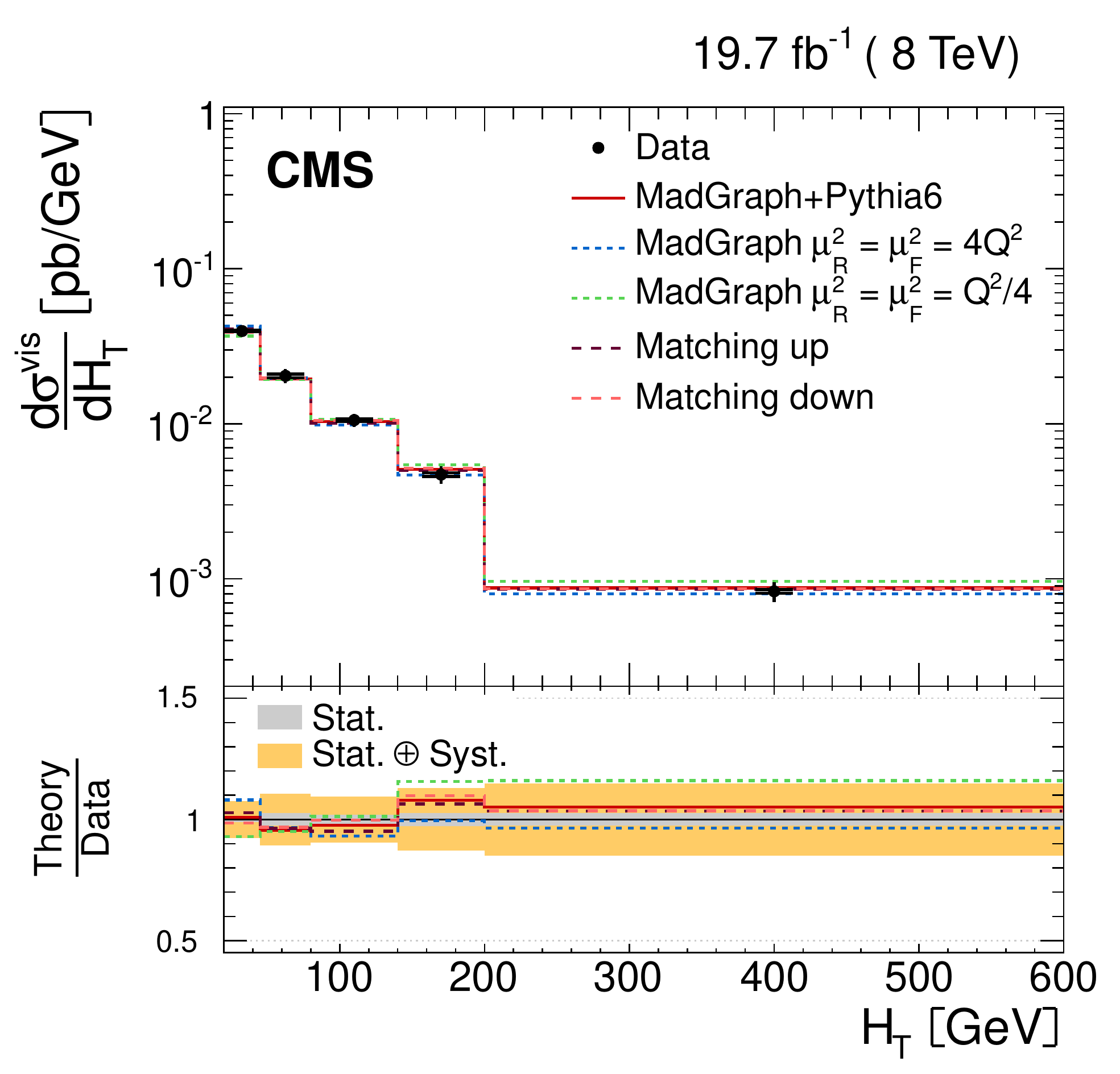}\\
      \caption{Absolute differential \ttbar cross section as a function of \pt of the leading additional jet (top) and the subleading additional jet (middle), and $\HT$ (bottom) in the visible phase space of the \ttbar system and the additional jets. Data are compared to predictions from \MADGRAPH{}+\PYTHIA{6}, \POWHEG{}+\PYTHIA{6}, \POWHEG{}+\HERWIG{6}, and \MCATNLO{}+\HERWIG{6} (left) and to \MADGRAPH with varied renormalization, factorization, and jet-parton matching scales (right). The inner (outer) vertical bars indicate the statistical (total) uncertainties. The lower part of each plot shows the ratio of the predictions to the data. }
\label{fig:inclusivept}
  \end{center}
\end{figure*}

The absolute differential cross sections as a function of the \pt of the leading and subleading additional jets and $\HT$, the scalar sum of the \pt of all additional jets in the event, are shown in Fig.~\ref{fig:inclusivept}. The total uncertainties in the absolute cross sections range from 8--14\% for the leading additional jet \pt and $\HT$, and up to 40\% for the subleading additional jet \pt, while the systematic uncertainties in the normalized cross sections for the bins with the larger number of events are about 3--4\%. The dominant sources of systematic uncertainties arise in both cases from model uncertainties, in particular the renormalization and factorization scales, and the parton shower modelling (up to 10\% for the absolute cross sections), and JES (3--6\% for the absolute cross sections). The typical contribution of other uncertainties such as the assumed top quark mass in the simulation, background contribution, etc., amounts to 1--3\% and 0.5--1.5\%, for the absolute and normalized cross sections, respectively.

In general, the simulation predictions describe the behaviour of the data for the leading additional jet momenta and $\HT$, although some predictions, in particular \POWHEG, favour a harder \pt spectrum for the leading jet. The \MCATNLO{}+\HERWIG{6} prediction yields the largest discrepancies. The varied \MADGRAPH samples provide similar descriptions of the shape of the data, except for \MADGRAPH with the lower $\mu_\mathrm{R} = \mu_\mathrm{F}$ scale, which worsens the agreement.

\begin{figure*}[htbp!]
  \begin{center}
      \includegraphics[width=0.40 \textwidth]{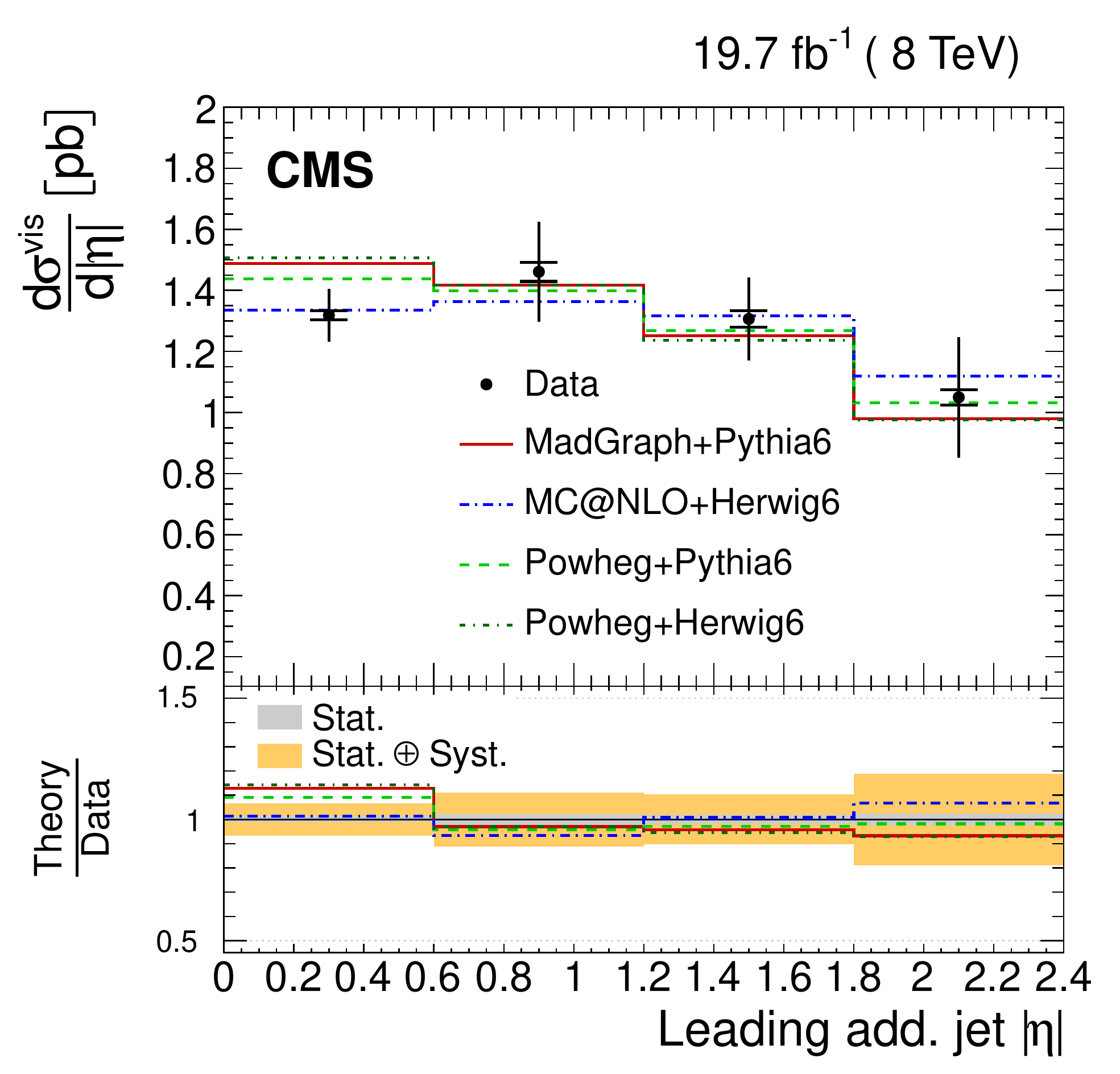}%
      \includegraphics[width=0.40 \textwidth]{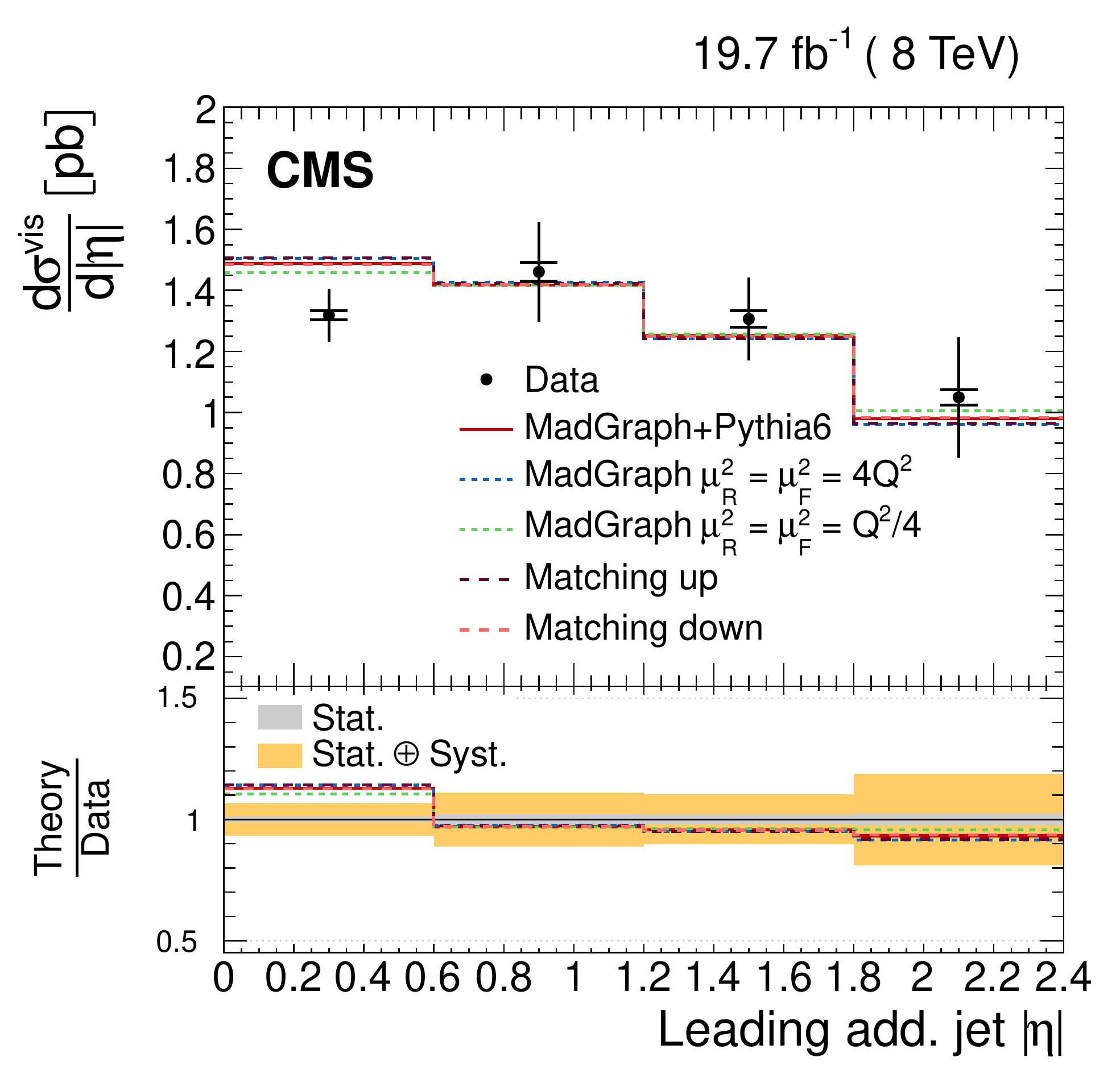}\\
      \includegraphics[width=0.40 \textwidth]{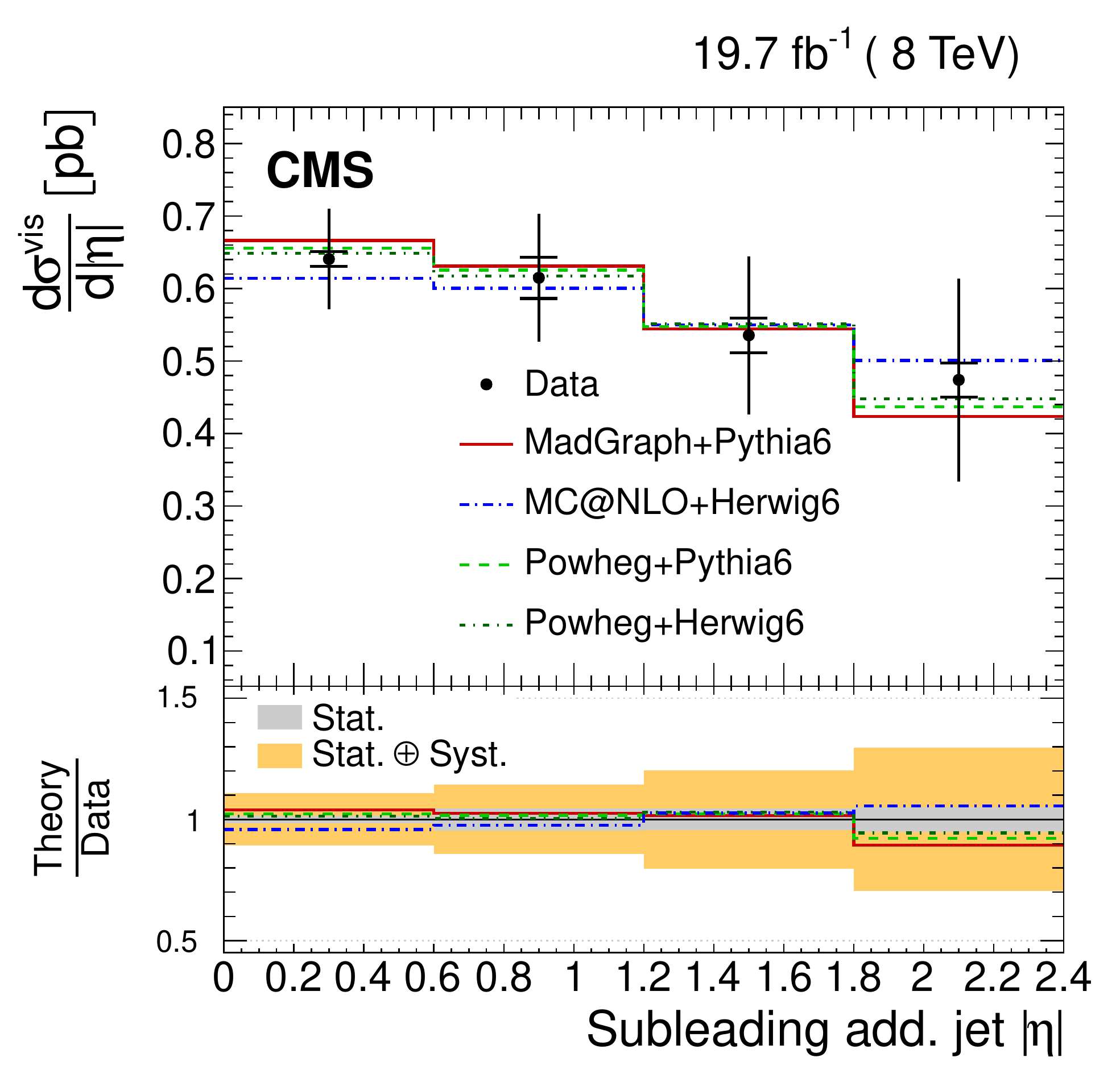}%
      \includegraphics[width=0.40 \textwidth]{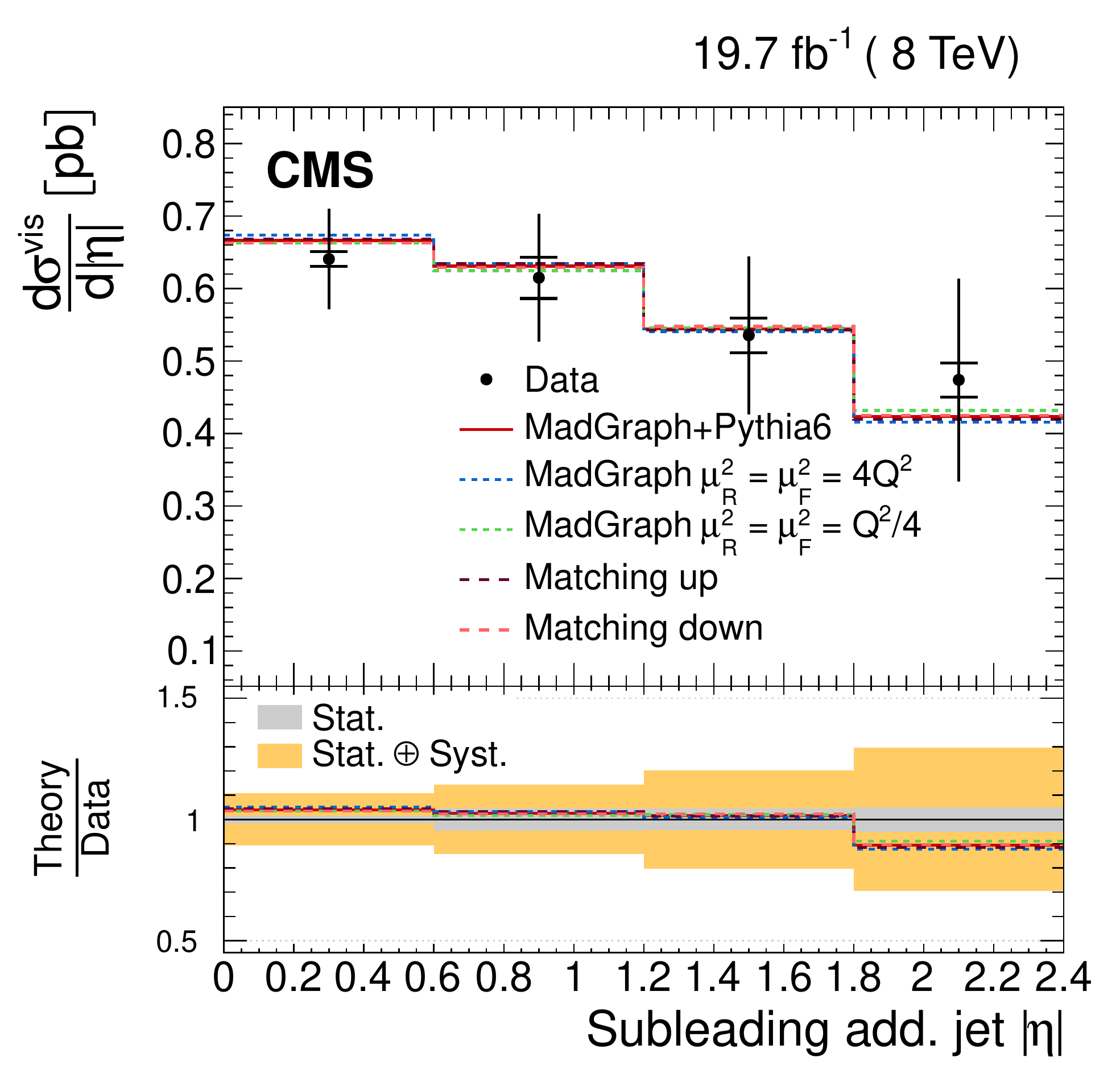}
\caption{Absolute differential \ttbar cross section as a function of the \abseta of the leading additional jet (top) and the subleading additional jet (bottom) in the visible phase space of the \ttbar system and the additional jets. Data are compared to predictions from \MADGRAPH{}+\PYTHIA{6}, \POWHEG{}+\PYTHIA{6}, \POWHEG{}+\HERWIG{6}, and \MCATNLO{}+\HERWIG{6} (left) and to \MADGRAPH with with varied renormalization, factorization, and jet-parton matching scales (right). The inner (outer) vertical bars indicate the statistical (total) uncertainties. The lower part of each plot shows the ratio of the predictions to the data.}
\label{fig:inclusiveeta}
  \end{center}
\end{figure*}

The results as a function of \abseta are presented in Fig.~\ref{fig:inclusiveeta}. The typical total systematic uncertainties in the absolute cross sections vary from 6.5--19\% for the leading additional jet and about 11--20\% for the subleading one. The uncertainty in the normalized cross section ranges from 1.5--9\% and 5--14\%, respectively. The shape of the \abseta distribution is well modelled by \MCATNLO{}+\HERWIG{6}. The distributions from \MADGRAPH and \POWHEG yield a similar description of the data, being slightly more central than \MCATNLO{}. Variations of the \MADGRAPH parameters have little impact on these distributions.

\begin{figure*}[htbp!]
  \begin{center}
      \includegraphics[width=0.40 \textwidth]{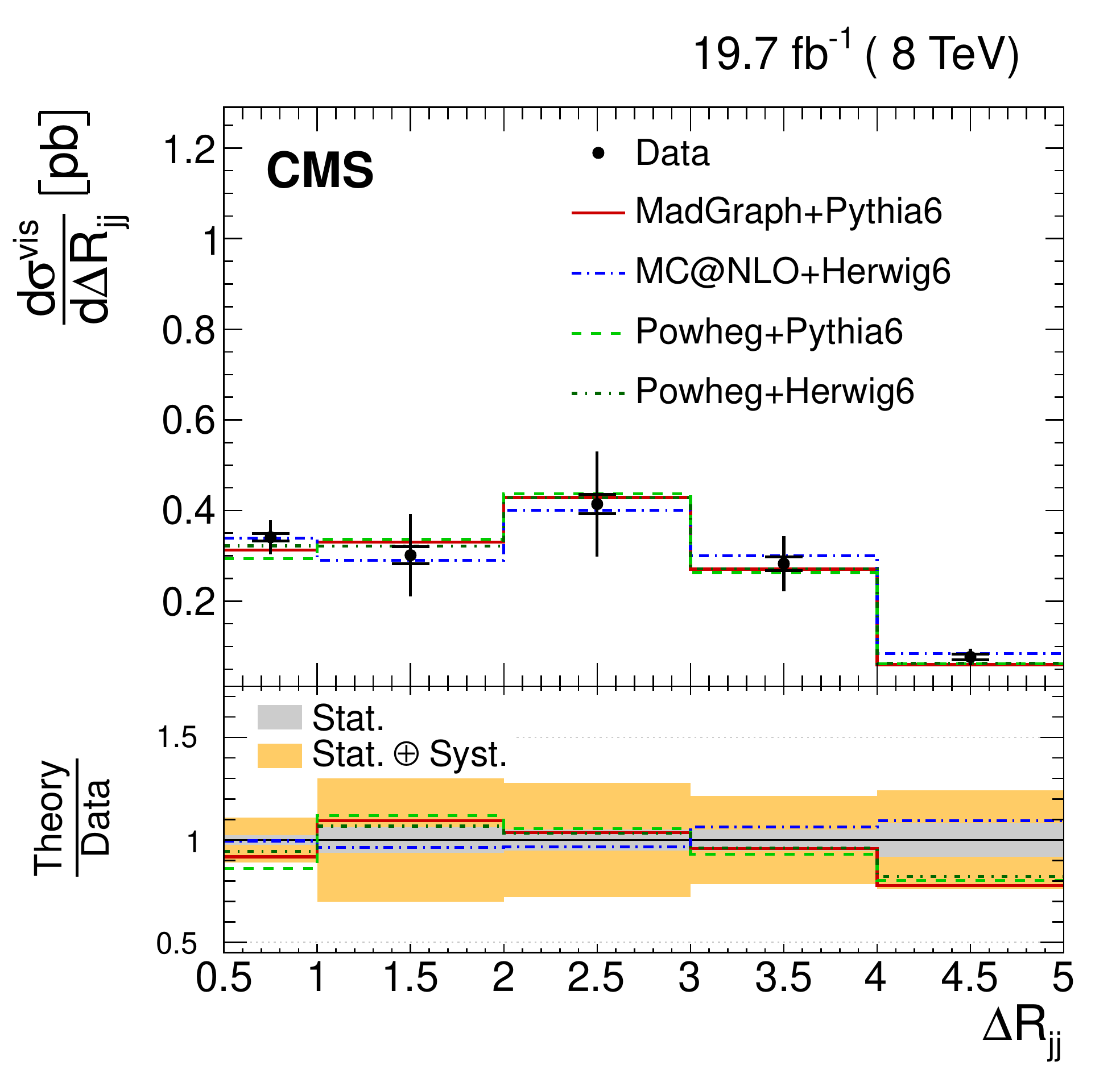}%
      \includegraphics[width=0.40 \textwidth]{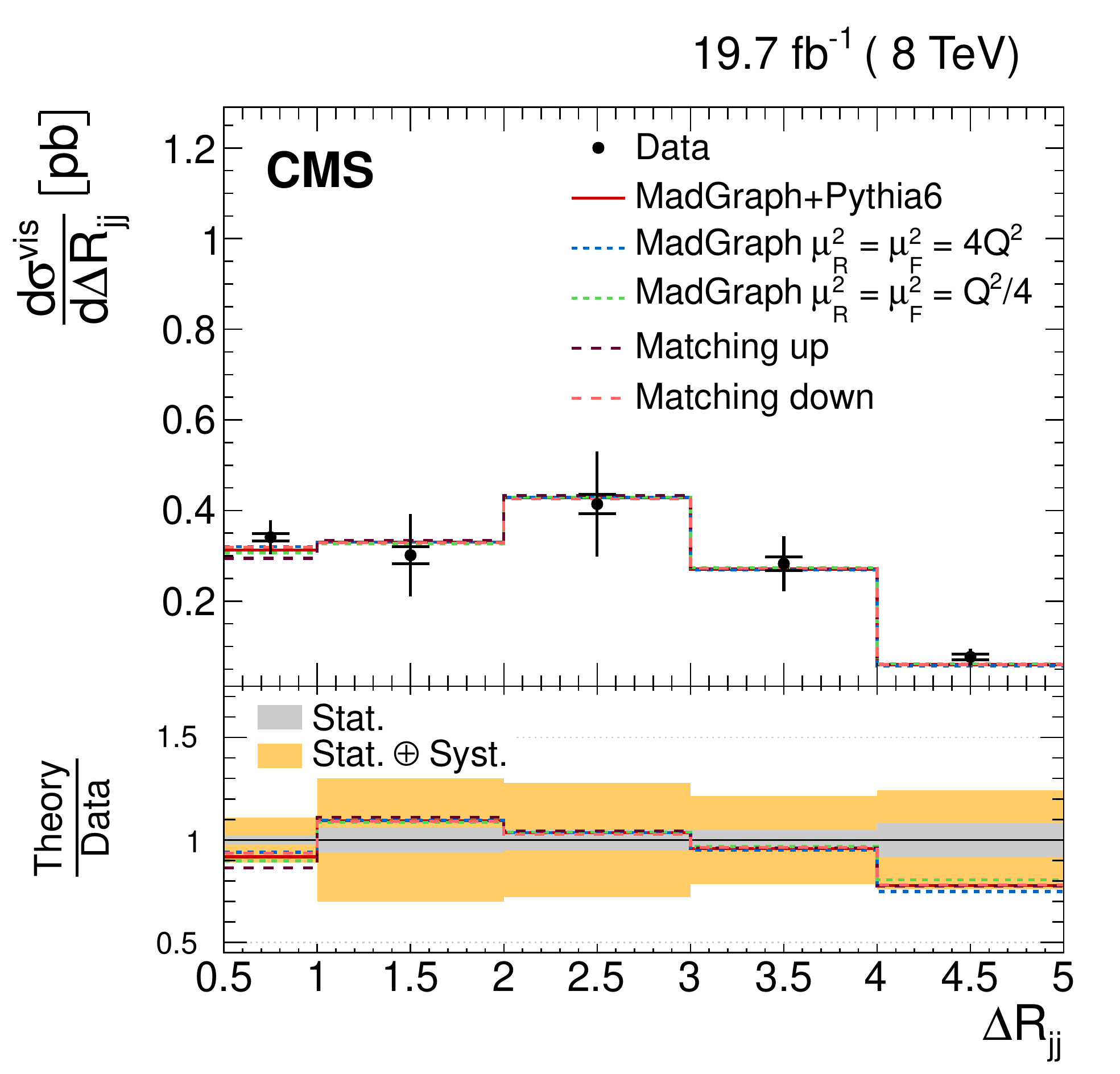}\\
      \includegraphics[width=0.40 \textwidth]{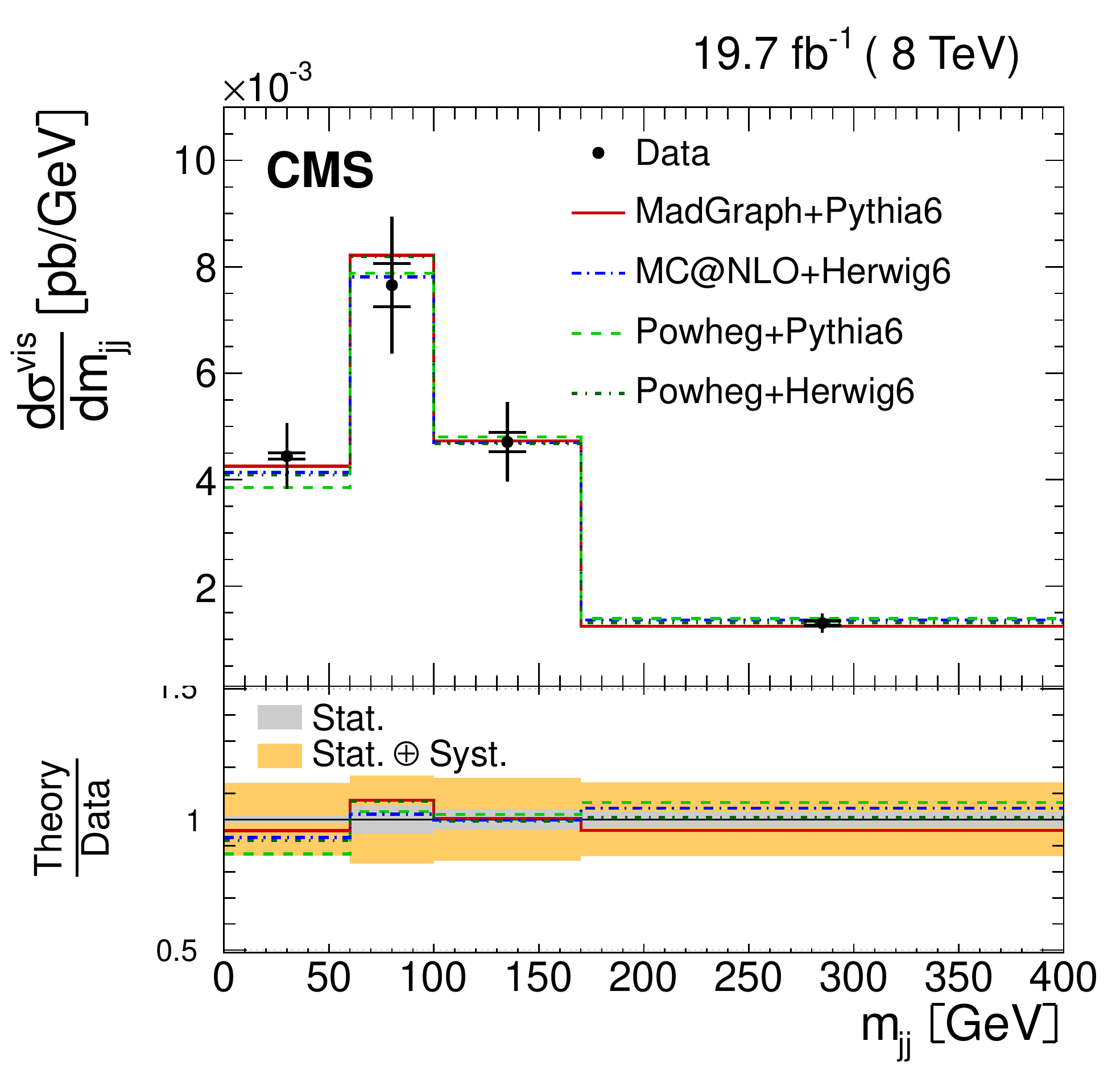}%
      \includegraphics[width=0.40 \textwidth]{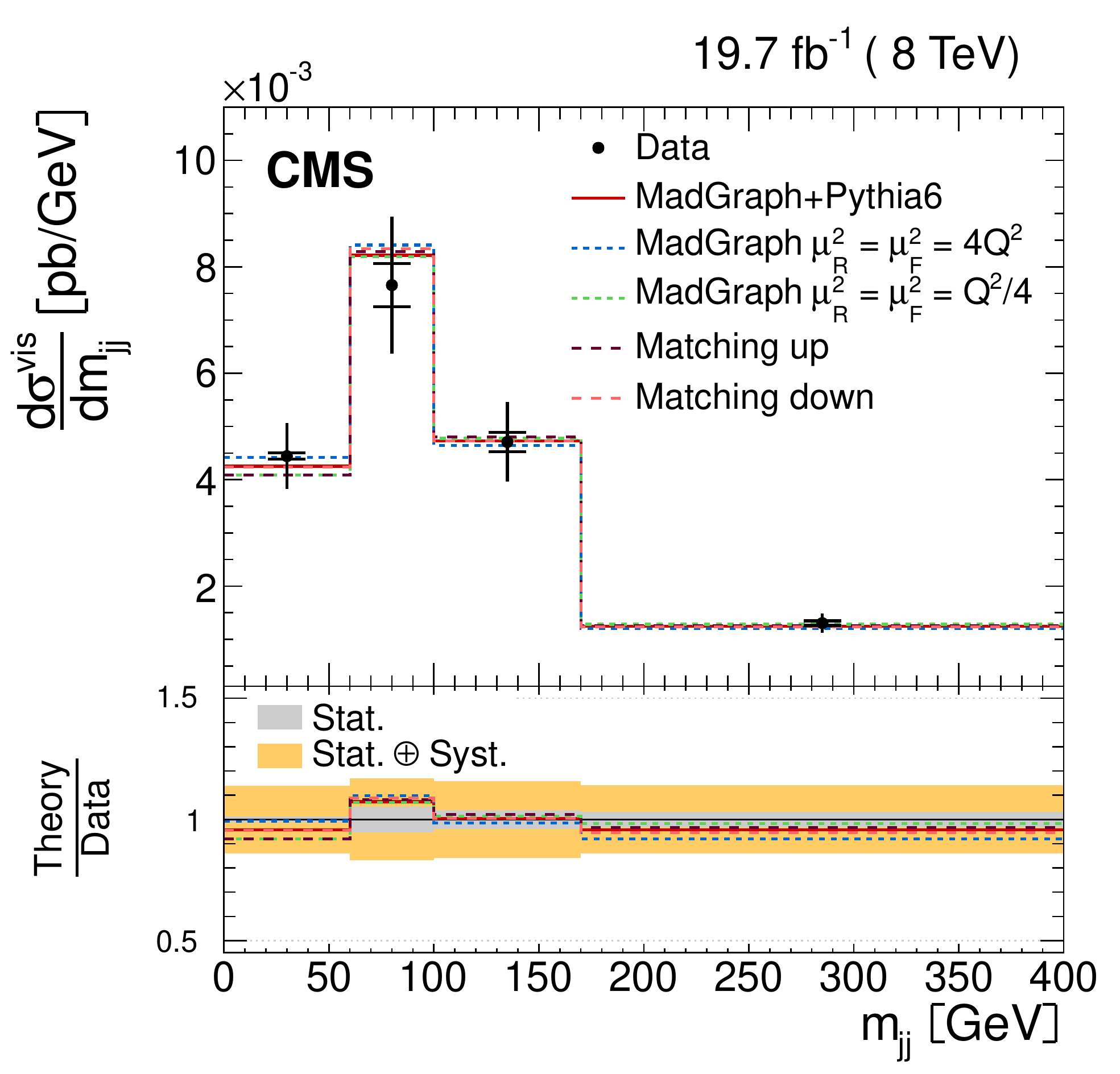}
      \caption{Absolute differential \ttbar cross section as a function of \Djj between the leading and subleading additional jets (top) and their invariant mass, \mjj (bottom). Data are compared to predictions from \MADGRAPH{}+\PYTHIA{6}, \POWHEG{}+\PYTHIA{6}, \POWHEG{}+\HERWIG{6}, and \MCATNLO{}+\HERWIG{6} (left) and to \MADGRAPH with varied renormalization, factorization, and jet-parton matching scales (right). The inner (outer) vertical bars indicate the statistical (total) uncertainties. The lower part of each plot shows the ratio of the predictions to the data.}
\label{fig:DeltaRmassjj}
  \end{center}
\end{figure*}

The differential cross section is also measured as a function of the dijet angular separation \Djj and invariant mass \mjj for the leading and subleading additional jets (Fig.~\ref{fig:DeltaRmassjj}). In general, all simulations provide a reasonable description of the distributions for both variables. All results are reported in Tables~\ref{tab:dilepton:SummaryResultsJet1}--\ref{tab:dilepton:SummaryResultsJet12} in Appendix~\ref{sec:summarytables}. Representative examples of the migration matrices are presented in Fig.~\ref{fig:migration} in Appendix~\ref{sec:migrationmatrix}.

The absolute and normalized differential cross sections are also measured as a function of the kinematic variables of the additional jets and \PQb jets in the event for the full phase space of the \ttbar system to facilitate comparison with theoretical calculations. In this case, the phase space is defined only by the kinematic requirements on the additional jets.

\begin{figure*}[htbp!]
  \begin{center}
      \includegraphics[width=0.40 \textwidth]{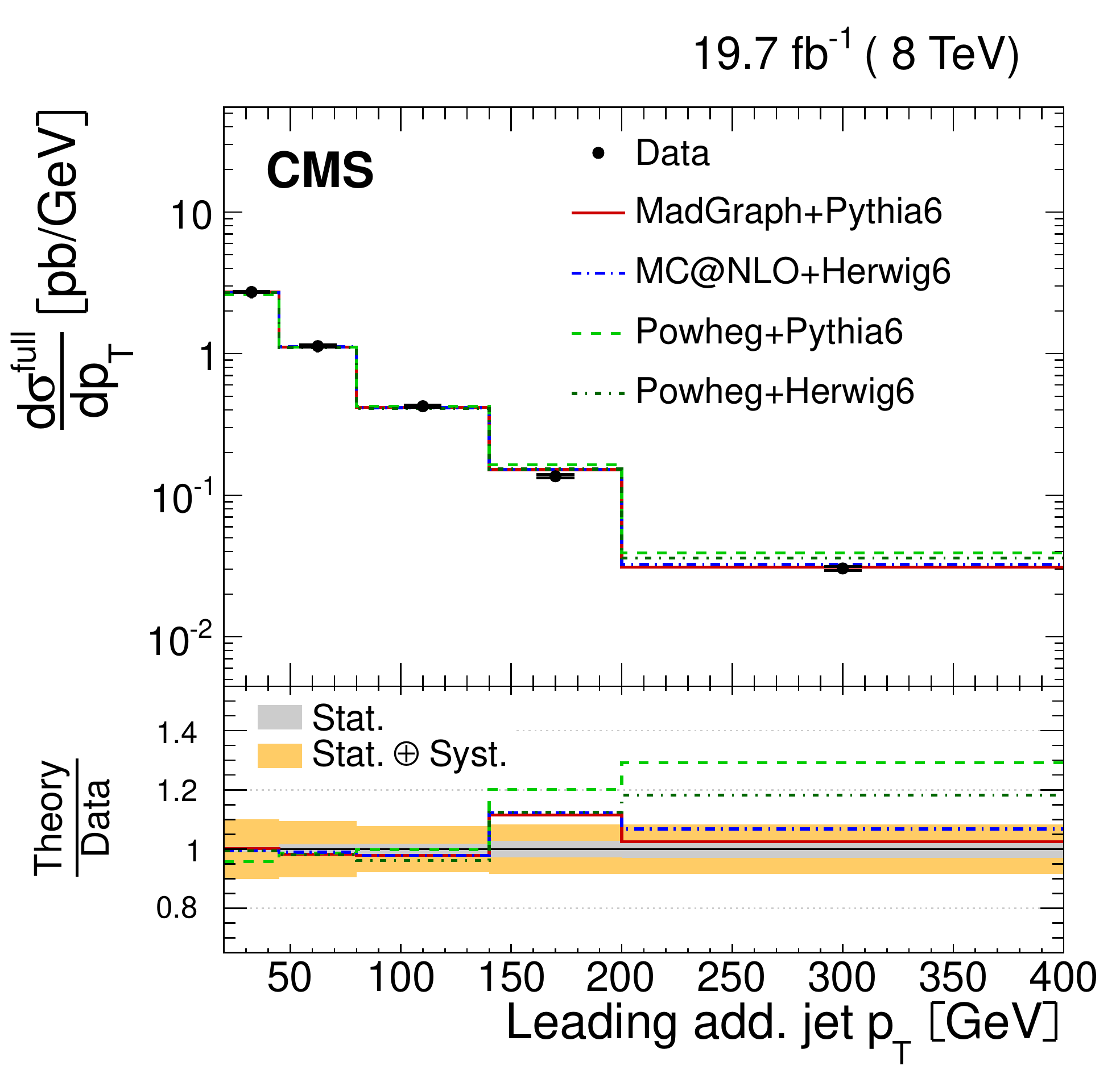}%
      \includegraphics[width=0.40 \textwidth]{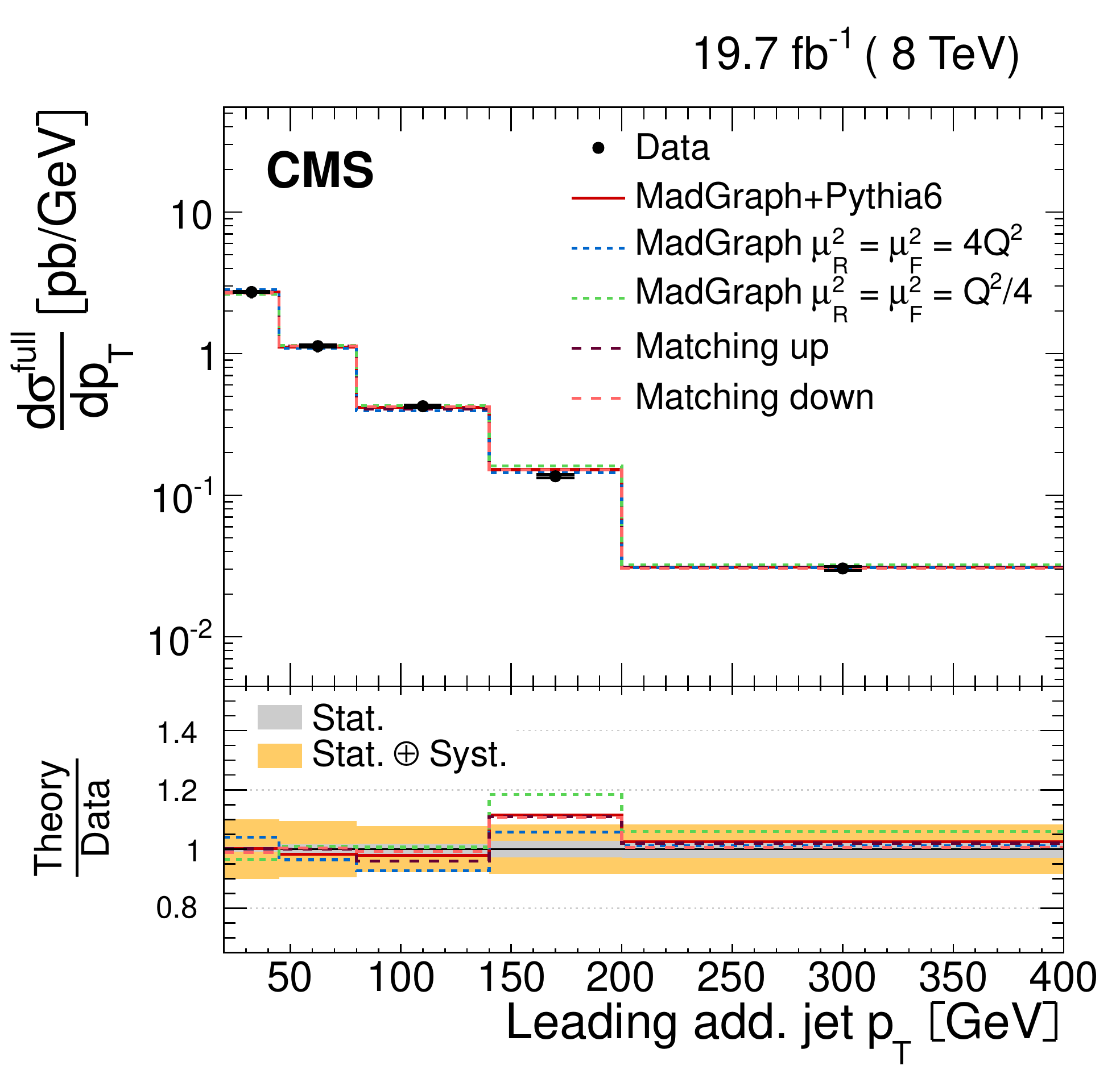}\\
      \includegraphics[width=0.40 \textwidth]{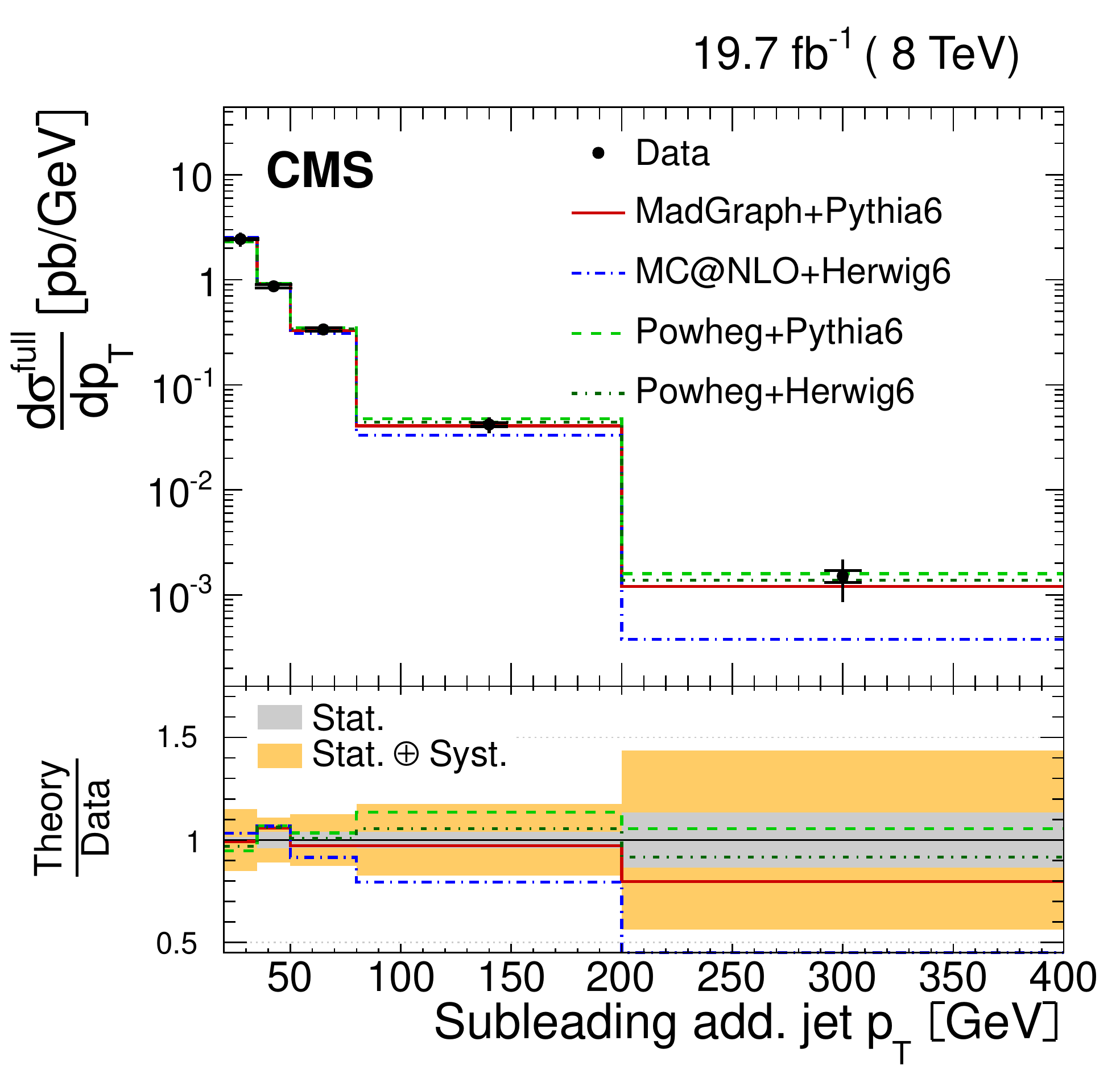}%
      \includegraphics[width=0.40 \textwidth]{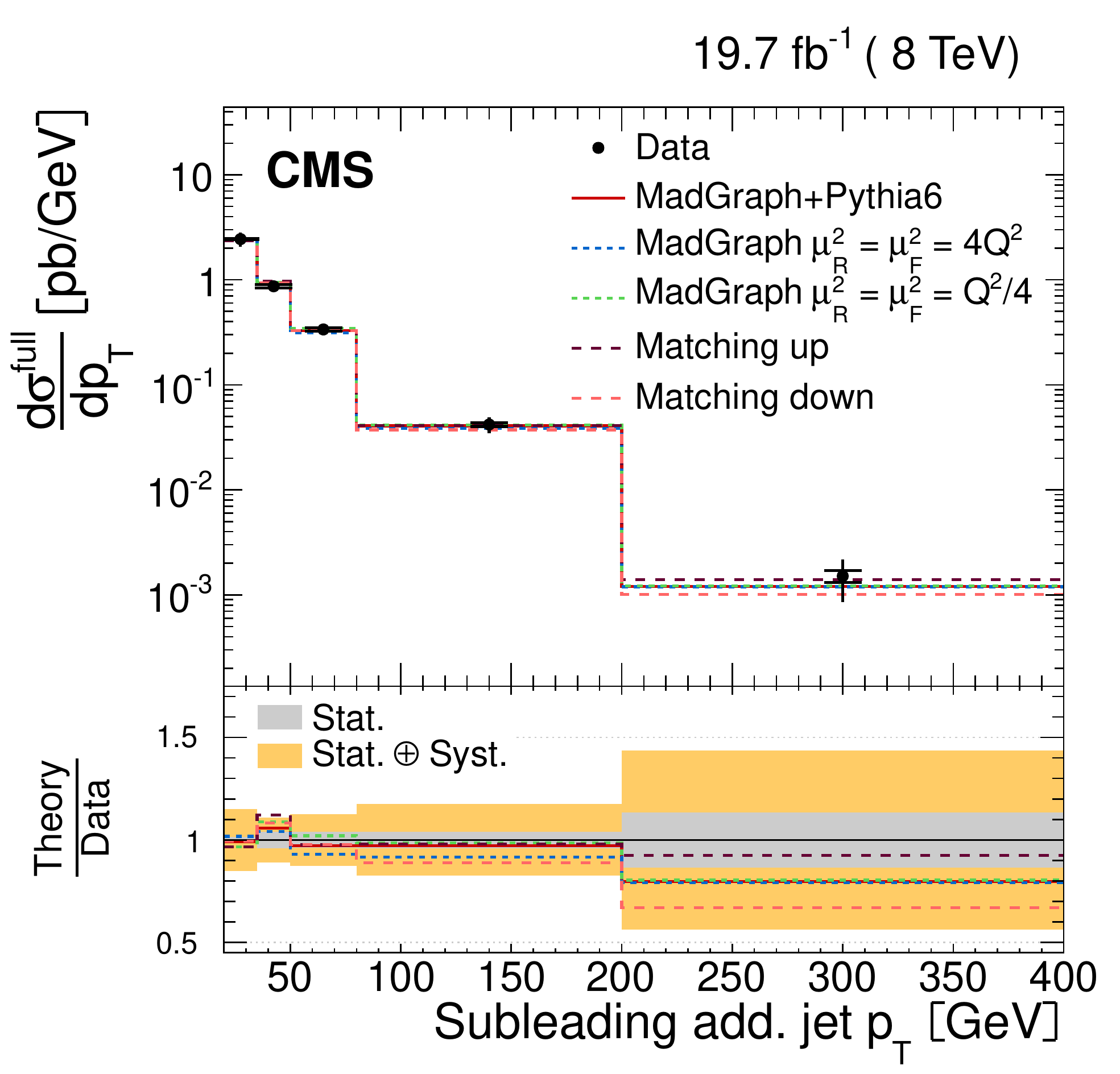}
      \includegraphics[width=0.40 \textwidth]{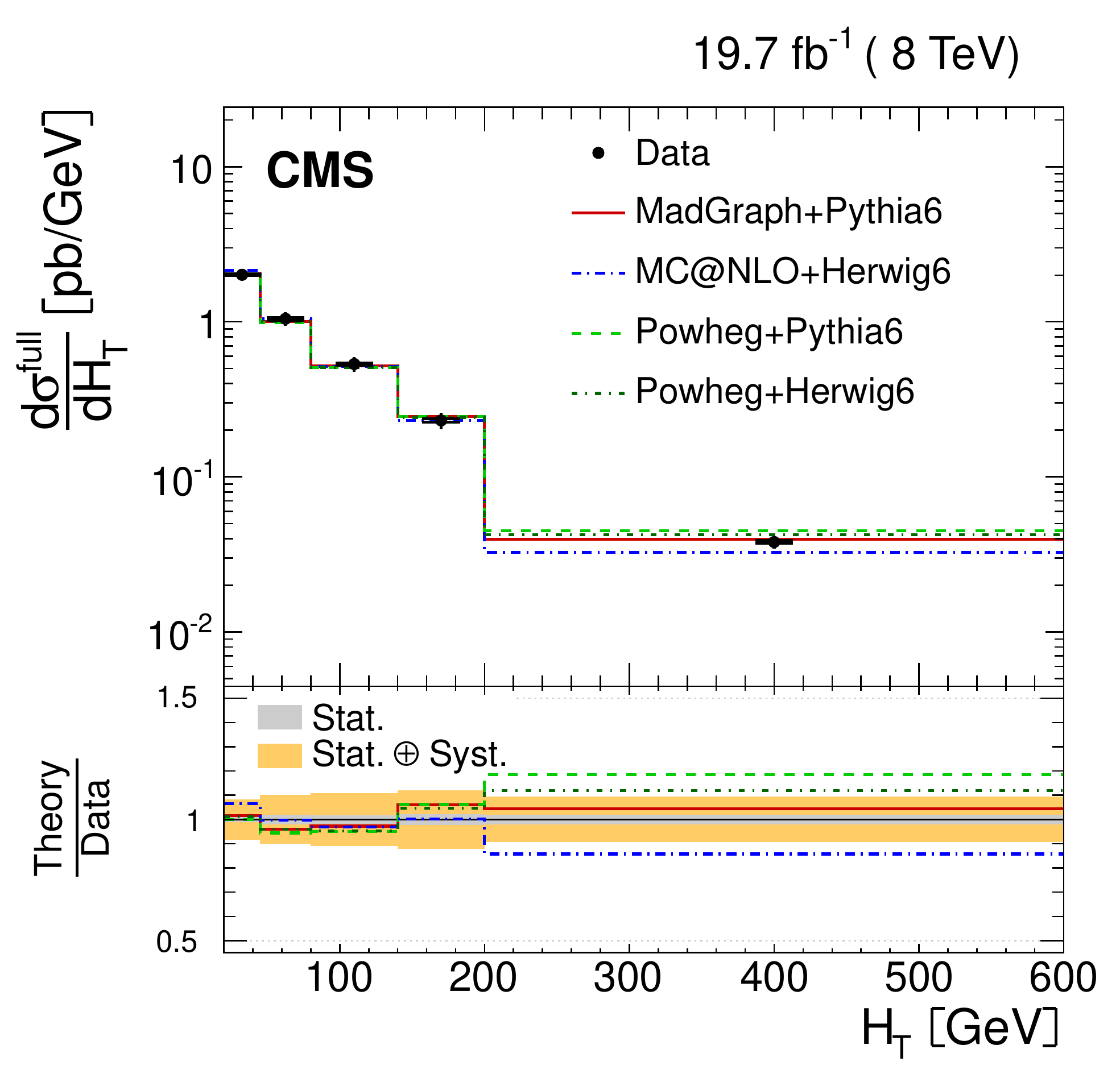}%
      \includegraphics[width=0.40 \textwidth]{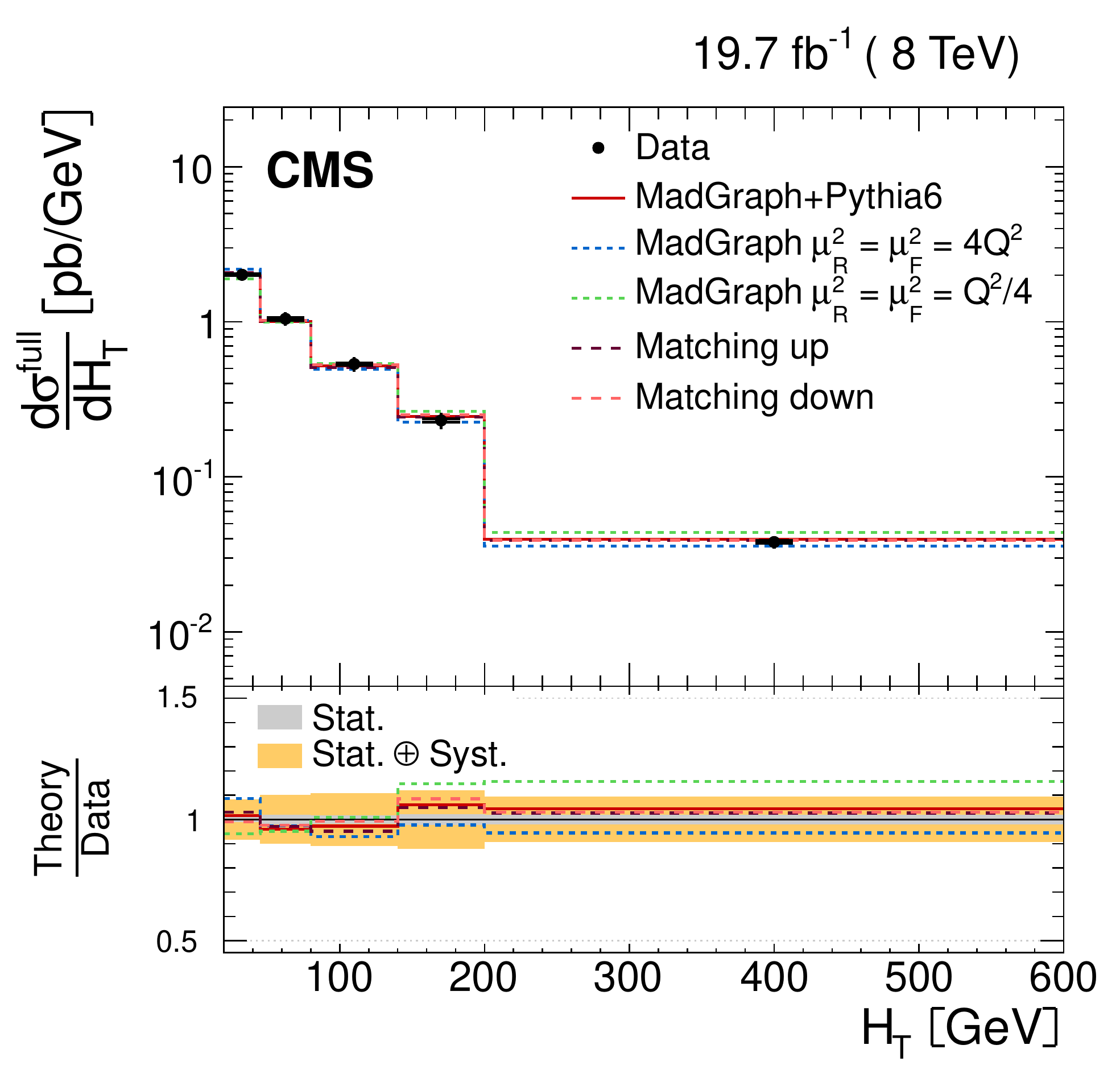}\\
      \caption{Absolute differential \ttbar cross section as a function of \pt of the leading additional jet (top) and the subleading additional jet (middle) and $\HT$ (bottom) measured in the full phase space of the \ttbar system, corrected for acceptance and branching fractions. Data are compared to predictions from \MADGRAPH{}+\PYTHIA{6}, \POWHEG{}+\PYTHIA{6}, \POWHEG{}+\HERWIG{6}, and \MCATNLO{}+\HERWIG{6} (left) and to \MADGRAPH with varied renormalization, factorization, and jet-parton matching scales (right). The inner (outer) vertical bars indicate the statistical (total) uncertainties. The lower part of each plot shows the ratio of the predictions to the data.}
\label{fig:inclusiveptFull}
  \end{center}
\end{figure*}

\begin{figure*}[htbp!]
  \begin{center}
      \includegraphics[width=0.40 \textwidth]{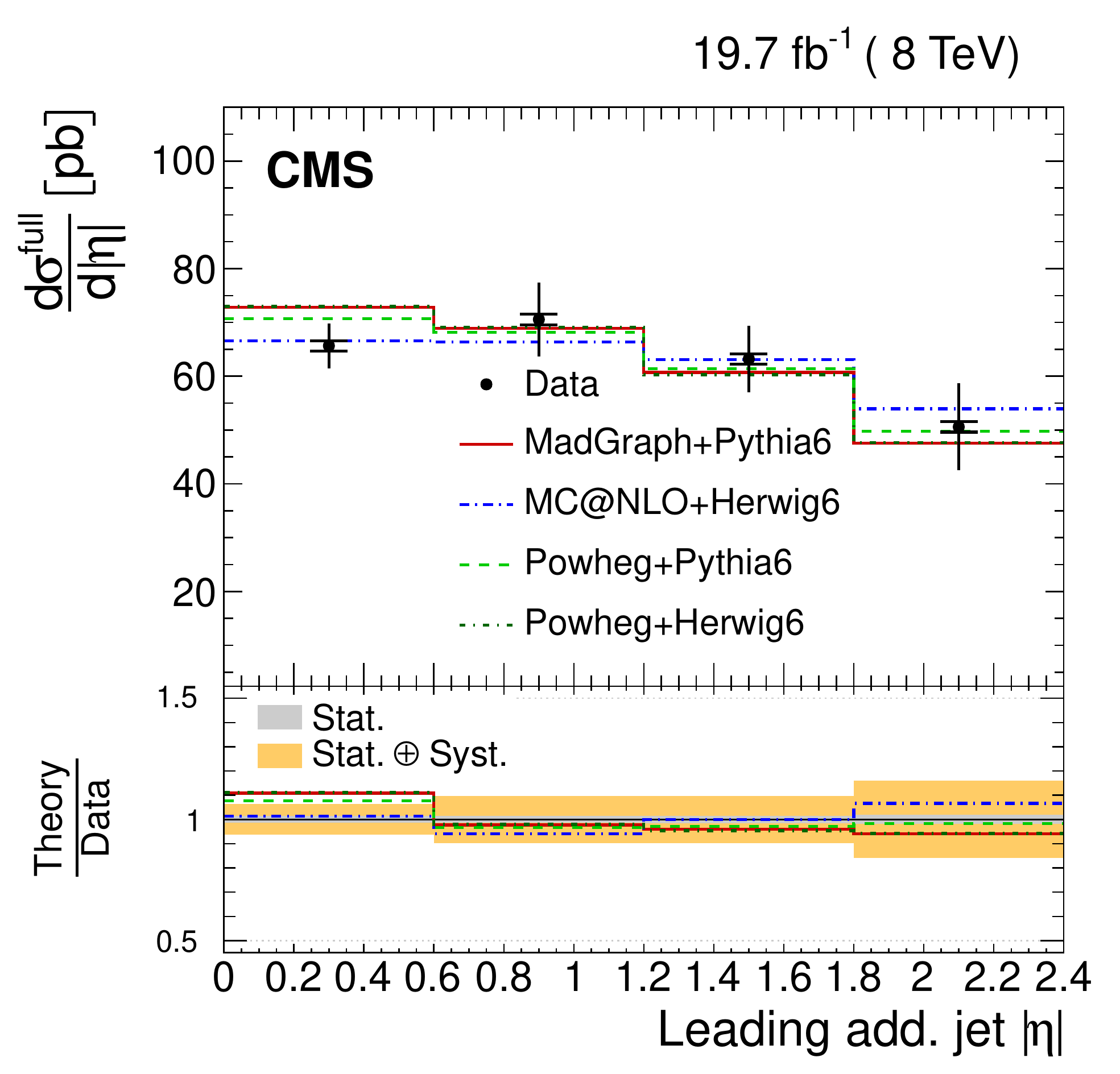}%
      \includegraphics[width=0.40 \textwidth]{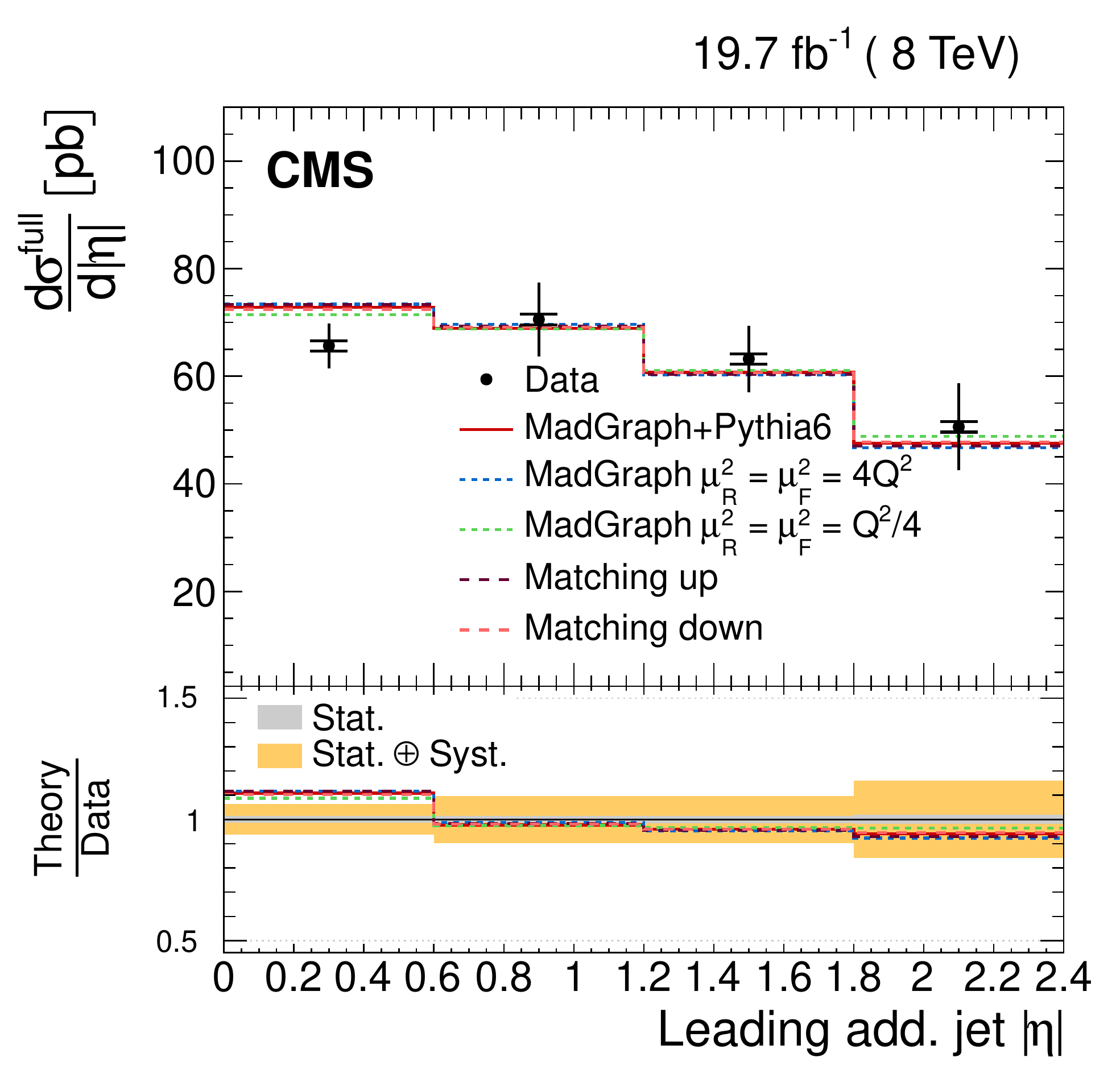}\\
      \includegraphics[width=0.40 \textwidth]{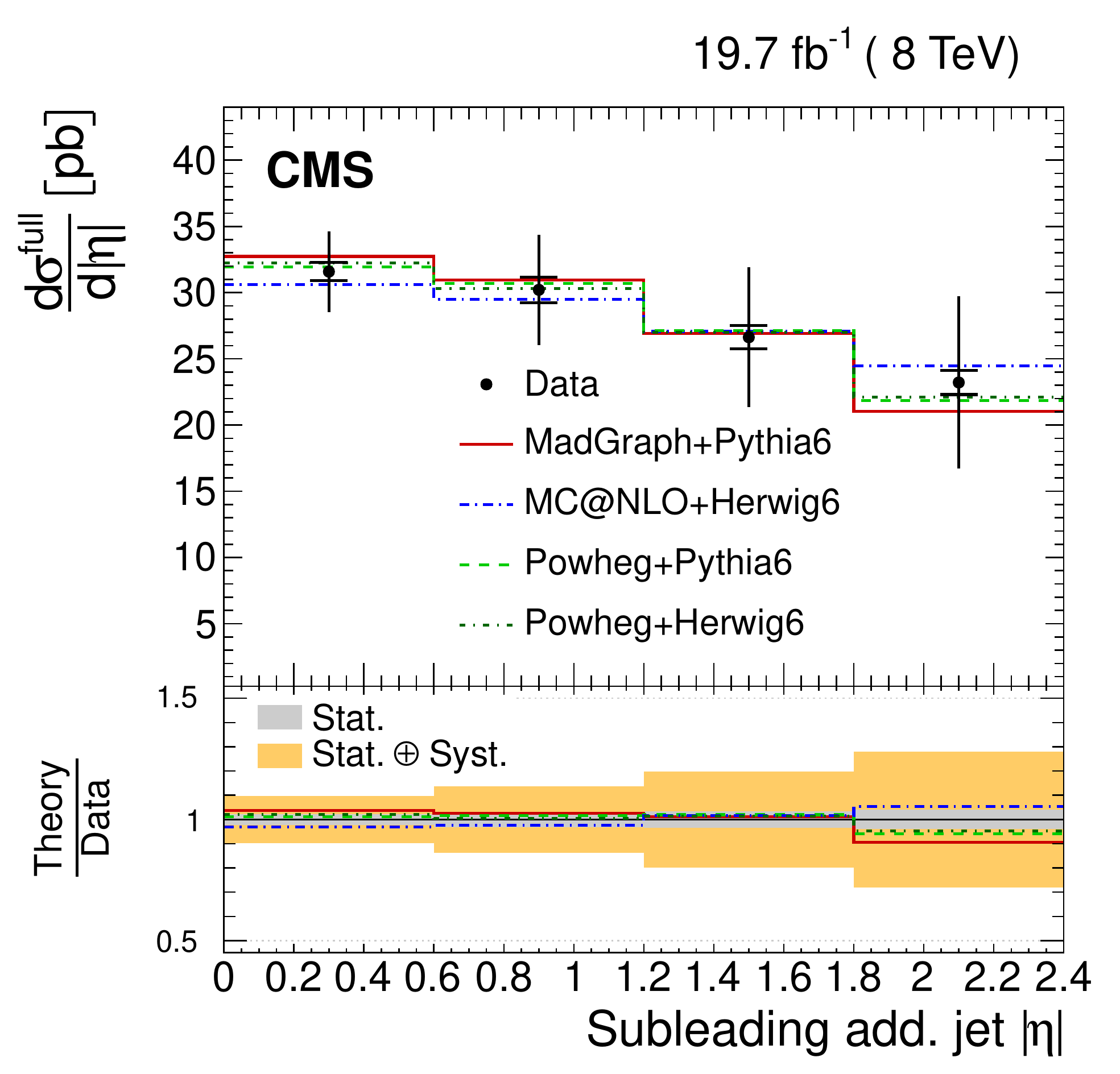}%
      \includegraphics[width=0.40 \textwidth]{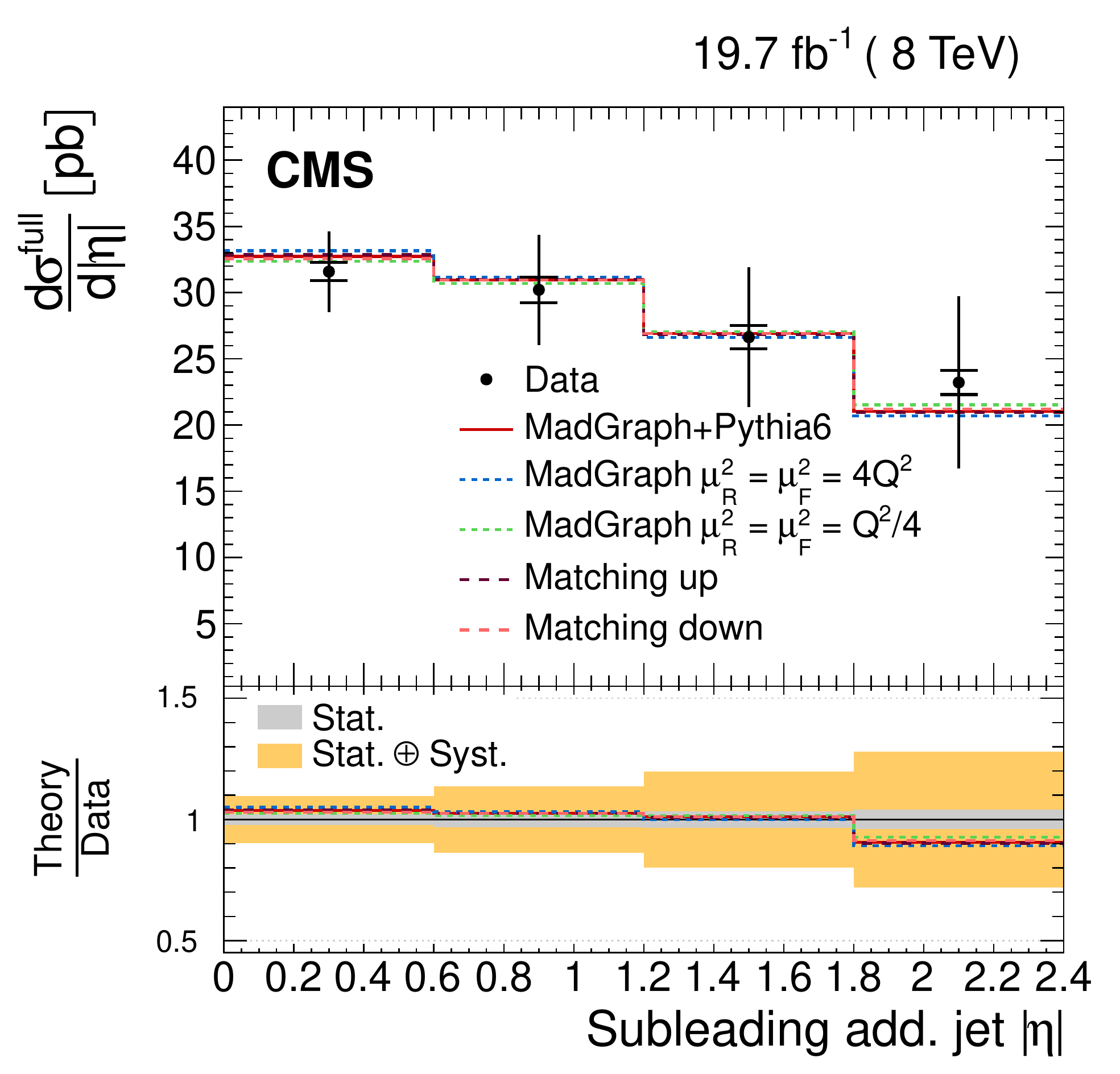}
\caption{Absolute differential \ttbar cross section as a function of the \abseta of the leading additional jet (top) and the subleading additional jet (bottom) measured in the full phase space of the \ttbar system, corrected for acceptance and branching fractions. Data are compared to predictions from \MADGRAPH{}+\PYTHIA{6}, \POWHEG{}+\PYTHIA{6}, \POWHEG{}+\HERWIG{6}, and \MCATNLO{}+\HERWIG{6} (left) and to \MADGRAPH with varied renormalization, factorization, and jet-parton matching scales (right). The inner (outer) vertical bars indicate the statistical (total) uncertainties. The lower part of each plot shows the ratio of the predictions to the data.}
\label{fig:inclusiveetaFull}
  \end{center}
\end{figure*}

\begin{figure*}[htbp!]
  \begin{center}
      \includegraphics[width=0.40 \textwidth]{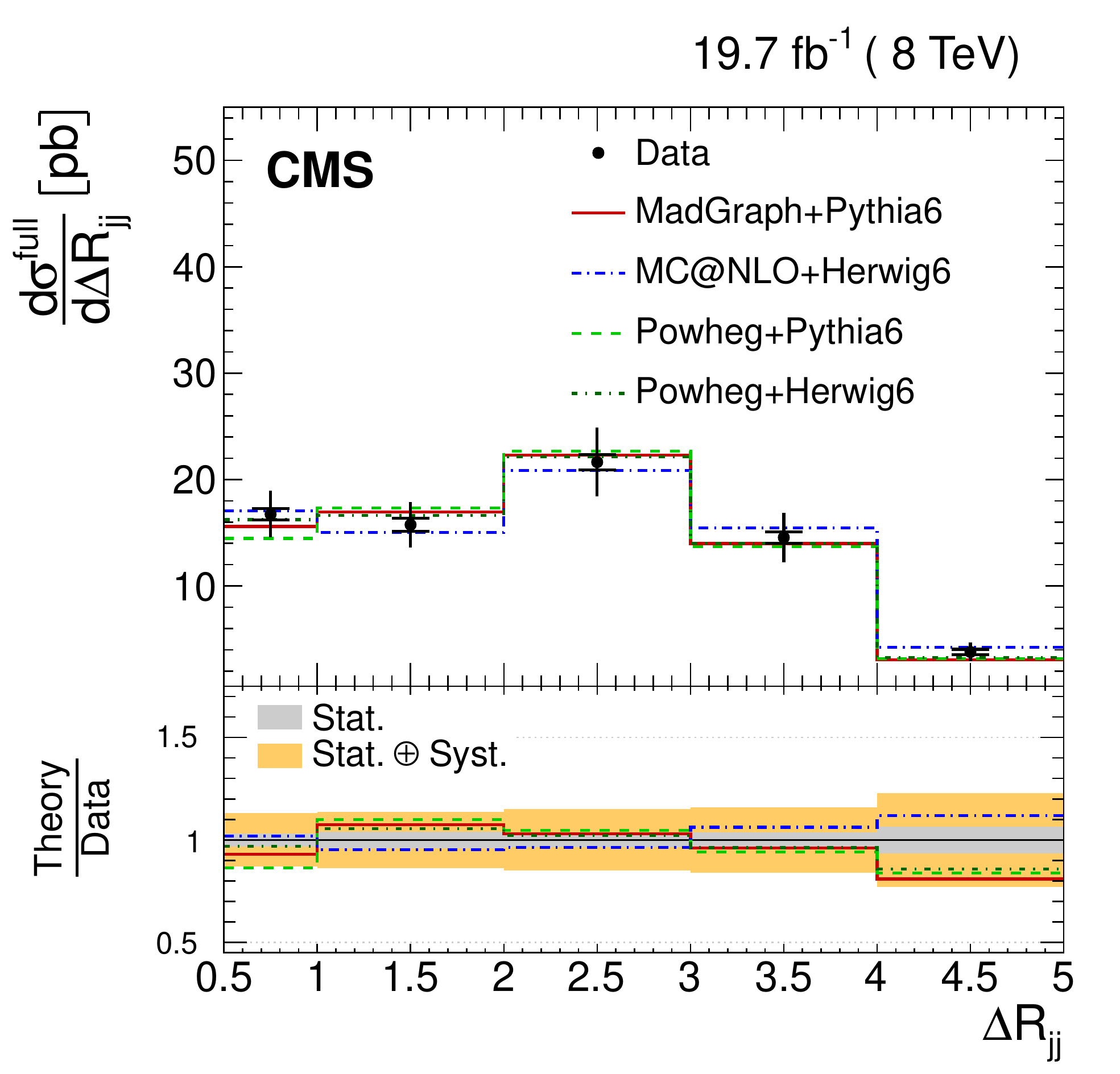}%
      \includegraphics[width=0.40 \textwidth]{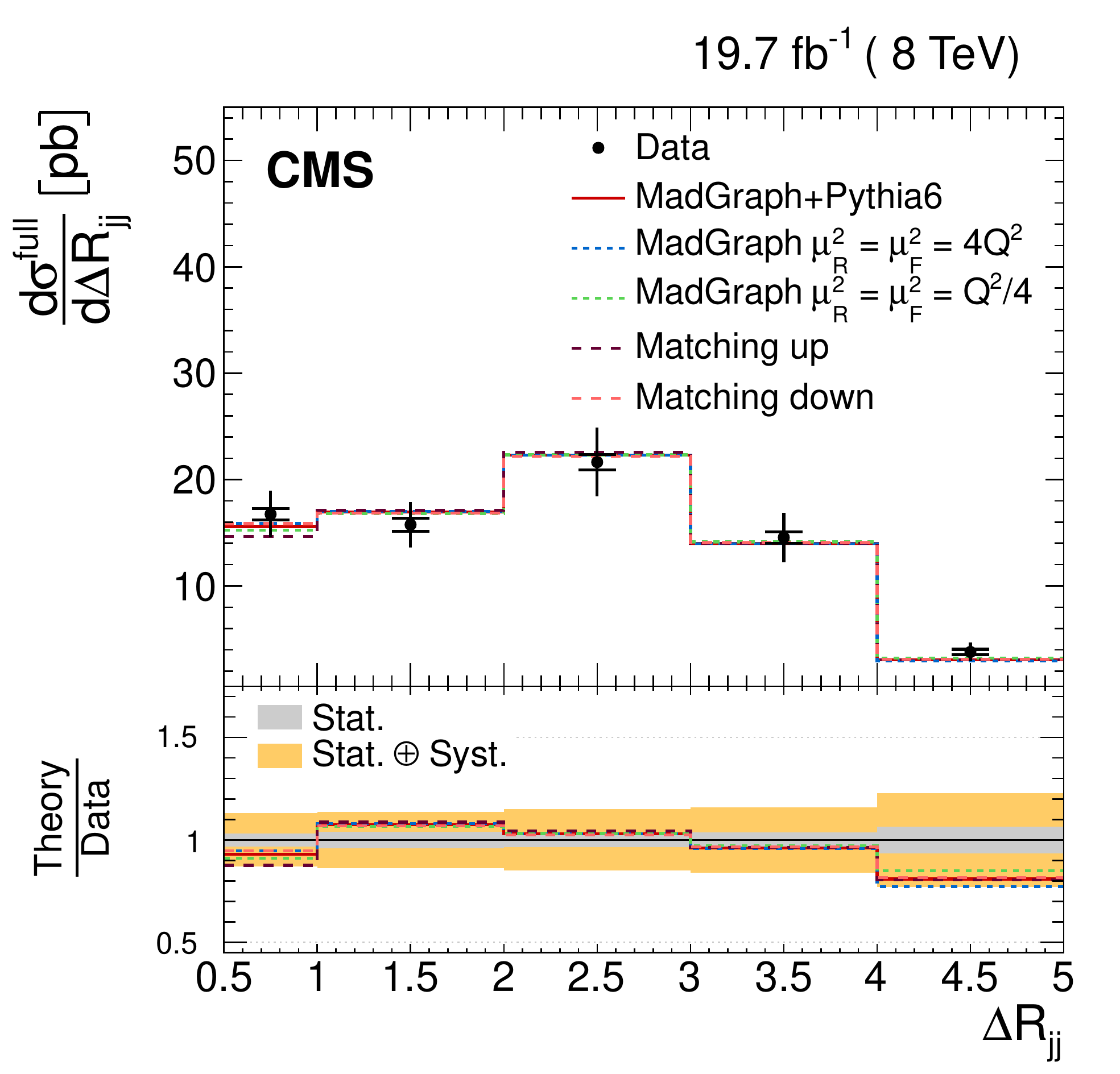}\\
      \includegraphics[width=0.40 \textwidth]{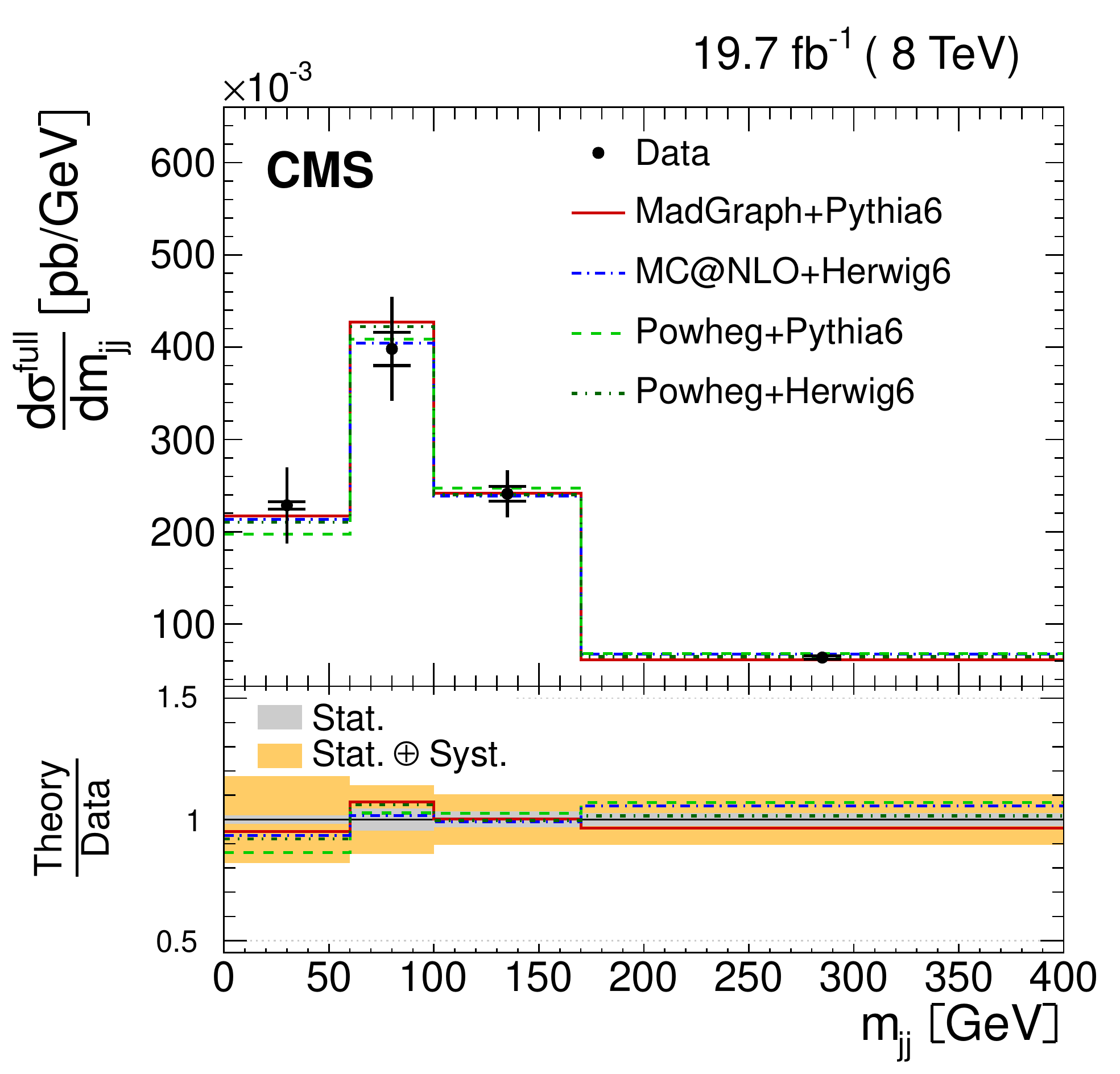}%
      \includegraphics[width=0.40 \textwidth]{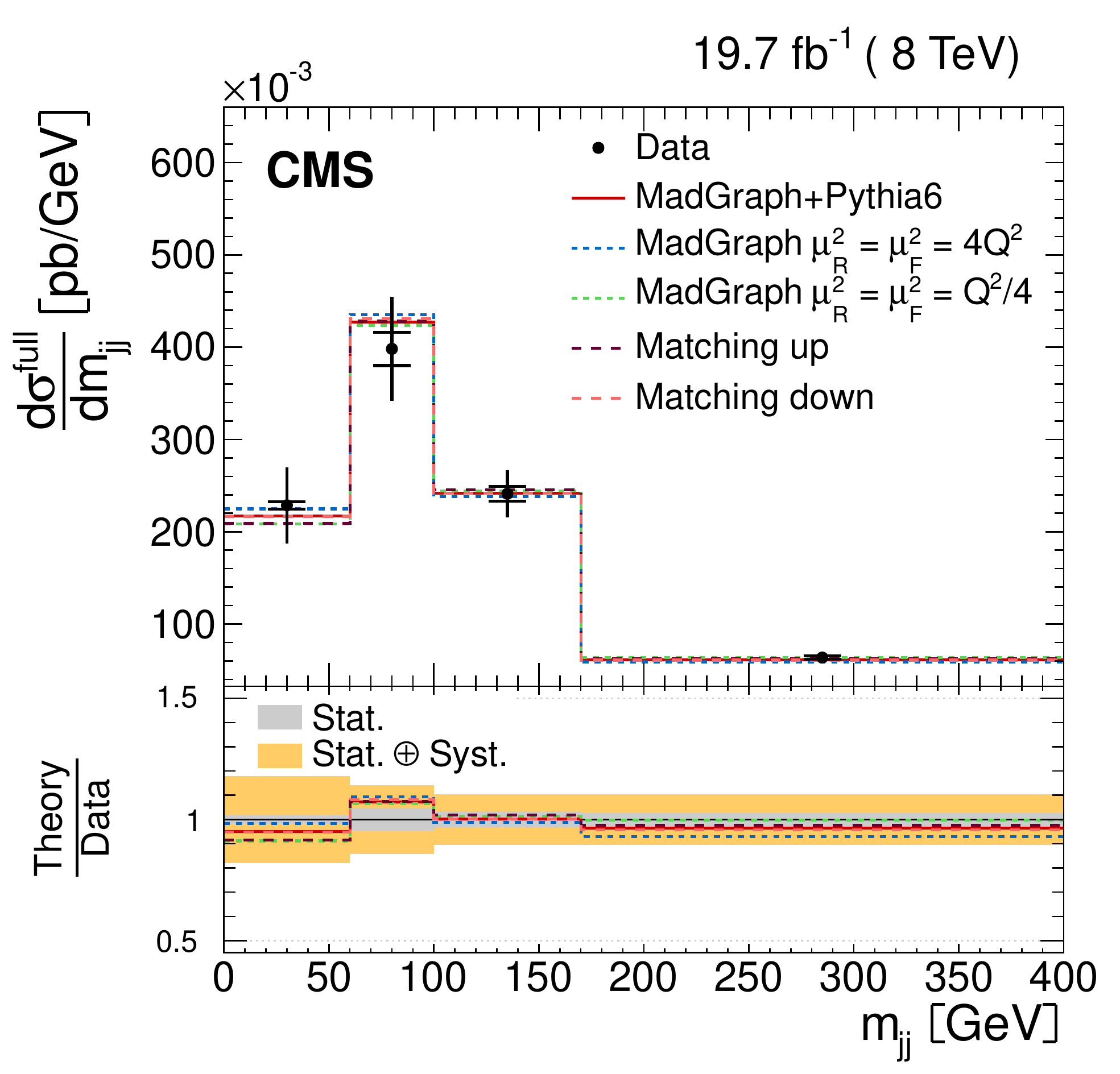}
      \caption{Absolute differential \ttbar cross section as a function of \Djj between the leading and subleading additional jets (top) and their invariant mass, \mjj (bottom) measured in the full phase space of the \ttbar system, corrected for acceptance and branching fractions. Data are compared to predictions from \MADGRAPH{}+\PYTHIA{6}, \POWHEG{}+\PYTHIA{6}, \POWHEG{}+\HERWIG{6}, and \MCATNLO{}+\HERWIG{6} (left) and to \MADGRAPH with varied renormalization, factorization, and jet-parton matching scales (right). The inner (outer) vertical bars indicate the statistical (total) uncertainties. The lower part of each plot shows the ratio of the predictions to the data.}
\label{fig:DeltaRmassjjFull}
  \end{center}
\end{figure*}

Figures~\ref{fig:inclusiveptFull} and~\ref{fig:inclusiveetaFull} show the absolute cross sections as a function of the \pt and \abseta of the leading and subleading additional jets and $\HT$, while the results as a function of \Djj and \mjj are presented in Fig.~\ref{fig:DeltaRmassjjFull}.

The total uncertainties range between 8--12\% for the leading jet \pt and $\HT$, 10\% at lower \pt and 40\% in the tails of distribution of the subleading jet \pt. The uncertainties for $|\eta|$ are 6--16\% and 10--30\% for the leading and subleading additional jets, respectively. The typical uncertainties in the cross section as a function of \Djj and \mjj are on the order of 10--20\%. The uncertainties are dominated by the JES, scale uncertainties, and shower modelling.

The numerical values are given in Tables~\ref{tab:dilepton:SummaryResultsJet1Full}--\ref{tab:dilepton:SummaryResultsJet12Full} of Appendix~\ref{sec:summarytables}, together with the normalized results. In the latter, the uncertainties are on average 2--3 times smaller than for the absolute cross sections, owing to the cancellation of uncertainties such as the integrated luminosity, lepton identification, and trigger efficiency, as well as a large fraction of the JES and model uncertainties, as discussed in Section~\ref{sec:diffxsecNJets}. The dominant systematic uncertainties are still the model uncertainties, although they are typically smaller than for the absolute cross sections.

The shapes of the distributions measured in the full and visible phase-space regions of the \ttbar system are similar, while the absolute differential cross sections are a factor of 2.2 larger than those in the visible phase space of the \ttbar system (excluding the factor due to the leptonic branching fraction correction $(4.54 \pm 0.10)\%$~\cite{PDG2014}).

\section{Differential \texorpdfstring{\ttbb(\ttb)}{t-tbar-b-bbar (t-tbar-b)} cross sections as a function of the kinematic variables of the additional \texorpdfstring{\PQb}{b} jets}
\sectionmark{\ttbb(\ttb) cross sections as a function of the kinematics of additional \PQb jets}
\label{sec:diffxsecAddbJets}

\begin{figure*}[htbp!]
  \begin{center}
    \includegraphics[width=0.40\textwidth]{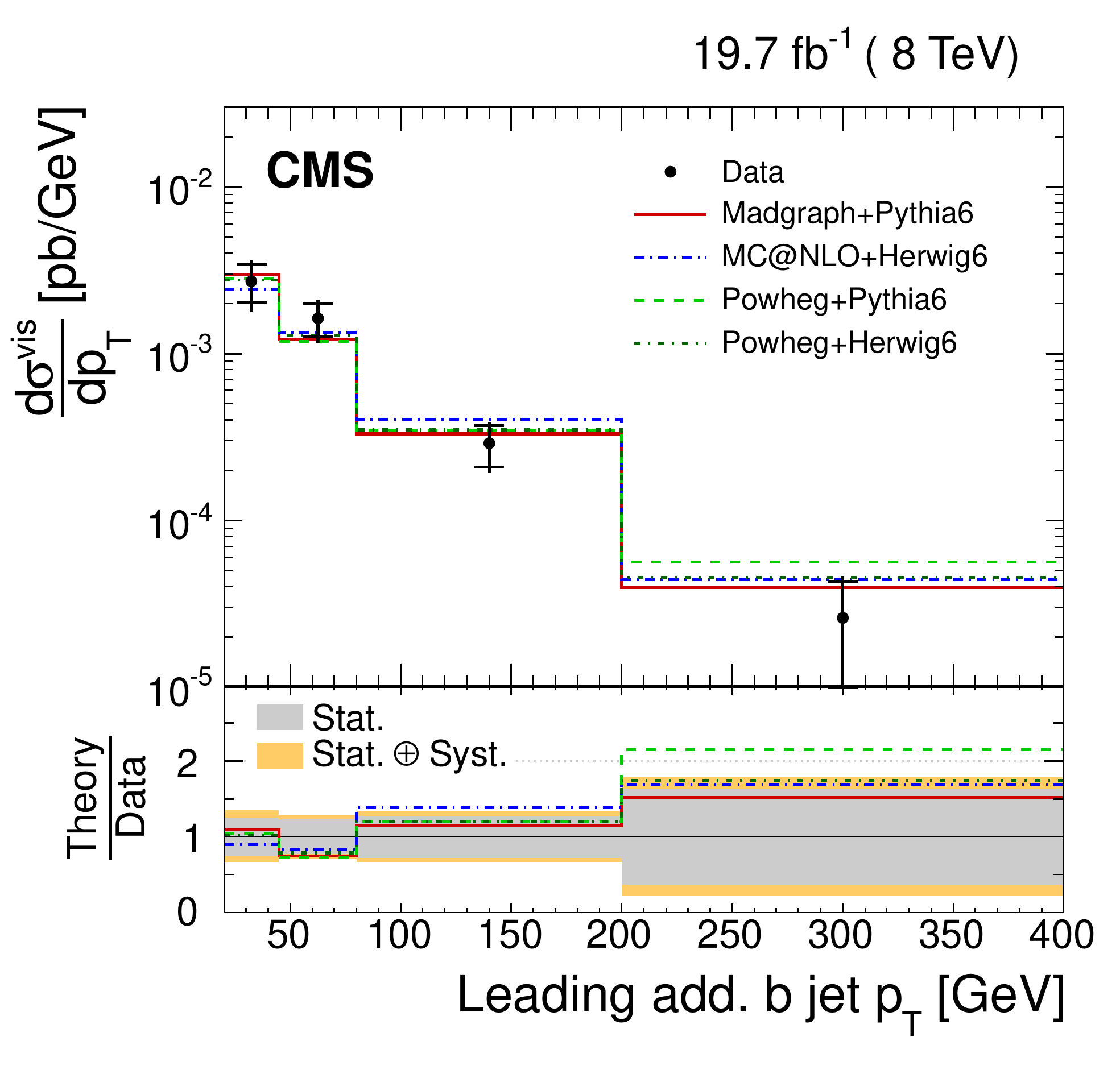}
    \includegraphics[width=0.40\textwidth]{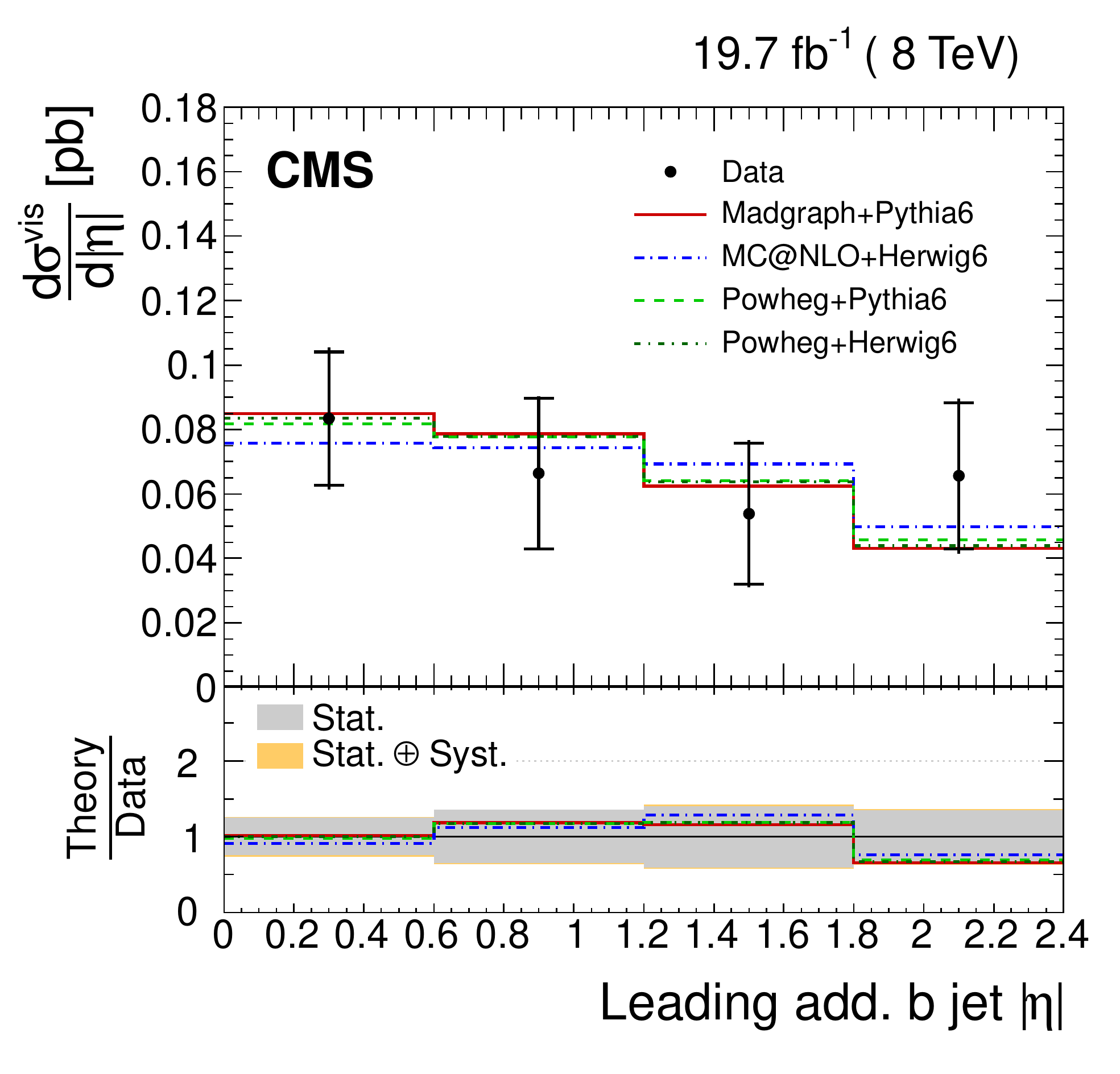}
    \includegraphics[width=0.40\textwidth]{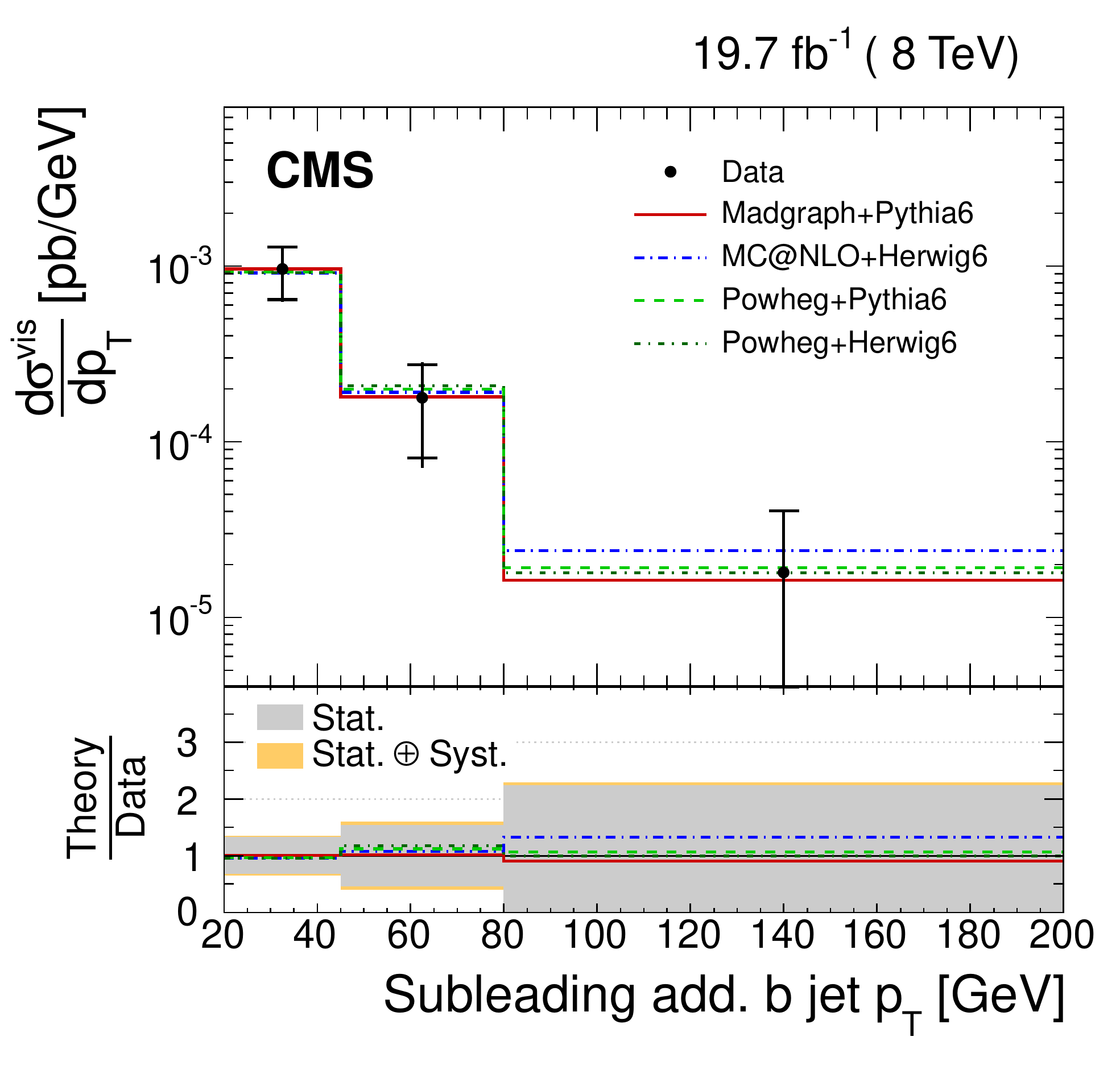}
    \includegraphics[width=0.40\textwidth]{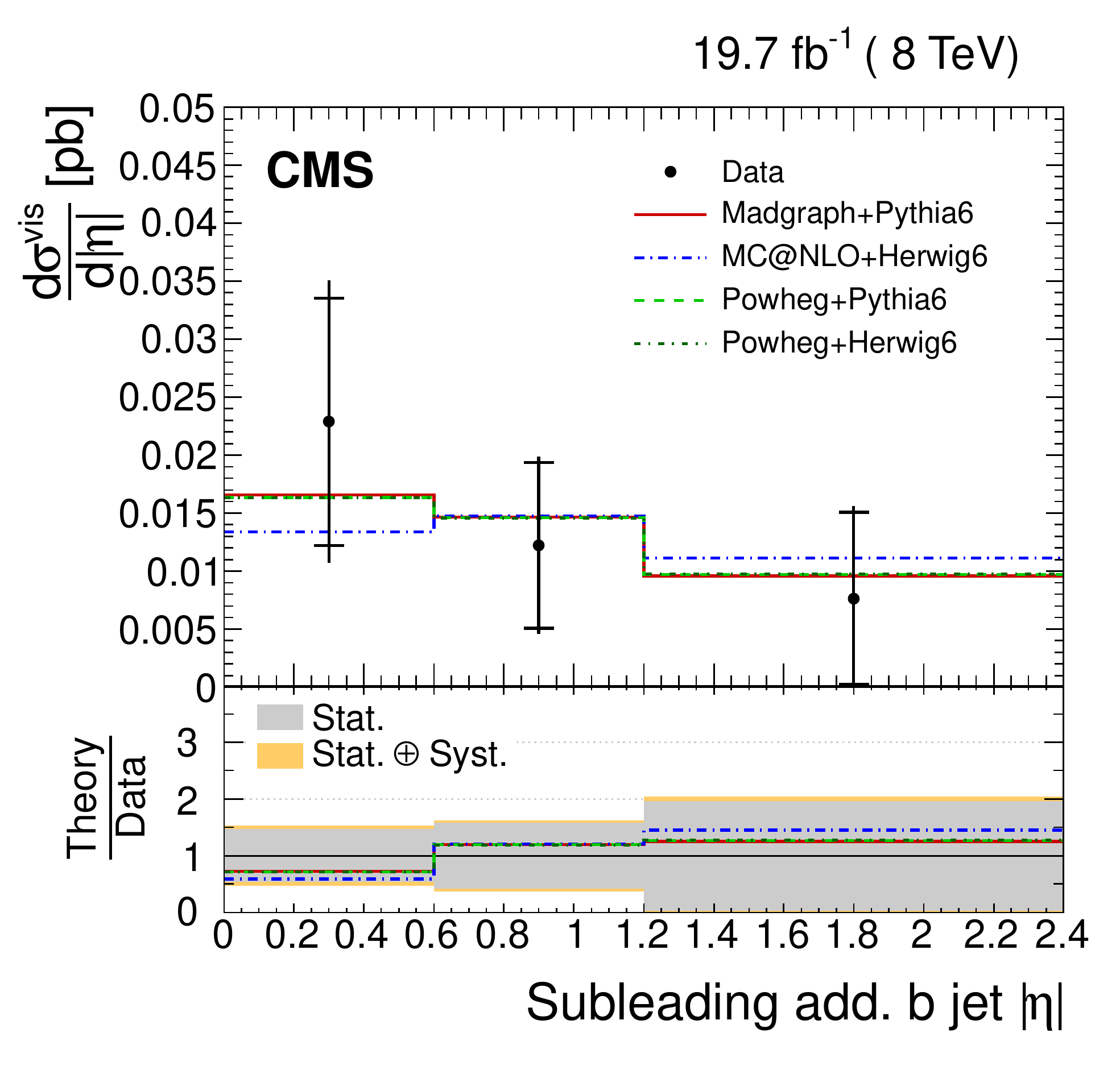}
    \includegraphics[width=0.40\textwidth]{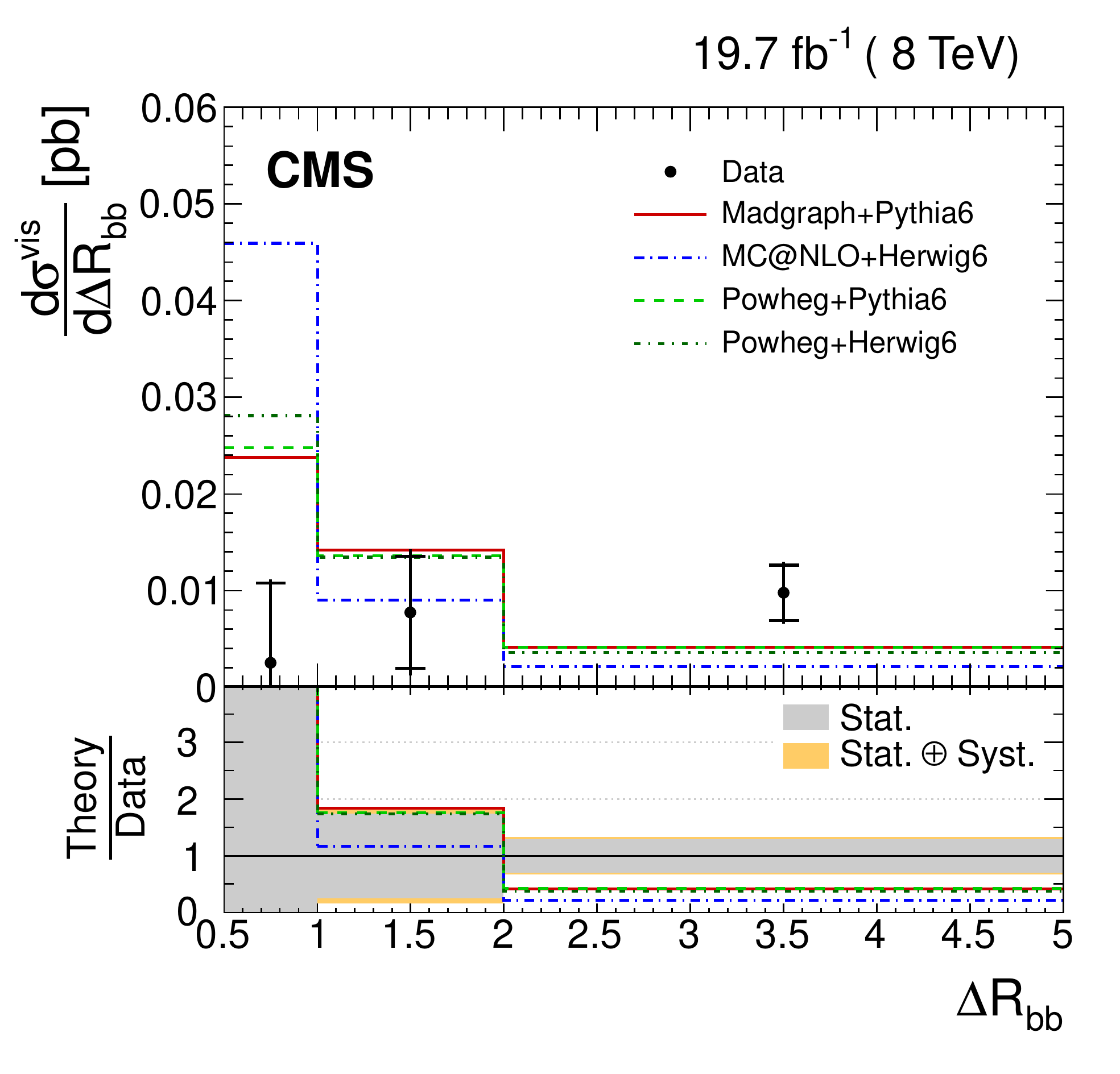}
    \includegraphics[width=0.40\textwidth]{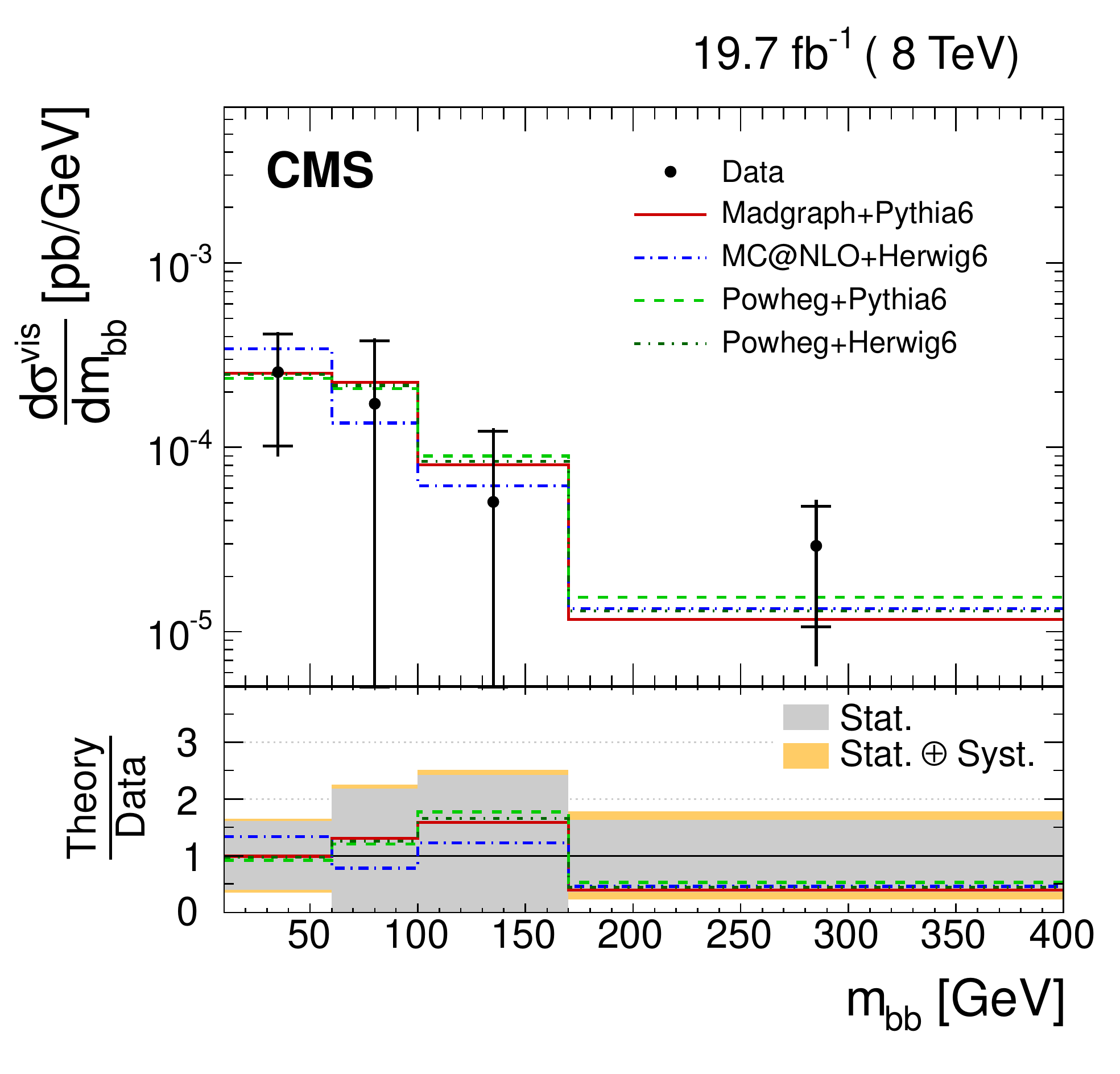}
    \caption{Absolute differential \ttbar cross section measured in the visible phase space of the \ttbar system and the additional \PQb jets, as a function of the leading additional \PQb jet \pt (top left) and \abseta (top right), subleading additional \PQb jet \pt (middle left) and \abseta (middle right), the angular separation $\Delta R_{\PQb\PQb}$ between the two leading additional \PQb jets (bottom left), and the invariant mass \mbb of the two \PQb jets (bottom right). Data are compared with predictions from \MADGRAPH interfaced with \PYTHIA{6}, \MCATNLO interfaced with \HERWIG{6}, and \POWHEG with \PYTHIA{6} and \HERWIG{6}, normalized to the measured inclusive cross section. The inner (outer) vertical bars indicate the statistical (total) uncertainties. The lower part of each plot shows the ratio of the predictions to the data.}
    \label{fig:xsec_bjets}
  \end{center}
\end{figure*}

Figure~\ref{fig:xsec_bjets} shows the absolute \ttbar differential cross sections in the visible phase space of the \ttbar system and the additional \PQb jets as a function of the \pt and \abseta of the leading and subleading additional \PQb jets, and $\Delta R_{\PQb\PQb}$ and \mbb of the two \PQb jets. The uncertainties in the measured cross sections as a function of the \PQb jet kinematic variables are dominated by the statistical uncertainties, with values varying from 20--100\%. The results are quantified in Tables~\ref{tab:dilepton:SummaryResultsBJet} and~\ref{tab:dilepton:SummaryResultsBJet12} in Appendix~\ref{sec:summarytables}, together with the normalized results. The corresponding migration matrices between the reconstructed and particle levels for the kinematic properties of the additional \PQb jets are presented in Fig.~\ref{fig:migrationttbb} in Appendix~\ref{sec:migrationmatrix} for illustration purposes.

The dominant systematic uncertainties are the \PQb tagging efficiency and JES, up to 20\% and 15\%, respectively. Other uncertainties have typical values on the order of or below 5\%. The experimental sources of systematic uncertainties affecting only the normalization, which are constrained in the fit, have a negligible impact. The largest model uncertainty corresponds to that from the renormalization and factorization scales of 8\%. The effect of the assumed top quark mass and the PDF uncertainties have typical values of 1--2\%. On average, the inclusion of all the systematic uncertainties increases the total uncertainties by 10\%.

{\tolerance=400
The measured distributions are compared with the \MADGRAPH{}+\PYTHIA{6} prediction, normalized to the corresponding measured inclusive cross section in the same phase space. The measurements are also compared to the predictions from \MCATNLO interfaced with \HERWIG{6} and from \POWHEG with \PYTHIA{6} and \HERWIG{6}. The normalization factors applied to the \MADGRAPH and \POWHEG predictions are found to be about 1.3 for results related to the leading additional \PQb jet. The predictions from both generators underestimate the \ttbb cross sections by a factor 1.8, in agreement with the results from Ref.~\cite{bib:ttbb_ratio:2014}. The normalization factors applied to \MCATNLO are approximately 2 and 4 for the leading and subleading additional \PQb jet quantities, respectively, reflecting the observation that the generator does not simulate sufficiently large jet multiplicities. All the predictions have slightly harder \pt spectra for the leading additional \PQb jet than the data, while they describe the behaviour of the \abseta and \mbb distributions within the current precision. The predictions favour smaller $\Delta R_{\PQb\PQb}$ values than the measurement, although the differences are in general within two standard deviations of the total uncertainty.
\par}

\begin{figure*}[htbp!]
  \begin{center}
    \includegraphics[width=0.40\textwidth]{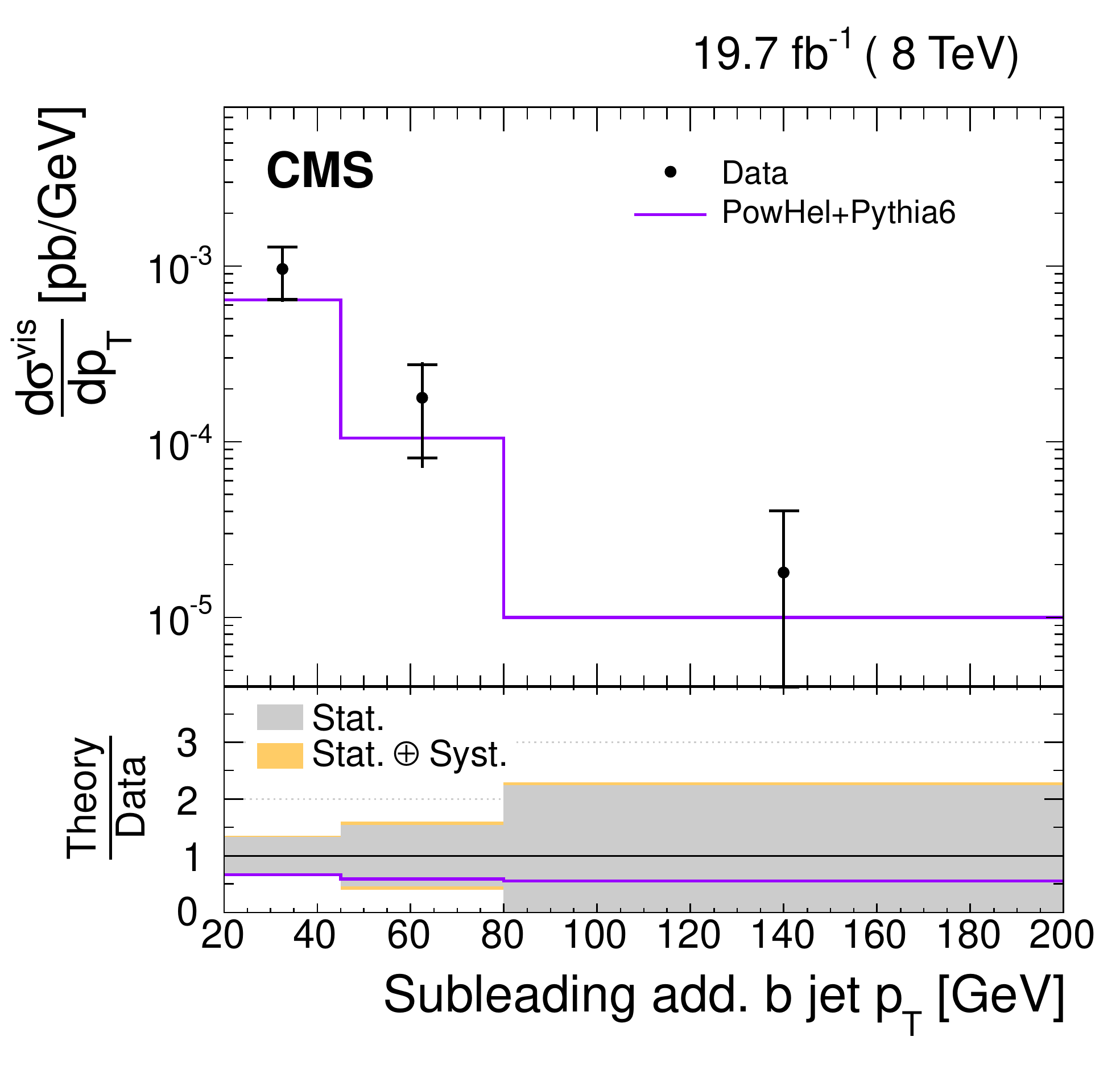}
    \includegraphics[width=0.40\textwidth]{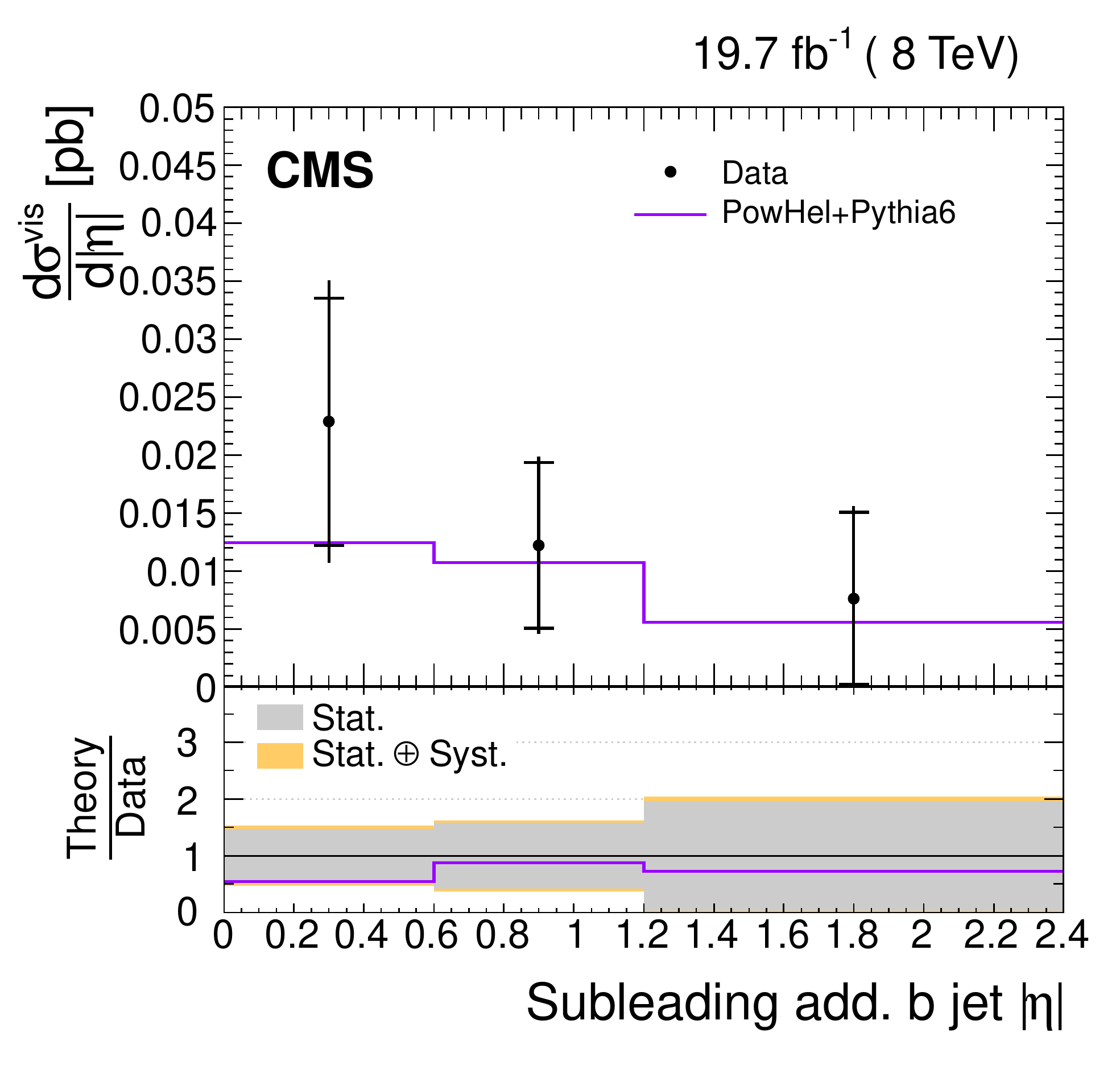}
    \includegraphics[width=0.40\textwidth]{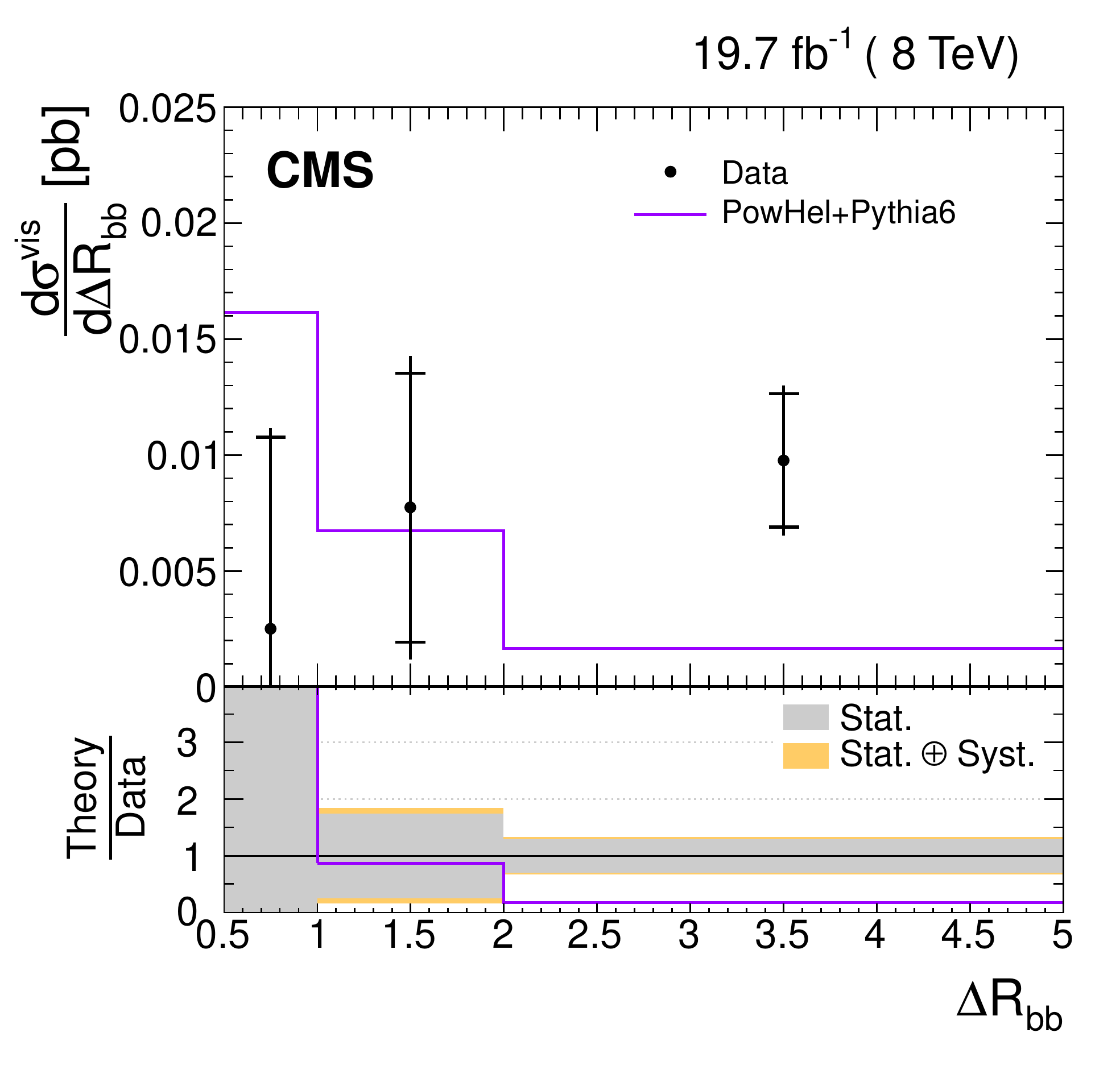}
    \includegraphics[width=0.40\textwidth]{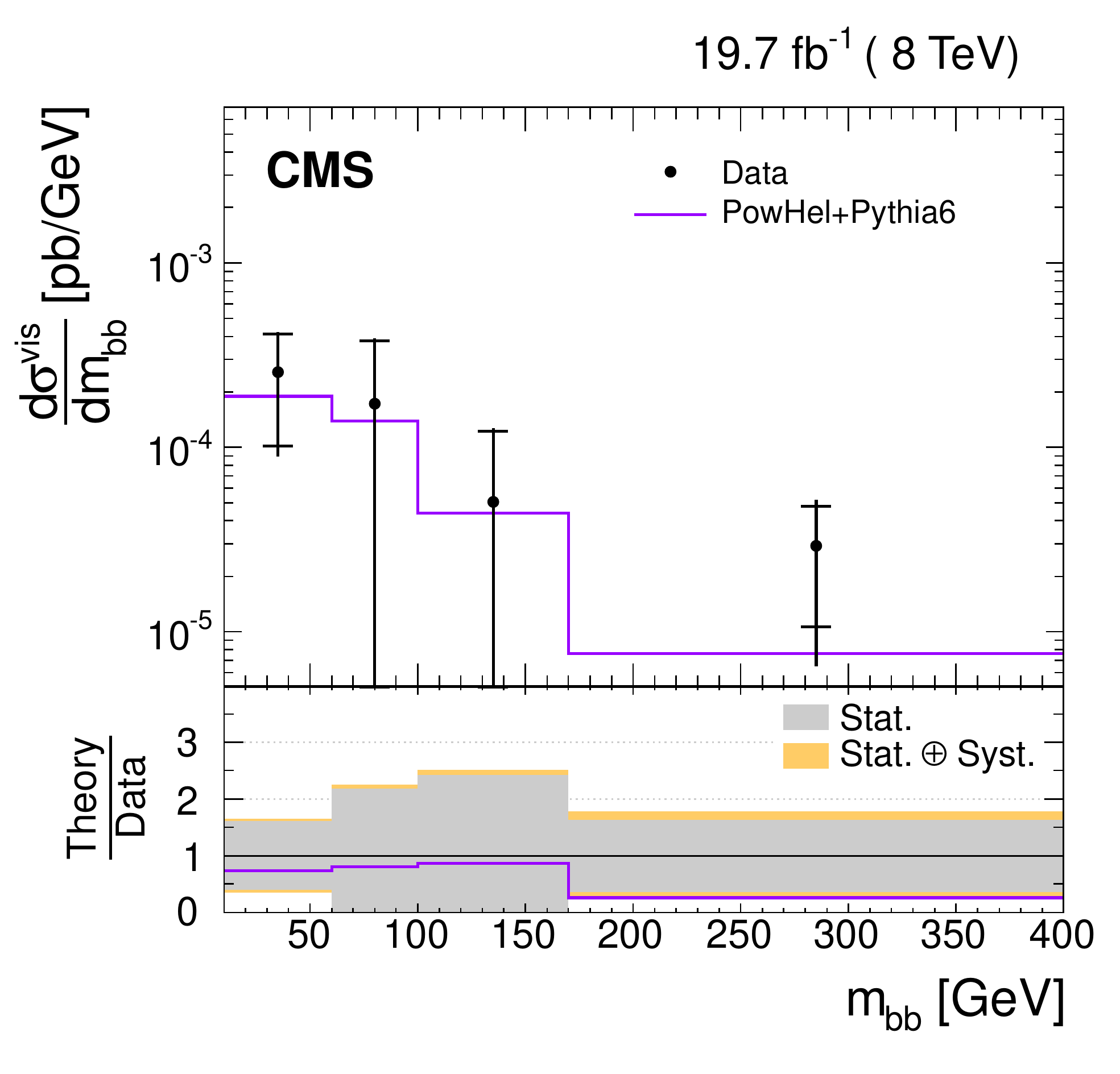}
    \caption{Absolute differential \ttbar cross section measured in the visible phase space of the \ttbar system and the additional \PQb jets, as a function of the second additional \PQb jet \pt (top left) and \abseta (top right), the angular separation $\Delta R_{\PQb\PQb}$ between the two leading additional \PQb jets (bottom left), and the invariant mass \mbb of the two \PQb jets (bottom right). Data are compared with predictions from \PowHel{}+\PYTHIA{6}. The inner (outer) vertical bars indicate the statistical (total) uncertainties. The lower part of each plot shows the ratio of the calculation to data.}
    \label{fig:xsec_bjetsNLO}
  \end{center}
\end{figure*}

{\tolerance=1200
The \ttbb production cross sections are compared to the NLO calculation by \PowHel{}+\PYTHIA{6} in Fig.~\ref{fig:xsec_bjetsNLO}. In the figure, the prediction is normalized to the absolute cross section given by the calculation of $20.8 \pm 0.6 \stat {}^{+7.9}_{-5.4} \text{(scale)}\unit{fb}$. The prediction describes well the shape of the different distributions, while the predicted absolute \ttbb cross section is about 30\% lower than the measured one, but compatible within the uncertainties.
\par}

\begin{figure*}[htbp!]
  \begin{center}
     \includegraphics[width=0.40\textwidth]{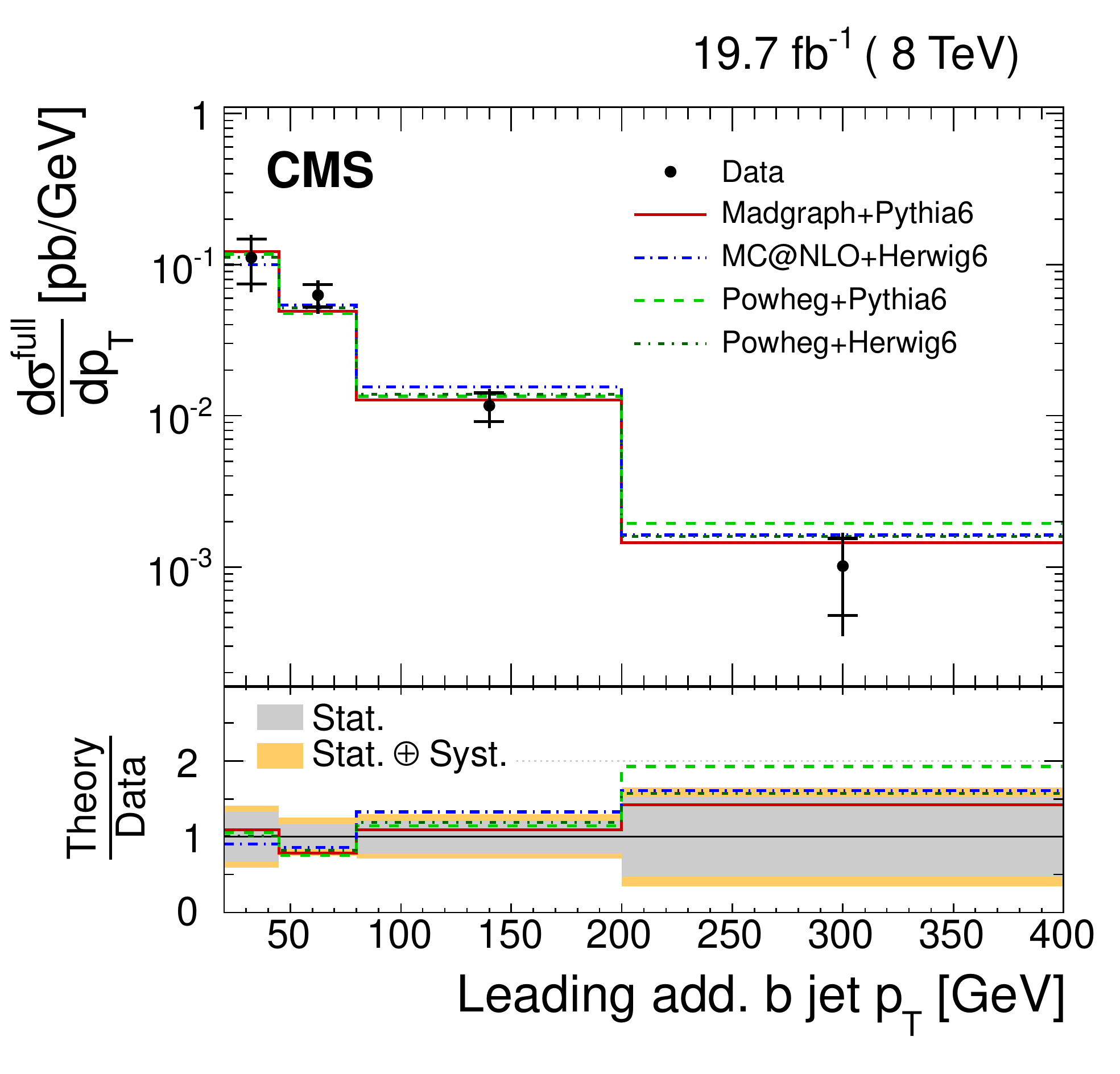}
    \includegraphics[width=0.40\textwidth]{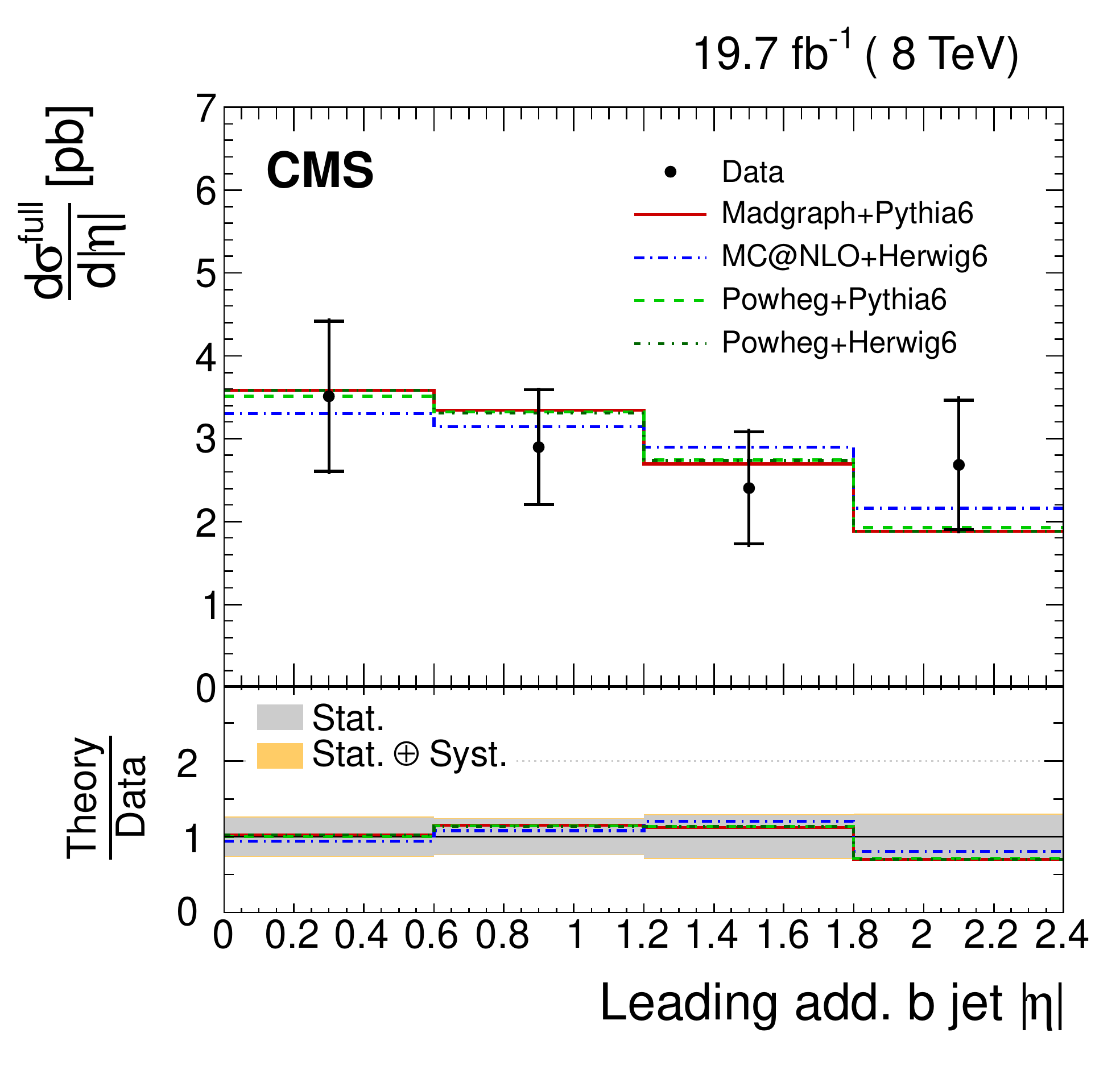}
    \includegraphics[width=0.40\textwidth]{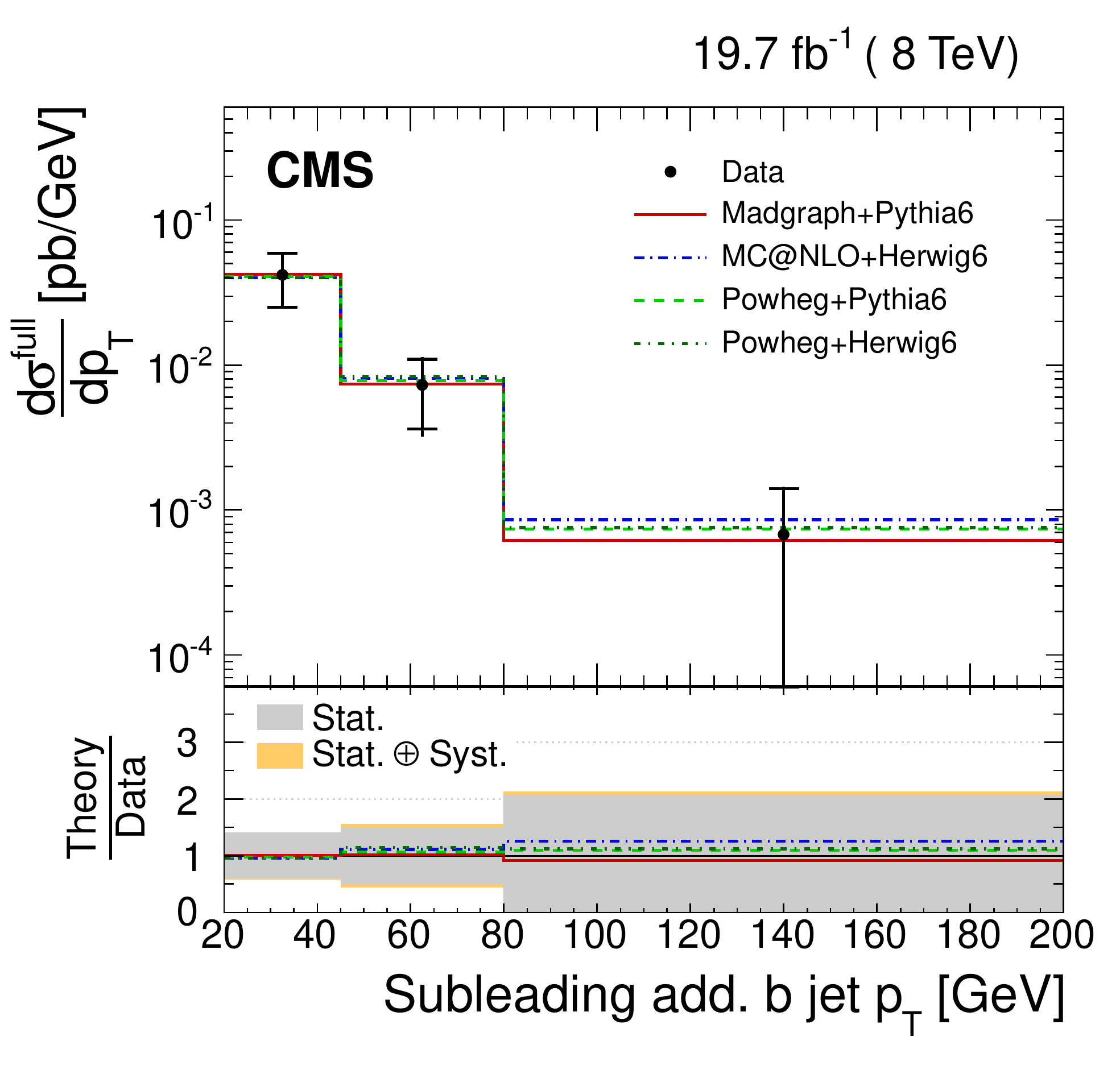}
    \includegraphics[width=0.40\textwidth]{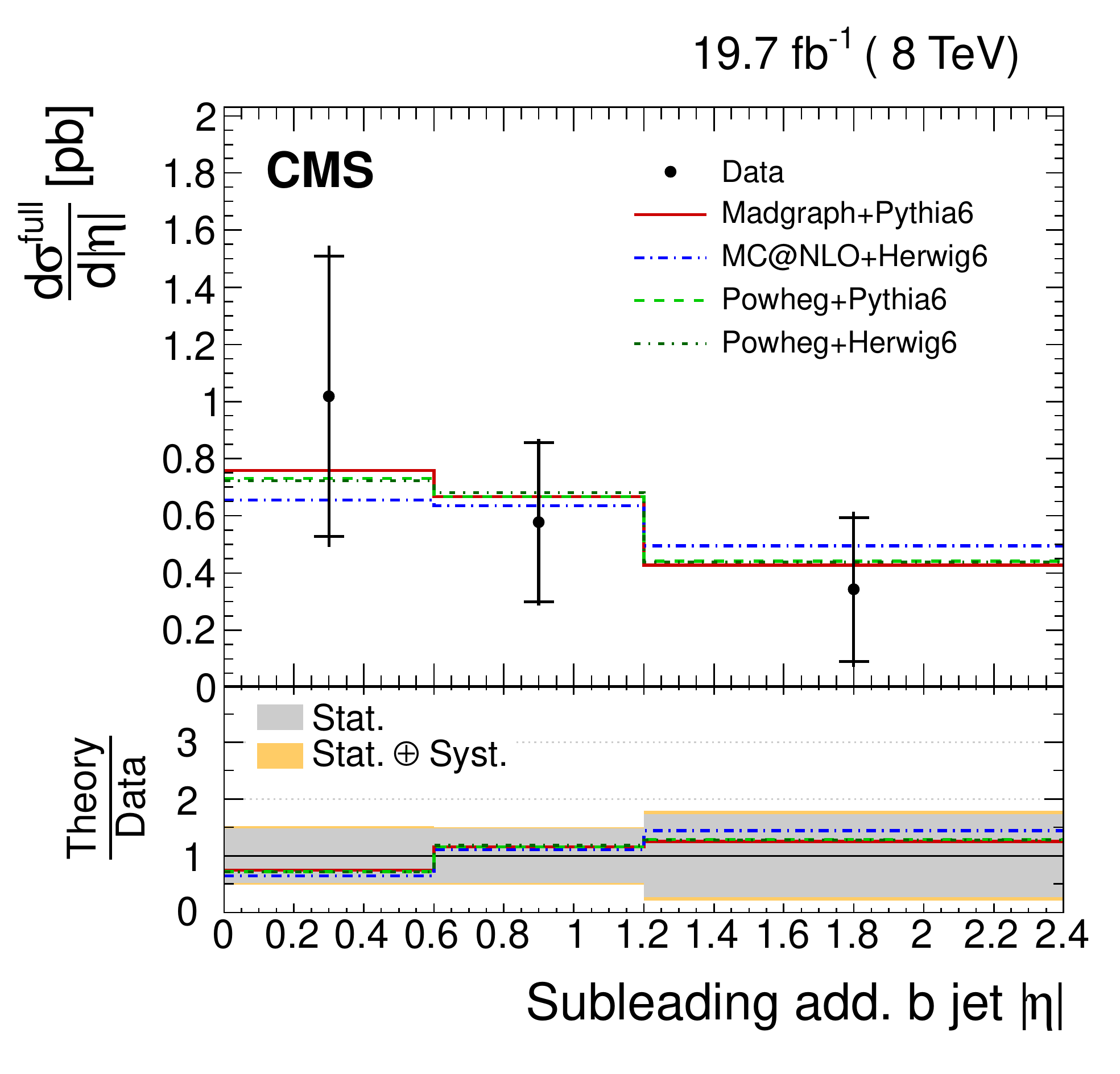}
    \includegraphics[width=0.40\textwidth]{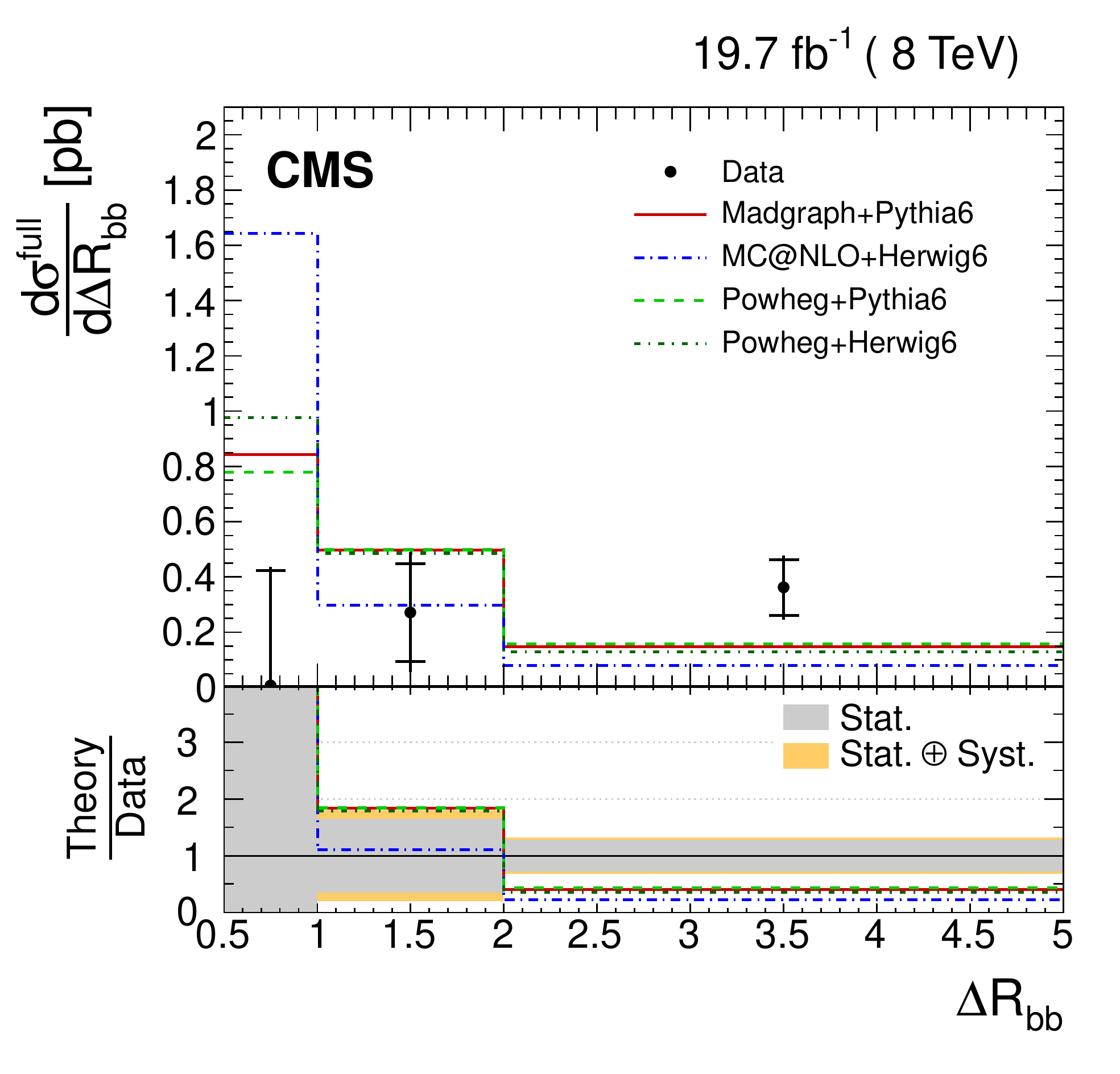}
    \includegraphics[width=0.40\textwidth]{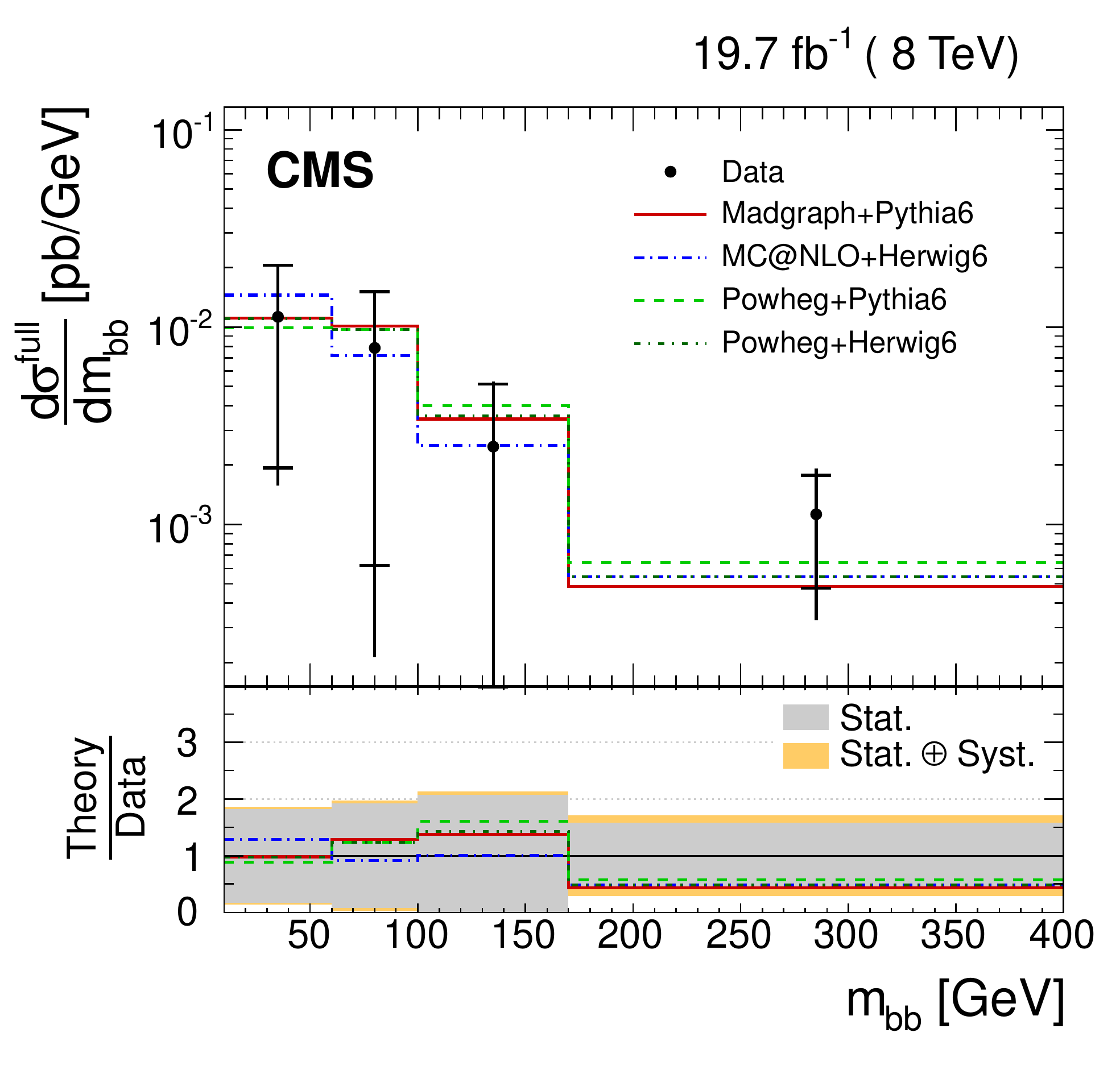}
   \caption{Absolute differential \ttbar cross section measured in the full phase space of the \ttbar system, corrected for acceptance and branching fractions, and the visible phase space of the additional \PQb jets, as a function of the leading additional \PQb jet \pt (top left) and \abseta (top right), subleading additional \PQb jet \pt (middle left) and \abseta (middle right), the angular separation $\Delta R_{\PQb\PQb}$ between the leading and subleading additional \PQb jets (bottom left), and the invariant mass \mbb of the two \PQb jets (bottom right). Data are compared with predictions from \MADGRAPH interfaced with \PYTHIA{6}, \MCATNLO interfaced with \HERWIG{6}, and \POWHEG intefarced with both \PYTHIA{6} and \HERWIG{6}, normalized to the measured inclusive cross section. The inner (outer) vertical bars indicate the statistical (total) uncertainties. The lower part of each plot shows the ratio of the predictions to the data.}
    \label{fig:xsec_bjetsFull}
  \end{center}
\end{figure*}

\begin{figure*}[htbp!]
  \begin{center}
    \includegraphics[width=0.40\textwidth]{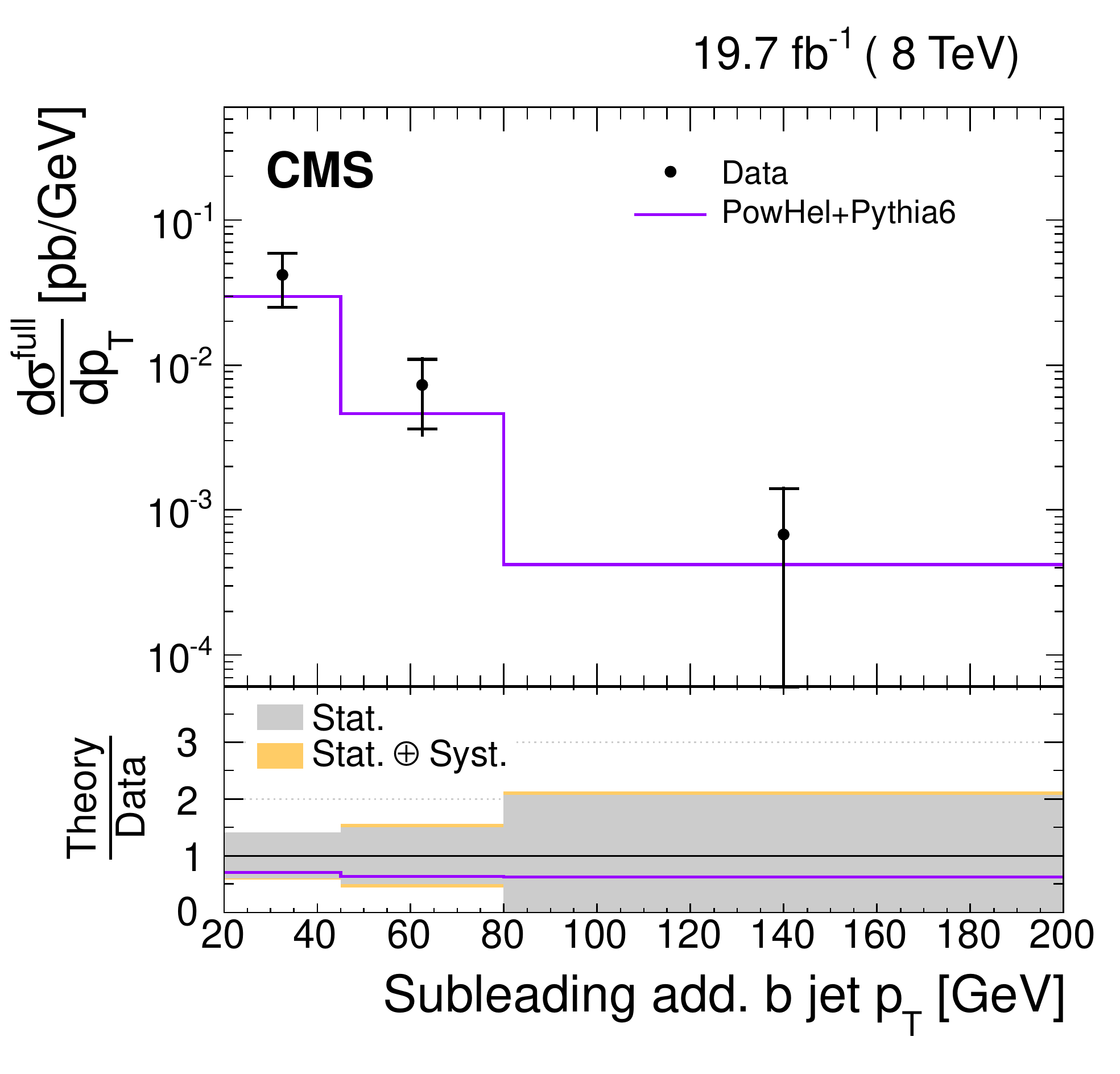}
    \includegraphics[width=0.40\textwidth]{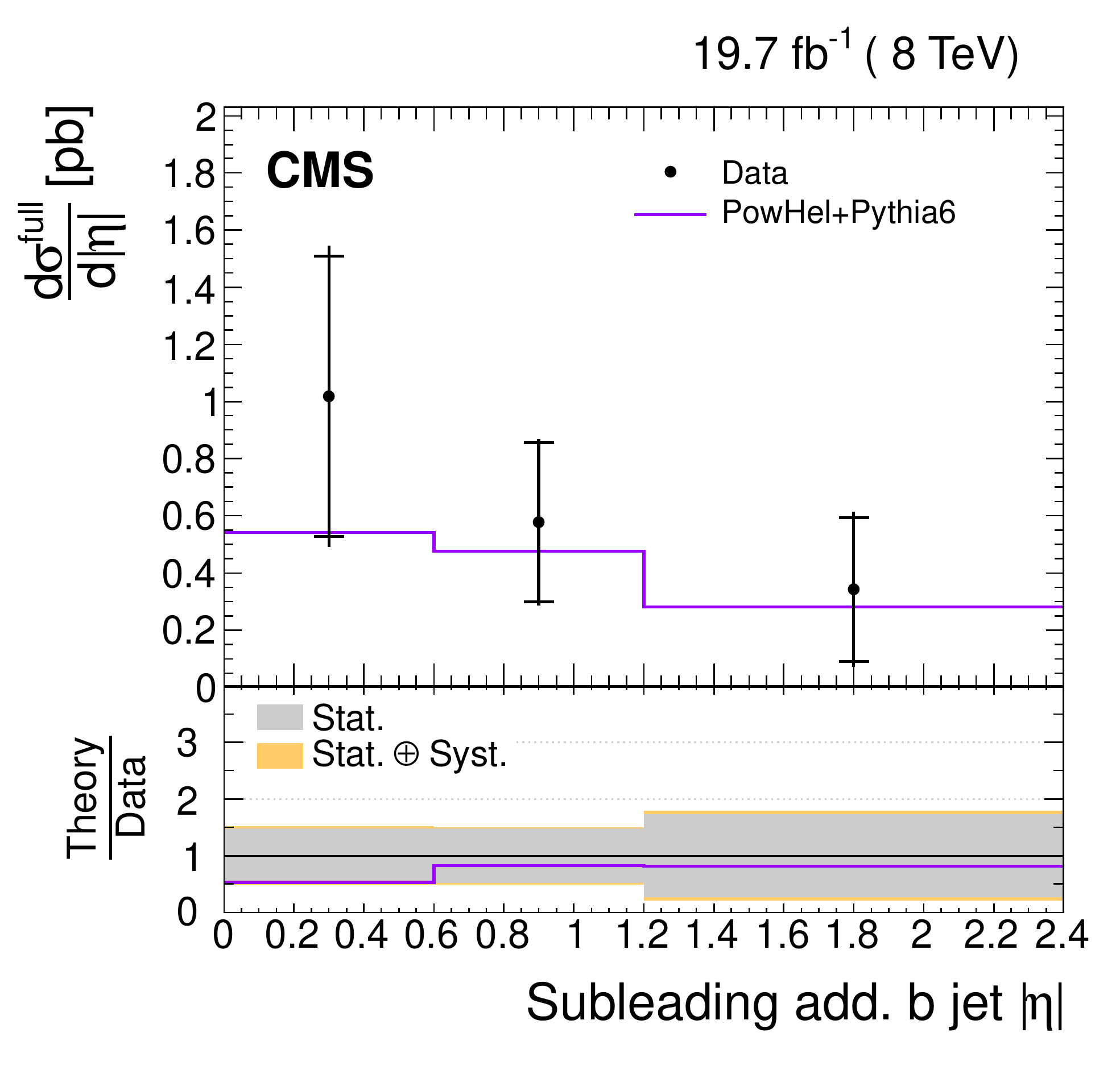}
    \includegraphics[width=0.40\textwidth]{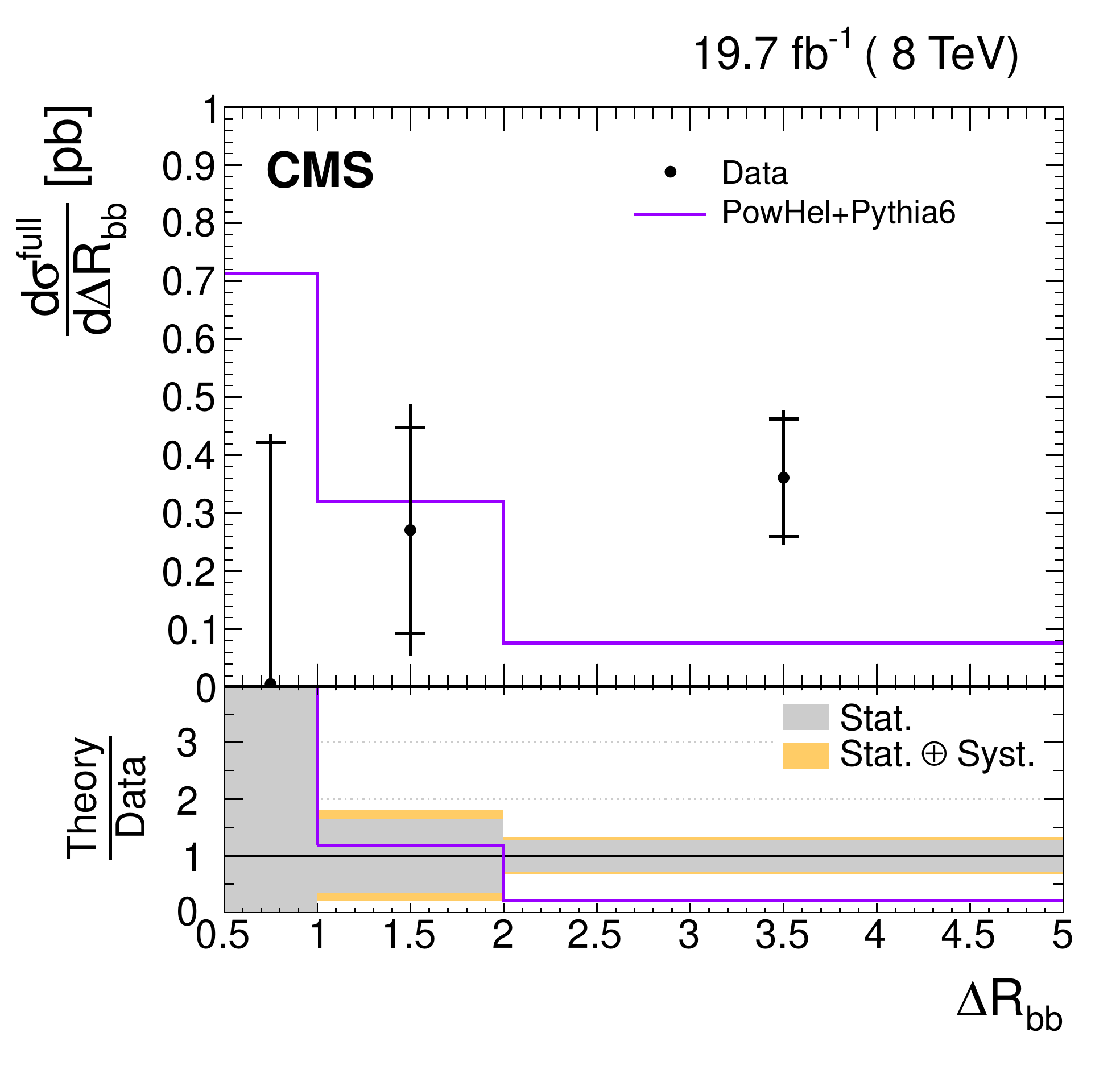}
    \includegraphics[width=0.40\textwidth]{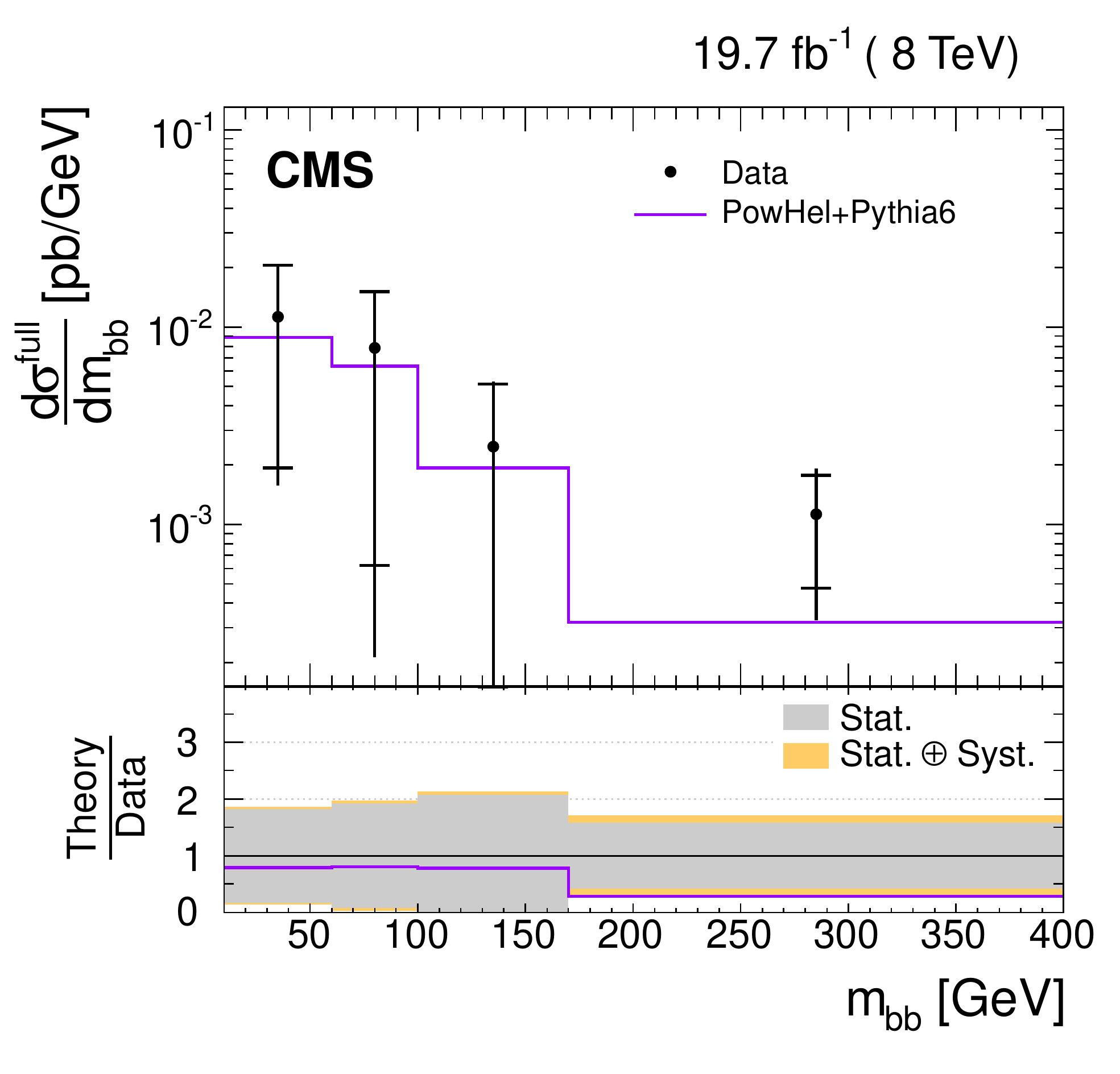}
    \caption{Absolute differential \ttbar cross section measured in the full phase space of the \ttbar system, corrected for acceptance and branching fractions, and the additional \PQb jets, as a function of the second additional \PQb jet \pt (top left) and \abseta (top right), the angular separation $\Delta R_{\PQb\PQb}$ between the leading and subleading additional \PQb jets (bottom left), and the invariant mass \mbb of the two \PQb jets (bottom right). Data are compared with predictions from \PowHel{}+\PYTHIA{6}. The inner (outer) vertical bars indicate the statistical (total) uncertainties. The lower part of each plot shows the ratio of the calculation to data.}
    \label{fig:xsec_bjetsNLOFull}
  \end{center}
\end{figure*}

The absolute differential cross sections measured in the visible phase space of the additional \PQb jets and the full phase space of the \ttbar system are presented in Fig.~\ref{fig:xsec_bjetsFull} and given in Tables~\ref{tab:dilepton:SummaryResultsBJetFullPS}--\ref{tab:dilepton:SummaryResultsBJet12FullPS} of Appendix~\ref{sec:summarytables}. The results are corrected for acceptance and dileptonic branching fractions including $\tau$ leptonic decays ($6.43 \pm 0.14$)\%~\cite{PDG2014}. The results are compared to the same predictions as in Fig.~\ref{fig:xsec_bjets}, which are scaled to the measured cross section, obtained by integrating all the bins of the corresponding distribution. The normalization factor applied to the simulations is similar to the previous one for the results in the visible phase space of the \ttbar system. The description of the data by the simulations is similar as well. The total measured $\sigma_{\ttbb}$, as well as the agreement between the data and the simulation, is in agreement with the result obtained in Ref.~\cite{bib:ttbb_ratio:2014}.
In the full phase space, the inclusive \ttbb cross section at NLO given by \PowHel{}+\PYTHIA{6} corresponds to $62 \pm 1 \stat {}^{+23}_{-17} \text{(scale)}\unit{fb}$ (excluding the dileptonic branching fraction correction).
The comparison of the differential \ttbb cross section with the NLO calculation is presented in Fig.~\ref{fig:xsec_bjetsNLOFull}.

Differences between the kinematic properties of the additional jets and b jets are expected owing to the different production mechanisms~\cite{Bevilacqua:2014qfa} of both processes. The dominant production mechanism of $\Pp\Pp\to\ttbb$ is gluon-gluon (gg) scattering, while in the case of $\Pp\Pp\to\ttjj$, the quark-gluon (qg) channel is equally relevant. The \abseta distributions of the additional b jets seem to be more central than the corresponding distributions of the additional jets, see Figs.~\ref{fig:inclusiveeta} and ~\ref{fig:inclusiveetaFull}. This difference can be attributed mainly to the contribution of the production via the qg channel, which favours the emission of jets at larger $\abseta$. The distributions of the differential cross section as a function of \mbb peak at smaller invariant masses than those as a function of \mjj, presented in Figs.~\ref{fig:DeltaRmassjj} and~\ref{fig:DeltaRmassjjFull}, because of the larger contribution of the gg channel. Given the large uncertainties in the \ttbb measurements, no statistically significant differences can be observed in the shape of the \pt distributions of the additional b jets compared to the additional jets, shown in Figs.~\ref{fig:inclusivept} and~\ref{fig:inclusiveptFull}.

\section{Additional jet gap fraction}
\label{sec:gap}

An alternative way to investigate the jet activity arising from quark and gluon radiation is to determine the fraction of events that do not contain additional jets above a given \pt threshold~\cite{bib:TOP-12-018,bib:atlas2}.
A threshold observable, referred to as the gap fraction, is defined as:
\begin{equation}
f(\pt^j)=\frac{N(\pt^j)}{N_{\text{total}}},
\end{equation}
where $N_{\text{total}}$ is the total number of selected events and $N(\pt^j)$ is the number of events that do not contain at least $j$ additional jets (apart from the two jets from the \ttbar solution hypothesis) above a \pt threshold, with $j$ corresponding to one or two jets. The measurements are presented as a function of the \pt of the leading and subleading additional jets, respectively.

A modified gap fraction can be defined as:
\begin{equation}
f(\HT)=\frac{N(\HT)}{N_{\text{total}}},
\end{equation}
where $N(\HT)$ is the number of events in which the sum of the scalar \pt of the additional jets $(\HT)$ is less than a certain threshold. In both cases, detector effects are unfolded using the \MADGRAPH simulation to obtain the results at the particle level.
The additional jets at the generator level are defined as all jets within the kinematic acceptance, excluding the two \PQb jets originating from the \PQb quarks from top quark decay (see~Section~\ref{sec:diffxsec}). For each value of the \pt and $\HT$ thresholds the gap fraction at the generator level is evaluated, along with the equivalent distributions after the detector simulation and analysis requirements.
Given the high purity of the selected events, above 70\% for any bin for the leading additional jet \pt and $\HT$, and above 85\% for any bin for the subleading additonal jets, a correction for detector effects is applied by following a simpler approach than the unfolding method used for other measurements presented here. The data are corrected to the particle level by applying the ratio of the generated distributions at particle level to the simulated ones at the reconstruction level, using the nominal \MADGRAPH simulation.

\begin{figure*}[htbp!]
  \begin{center}
      \includegraphics[width=0.40 \textwidth]{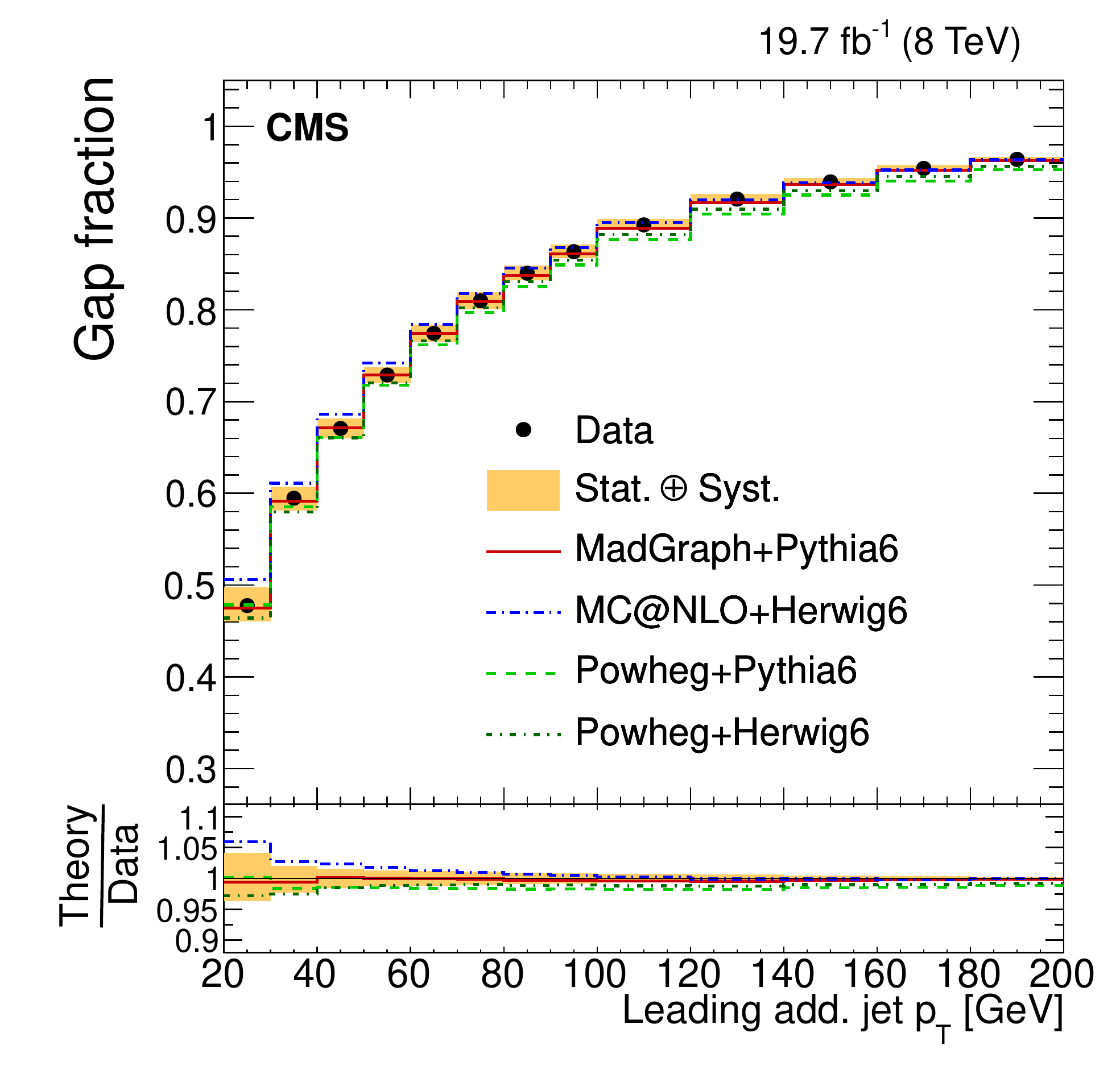}%
      \includegraphics[width=0.40 \textwidth]{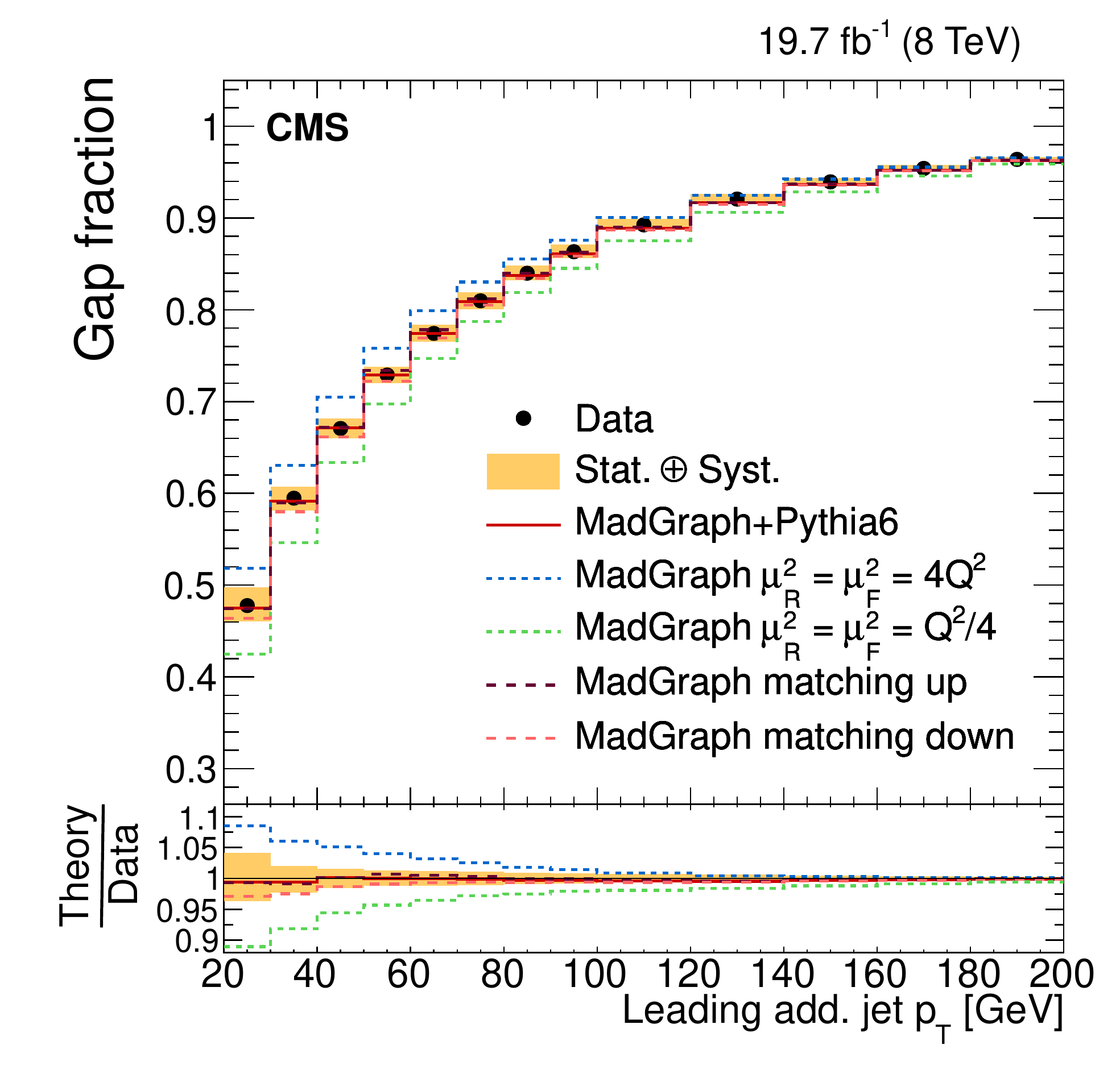}\\
      \includegraphics[width=0.40 \textwidth]{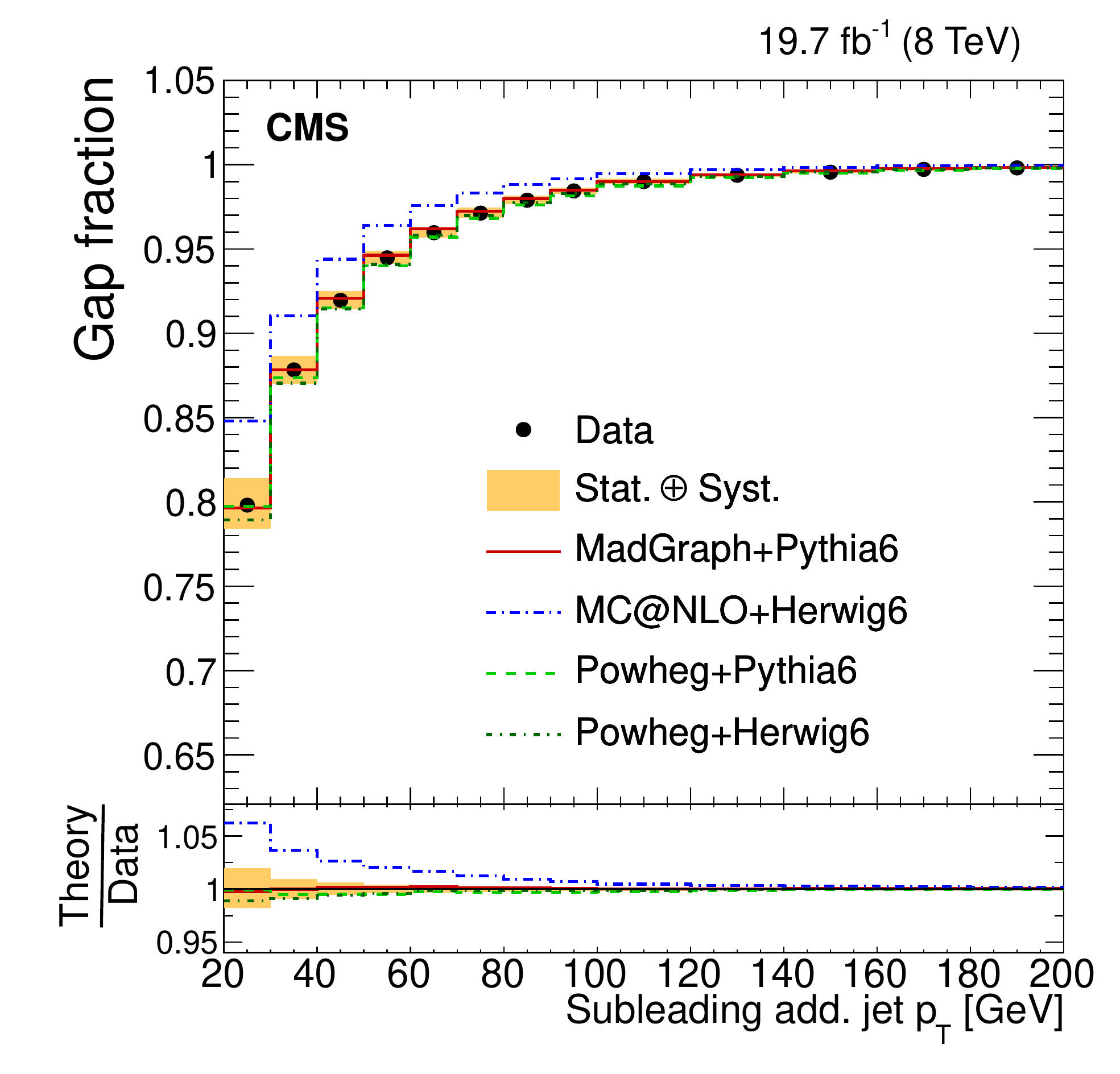}%
      \includegraphics[width=0.40\textwidth]{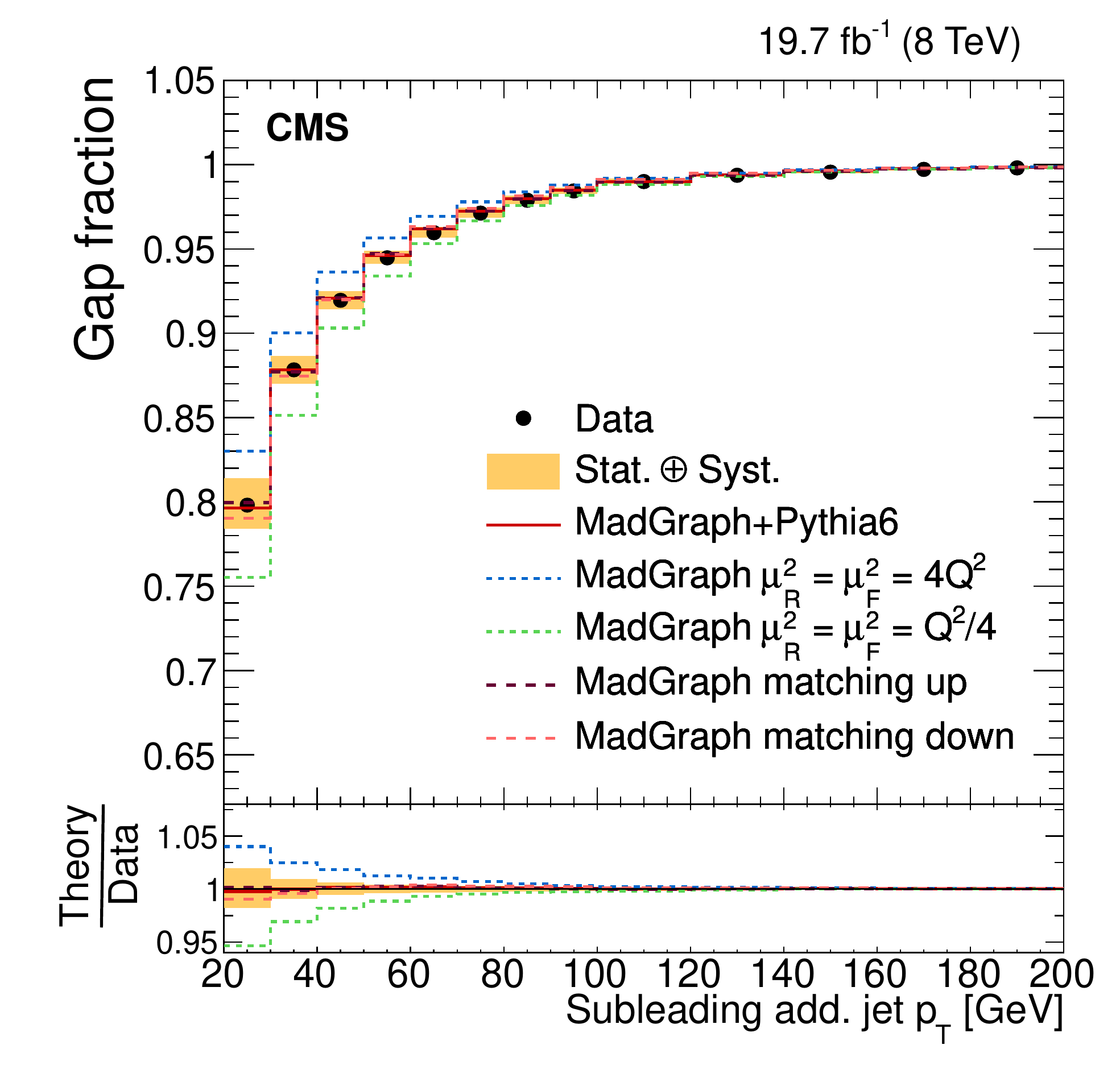}\\
      \includegraphics[width=0.40 \textwidth]{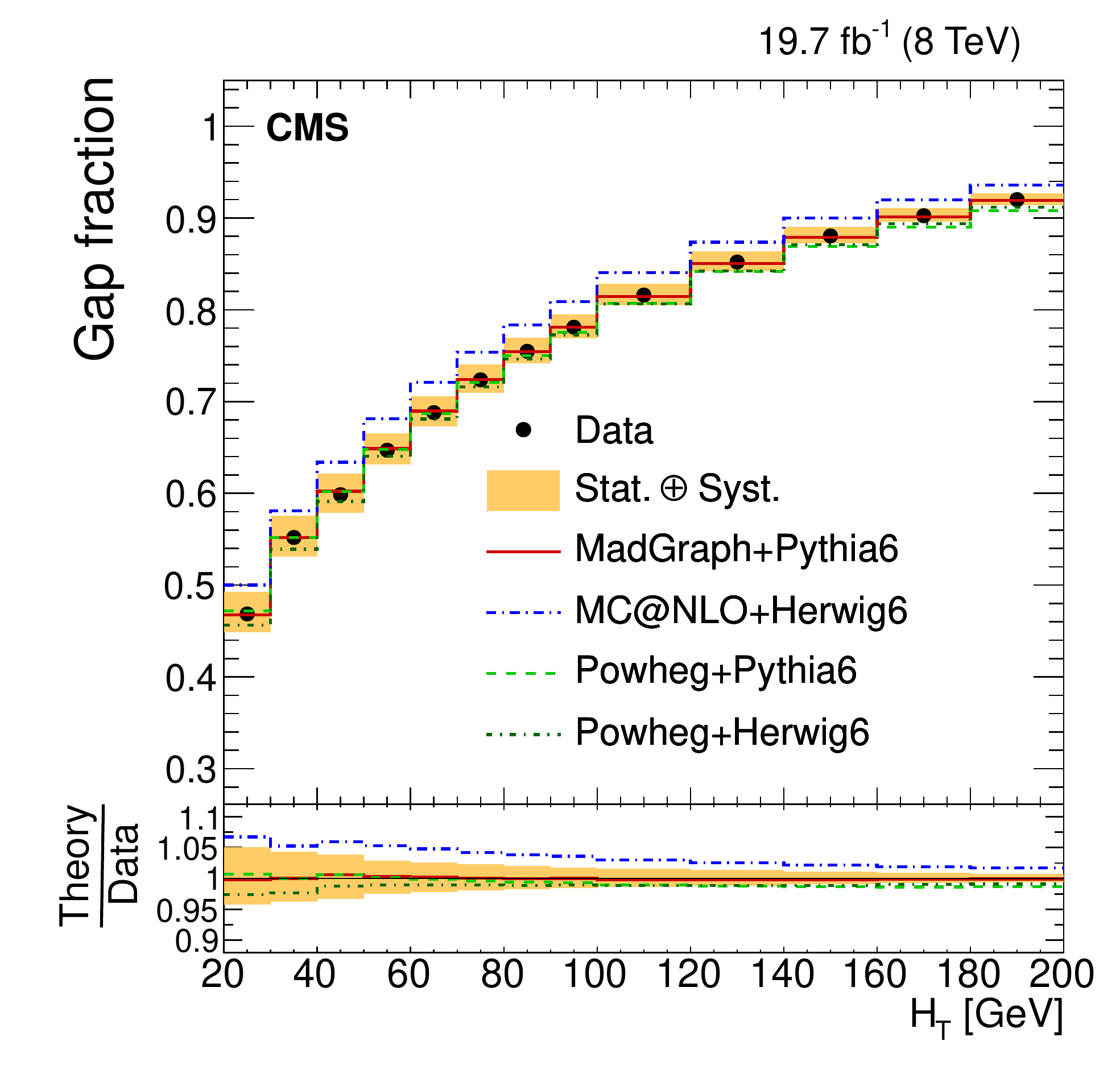}%
      \includegraphics[width=0.40 \textwidth]{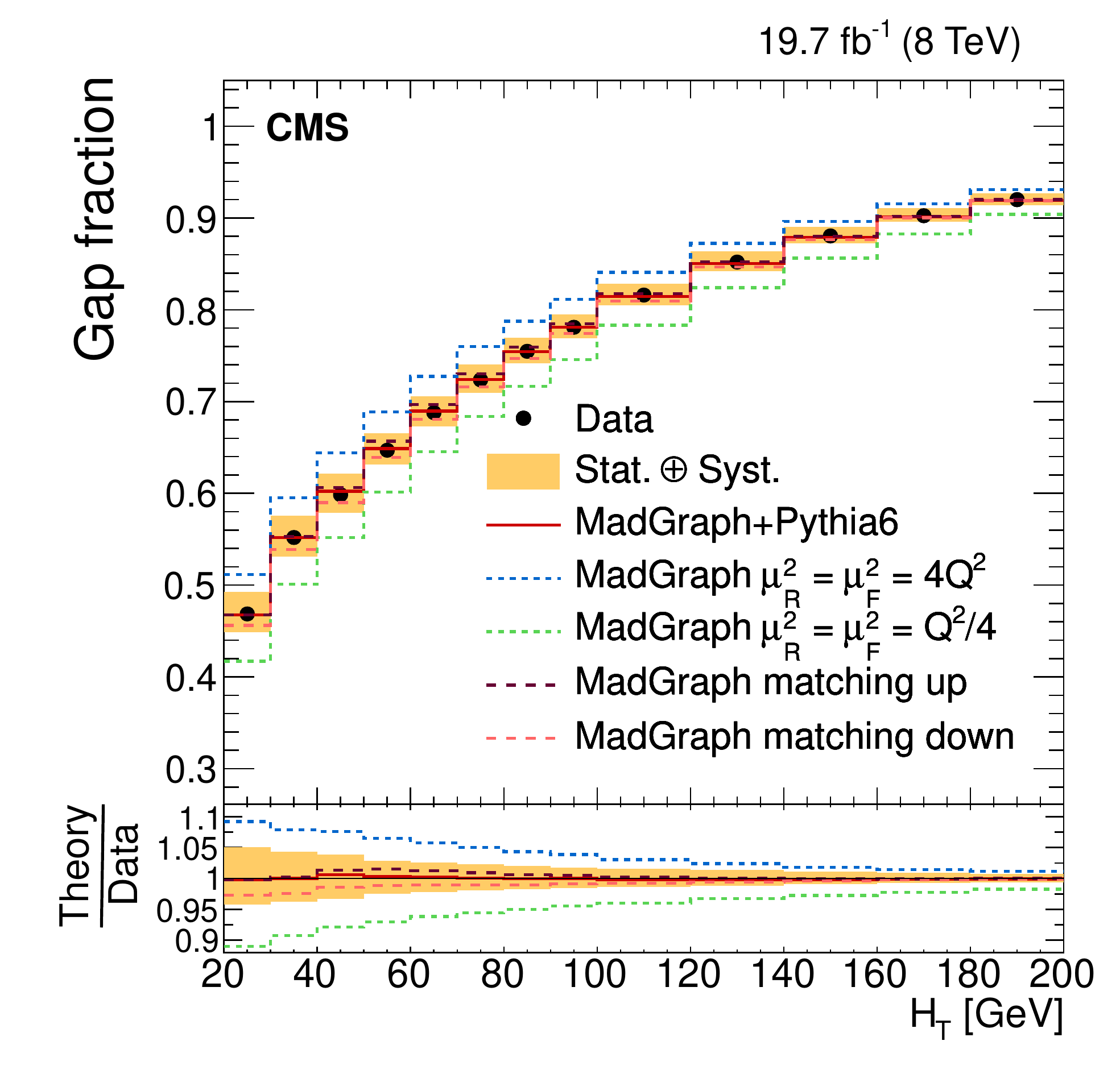}\\
\caption{Measured gap fraction as a function of the leading additional jet \pt (top row), subleading additional jet \pt (middle row), and of $\HT$ (bottom row). Data are compared to predictions from \MADGRAPH, \POWHEG interfaced with \PYTHIA and \HERWIG, and \MCATNLO interfaced with \HERWIG (left), and to \MADGRAPH with varied renormalization, factorization, and jet-parton matching scales (right). For each bin the threshold is defined at the value where the data point is placed. The vertical bars on the data points indicate the statistical uncertainty. The shaded band corresponds to the statistical and the total systematic uncertainty added in quadrature. The lower part of each plot shows the ratio of the predictions to the data.}
\label{fig:gap}
  \end{center}
\end{figure*}

The measured gap fraction distributions are compared to predictions from \MADGRAPH interfaced with \PYTHIA{6}, \POWHEG{6} interfaced with \PYTHIA{6} and \HERWIG{6}, \MCATNLO interfaced with \HERWIG{6}, and to the \MADGRAPH predictions with varied renormalization, factorization, and jet-parton matching scales. Figure~\ref{fig:gap} displays the gap fraction distribution as a function of the \pt of the leading and subleading additional jets, and $\HT$. The lower part of the figures shows the ratio of the predictions to the data. The light band indicates the total uncertainty in the data in each bin. The threshold, defined at the value where the data point is shown, is varied from 25\GeV (lower value compared to previous measurements~\cite{bib:TOP-12-018}) to 190\GeV. In general, \MADGRAPH interfaced with \PYTHIA{6} agrees with the data distributions of the three variables, while \POWHEG interfaced with \PYTHIA{6} and \HERWIG{6} also provide a good description of the data, though they tend to predict a lower gap fraction than the measured ones. The \MCATNLO generator interfaced with \HERWIG{6} describes the data well as a function of the leading additional jet \pt. However, it predicts higher values of the gap fraction as a function of the subleading jet \pt and $\HT$. Modifying the renormalization and factorization scales in \MADGRAPH worsens the agreement with data, while variations of the jet-parton matching threshold provide similar predictions as the nominal \MADGRAPH simulation, in agreement with the results shown before.

\begin{figure*}[htbp!]
  \begin{center}
      \includegraphics[width=0.40 \textwidth]{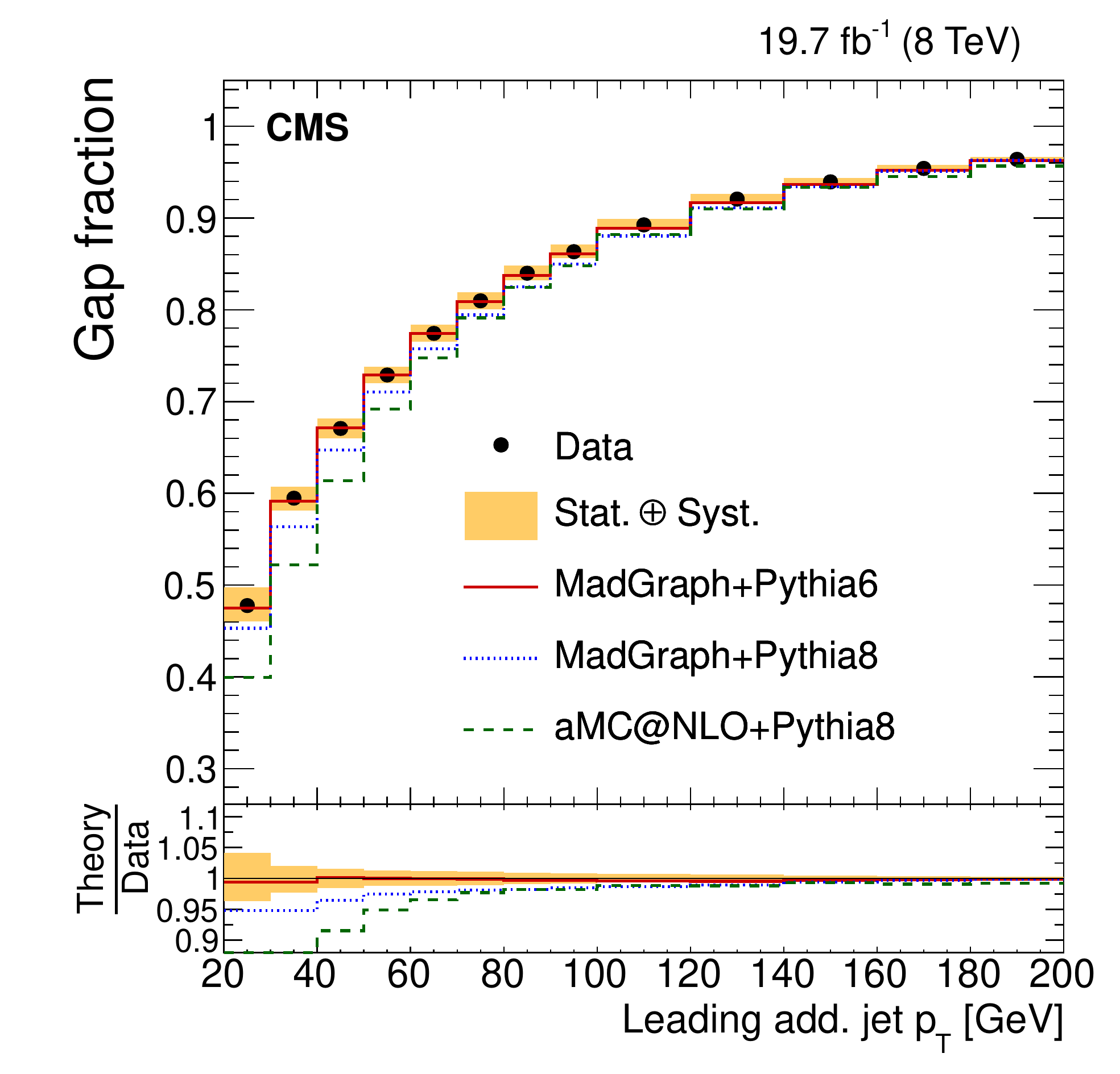}%
      \includegraphics[width=0.40 \textwidth]{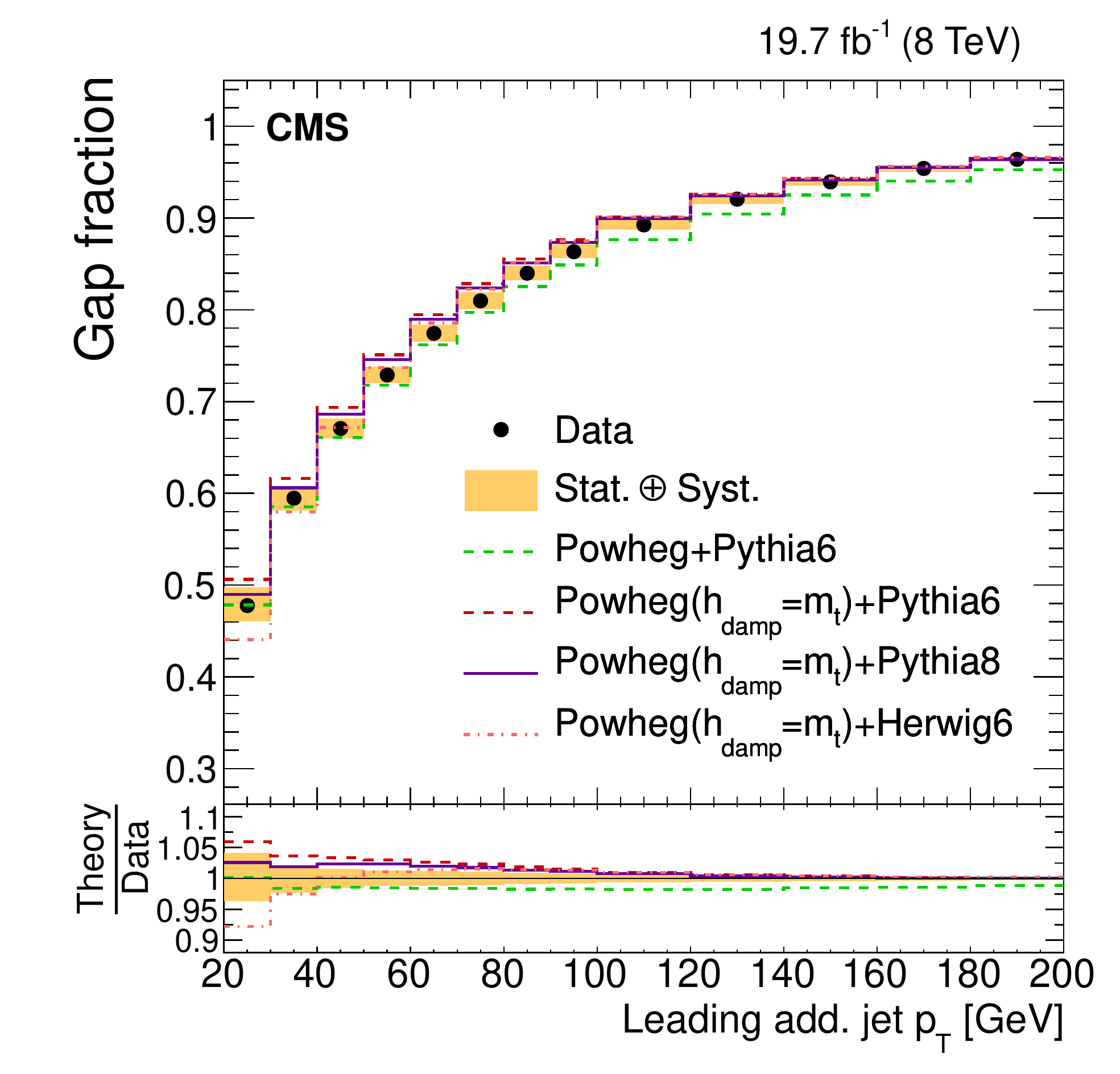}\\
      \includegraphics[width=0.40 \textwidth]{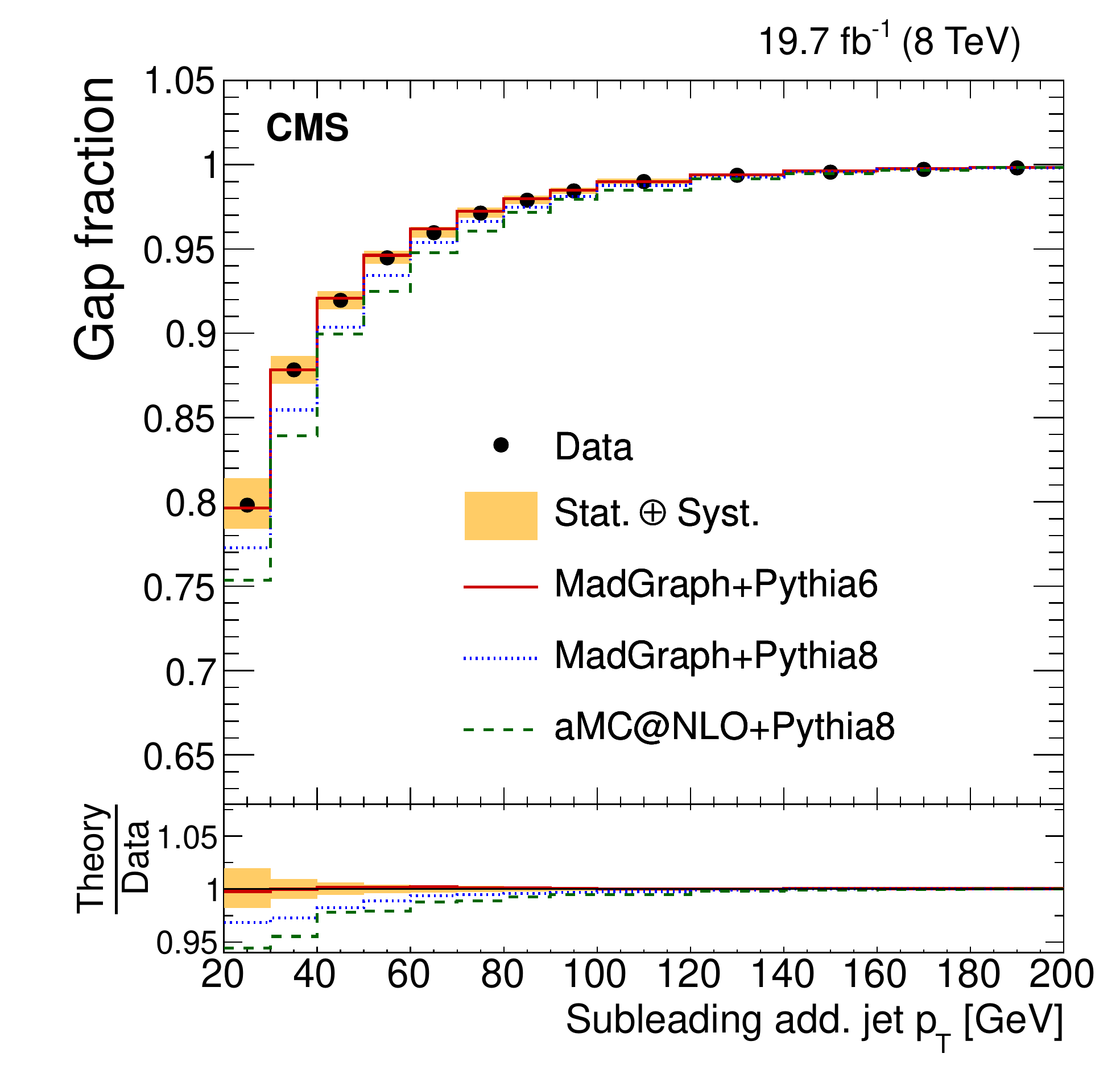}%
      \includegraphics[width=0.40\textwidth]{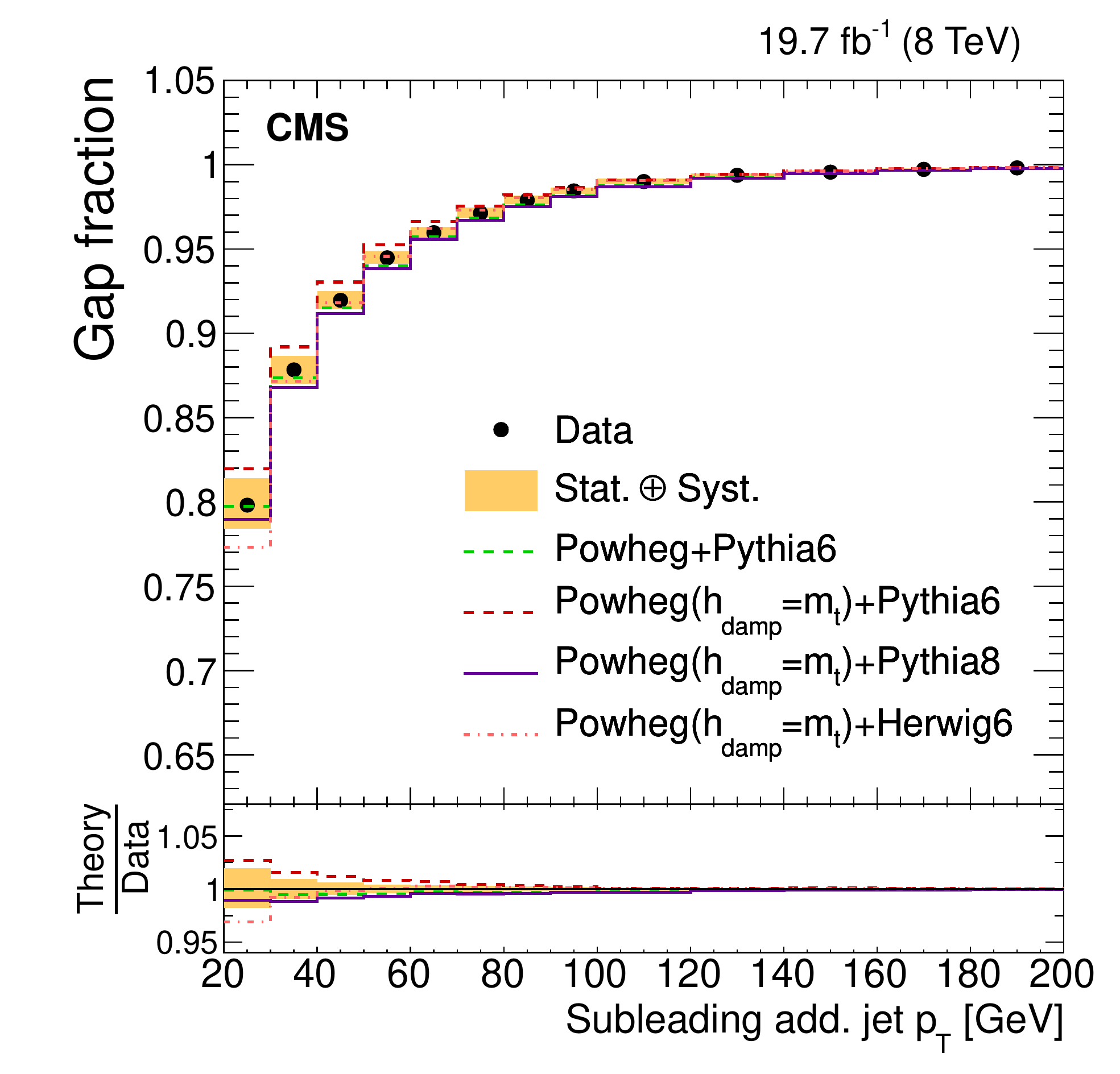}\\
      \includegraphics[width=0.40 \textwidth]{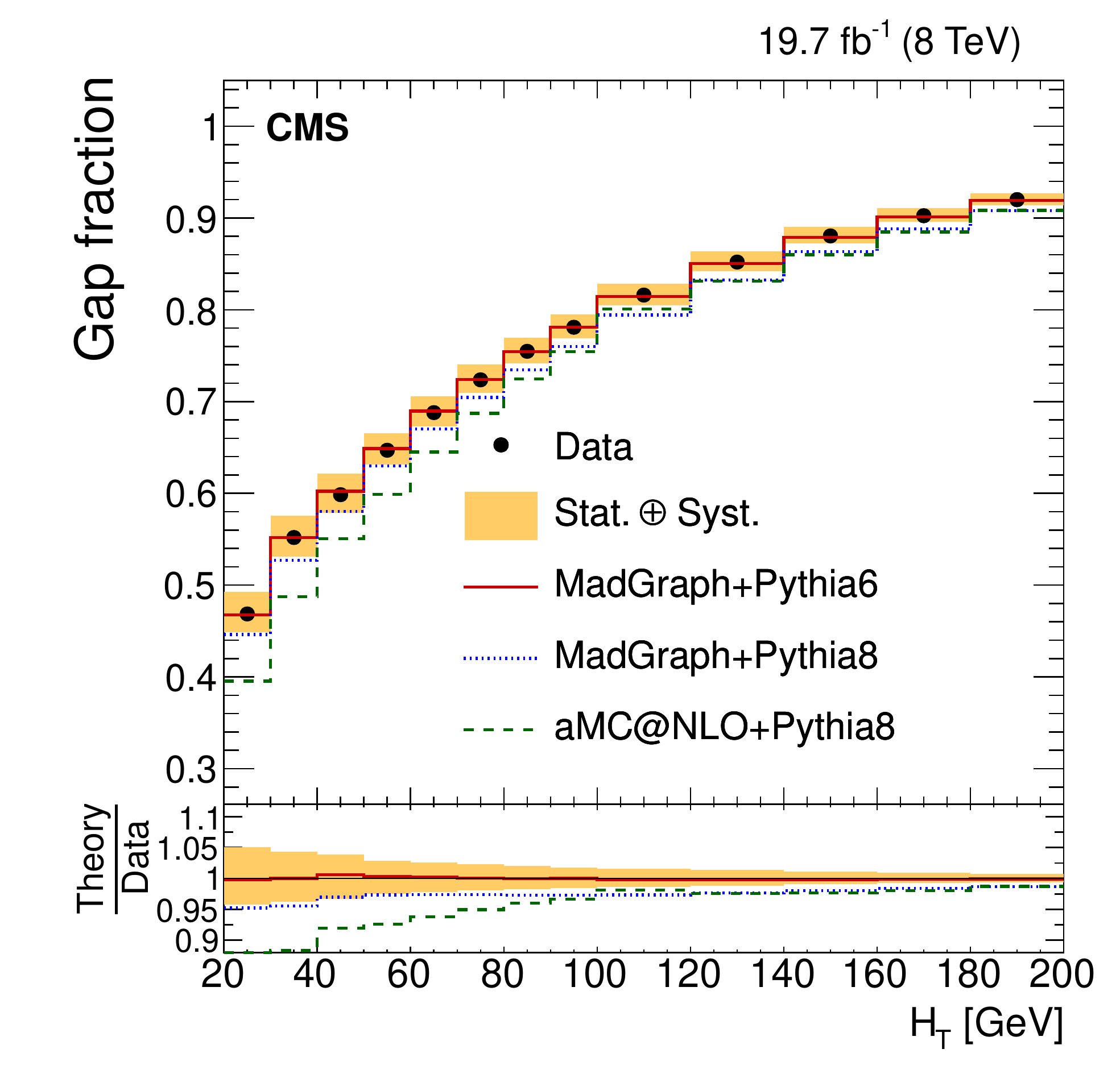}%
      \includegraphics[width=0.40 \textwidth]{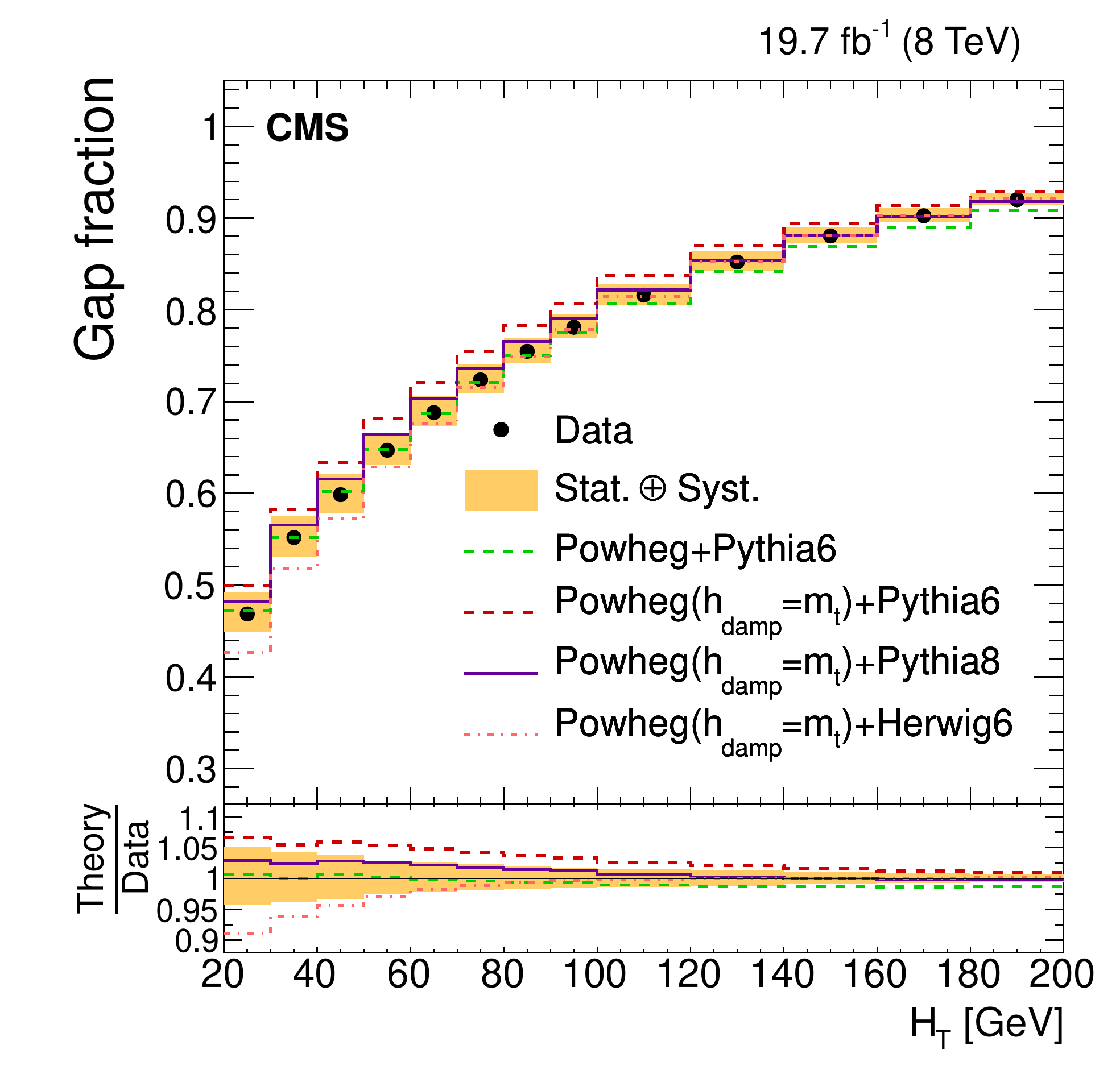}\\
\caption{Measured gap fraction as a function of the leading additional jet \pt (top row), subleading additional jet \pt (middle row), and of $\HT$ (bottom row). Data are compared to predictions from \MADGRAPH, interfaced with \PYTHIA{6} and \PYTHIA{8}, and \amcatnlo interfaced with \HERWIG{6} (left), and to \POWHEG interfaced with different versions of \PYTHIA and \HERWIG{6} (right). For each bin the threshold is defined at the value where the data point is placed. The vertical bars on the data points indicate the statistical uncertainty. The shaded band corresponds to the statistical and the total systematic uncertainty added in quadrature. The lower part of each plot shows the ratio of the predictions to the data.}
\label{fig:gapNewMC}
  \end{center}
\end{figure*}

The results are also compared in Fig.~\ref{fig:gapNewMC} with the recently available simulations, described in Section~\ref{sec:theory}, matched to different versions of the parton showering models. The \MADGRAPH and \amcatnlo generators interfaced with \PYTHIA{8} predict up to 10\% lower values of the gap fraction for all the variables, which reflects the fact that those simulations generate larger jet multiplicities, as discussed in Section~\ref{sec:diffxsecNJets}. Within the uncertainties, the predictions of the \POWHEG{}+\PYTHIA{8} simulation agree well with data, while the \POWHEG{} generator (with $\hdamp = m_{\PQt}$) interfaced with \PYTHIA{6} and \HERWIG{6} tends to overestimate and underestimate the measured values, respectively.

\begin{figure*}[htbp!]
  \begin{center}
      \includegraphics[width=0.40 \textwidth]{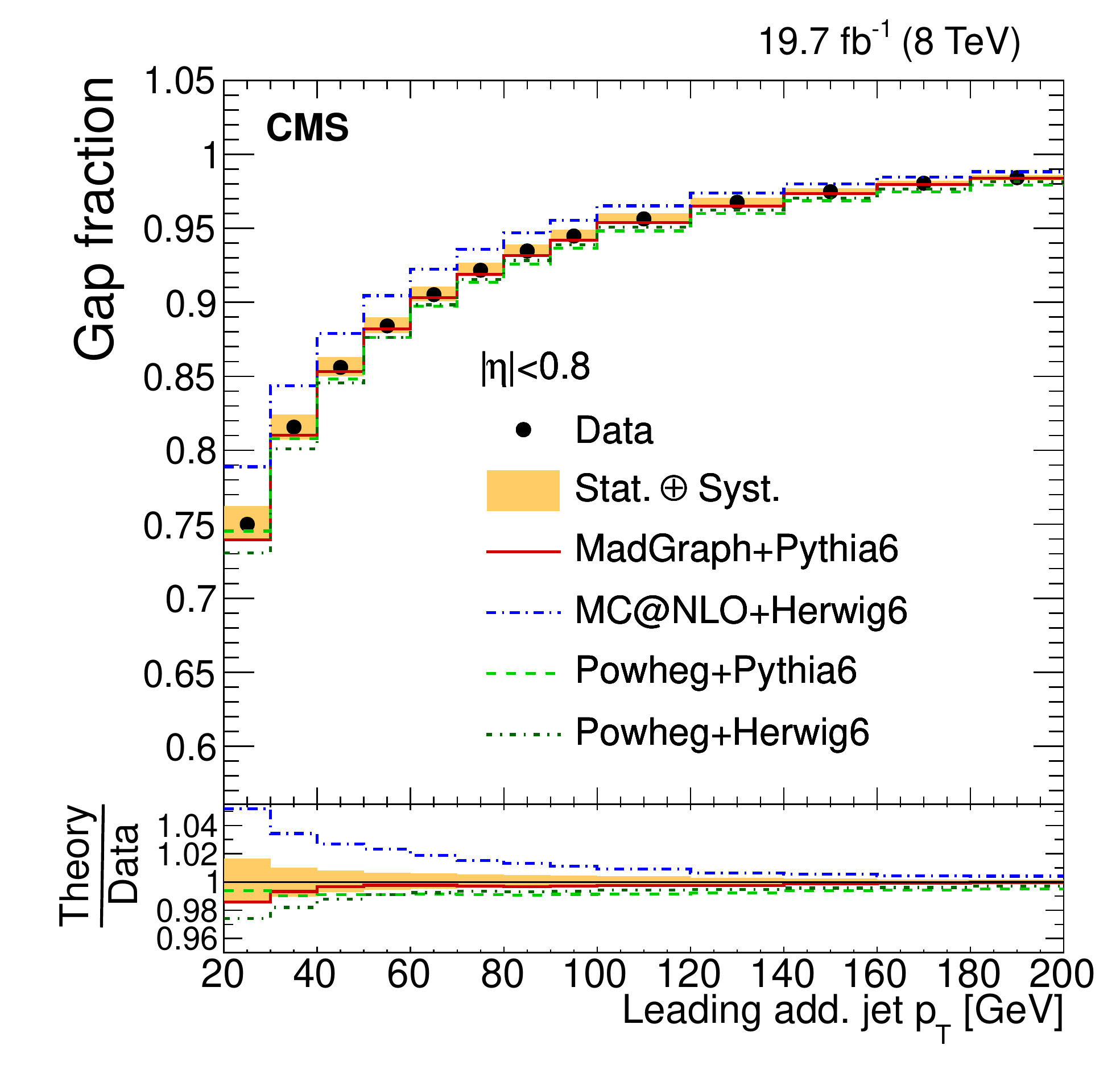}%
      \includegraphics[width=0.40 \textwidth]{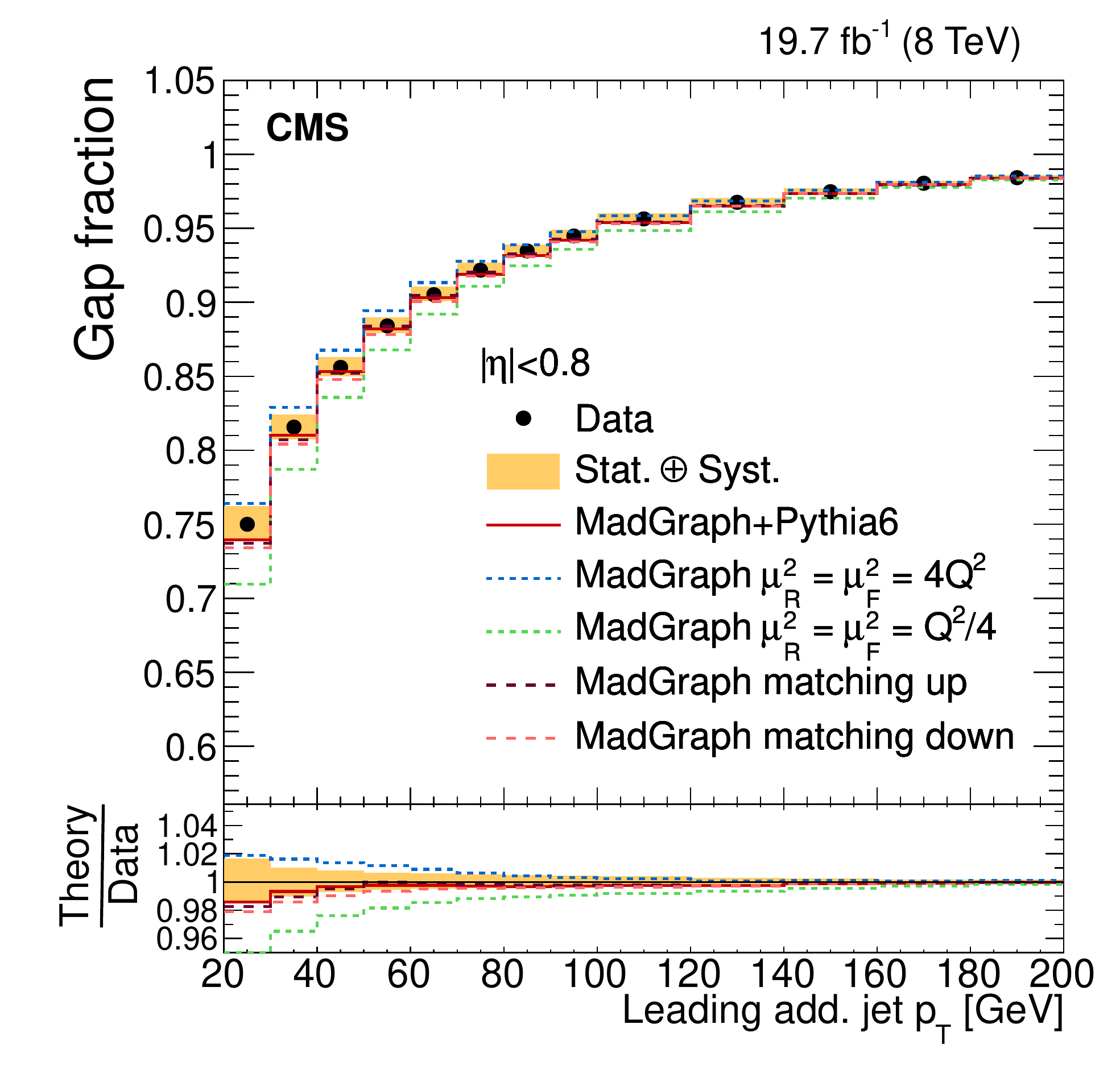}\\
      \includegraphics[width=0.40 \textwidth]{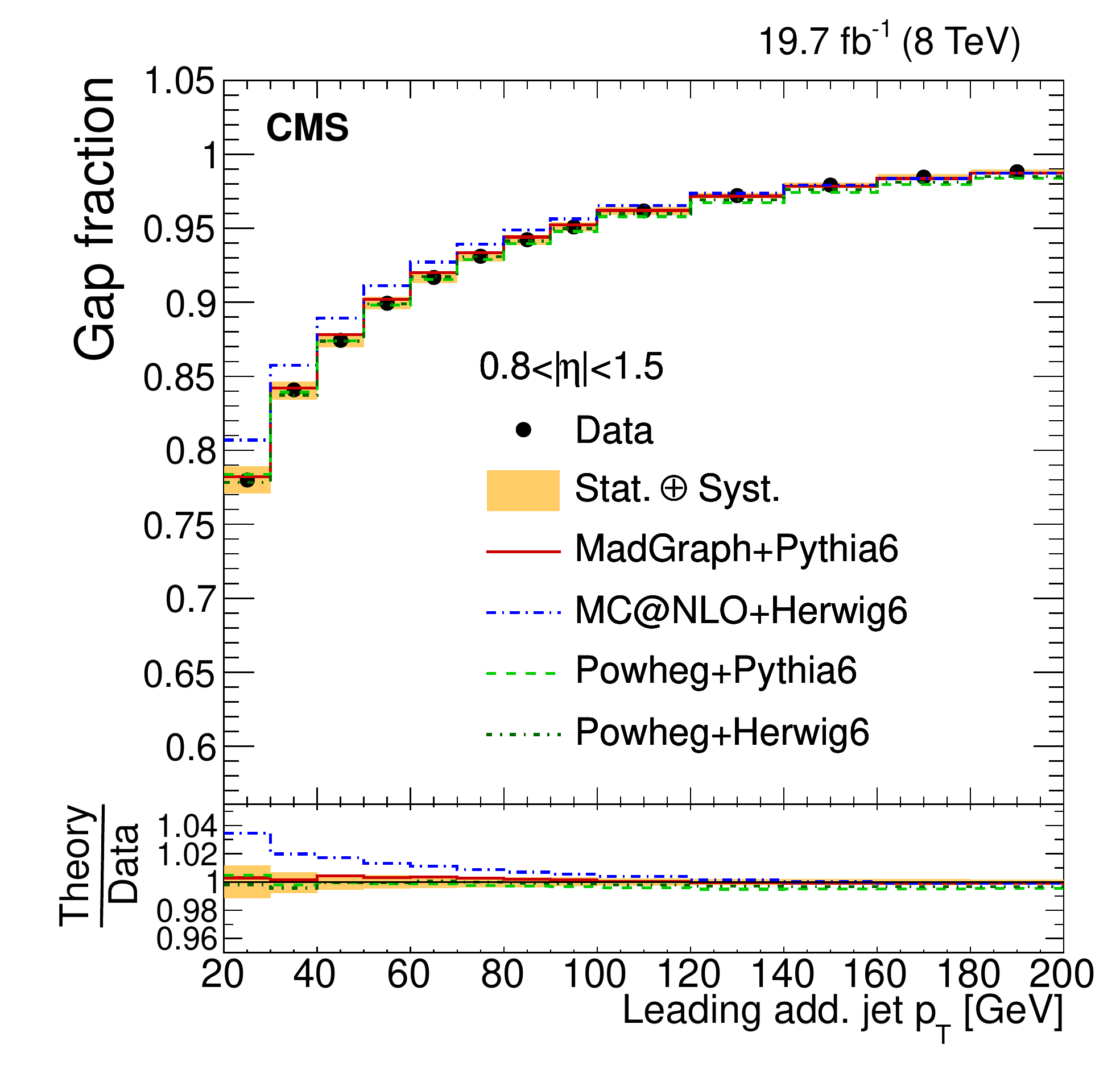}%
      \includegraphics[width=0.40 \textwidth]{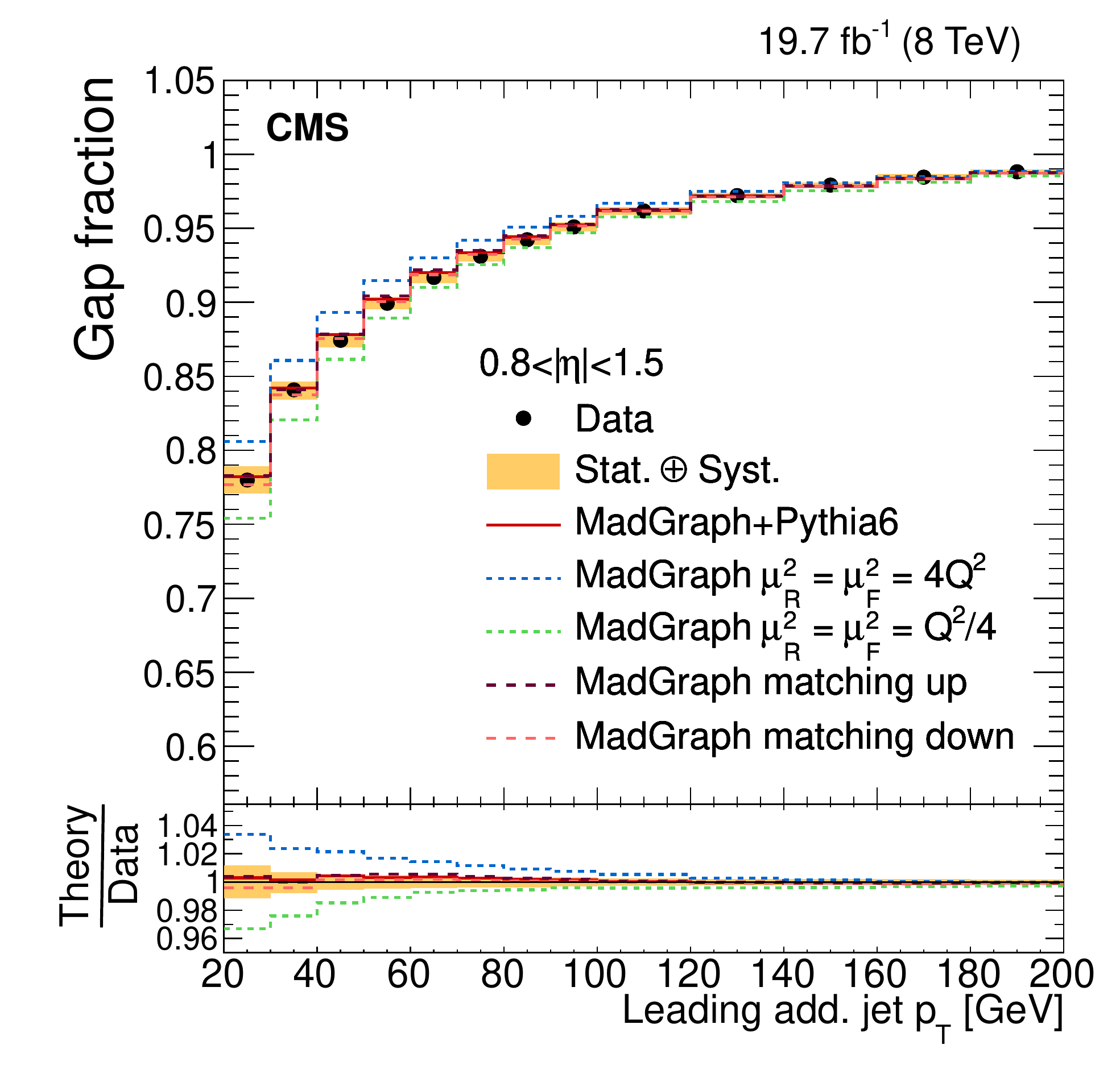}\\
      \includegraphics[width=0.40 \textwidth]{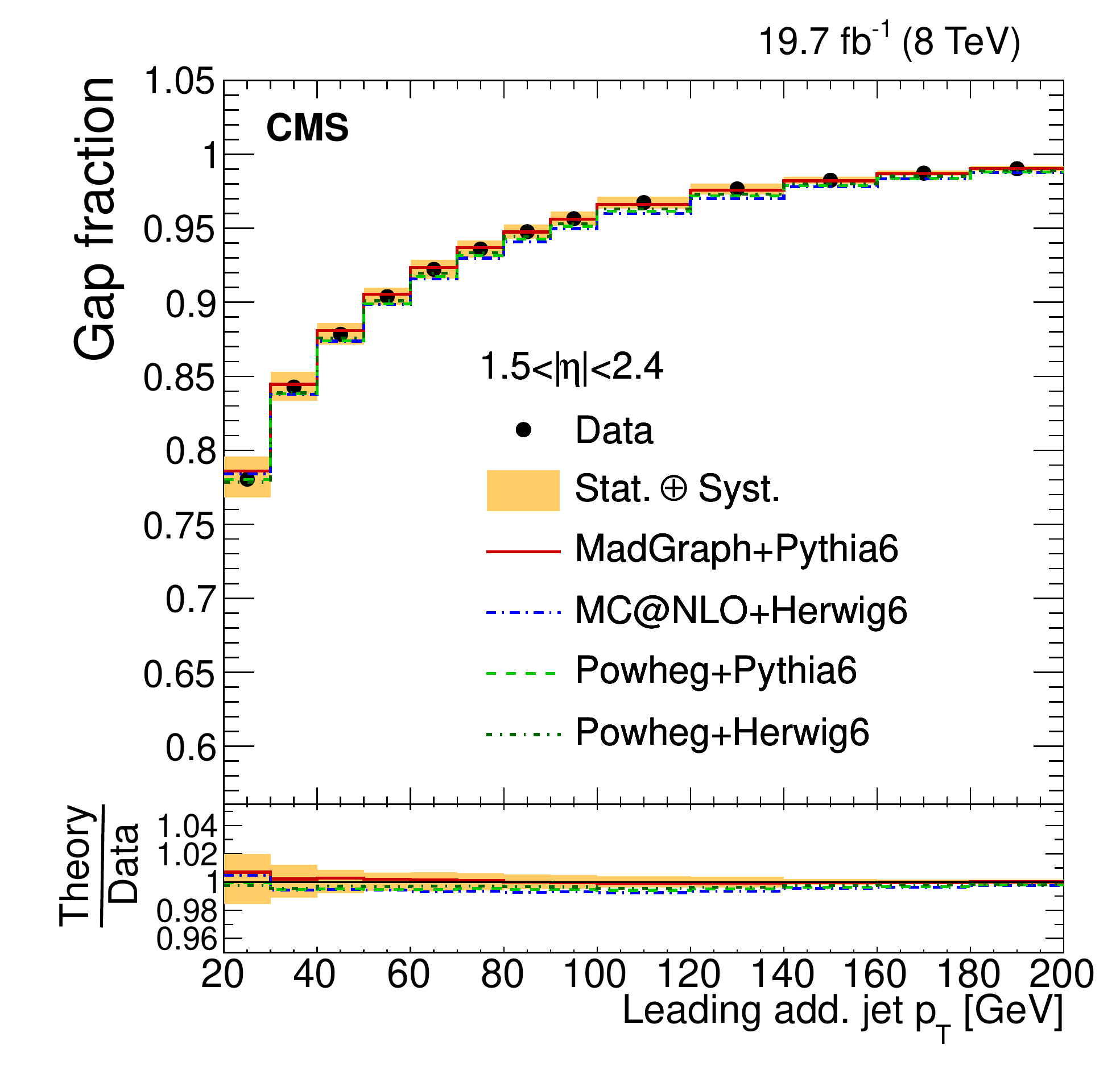}%
      \includegraphics[width=0.40 \textwidth]{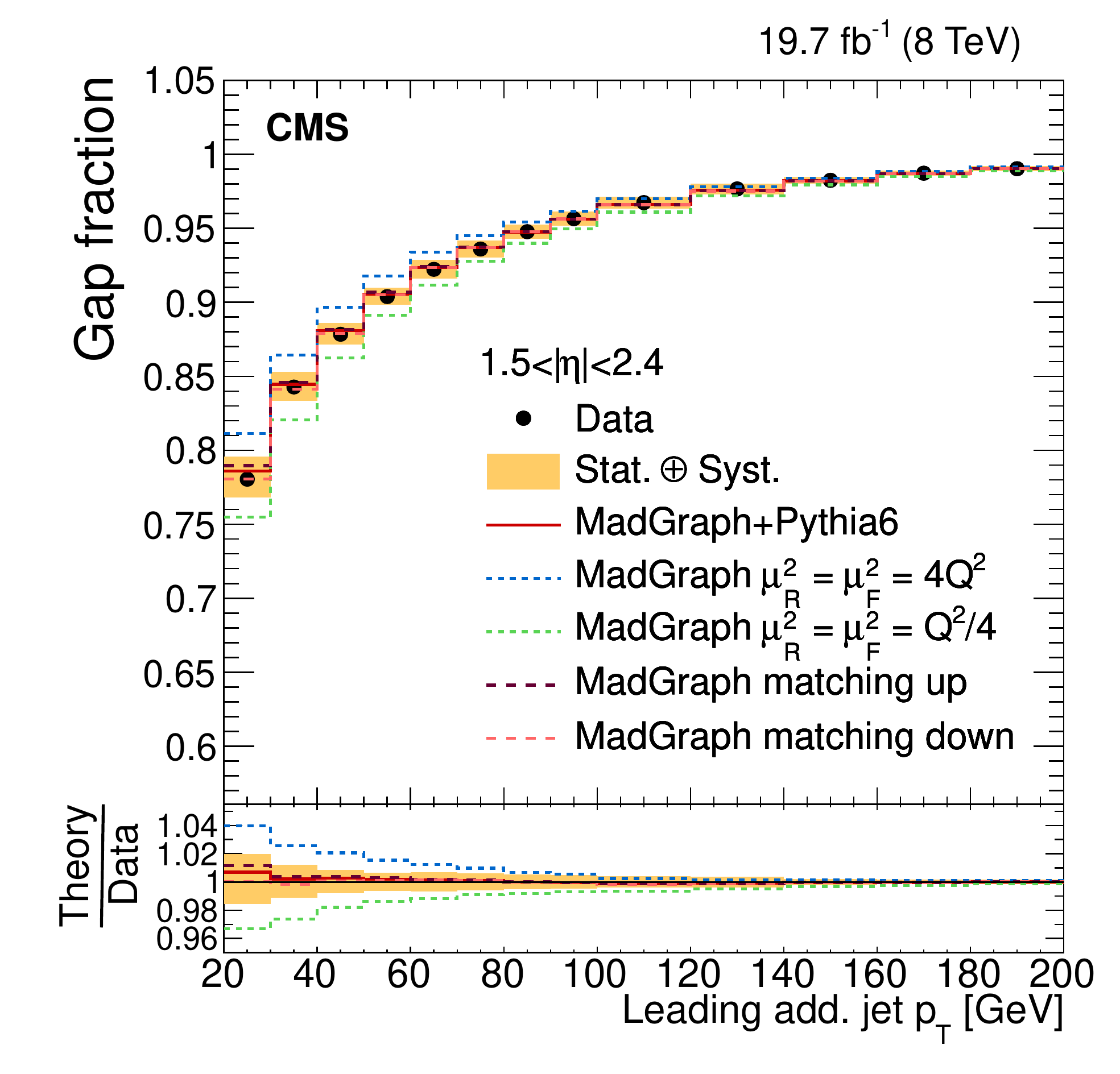}
\caption{Measured gap fraction as a function of the leading additional jet \pt in different $\eta$ regions. Data are compared to predictions from \MADGRAPH, \POWHEG interfaced with \PYTHIA{6} and \HERWIG{6}, and \MCATNLO interfaced with \HERWIG{6} (left) and to \MADGRAPH with varied renormalization, factorization, and jet-parton matching scales (right). For each bin the threshold is defined at the value where the data point is placed. The vertical bars on the data points indicate the statistical uncertainty. The shaded band corresponds to the statistical uncertainty and the total systematic uncertainty added in quadrature. The lower part of each plot shows the ratio of the predictions to the data.}
 \label{fig:Gap1eta}
    \end{center}
\end{figure*}

\begin{figure*}[htbp!]
  \begin{center}
      \includegraphics[width=0.40 \textwidth]{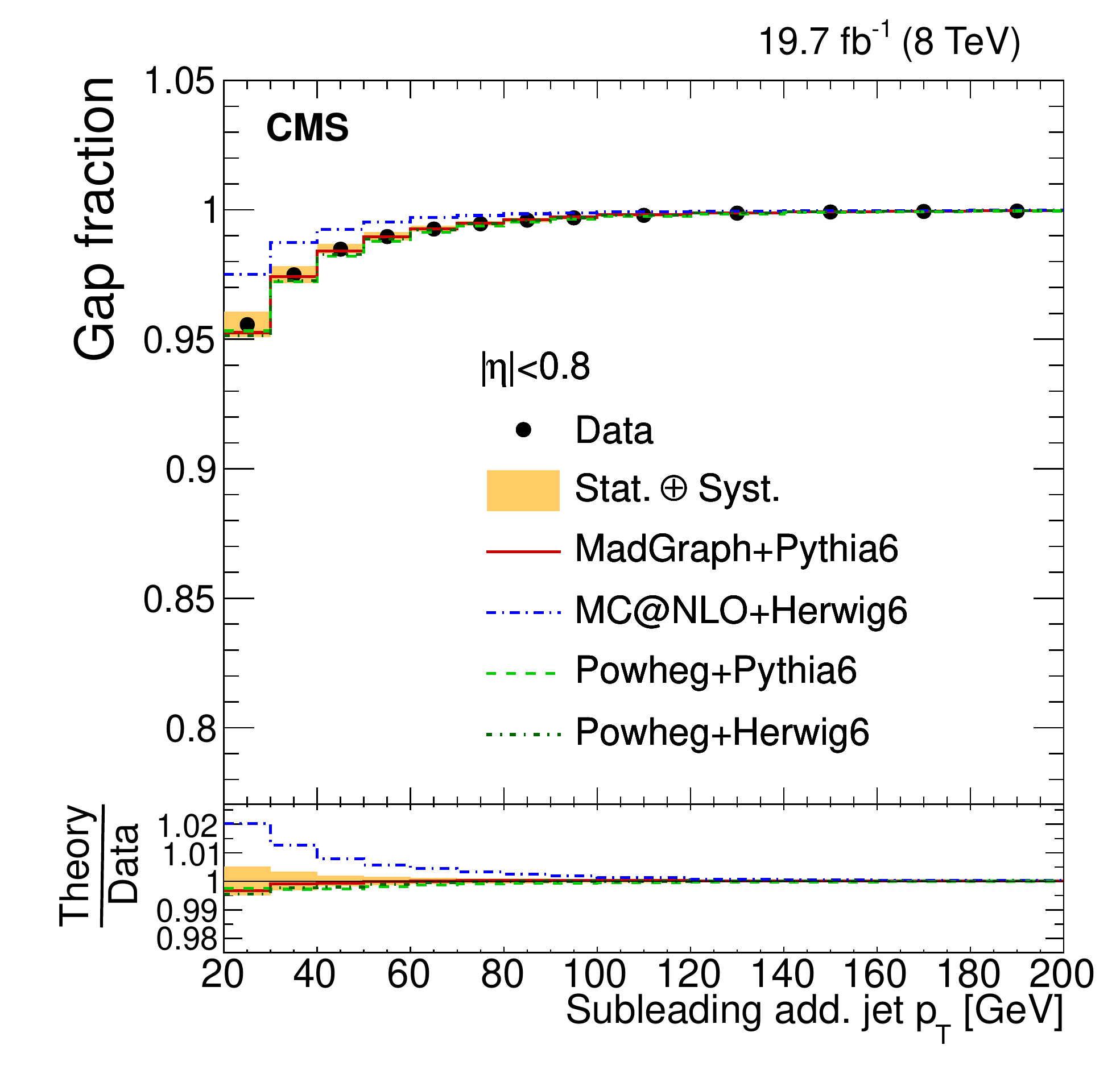}%
      \includegraphics[width=0.40 \textwidth]{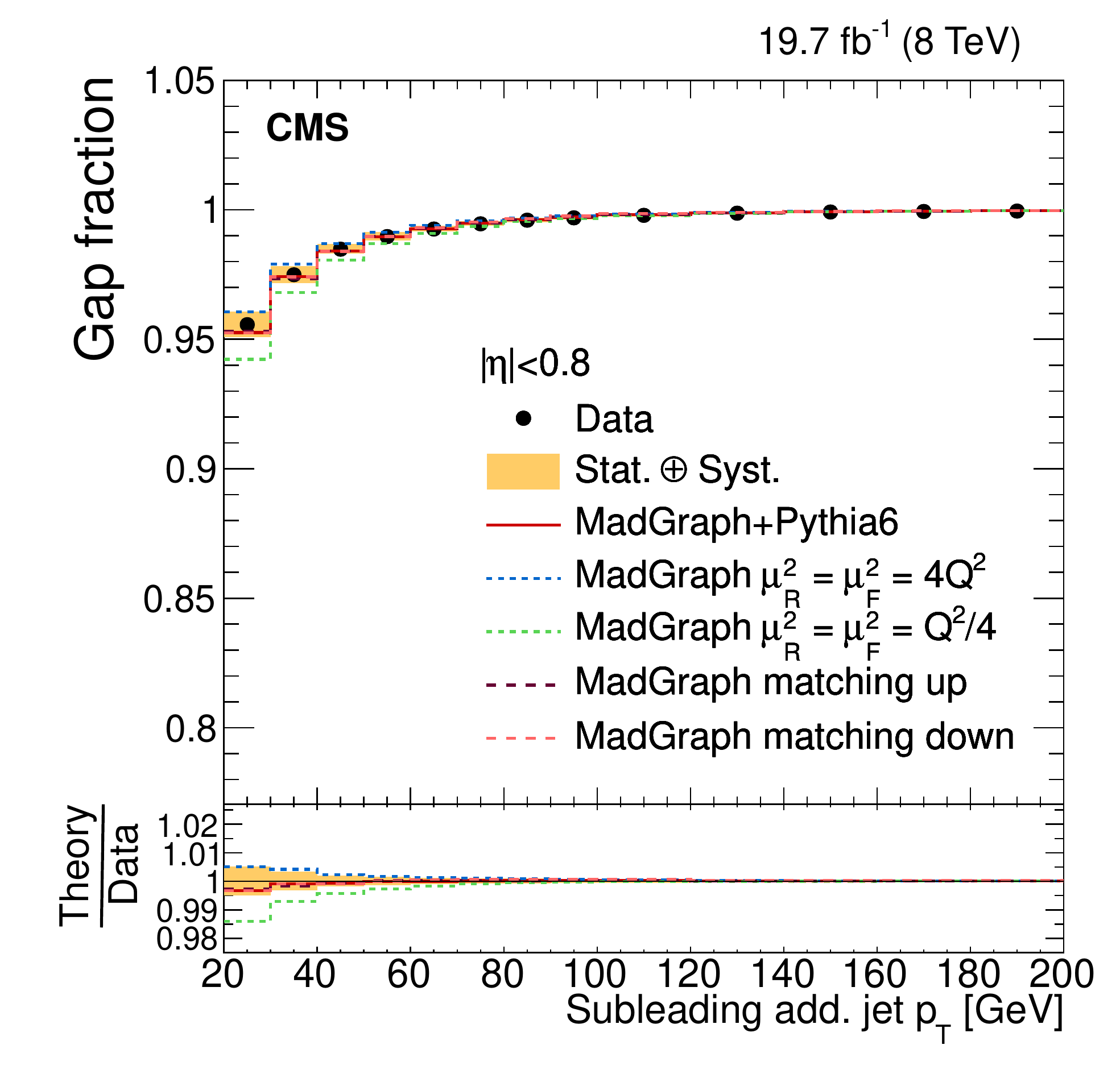}\\
      \includegraphics[width=0.40 \textwidth]{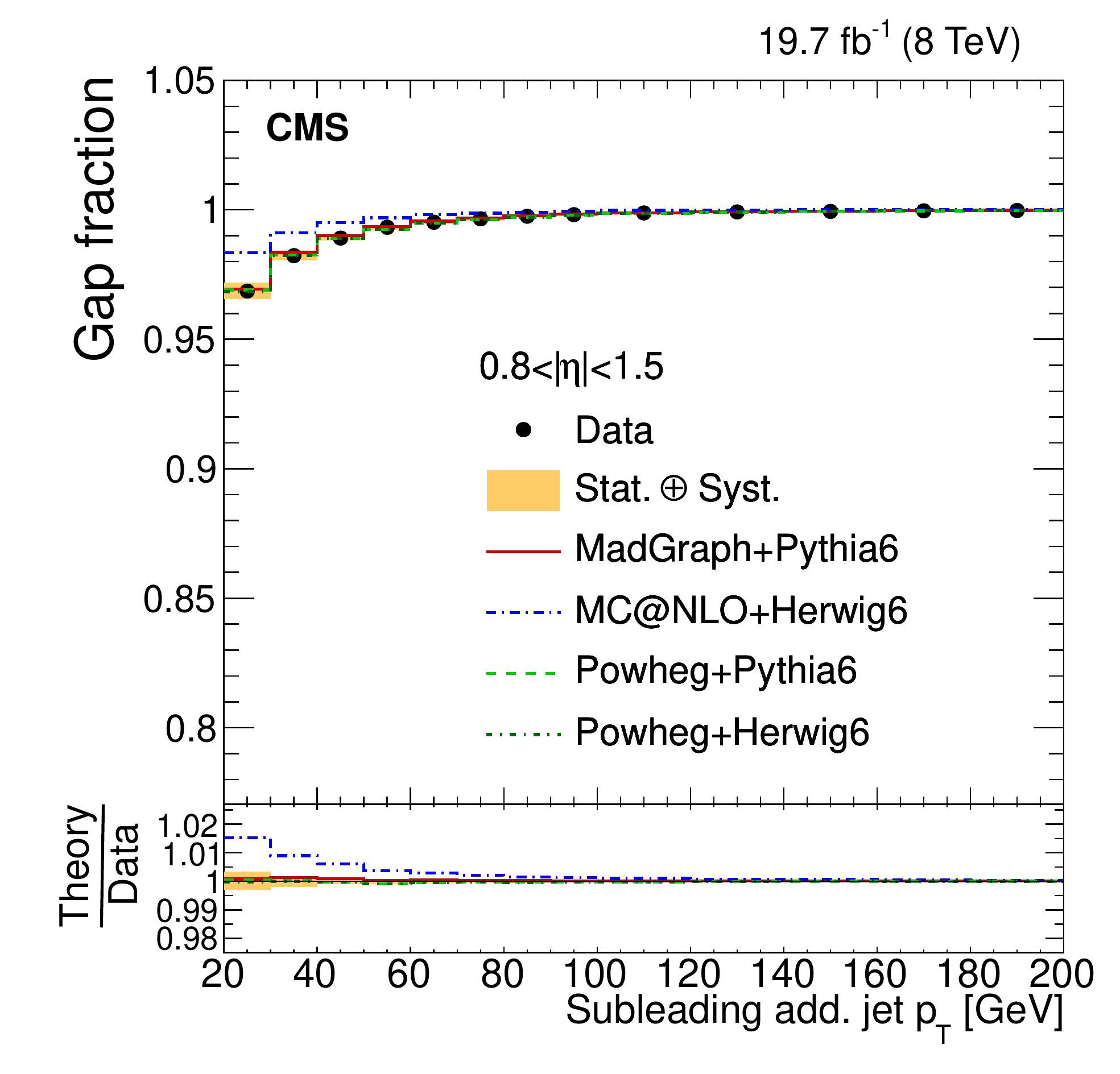}%
      \includegraphics[width=0.40 \textwidth]{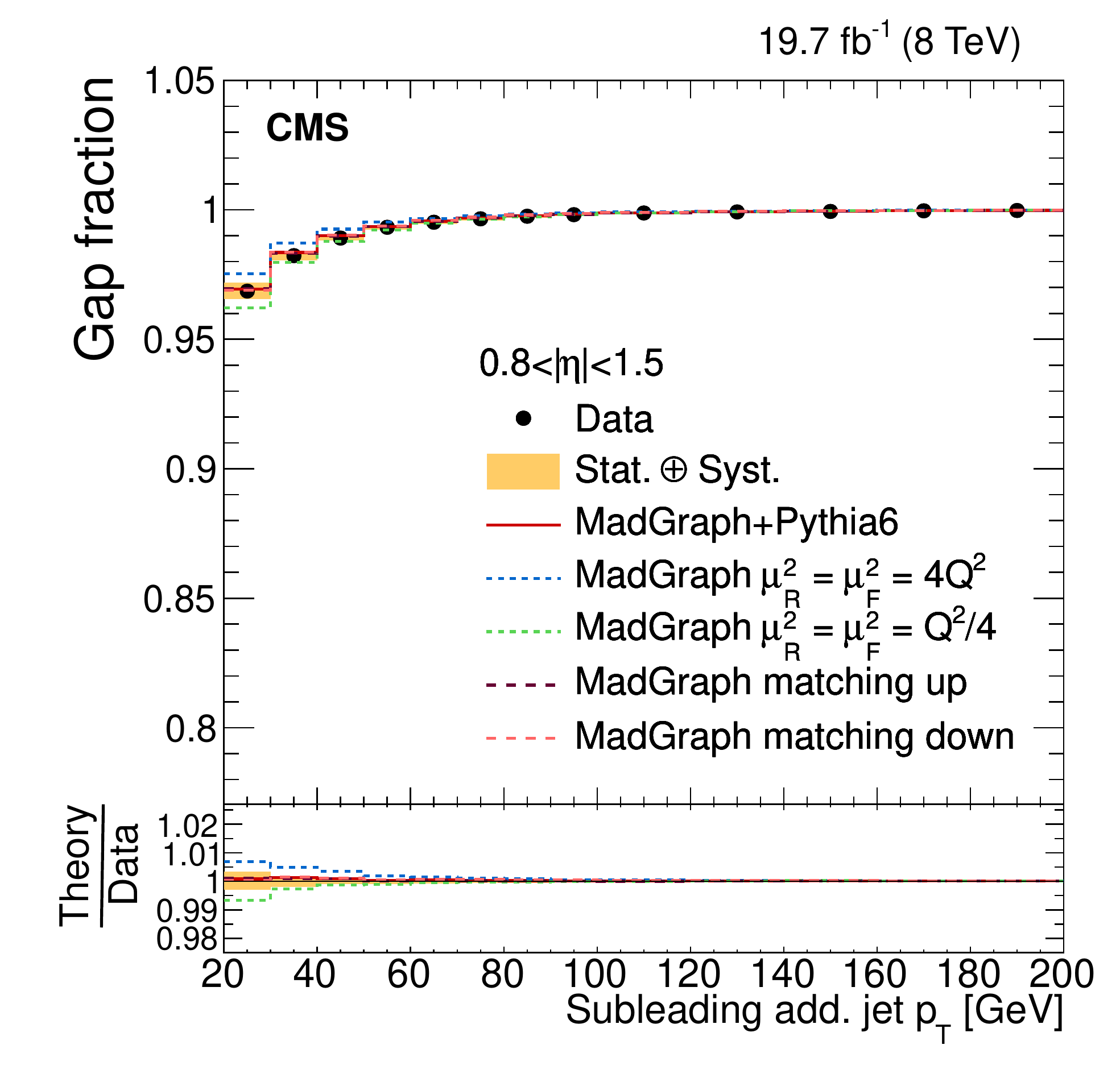}\\
      \includegraphics[width=0.40 \textwidth]{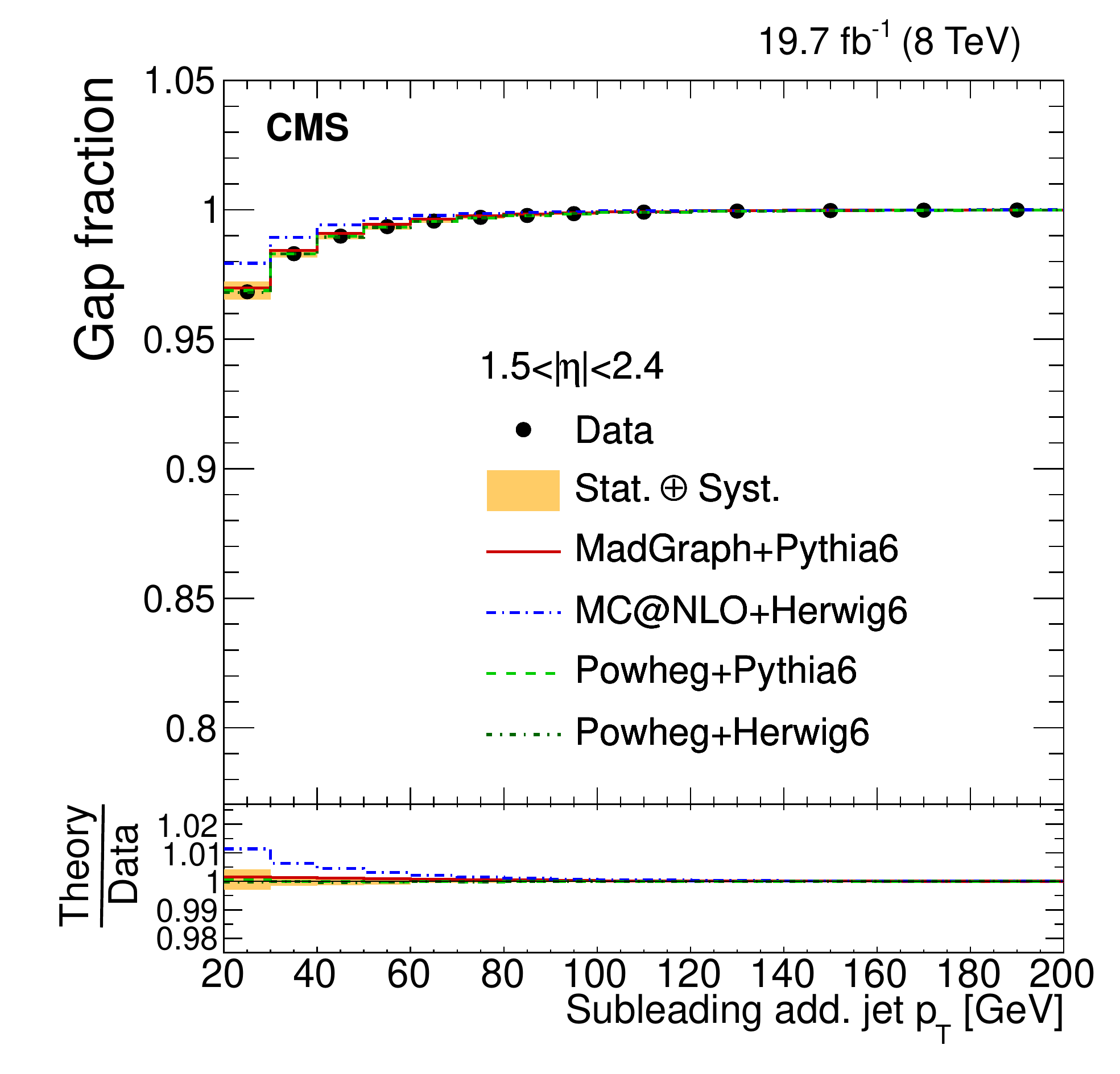}%
      \includegraphics[width=0.40 \textwidth]{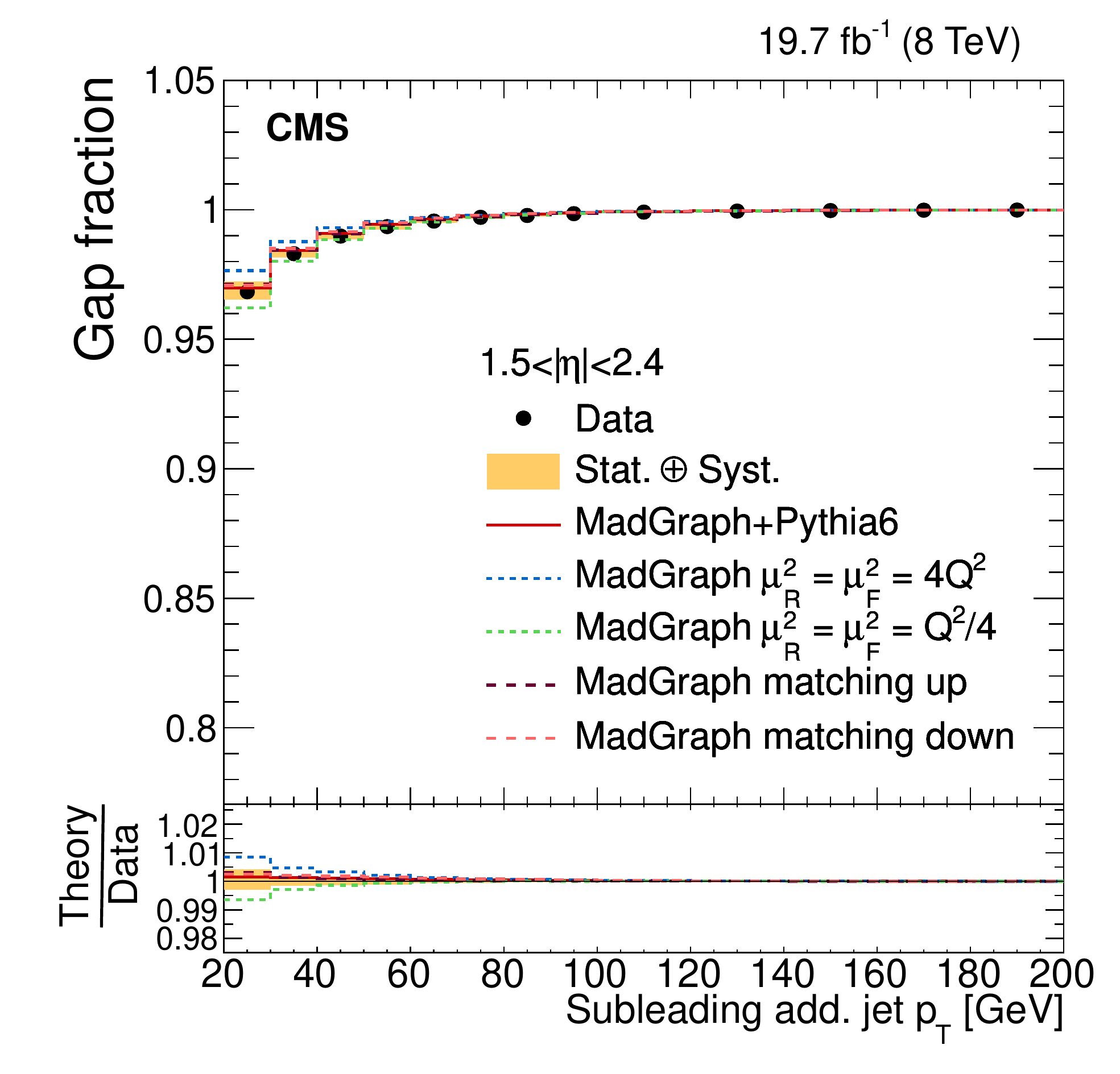}
\caption{Measured gap fraction as a function of the subleading additional jet \pt in different \abseta regions. Data are compared to predictions from \MADGRAPH, \POWHEG interfaced with \PYTHIA{6} and \HERWIG{6}, and \MCATNLO interfaced with \HERWIG{6} (left) and to \MADGRAPH with varied with varied renormalization, factorization, and jet-parton matching scales (right). For each bin the threshold is defined at the value where the data point is placed. The vertical bars on the data points indicate the statistical uncertainty. The shaded band corresponds to the statistical uncertainty and the total systematic uncertainty added in quadrature. The lower part of each plot shows the ratio of the predictions to the data.}
 \label{fig:Gap2eta}
   \end{center}
\end{figure*}

\begin{figure*}[htbp!]
  \begin{center}
      \includegraphics[width=0.40 \textwidth]{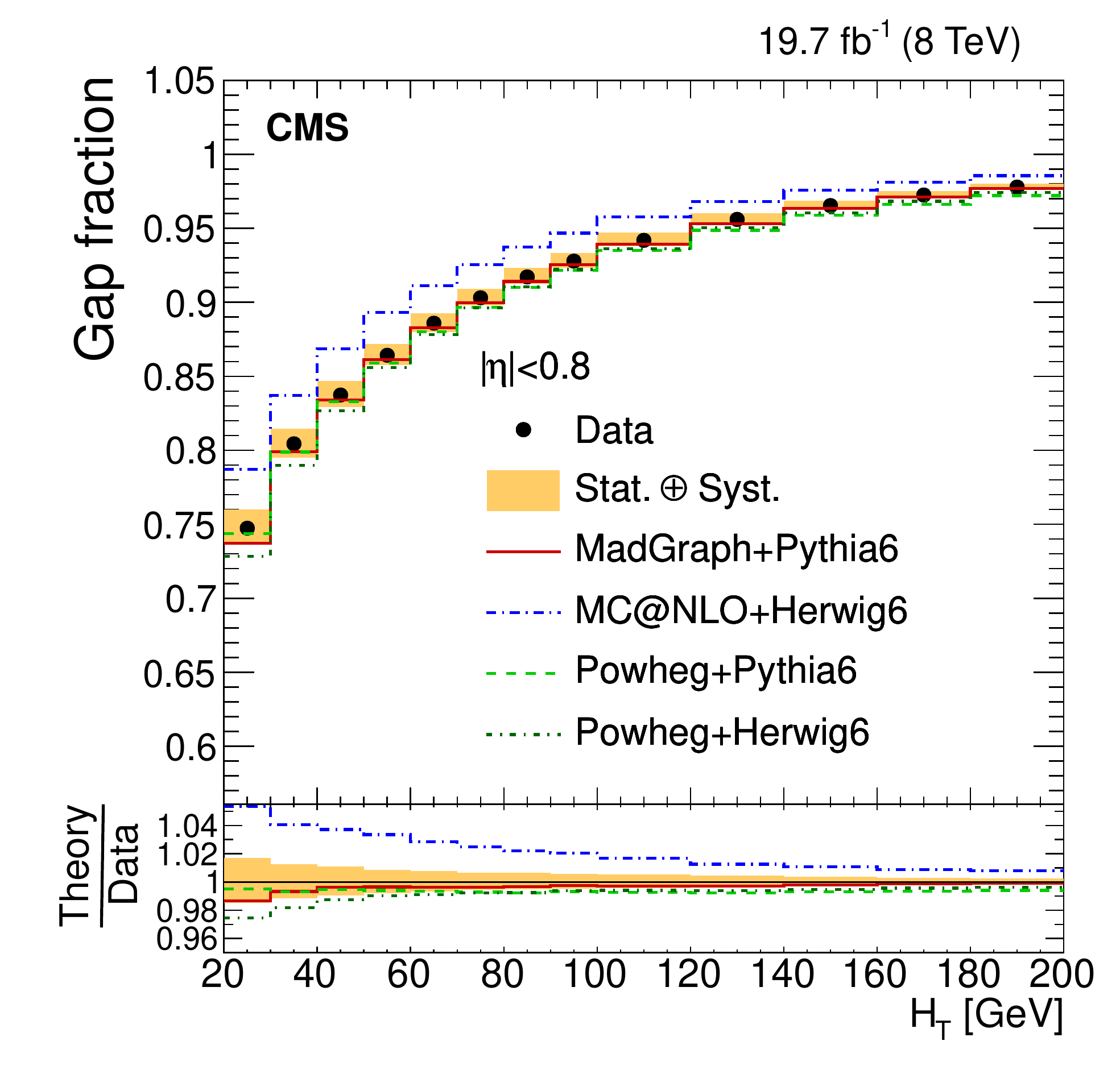}%
      \includegraphics[width=0.40 \textwidth]{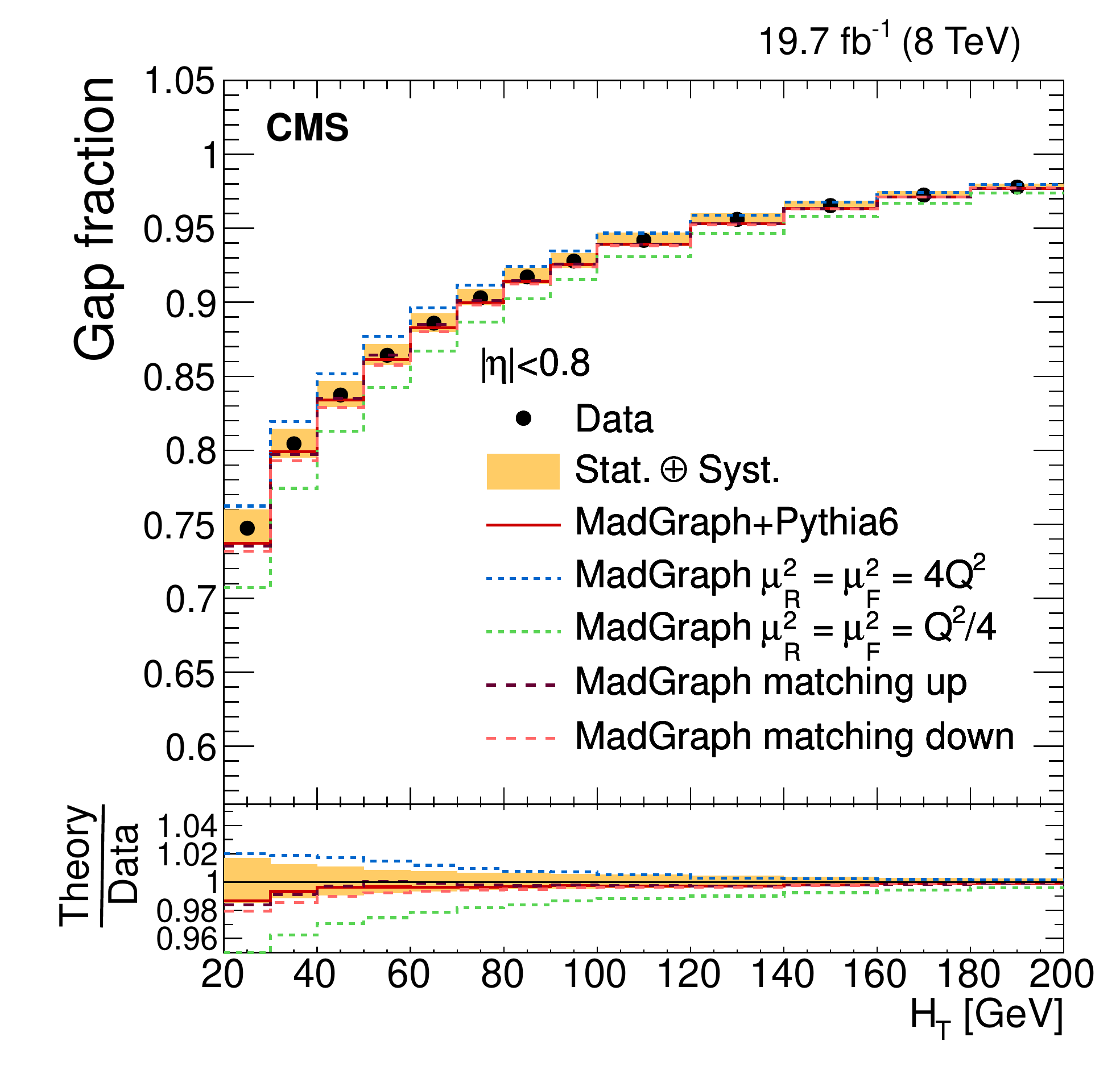}\\
      \includegraphics[width=0.40 \textwidth]{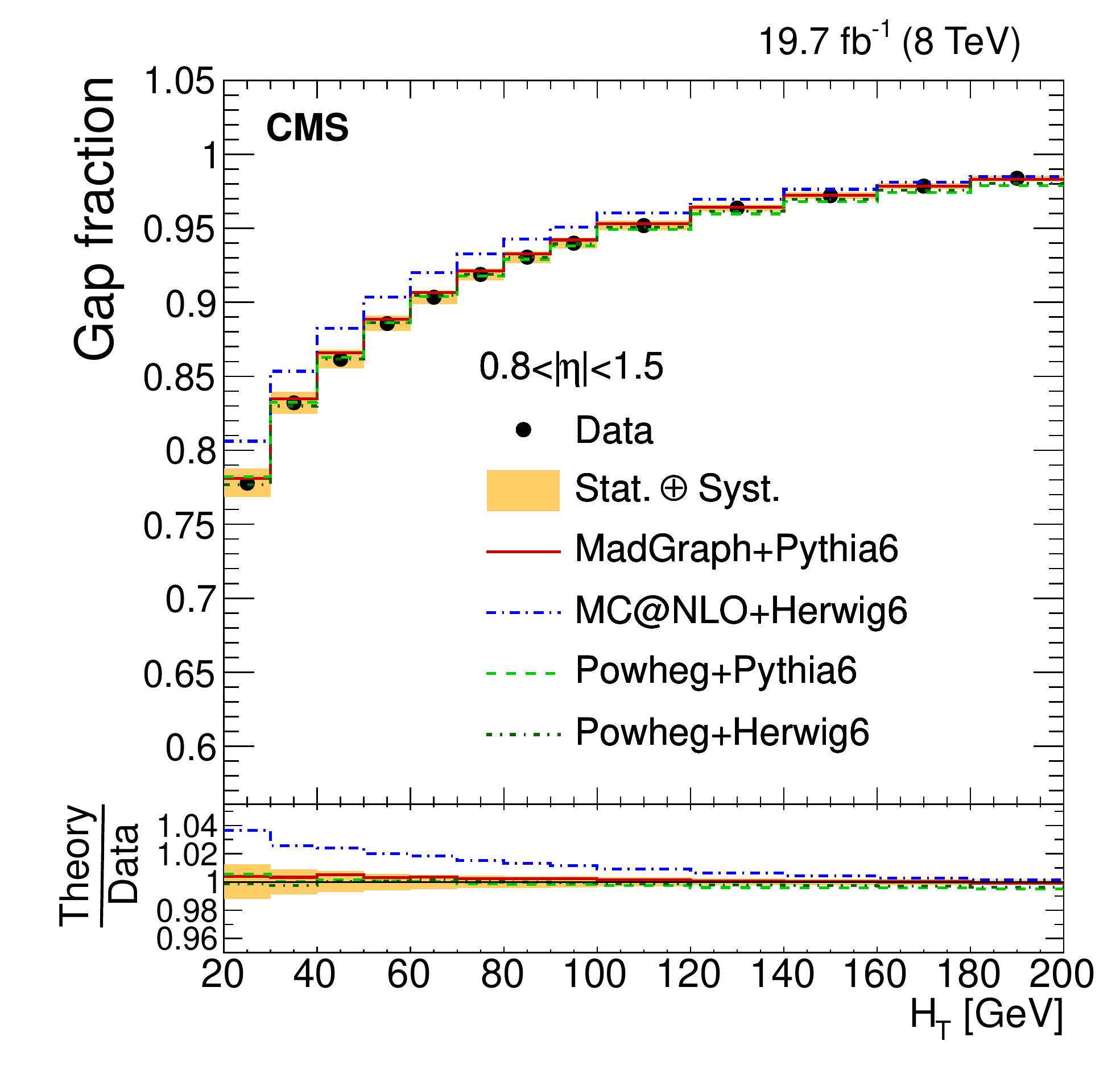}%
      \includegraphics[width=0.40 \textwidth]{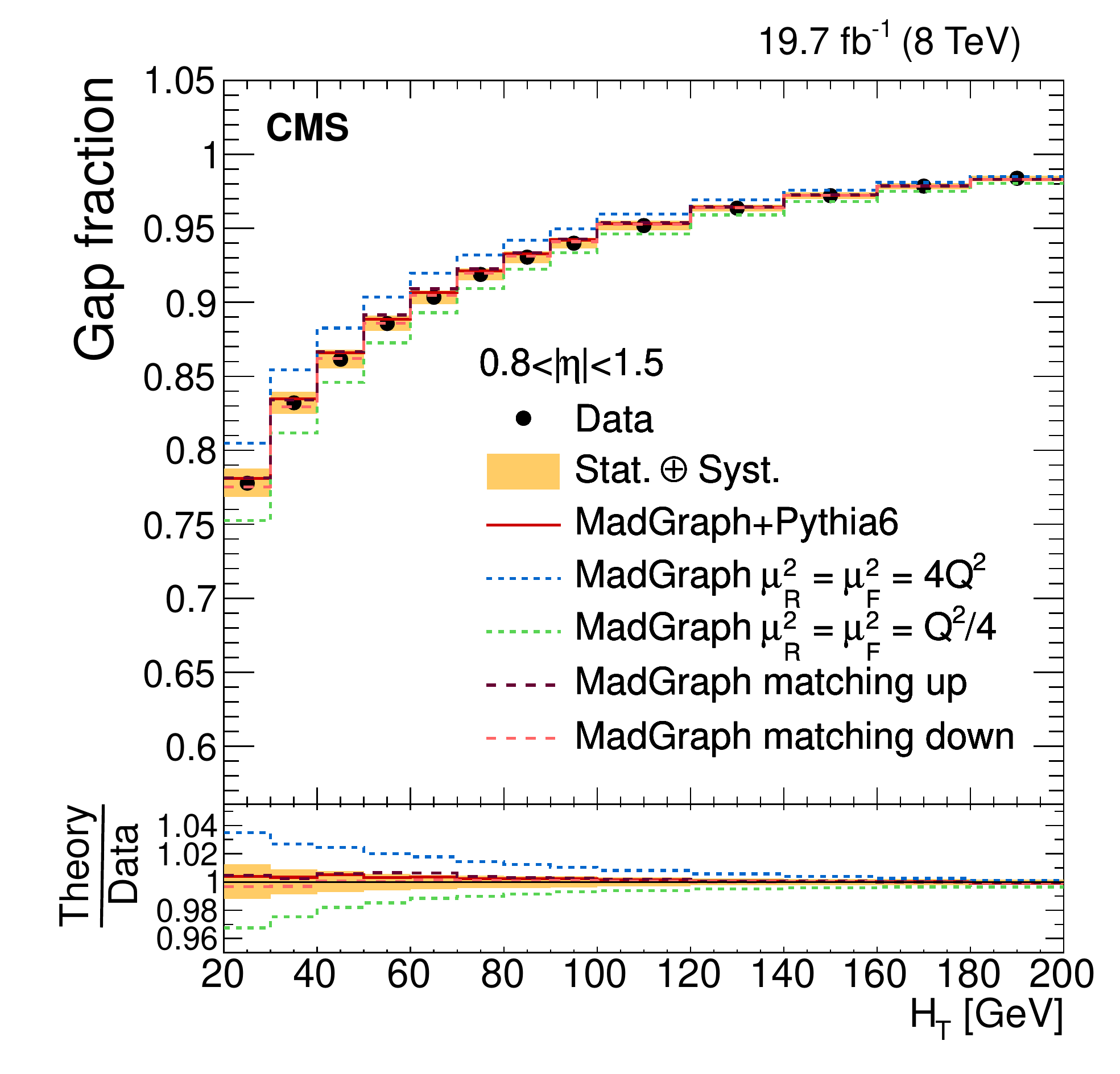}\\
      \includegraphics[width=0.40 \textwidth]{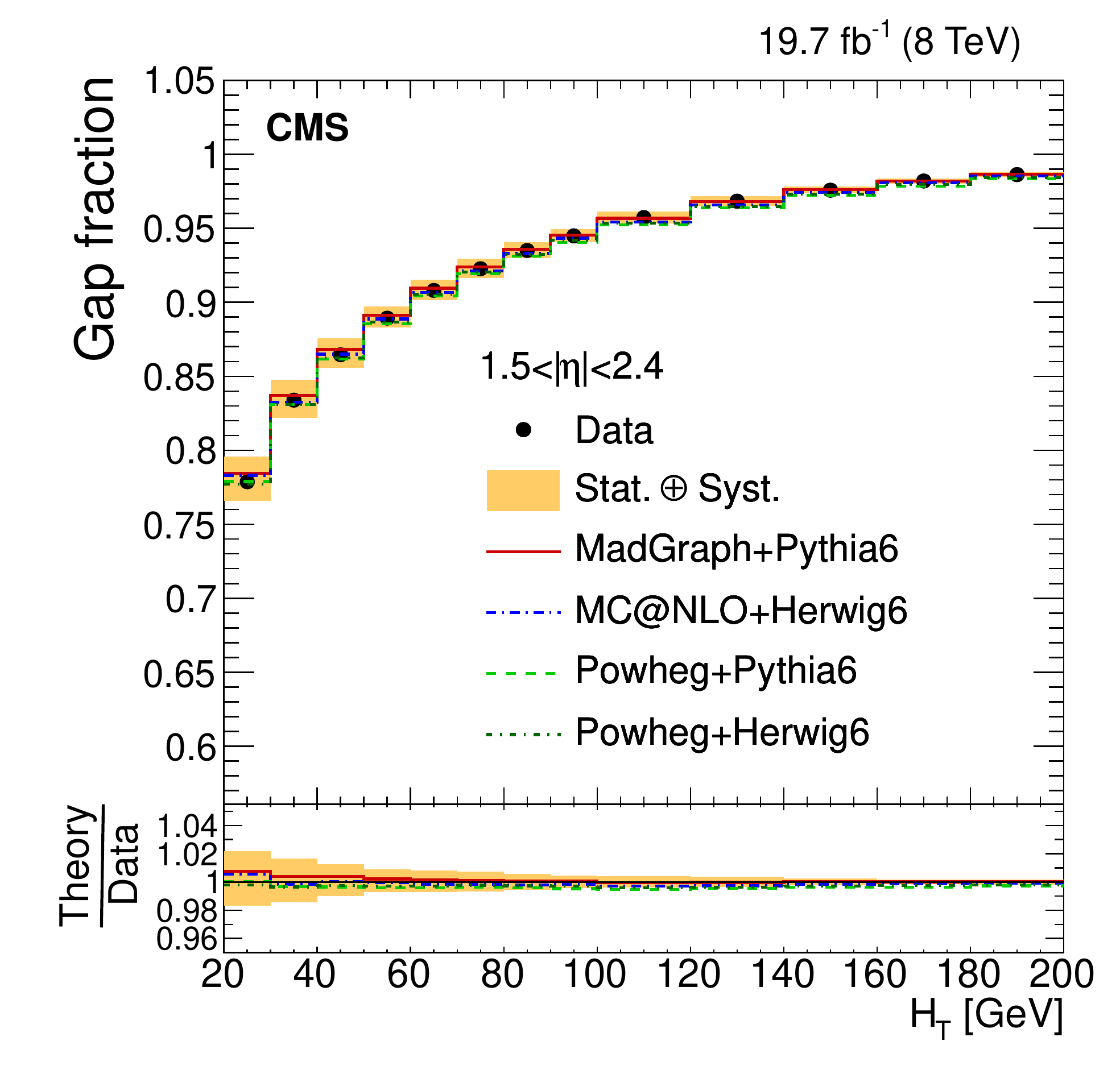}%
      \includegraphics[width=0.40 \textwidth]{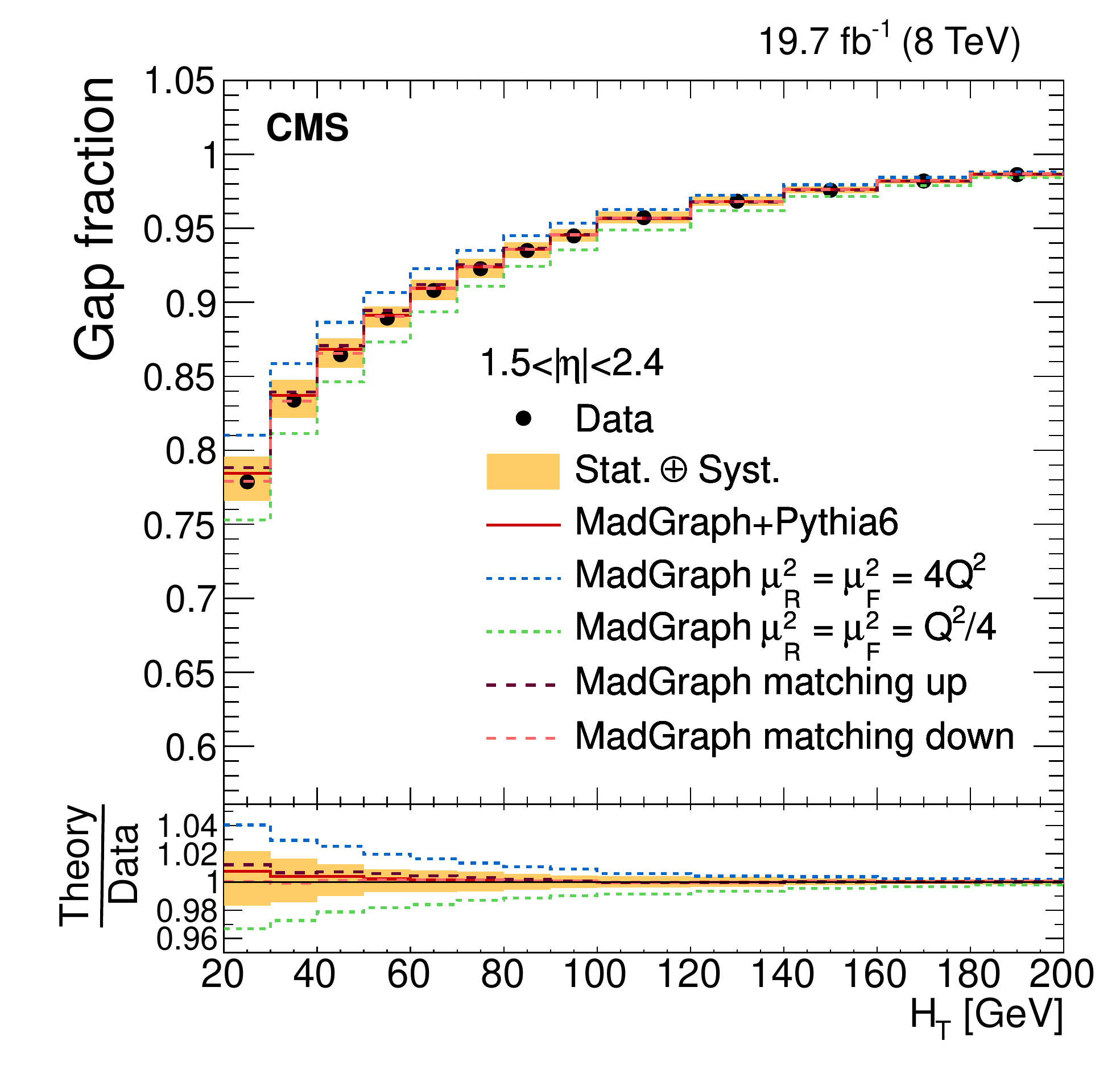}
 \caption{Measured gap fraction as a function of $\HT$ in different $\eta$ regions. Results in data are compared to the nominal \MADGRAPH signal sample, \POWHEG and \MCATNLO (left) and to the samples with varied renormalization, factorization, and jet-parton matching scales (right). For each bin the threshold is defined at the value where the data point is placed. The vertical bars on the data points indicate the statistical uncertainty. The shaded band corresponds to the statistical uncertainty and the total systematic uncertainty added in quadrature. The lower part of each plot shows the ratio of the predictions to the data.}
 \label{fig:GapHTeta}
    \end{center}
\end{figure*}

The gap fraction is also measured in different \abseta regions of the additional jets, with the results presented in Figs.~\ref{fig:Gap1eta}--~\ref{fig:GapHTeta} as a function of the leading additional jet \pt, subleading additional jet \pt, and $\HT$, respectively. In general, the gap fraction values predicted by the simulations describe the data better in the higher \abseta ranges. The values given by \MADGRAPH and \POWHEG interfaced with \PYTHIA{6} are slightly below the measured ones in the central region for the leading \pt jet and $\HT$, while \MCATNLO{}+\HERWIG{6} yields higher values of the gap fraction. In the case of the subleading jet \pt, all predictions agree with the data within the uncertainties, except for \MCATNLO{}+\HERWIG{6} in the more central regions. Variations of the jet-parton matching threshold do not have a noticeable impact on the gap fraction, while \MADGRAPH with the varied renormalization and factorization scales provides a poorer description of the data.

The total systematic uncertainty in the gap fraction distributions is about 5\% for low values of the threshold (\pt or $\HT$) and decreases to ${<}0.5\%$ for the highest values. The measurement of the gap fraction as a function of $\HT$ has larger uncertainties because of the impact of the lower-momentum jets that have a significantly larger uncertainty, as discussed in Section~\ref{sec:diffxsecJets}. The uncertainty in JES is the dominant source of systematic uncertainty, corresponding to approximately 4\% for the smallest \pt and $\HT$ values. Other sources with a smaller impact on the total uncertainty are the \PQb tagging efficiency, JER, pileup, and the simulated sample used to correct the data to the particle level.

\section{Summary}
\label{sec:summary}
Measurements of the absolute and normalized differential top quark pair production cross sections have been presented using pp collisions at a centre-of-mass energy of 8\TeV, corresponding to an integrated luminosity of 19.7\fbinv, in the dilepton decay channel as a function of the number of jets in the event, for three different jet \pt thresholds, and as a function of the kinematic variables of the leading and subleading additional jets. The results have been compared to the predictions from \MADGRAPH interfaced with \PYTHIA{6}, \POWHEG interfaced with both \PYTHIA{6} and \HERWIG{6}, \MCATNLO interfaced with \HERWIG{6}, and \MADGRAPH samples with varied renormalization, factorization, and jet-parton matching scales. In general, all these generators are found to give a reasonable description of the data.

The \MADGRAPH and \POWHEG generators interfaced with \PYTHIA{6} describe the data well for all measured jet multiplicities; while \MCATNLO interfaced with \HERWIG{6} generates lower multiplicities than observed for the lower-\pt thresholds. The prediction from \MADGRAPH with varied renormalization and factorization scales does not provide an improved description of the data compared to the nominal simulation.

These results are also compared to the predictions from \POWHEG with the \hdamp parameter set to the top quark mass interfaced with \PYTHIA{6}, \PYTHIA{8}, and \HERWIG{6}, which provide a reasonable description of the data within the uncertainties, and the predictions from \MADGRAPH and \amcatnlo interfaced with \PYTHIA{8}, which generate higher jet multiplicities for all the \pt thresholds.

The measured kinematic variables of the leading and subleading additional jets are consistent with the various predictions. The simulations also describe well the data distributions of the leading additional jet \pt and $\HT$, although they tend to predict higher \pt values and more central values in $\eta$.
\MADGRAPH with varied parameters yields similar predictions, except for varying the renormalization and factorization scales, which tends to give higher $\HT$ values. The \MCATNLO generator predicts lower yields than observed for the subleading additional jet \pt.

The uncertainties in the measured \ttbb(\ttb) absolute and normalized differential cross sections as a function of the \PQb jet kinematic variables are dominated by the statistical uncertainties. In general, the predictions describe well the shape of the measured cross sections as a function of the variables studied, except for $\Delta R_{\PQb\PQb}$, where they favour smaller values than the measurement. The predictions underestimate the total \ttbb cross section by approximately a factor of 2, in agreement with previous measurements~\cite{bib:ttbb_ratio:2014}. The calculation by \PowHel~\cite{Garzelli:2014aba} describes well the shape of the distributions, while the predicted absolute cross section is about 30\% lower, but compatible with the measurements within the uncertainties.

The gap fraction has been measured as a function of the \pt of the leading and subleading additional jets and $\HT$ of the additional jets in different $\eta$ ranges. For a given threshold value, the gap fraction as a function of $\HT$ is lower than the gap fraction as a function of the \pt of the leading additional jet, showing that the measurement is probing multiple quark and gluon emission. Within the uncertainties, all predictions describe the gap fraction well as a function of the momentum of the first additional jet, while \MCATNLO interfaced with \HERWIG fails to describe the gap fraction as a function of the subleading additional jet \pt and $\HT$. In general, \MADGRAPH with decreased renormalization and factorization scales more poorly describes the observed gap fraction, while varying the jet-parton matching threshold provides a similar description of the data. The \MADGRAPH and \amcatnlo generators interfaced with \PYTHIA{8} predict lower values than measured. The \POWHEG simulation with $\hdamp = m_{\PQt}$ interfaced with \PYTHIA{8} is consistent with the data, while the simulation interfaced with \HERWIG{6} and \PYTHIA{6} tends to worsen the comparison with the measurement.

In general, the different measurements presented are in agreement with the SM predictions as formulated by the various event generators, within their uncertainties. The correct description of \ttbar{}+jets production is important since it constitutes a major background in searches for new particles in several supersymmetric models and in \ttH processes, where the Higgs boson decays into $\bbbar$. The \ttbb(\ttb) differential cross sections, measured here for the first time, also provide important information about the main irreducible background in the search for \ttHtobb.

\section*{Acknowledgements}
\hyphenation{Bundes-ministerium Forschungs-gemeinschaft Forschungs-zentren}

{\tolerance=500
We thank M. V. Garzelli for providing the theoretical predictions from \PowHel{}+\PYTHIA{6}.
We congratulate our colleagues in the CERN accelerator departments for the excellent performance of the LHC and thank the technical and administrative staffs at CERN and at other CMS institutes for their contributions to the success of the CMS effort. In addition, we gratefully acknowledge the computing centres and personnel of the Worldwide LHC Computing Grid for delivering so effectively the computing infrastructure essential to our analyses. Finally, we acknowledge the enduring support for the construction and operation of the LHC and the CMS detector provided by the following funding agencies: the Austrian Federal Ministry of Science, Research and Economy and the Austrian Science Fund; the Belgian Fonds de la Recherche Scientifique, and Fonds voor Wetenschappelijk Onderzoek; the Brazilian Funding Agencies (CNPq, CAPES, FAPERJ, and FAPESP); the Bulgarian Ministry of Education and Science; CERN; the Chinese Academy of Sciences, Ministry of Science and Technology, and National Natural Science Foundation of China; the Colombian Funding Agency (COLCIENCIAS); the Croatian Ministry of Science, Education and Sport, and the Croatian Science Foundation; the Research Promotion Foundation, Cyprus; the Ministry of Education and Research, Estonian Research Council via IUT23-4 and IUT23-6 and European Regional Development Fund, Estonia; the Academy of Finland, Finnish Ministry of Education and Culture, and Helsinki Institute of Physics; the Institut National de Physique Nucl\'eaire et de Physique des Particules~/~CNRS, and Commissariat \`a l'\'Energie Atomique et aux \'Energies Alternatives~/~CEA, France; the Bundesministerium f\"ur Bildung und Forschung, Deutsche Forschungsgemeinschaft, and Helmholtz-Gemeinschaft Deutscher Forschungszentren, Germany; the General Secretariat for Research and Technology, Greece; the National Scientific Research Foundation, and National Innovation Office, Hungary; the Department of Atomic Energy and the Department of Science and Technology, India; the Institute for Studies in Theoretical Physics and Mathematics, Iran; the Science Foundation, Ireland; the Istituto Nazionale di Fisica Nucleare, Italy; the Ministry of Science, ICT and Future Planning, and National Research Foundation (NRF), Republic of Korea; the Lithuanian Academy of Sciences; the Ministry of Education, and University of Malaya (Malaysia); the Mexican Funding Agencies (CINVESTAV, CONACYT, SEP, and UASLP-FAI); the Ministry of Business, Innovation and Employment, New Zealand; the Pakistan Atomic Energy Commission; the Ministry of Science and Higher Education and the National Science Centre, Poland; the Funda\c{c}\~ao para a Ci\^encia e a Tecnologia, Portugal; JINR, Dubna; the Ministry of Education and Science of the Russian Federation, the Federal Agency of Atomic Energy of the Russian Federation, Russian Academy of Sciences, and the Russian Foundation for Basic Research; the Ministry of Education, Science and Technological Development of Serbia; the Secretar\'{\i}a de Estado de Investigaci\'on, Desarrollo e Innovaci\'on and Programa Consolider-Ingenio 2010, Spain; the Swiss Funding Agencies (ETH Board, ETH Zurich, PSI, SNF, UniZH, Canton Zurich, and SER); the Ministry of Science and Technology, Taipei; the Thailand Center of Excellence in Physics, the Institute for the Promotion of Teaching Science and Technology of Thailand, Special Task Force for Activating Research and the National Science and Technology Development Agency of Thailand; the Scientific and Technical Research Council of Turkey, and Turkish Atomic Energy Authority; the National Academy of Sciences of Ukraine, and State Fund for Fundamental Researches, Ukraine; the Science and Technology Facilities Council, UK; the US Department of Energy, and the US National Science Foundation.

Individuals have received support from the Marie-Curie programme and the European Research Council and EPLANET (European Union); the Leventis Foundation; the A. P. Sloan Foundation; the Alexander von Humboldt Foundation; the Belgian Federal Science Policy Office; the Fonds pour la Formation \`a la Recherche dans l'Industrie et dans l'Agriculture (FRIA-Belgium); the Agentschap voor Innovatie door Wetenschap en Technologie (IWT-Belgium); the Ministry of Education, Youth and Sports (MEYS) of the Czech Republic; the Council of Science and Industrial Research, India; the HOMING PLUS programme of the Foundation for Polish Science, cofinanced from European Union, Regional Development Fund; the OPUS programme of the National Science Center (Poland); the Compagnia di San Paolo (Torino); the Consorzio per la Fisica (Trieste); MIUR project 20108T4XTM (Italy); the Thalis and Aristeia programmes cofinanced by EU-ESF and the Greek NSRF; the National Priorities Research Program by Qatar National Research Fund; the Rachadapisek Sompot Fund for Postdoctoral Fellowship, Chulalongkorn University (Thailand); and the Welch Foundation, contract C-1845.
\par}

\bibliography{auto_generated}
\newpage

\numberwithin{table}{section}

\appendix

\section{BDT variables}
\label{ap:mvaVariables}

The variables used for the BDT are listed below. The candidate \PQb jet is denoted with the superscript \PQb in the following equations, while the candidate anti-\PQb jet is denoted as~$\PAQb$.
Combinations of particles that are treated as a system by adding their four-momentum vectors are denoted without a comma, \eg $\PQb\ell^+$ represents the \PQb jet and the antilepton system. The angular separation $\Delta R = \sqrt{\smash[b]{(\Delta \eta)^2 + (\Delta \phi)^2}}$ and the azimuthal angular difference $\Delta \phi$ between the directions of two particles is designated using the two particle abbreviations in a superscript, separated by a comma.

One variable is the difference in the jet charges, $c_\text{rel}$, of the \PQb and \PAQb jets:
\begin{itemize}
\item[$\bullet$]  $c_\text{rel}^{\bar{\text{b}}} - c_\text{rel}^{\text{b}}$
\end{itemize}
It is the only variable not directly related to the kinematical  properties of the \ttbar decay and the additional radiation. The values are by definition positive, as the jet with the highest charge is always assigned as the anti-\PQb jet.

There are three angular variables:
\begin{itemize}
\item[$\bullet$]  $0.5\, \left( \lvert \Delta\phi^{\mathrm{b},\Vptmiss} \rvert + \lvert \Delta\phi^{\mathrm{\bar{b}},\Vptmiss} \rvert \right)$
\item[$\bullet$]  $ \lvert \Delta\phi^{\mathrm{b\ell}^{+},\mathrm{\bar{b}\ell^{-}}} \rvert $
\item[$\bullet$]  $\Delta{R}^{\mathrm{b,\ell^{+}}}$ \quad and \quad $\Delta{R}^{\mathrm{\bar{b},\ell^{-}}}$
\end{itemize}
Here, \Vptmiss denotes the missing transverse momentum in an event. The angles are defined such that $-\pi\leq\Delta\phi\leq\pi$, and consequently the absolute values are within $[0, \pi]$.

Two variables are the \pt of the \PQb jet (\PAQb jet) and charged antilepton (lepton) systems:
\begin{itemize}
\item[$\bullet$]  $\pt^{\PQb\ell^+}$ \quad and \quad $\pt^{\PAQb\ell^-}$
\end{itemize}

The remaining variables are based on the invariant or transverse masses of several particle combinations:
\begin{itemize}
\item[$\bullet$]  $m^{\PQb\ell^+} + m^{\PAQb\ell^-}$
\item[$\bullet$]  $m^{\PQb\ell^+} - m^{\PAQb\ell^-}$
\item[$\bullet$]  $m^{\PQb\PAQb\ell^+\ell^-} - m^{\bbbar}$
\item[$\bullet$]  $m^\mathrm{jets}_{\text{recoil}} - m^{\bbbar}$
\item[$\bullet$]  $0.5\, \left( m^{\PQb\Vptmiss}_{\text{T}} + m^{\PAQb\Vptmiss}_{\text{T}} \right) $
\end{itemize}
For any pair of jets, the variable $m^\mathrm{jets}_{\text{recoil}}$ is the invariant mass of all the other selected jets recoiling against this pair, \ie all selected jets except these two.

\clearpage
\section{Summary tables of absolute and normalized cross section measurements}
\label{sec:summarytables}
\begin{table*}[h]
 \renewcommand{\arraystretch}{1.2}
  \begin{center}
    \topcaption{Absolute (left) and normalized (right) differential \ttbar cross sections as a function of the jet multiplicity ($N_{\text{jets}}$) for jets with $\pt > 30\GeV$ (top), $\pt > 60\GeV$ (middle), and $\pt > 100\GeV$ (bottom), along with their statistical, systematic, and total uncertainties. The results are presented at the particle level in the visible phase space of the \ttbar decay products and the additional jets.}
    \label{tab:dilepton:SummaryResultsJetMult}
   \resizebox{\textwidth}{!}{
    \begin{tabular}{c|x{4}ccc|x{-1}ccc}
    \multicolumn{9}{c}{ $\pt>30\GeV$ } \\
    \hline
     \multirow{2}{*}{$N_{\text{jets}}$} &
     \multicolumn{1}{c}{ $ { \rd \sigma^{\text{vis}} }/{ \rd N_{\text{jets}} }  $ } &
     stat. & syst. & tot. &
     \multicolumn{1}{c}{ \multirow{2}{*}{ $({1}/{\sigma^{\text{vis}}})({\rd \sigma^{\text{vis}}}/{\rd N_{\text{jets}}})$ } }  &
     stat. & syst. & tot. \\ [-2pt]
     &  \multicolumn{1}{c}{ (pb) } & (\%) &(\%) &(\%) & &  (\%) & (\%) &(\%)
      \\
\hline
$ 2$ & \multicolumn{1}{c}{$2.59$} & 0.6 & 5.8 & 5.8   & 5.38 , 10^{-1} & 0.6 & 3.6 & 3.6  \\
$ 3$ &\multicolumn{1}{c}{$1.43$} & 1.5 & 10 & 10 & 2.95 , 10^{-1} & 1.2 & 3.8 & 4.0  \\
$ 4$ & 5.1 , 10^{-1} & 2.2 & 14 & 14 & 1.05 , 10^{-1} & 2.1 & 9.3 & 9.5  \\
$ 5$ & 1.5 , 10^{-1} & 3.6 & 28 & 28 & 3.1 , 10^{-2} & 3.5 & 28 & 29\\
$ \geq 6$ & 5.0 , 10^{-2} & 6.4 & 20 & 21 & 1.1 , 10^{-2} & 6.2 & 16 & 17\\
    \hline
   \multicolumn{9}{c}{ $\pt>60\GeV$ } \\
   \hline
     \multirow{2}{*}{$N_{\text{jets}}$} &
     \multicolumn{1}{c}{ $ { \rd \sigma^{\text{vis}} }/{ \rd N_{\text{jets}} }  $ } &
     stat. & syst. & tot. &
     \multicolumn{1}{c}{ \multirow{2}{*}{ $({1}/{\sigma^{\text{vis}}})({\rd \sigma^{\text{vis}}}/{\rd N_{\text{jets}}})$ } }  &
     stat. & syst. & tot. \\ [-2pt]
     &  \multicolumn{1}{c}{ (pb) } & (\%) &(\%) &(\%) & &  (\%) & (\%) &(\%)
      \\
\hline
$ 0$ & 5.56 , 10^{-1} & 1.5 & 7.4 & 7.5  & 1.17 , 10^{-1} & 2.0 & 5.9 & 6.2 \\
$ 1$ & \multicolumn{1}{c}{$1.73$} & 2.0 & 6.8 & 7.1  & 3.67 , 10^{-1} & 1.4 & 1.9 & 2.3 \\
$ 2$ & \multicolumn{1}{c}{$1.87$} & 1.2 & 5.9 & 6.1  & 3.93 , 10^{-1} & 1.0 & 1.8 & 2.1 \\
$ 3$ & 4.73 , 10^{-1} & 2.2 & 8.4 & 8.6  & 9.85 , 10^{-2} & 2.1 & 3.7 & 4.3 \\
$ 4$ & 9.2 , 10^{-2} & 4.6 & 19 & 19 & 2.0 , 10^{-2} & 4.4 & 18 & 18 \\
$ \geq 5$ & 1.9 , 10^{-2} & 9.1 & 13 & 16& 4.2 , 10^{-3} & 8.7 & 9.2 & 13 \\
 \hline
    \multicolumn{9}{c}{ $\pt>100\GeV$ } \\
 \hline
      \multirow{2}{*}{$N_{\text{jets}}$} &
     \multicolumn{1}{c}{ $ { \rd \sigma^{\text{vis}} }/{ \rd N_{\text{jets}} }  $ } &
     stat. & syst. & tot. &
     \multicolumn{1}{c}{ \multirow{2}{*}{ $({1}/{\sigma^{\text{vis}}})({\rd \sigma^{\text{vis}}}/{\rd N_{\text{jets}}})$ } }  &
     stat. & syst. & tot. \\ [-2pt]
     &  \multicolumn{1}{c}{ (pb) } & (\%) &(\%) &(\%) & &  (\%) & (\%) &(\%)
      \\
\hline
$ 0 $ & \multicolumn{1}{c}{$2.66$} & 0.6 & 6.1 & 6.2 & 5.59 , 10^{-1} & 0.6 & 2.9 & 2.9 \\
$ 1 $ & \multicolumn{1}{c}{$1.37$} & 1.6 & 7.6 & 7.7 & 2.92 , 10^{-1} & 1.3 & 4.5 & 4.6 \\
$ 2 $ & 6.00 , 10^{-1} & 2.0 & 6.5 & 6.8 & 1.25 , 10^{-1} & 1.8 & 2.0 & 2.7 \\
$ 3 $ & 9.29 , 10^{-2} & 4.5 & 13 & 14 & 2.0 , 10^{-2} & 4.4 & 13 & 14 \\
$ \geq 4 $ & 1.37 , 10^{-2} & 12 & 14 & 18& 2.9 , 10^{-3} & 11 & 13 & 17 \\
\hline
     \end{tabular}
     }
  \end{center}
\end{table*}

\begin{table*}[h]
  \renewcommand{\arraystretch}{1.2}
  \begin{center}
    \topcaption{Absolute (left) and normalized (right) differential \ttbar cross sections as a function of the \pt (${\pt}^{\,\mathrm{j1}}$) and the $|\eta|$ ($|\eta^{\,\mathrm{j1}}|$) of the leading additional jet in the event (not coming from the top quark decay products), along with their statistical, systematic, and total uncertainties. The results are presented at the particle level in the visible phase space.}
    \label{tab:dilepton:SummaryResultsJet1}
   \resizebox{\textwidth}{!}{
    \begin{tabular}{y|x{4}ccc|x{-1}ccc}
    \hline
    \multicolumn{1}{c|}{   $ {\pt}^{\,\mathrm{j1} } $ bin range  }
      & \multicolumn{1}{c}{ $ { \rd\sigma^{\text{vis}} } / { \rd {\pt}^{\,\mathrm{j1}} } $ }
      & stat. & syst. & tot.
      & \multicolumn{1}{c}{  $( 1/{\sigma^{\text{vis}}} )( {\rd \sigma^{\text{vis}} }/{\rd {\pt}^{\,\mathrm{j1}} } )$  }
      & stat. & syst. & tot. \\ [-2pt]
    \multicolumn{1}{c|}{  (\GeVns{})  }
      & \multicolumn{1}{c}{ (pb/\GeVns{}) } & (\%) & (\%) & (\%)
      & \multicolumn{1}{c}{ ($\GeVns^{-1}$) } & (\%) & (\%) & (\%) \\
\hline
\ \ \, 20,\, 45 \ \
& 5.30 , 10^{-2} & 0.8 & 8.2 & 8.2  & 1.82 , 10^{-2} & 0.8 & 2.8 & 2.9 \\
\ \ \, 45,\, 80 \ \
& 2.17 , 10^{-2} & 2.2 & 7.7 & 8.0  & 7.44 , 10^{-3} & 1.4 & 3.9 & 4.1 \\
\ \ \, 80,\, 140 \,
& 8.64 , 10^{-3} & 2.2 & 7.9 & 8.2  & 2.96 , 10^{-3} & 2.1 & 4.9 & 5.3 \\
 \, 140,\, 200 \,
& 2.8 , 10^{-3} & 3.4 & 9.3 & 10 & 9.78 , 10^{-4} & 3.3 & 6.7 & 7.4   \\
 \, 200,\, 400 \,
& 6.9 , 10^{-4} & 3.8 & 14 & 14 & 2.4 , 10^{-4} & 3.5 & 14 & 14\\
    \hline
    \multicolumn{9}{c}{} \\ [-10pt]
    \hline
    \multicolumn{1}{c|}{ $ \abs{\eta^{\,\mathrm{j1} } } $ bin range }
      & \multicolumn{1}{c}{ $ { \rd\sigma^{\text{vis}} } / { \rd | {\eta^{\,\mathrm{j1}} } | } $ }
      & stat. & syst. & tot.
      & \multicolumn{1}{c}{ \multirow{2}{*}{ $( 1/{\sigma^{\text{vis}}} )( {\rd \sigma^{\text{vis}} }/{\rd | { \eta^{\,\mathrm{j1} } } | } )$ } }
      & stat. & syst. & tot. \\ [-2pt]
    \multicolumn{1}{c|}{ (\GeVns{}) }
      & \multicolumn{1}{c}{ (pb) } & (\%) & (\%) & (\%)
      & & (\%) & (\%) & (\%)   \\
\hline
\ \ \, 0,\,  0.6 \,
& \multicolumn{1}{c}{$1.32$} & 1.2 & 6.5 & 6.6   & 4.27 , 10^{-1} & 1.7 & 6.4 & 6.6 \\
\, 0.6,\,  1.2 \,
& \multicolumn{1}{c}{$1.5$} & 2.2 & 11 & 11 & 4.77 , 10^{-1} & 1.4 & 2.3 & 2.7 \\
\, 1.2,\,  1.8 \,
& \multicolumn{1}{c}{$1.3$} & 2.0 & 10 & 10 & 4.20 , 10^{-1} & 1.6 & 1.4 & 2.1 \\
\, 1.8,\,  2.4 \,
& \multicolumn{1}{c}{$1.1$} & 2.4 & 19 & 19 & 3.42 , 10^{-1} & 1.9 & 9.3 & 9.5 \\
\hline
    \end{tabular}
    }
  \end{center}
\end{table*}

\begin{table*}[htb]
  \renewcommand{\arraystretch}{1.2}
  \begin{center}
    \topcaption{Absolute (left) and normalized (right) differential \ttbar cross sections as a function of the \pt (${\pt}^{\,\mathrm{j2}}$) and the $|\eta|$ ($|\eta^{\mathrm{j2}}|$) of the subleading additional jet, along with their statistical, systematic, and total uncertainties. The results are presented at particle level in the visible phase space.}
    \label{tab:dilepton:SummaryResultsJet2}
   \resizebox{\textwidth}{!}{
    \begin{tabular}{y|x{4}ccc|x{-1}ccc}
    \hline
    \multicolumn{1}{c|}{   $ {\pt}^{\,\mathrm{j2} } $ bin range  }
      & \multicolumn{1}{c}{ $ { \rd\sigma^{\text{vis}} } / { \rd {\pt}^{\,\mathrm{j2}} } $ }
      & stat. & syst. & tot.
      & \multicolumn{1}{c}{ $( 1/{\sigma^{\text{vis}}} )( {\rd \sigma^{\text{vis}} }/{\rd {\pt}^{\,\mathrm{j2}} } )$ }
      & stat. & syst. & tot. \\ [-2pt]
    \multicolumn{1}{c|}{  (\GeVns{})  }
      & \multicolumn{1}{c}{ (pb/GeVns{}) } & (\%) & (\%) & (\%)
      & \multicolumn{1}{c}{ ($\GeVns^{-1}$) } & (\%) & (\%) & (\%) \\
\hline
\ \ \, 20,\, 35 \ \
& 4.7 , 10^{-2} & 2.6 & 12 & 12 & 3.68 , 10^{-2} & 1.1 & 4.5 & 4.7 \\
\ \ \, 35,\, 50 \ \
& 1.7 , 10^{-2} & 4.7 & 8.8 & 10  & 1.32 , 10^{-2} & 2.7 & 5.6 & 6.3 \\
\ \ \, 50,\, 80 \ \
& 6.82 , 10^{-3} & 4.3 & 8.5 & 9.6   & 5.30 , 10^{-3} & 5.2 & 7.1 & 8.7 \\
\ \ \, 80,\, 200 \,
& 9.0 , 10^{-4} & 4.9 & 27 & 27 & 7.1 , 10^{-4} & 4.6 & 25 & 26 \\
 \, 200,\, 400 \,
& 4.0 , 10^{-5} & 15 & 35 & 38 & 2.7 , 10^{-5} & 16 & 49 & 51 \\
\hline
    \multicolumn{9}{c}{} \\ [-10pt]
    \hline
    \multicolumn{1}{c|}{ $ \abs{\eta^{\,\mathrm{j2} } } $ bin range }
      & \multicolumn{1}{c}{ $ { \rd\sigma^{\text{vis}} } / { \rd | {\eta^{\,\mathrm{j2}} } | } $ }
      & stat. & syst. & tot.
      & \multicolumn{1}{c}{ \multirow{2}{*}{ $( 1/{\sigma^{\text{vis}}} )( {\rd \sigma^{\text{vis}} }/{\rd | { \eta^{\,\mathrm{j2} } } | } )$ } }
      & stat. & syst. & tot. \\ [-2pt]
    \multicolumn{1}{c|}{ (\GeVns{}) }
      & \multicolumn{1}{c}{ (pb) } & (\%) & (\%) & (\%)
      & & (\%) & (\%) & (\%)   \\
\hline
\ \ \, 0,\,  0.6 \,
& 6.4 , 10^{-1} & 1.6 & 11 & 11 & 4.69 , 10^{-1} & 3.2 & 8.6 & 9.2 \\
\, 0.6,\,  1.2 \,
& 6.2 , 10^{-1} & 4.6 & 14 & 14 & 4.50 , 10^{-1} & 2.9 & 5.2 & 6.0 \\
\, 1.2,\,  1.8 \,
& 5.3 , 10^{-1} & 4.5 & 20 & 20 & 3.99 , 10^{-1} & 3.2 & 6.0 & 6.8 \\
\, 1.8,\,  2.4 \,
& 4.7 , 10^{-1} & 5.0 & 29 & 30 & 3.5 , 10^{-1} & 3.8 & 14 & 14 \\
    \hline
    \end{tabular}
  }
  \end{center}
\end{table*}

\begin{table*}[htb]
  \renewcommand{\arraystretch}{1.2}
  \begin{center}
    \topcaption{Absolute (left) and normalized (right) differential \ttbar cross sections as a function of the invariant mass ($\mjj$) of the two leading additional jets in the event, the angle \DR\ between them ($\Djj$), and $\HT$, along with their statistical, systematic, and total uncertainties. The results are presented at the particle level in the visible phase space.}
    \label{tab:dilepton:SummaryResultsJet12}
    \resizebox{\textwidth}{!}{
    \begin{tabular}{y|x{4}ccc|x{-1}ccc}
    \hline
    \multicolumn{1}{c|}{   $\mjj$ bin range  }
     & \multicolumn{1}{c}{ $ { \rd\sigma^{\text{vis}} } / { \rd \mjj } $ }
      & stat. & syst. & tot.
      & \multicolumn{1}{c}{ $( 1/{\sigma^{\text{vis}}} )( {\rd \sigma^{\text{vis}} }/{\rd \mjj } )$ }
      & stat. & syst. & tot. \\ [-2pt]
    \multicolumn{1}{c|}{  (\GeVns{})  }
      & \multicolumn{1}{c}{ (pb/GeV) } & (\%) & (\%) & (\%)
      & \multicolumn{1}{c}{ $(\GeVns^{-1})$ } & (\%) & (\%) & (\%) \\
    \hline
0,\, 60 &
4.4 , 10^{-3} & 1.3 & 14 & 14   & 3.7 , 10^{-3} & 2.4 & 13 & 13 \\
60,\, 100 &
7.6 , 10^{-3} & 5.3 & 16 & 17  & 6.33 , 10^{-3} & 3.6 & 4.9 & 6.0   \\
100,\, 170 &
4.7 , 10^{-3} & 3.9 & 15 & 16 & 3.96 , 10^{-3} & 2.8 & 4.9 & 5.6   \\
170,\, 400 &
1.3 , 10^{-3} & 3.2 & 14 & 14 & 1.08 , 10^{-3} & 2.4 & 4.3 & 5.2   \\
    \hline
    \multicolumn{9}{c}{} \\ [-10pt]
    \hline
    \multicolumn{1}{c|}{ \multirow{2}{*}{ $\Djj$ bin range } }
      & \multicolumn{1}{c}{ $ { \rd\sigma^{\text{vis}} } / { \rd \Djj } $ }
      & stat. & syst. & tot.
      & \multicolumn{1}{c}{ \multirow{2}{*}{ $( 1/{\sigma^{\text{vis}}} )( {\rd \sigma^{\text{vis}} }/{\rd \Djj } )$ } }
      & stat. & syst. & tot. \\ [-2pt]
      & \multicolumn{1}{c}{ (pb) } & (\%) & (\%) & (\%)
      & & (\%) & (\%) & (\%)   \\
    \hline	
0.5,\, 1.0 &
3.4 , 10^{-1} & 2.4 & 11 & 11  & 2.8 , 10^{-1} & 5.4 & 18 & 19 \\
1.0,\, 2.0 &
3.0 , 10^{-1} & 6.2 & 29 & 30 & 2.4 , 10^{-1} & 3.8 & 9.2 & 10  \\
2.0,\, 3.0 &
4.1 , 10^{-1} & 5.1 & 28 & 28 & 3.29 , 10^{-1} & 3.0 & 7.5 & 8.1   \\
3.0,\, 4.0 &
2.8 , 10^{-1} & 5.2 & 21 & 21 & 2.28 , 10^{-1} & 3.5 & 7.2 & 8.0   \\
4.0,\, 5.0 &
7.7 , 10^{-2} & 8.1 & 23 & 24 & 6.0 , 10^{-2} & 7.3 & 19 & 20 \\
    \hline
    \multicolumn{9}{c}{} \\ [-10pt]
    \hline
    \multicolumn{1}{c|}{ \multirow{2}{*}{ $\HT$ bin range } }
      & \multicolumn{1}{c}{ $ { \rd\sigma^{\text{vis}} } / { \rd \HT } $ }
      & stat. & syst. & tot.
      & \multicolumn{1}{c}{ \multirow{2}{*}{ $( 1/{\sigma^{\text{vis}}} )( {\rd \sigma^{\text{vis}} }/{\rd \HT } )$ } }
      & stat. & syst. & tot. \\ [-2pt]
      & \multicolumn{1}{c}{ (pb) } & (\%) & (\%) & (\%)
      & & (\%) & (\%) & (\%)   \\
    \hline	
20,\,45 &
3.96 , 10^{-2} & 1.0 & 7.6 & 7.7      & 1.35 , 10^{-2} & 0.9 & 3.6 & 3.7 \\
45,\,80 &
2.0 , 10^{-2} & 2.6 & 10 & 11    & 6.91 , 10^{-3} & 1.7 & 3.2 & 3.6 \\
80,\,140 &
1.06 , 10^{-2} & 2.0 & 9.3 & 9.5    & 3.53 , 10^{-3} & 1.9 & 2.6 & 3.3 \\
140,\,200 &
4.7 , 10^{-3} & 2.7 & 13 & 13 & 1.62 , 10^{-3} & 2.6 & 6.6 & 7.1 \\
200,\,600 &
8.3 , 10^{-4} & 2.6 & 15 & 15 & 2.8 , 10^{-4} & 2.3 & 11 & 12 \\
    \hline
    \end{tabular}
    }
  \end{center}
\end{table*}

\begin{table*}[htb]
  \renewcommand{\arraystretch}{1.2}
  \begin{center}
    \topcaption{Absolute (left) and normalized (right) differential \ttbar cross sections as a function of the \pt (${\pt}^{\,\mathrm{j1}}$) and the $|\eta|$ ($|\eta^{\,\mathrm{j1}}|$) of the leading additional jet in the event (not coming from the top quark decay products), along with their statistical, systematic, and total uncertainties. The results are presented at the particle level in the full phase space of the \ttbar system, corrected for acceptance and branching fractions.}
    \label{tab:dilepton:SummaryResultsJet1Full}
   \resizebox{\textwidth}{!}{
    \begin{tabular}{y|x{4}ccc|x{-1}ccc}
    \hline
    \multicolumn{1}{c|}{   $ {\pt}^{\,\mathrm{j1} } $ bin range  }
      & \multicolumn{1}{c}{ $ { \rd\sigma^{\text{full}} } / { \rd {\pt}^{\,\mathrm{j1}} } $ }
      & stat. & syst. & tot.
      & \multicolumn{1}{c}{  $( 1/{\sigma^{\text{full}}} )( {\rd \sigma^{\text{full}} }/{\rd {\pt}^{\,\mathrm{j1}} } )$  }
      & stat. & syst. & tot. \\ [-2pt]
    \multicolumn{1}{c|}{  (\GeVns{})  }
      & \multicolumn{1}{c}{ (pb/\GeVns{}) } & (\%) & (\%) & (\%)
      & \multicolumn{1}{c}{ $(\GeVns^{-1})$ } & (\%) & (\%) & (\%) \\
\hline
20,\, 45 &
\multicolumn{1}{c}{2.7}  & 0.9 & 10 & 10& 1.85 , 10^{-2} & 0.7 & 2.3 & 2.4 \\
45,\, 80 &
\multicolumn{1}{c}{1.13} & 1.7 & 9.3 & 9.4  & 7.66 , 10^{-3} & 1.3 & 3.4 & 3.6 \\
80,\, 140 &
4.25 , 10^{-1} & 1.8 & 7.6 & 7.8 & 2.88 , 10^{-3} & 1.7 & 3.2 & 3.6 \\
140,\, 200 &
1.36 , 10^{-1} & 2.7 & 7.8 & 8.3 & 9.26 , 10^{-4} & 2.6 & 4.4 & 5.1 \\
200,\, 400 &
3.04 , 10^{-2} & 3.0 & 7.8 & 8.4 & 2.07 , 10^{-4} & 2.9 & 8.0 & 8.5 \\
    \hline
    \multicolumn{9}{c}{} \\ [-10pt]
    \hline
    \multicolumn{1}{c|}{ $ |\eta^{\,\mathrm{j1} } | $ bin range }
      & \multicolumn{1}{c}{ $ { \rd\sigma^{\text{full}} } / { \rd | {\eta^{\,\mathrm{j1}} } | } $ }
      & stat. & syst. & tot.
      & \multicolumn{1}{c}{ \multirow{2}{*}{ $( 1/{\sigma^{\text{full}}} )( {\rd \sigma^{\text{full}} }/{\rd | { \eta^{\,\mathrm{j1} } } | } )$ } }
      & stat. & syst. & tot. \\ [-2pt]
      \multicolumn{1}{c|}{ (\GeVns{}) }
      & \multicolumn{1}{c}{ (pb) } & (\%) & (\%) & (\%)
      & & (\%) & (\%) & (\%)   \\
    \hline
0,\, 0.6 &
\multicolumn{1}{c}{65.7} & 1.4 & 6.2 & 6.4 & 4.37 , 10^{-1} & 1.5 & 5.8 & 5.9\\
0.6,\, 1.2 &
\multicolumn{1}{c}{70.6} & 1.4 & 9.6 & 9.8 & 4.72 , 10^{-1} & 1.2 & 2.2 & 2.5\\
1.2,\, 1.8 &
\multicolumn{1}{c}{63.2} & 1.5 & 9.6 & 9.8 & 4.19 , 10^{-1} & 1.3 & 0.8 & 1.5\\
1.8,\, 2.4 &
\multicolumn{1}{c}{51}    & 1.9 & 16 & 16& 3.38 , 10^{-1} & 1.7 & 7.4 & 7.6 \\
   \hline
    \end{tabular}
  }
  \end{center}
\end{table*}

\begin{table*}[htb]
  \renewcommand{\arraystretch}{1.2}
  \begin{center}
    \topcaption{Absolute (left) and normalized (right) differential \ttbar cross sections as a function of the \pt (${\pt}^{\,\mathrm{j2}}$) and the $|\eta|$ ($| {\eta^{\,\mathrm{j2}} } |$) of the subleading additional jet in the event (not coming from the top quark decay products), along with their statistical, systematic, and total uncertainties. The results are presented at the particle level in the full phase space of the \ttbar system, corrected for acceptance and branching fractions.}
    \label{tab:dilepton:SummaryResultsJet2Full}
   \resizebox{\textwidth}{!}{
    \begin{tabular}{y|x{4}ccc|x{-1}ccc}
    \hline
    \multicolumn{1}{c|}{   $ {\pt}^{\,\mathrm{j2} } $ bin range  }
      & \multicolumn{1}{c}{ $ { \rd\sigma^{\text{full}} } / { \rd {\pt}^{\,\mathrm{j2}} } $ }
      & stat. & syst. & tot.
      & \multicolumn{1}{c}{ $( 1/{\sigma^{\text{full}}} )( {\rd \sigma^{\text{full}} }/{\rd {\pt}^{\,\mathrm{j2}} } )$ }
      & stat. & syst. & tot. \\ [-2pt]
    \multicolumn{1}{c|}{  (\GeVns{})  }
      & \multicolumn{1}{c}{ (pb/GeV) } & (\%) & (\%) & (\%)
      & \multicolumn{1}{c}{ $(\GeVns^{-1})$ } & (\%) & (\%) & (\%) \\
\hline
\ \ \, 20,\, 35 \ \  &
\multicolumn{1}{c}{$2.4$} & 1.6 & 15 & 15 & 3.76 , 10^{-2}  & 0.9 & 3.9 & 4.0 \\
\ \ \, 35,\, 50 \ \ &
8.7 , 10^{-1} & 4.0 & 10 & 11 & 1.33 , 10^{-2}  & 2.8 & 5.8 & 6.5 \\
\ \ \, 50,\, 80 \ \ &
3.4 , 10^{-1} & 3.9 & 12 & 13 & 5.18 , 10^{-3}  & 4.3 & 5.5 & 7.0 \\
\ \ \, 80,\, 200 \, &
4.2 , 10^{-2} & 4.0 & 17 & 18 & 6.5 , 10^{-4} & 3.8 & 21 & 21 \\
 \, 200,\, 400 \,  &
1.5 , 10^{-3} & 13 & 42 & 44 & 2.2 , 10^{-5} & 14 & 52 & 54\\
\hline
    \multicolumn{9}{c}{} \\ [-10pt]
    \hline
    \multicolumn{1}{c|}{ $ |\eta^{\,\mathrm{j2} } | $ bin range }
      & \multicolumn{1}{c}{ $ { \rd\sigma^{\text{full}} } / { \rd | {\eta^{\,\mathrm{j2}} } | } $ }
      & stat. & syst. & tot.
      & \multicolumn{1}{c}{ \multirow{2}{*}{ $( 1/{\sigma^{\text{full}}} )( {\rd \sigma^{\text{full}} }/{\rd | { \eta^{\,\mathrm{j2} } } | } )$ } }
      & stat. & syst. & tot. \\ [-2pt]
    \multicolumn{1}{c|}{ (\GeVns{}) }
      & \multicolumn{1}{c}{ (pb) } & (\%) & (\%) & (\%)
      & & (\%) & (\%) & (\%)   \\
\hline
\ \ \, 0,\,  0.6 \,
& \multicolumn{1}{c}{31.6} & 2.2 & 9.4 & 9.7  & 4.69 , 10^{-1} & 2.9 & 9.1 & 9.5 \\
\, 0.6,\,  1.2 \,
& \multicolumn{1}{c}{30} & 3.2 & 13 & 14 & 4.50 , 10^{-1} & 2.4 & 4.4 & 5.0 \\
\, 1.2,\,  1.8 \,
& \multicolumn{1}{c}{27}  & 3.3 & 20 & 20 & 4.02 , 10^{-1} & 2.7 & 5.7 & 6.3 \\
\, 1.8,\,  2.4 \,
& \multicolumn{1}{c}{23}  & 4.0 & 28 & 28 & 3.5 , 10^{-1} & 3.4 & 13 & 13 \\
    \hline
    \end{tabular}
  }
    \end{center}
\end{table*}

\begin{table*}[htb]
  \renewcommand{\arraystretch}{1.2}
  \begin{center}
    \topcaption{Absolute (left) and normalized (right) differential \ttbar cross sections as a function of the invariant mass of the two first leading additional jets in the event ($\mjj$), the angle \DR\ between them ($\Djj$), and $\HT$, along with their statistical, systematic, and total uncertainties. The results are presented at the particle level in the full phase space of \ttbar system, corrected for acceptance and branching fractions.}
    \label{tab:dilepton:SummaryResultsJet12Full}
    \resizebox{\textwidth}{!}{
    \begin{tabular}{y|x{4}ccc|x{-1}ccc}
    \hline
    \multicolumn{1}{c|}{   $\mjj$ bin range  }
     & \multicolumn{1}{c}{ $ { \rd\sigma^{\text{full}} } / { \rd \mjj } $ }
      & stat. & syst. & tot.
      & \multicolumn{1}{c}{ $( 1/{\sigma^{\text{full}}} )( {\rd \sigma^{\text{full}} }/{\rd \mjj } )$ }
      & stat. & syst. & tot. \\ [-2pt]
    \multicolumn{1}{c|}{  (\GeVns{})  }
      & \multicolumn{1}{c}{ (pb/GeVns{}) } & (\%) & (\%) & (\%)
      & \multicolumn{1}{c}{ $(\GeVns^{-1})$ } & (\%) & (\%) & (\%) \\
    \hline
0,\, 60    & 2.3 , 10^{-1} & 1.7 & 18 & 18 & 3.7 , 10^{-3} & 2.4 & 13 & 13 \\
60,\, 100  & 4.0 , 10^{-1} & 5.0 & 13 & 14 & 6.47 , 10^{-3} & 3.5 & 4.3 & 5.5 \\
100,\, 170 & 2.4 , 10^{-1} & 3.3 & 10 & 12 & 3.98 , 10^{-3} & 2.9 & 4.2 & 5.1 \\
170,\, 400 & 6.4 , 10^{-2} & 2.7 & 10  & 10 & 1.04 , 10^{-3} & 2.5 & 5.4 & 6.0 \\
    \hline
    \multicolumn{9}{c}{} \\ [-10pt]
    \hline
    \multicolumn{1}{c|}{ \multirow{2}{*}{ $\Djj$ bin range } }
      & \multicolumn{1}{c}{ $ { \rd\sigma^{\text{full}} } / { \rd \Djj } $ }
      & stat. & syst. & tot.
      & \multicolumn{1}{c}{ \multirow{2}{*}{ $( 1/{\sigma^{\text{full}}} )( {\rd \sigma^{\text{full}} }/{\rd \Djj } )$ } }
      & stat. & syst. & tot. \\ [-2pt]
      & \multicolumn{1}{c}{ (pb) } & (\%) & (\%) & (\%)
      & & (\%) & (\%) & (\%)   \\
    \hline	
0.5,\, 1.0 & \multicolumn{1}{c}{17} & 3.2 & 13 & 13 & 2.6 , 10^{-1} & 4.5 & 11.6 & 12\\
1.0,\, 2.0 & \multicolumn{1}{c}{16} & 4.0 & 13 & 14 & 2.45 , 10^{-1} & 3.0 & 5.4 & 6.2 \\
2.0,\, 3.0 & \multicolumn{1}{c}{22} & 3.4 & 15 & 15 & 3.35 , 10^{-1} & 2.4 & 5.7 & 6.2 \\
3.0,\, 4.0 & \multicolumn{1}{c}{15} & 3.6 & 16 & 16 & 2.27 , 10^{-1} & 2.8 & 6.0 & 6.7 \\
4.0,\, 5.0 & \multicolumn{1}{c}{3.8} & 6.5 & 22 & 23 & 5.8 , 10^{-2} & 6.0 & 15 & 16\\
    \hline
    \multicolumn{9}{c}{} \\ [-10pt]
    \hline
    \multicolumn{1}{c|}{ \multirow{2}{*}{ $\HT$ bin range } }
      & \multicolumn{1}{c}{ $ { \rd\sigma^{\text{full}} } / { \rd \HT } $ }
      & stat. & syst. & tot.
      & \multicolumn{1}{c}{ \multirow{2}{*}{ $( 1/{\sigma^{\text{full}}} )( {\rd \sigma^{\text{full}} }/{\rd \HT } )$ } }
      & stat. & syst. & tot. \\ [-2pt]
      & \multicolumn{1}{c}{ (pb) } & (\%) & (\%) & (\%)
      & & (\%) & (\%) & (\%)   \\
    \hline	
20,\, 45 & \multicolumn{1}{c}{2.01} & 1.0 & 8.2 & 8.3      & 1.36 , 10^{-2} & 0.9 & 2.7 & 2.8 \\
45,\, 80 & \multicolumn{1}{c}{1.1} & 2.0 & 9.9 & 10     & 7.08 , 10^{-3} & 1.5 & 2.0 & 2.5 \\
80,\, 140 & 5.3 , 10^{-1} & 1.7 & 11 & 11  & 3.56 , 10^{-3} & 1.6 & 3.0 & 3.5 \\
140,\, 200 & 2.3 , 10^{-1} & 2.3 & 12 & 12 & 1.58 , 10^{-3} & 2.2 & 4.7 & 5.1 \\
200,\, 600 & 3.80 , 10^{-2} & 2.0 & 9.2 & 9.4   & 2.56 , 10^{-4} & 1.9 & 5.8 & 6.1 \\
    \hline
    \end{tabular}
    }
  \end{center}
\end{table*}

\begin{table*}[htb]
  \renewcommand{\arraystretch}{1.2}
  \begin{center}
    \topcaption{Absolute (left) and normalized (right) differential \ttbar cross sections as a function of the \pt and the $|\eta|$ of the leading (${\pt}^{\, \mathrm{b1}}$, $|\eta^{\, \mathrm{b1}}|$) and subleading (${\pt}^{\, \mathrm{b2}}$, $|\eta^{\, \mathrm{b2}}|$) additional \PQb jet in the event (not coming from the top quark decay products), along with their statistical, systematic, and total uncertainties. The results are presented at particle level in the visible phase space.}
    \label{tab:dilepton:SummaryResultsBJet}
   \resizebox{\textwidth}{!}{
    \begin{tabular}{y|x{4}ccc|x{10}ccc}
    \hline
    \multicolumn{1}{c|}{   $ {\pt}^{\,\mathrm{b1} } $ bin range  }
      & \multicolumn{1}{c}{ $ { \rd\sigma^{\text{vis}} } / { \rd {\pt}^{\,\mathrm{b1}} } $ }
      & stat. & syst. & tot.
      & \multicolumn{1}{c}{  $( 1/{\sigma^{\text{vis}}} )( {\rd \sigma^{\text{vis}} }/{\rd {\pt}^{\,\mathrm{b1}} } )$  }
      & stat. & syst. & tot. \\ [-2pt]
    \multicolumn{1}{c|}{  (\GeVns{})  }
      & \multicolumn{1}{c}{ (pb/GeVns{}) } & (\%) & (\%) & (\%)
      & \multicolumn{1}{c}{ $(\GeVns^{-1})$ } & (\%) & (\%) & (\%) \\
\hline
20,\, 45   & 2.7 , 10^{-3}   & 25   & 23   & 35 & 1.6 , 10^{-2}  & 26  & 25  & 36 \\
45,\, 80   & 1.6 , 10^{-3}   & 23   & 18   & 29 & 9.8 , 10^{-3}  & 23  & 19  & 30 \\
80,\, 200  & 2.9 , 10^{-4}   & 28   & 19   & 34 & 1.8 , 10^{-3}  & 28  & 21  & 35 \\
200,\, 400 & 2.6 , 10^{-5}   & 64   & 46   & 78 & 1.6 , 10^{-4}  & 62  & 46  & 78 \\
    \hline
    \multicolumn{9}{c}{} \\ [-10pt]
    \hline
    \multicolumn{1}{c|}{ $ \abs{\eta^{\,\mathrm{b1} } } $ bin range }
      & \multicolumn{1}{c}{ $ { \rd\sigma^{\text{vis}} } / { \rd \abs{ {\eta^{\,\mathrm{b1}} } } } $ }
      & stat. & syst. & tot.
      & \multicolumn{1}{c}{ \multirow{2}{*}{ $( 1/{\sigma^{\text{vis}}} )( {\rd \sigma^{\text{vis}} }/{\rd | { \eta^{\,\mathrm{b1} } } | } )$ } }
      & stat. & syst. & tot. \\ [-2pt]
    \multicolumn{1}{c|}{ (\GeVns{}) }
      & \multicolumn{1}{c}{ (pb) } & (\%) & (\%) & (\%)
      & & (\%) & (\%) & (\%)   \\
\hline
0,\, 0.6 & 8.3 , 10^{-2}  & 25   & 8   & 26 & \multicolumn{1}{c}{0.5}  & 32   & 8   & 33 \\
0.6,\, 1.2 & 6.6 , 10^{-2}  & 35   & 7   & 36 & \multicolumn{1}{c}{0.4}  & 30   & 7   & 30 \\
1.2,\, 1.8 & 5.4 , 10^{-2}  & 41   & 12   & 42 & \multicolumn{1}{c}{0.3}  & 34   & 12   & 36 \\
1.8,\, 2.4 & 6.6 , 10^{-2}  & 35   & 12   & 37 & \multicolumn{1}{c}{0.4}  & 29   & 12   & 32 \\
\hline
    \multicolumn{9}{c}{} \\ [-10pt]
    \hline
    \multicolumn{1}{c|}{   $ {\pt}^{\,\mathrm{b2} } $ bin range  }
      & \multicolumn{1}{c}{ $ { \rd\sigma^{\text{vis}} } / { \rd {\pt}^{\,\mathrm{b2}} } $ }
      & stat. & syst. & tot.
      & \multicolumn{1}{c}{  $( 1/{\sigma^{\text{vis}}} )( {\rd \sigma^{\text{vis}} }/{\rd {\pt}^{\,\mathrm{b2}} } )$  }
      & stat. & syst. & tot. \\ [-2pt]
    \multicolumn{1}{c|}{  (\GeVns{})  }
      & \multicolumn{1}{c}{ (pb/\GeVns{}) } & (\%) & (\%) & (\%)
      & \multicolumn{1}{c}{ $(\GeVns^{-1})$ } & (\%) & (\%) & (\%) \\
\hline
20,\, 45  & 9.6 , 10^{-4}  & 33  & 11 & 34  & 3.0 , 10^{-2}  & 18 & 8  & 20 \\
45,\, 80  & 1.8 , 10^{-4}  & 54  & 24 & 60  & 5.5 , 10^{-3}  & 51 & 24  & 56 \\
80,\, 200 & 1.8 , 10^{-5}  & 124  & 35 & 129  & 5.5 , 10^{-4}  & 128 & 35  & 132 \\
    \hline
    \multicolumn{9}{c}{} \\ [-10pt]
    \hline
    \multicolumn{1}{c|}{ $ |\eta^{\,\mathrm{b2} } | $ bin range }
      & \multicolumn{1}{c}{ $ { \rd\sigma^{\text{vis}} } / { \rd | {\eta^{\,\mathrm{b2}} } | } $ }
      & stat. & syst. & tot.
      & \multicolumn{1}{c}{ \multirow{2}{*}{ $( 1/{\sigma^{\text{vis}}} )( {\rd \sigma^{\text{vis}} }/{\rd | { \eta^{\,\mathrm{b2} } } | } )$ } }
      & stat. & syst. & tot. \\ [-2pt]
    \multicolumn{1}{c|}{ (\GeVns{}) }
      & \multicolumn{1}{c}{ (pb) } & (\%) & (\%) & (\%)
      & & (\%) & (\%) & (\%)   \\
\hline
0,\, 0.6 & 2.3 , 10^{-2}  & 47  & 25  & 53 & \multicolumn{1}{c}{0.8}  & 57  & 25  & 62 \\
0.6,\, 1.2 & 1.2 , 10^{-2}  & 58  & 18  & 61 & \multicolumn{1}{c}{0.4}  & 47  & 14  & 49 \\
1.2,\, 2.4 & 7.6 , 10^{-3}  & 97  & 38  & 104 & \multicolumn{1}{c}{0.3}  & 79  & 37  & 87 \\
\hline
     \end{tabular}
     }
  \end{center}
\end{table*}

\begin{table*}[h]
  \renewcommand{\arraystretch}{1.2}
  \begin{center}
    \topcaption{Absolute (left) and normalized (right) differential \ttbar cross sections as a function of the invariant mass of the two leading additional \PQb jets in the event ($\mbb$) and the angle $\DR_{\PQb\PQb}$, along with their statistical, systematic, and total uncertainties. The results are presented at particle level in the visible phase space.}
    \label{tab:dilepton:SummaryResultsBJet12}
    \resizebox{\textwidth}{!}{
    \begin{tabular}{y|x{4}ccc|x{-1}ccc}
    \hline
    \multicolumn{1}{c|}{   $\mbb$ bin range  }
     & \multicolumn{1}{c}{ $ { \rd\sigma^{\text{vis}} } / { \rd \mbb } $ }
      & stat. & syst. & tot.
      & \multicolumn{1}{c}{ $( 1/{\sigma^{\text{vis}}} )( {\rd \sigma^{\text{vis}} }/{\rd \mbb } )$ }
      & stat. & syst. & tot. \\ [-2pt]
    \multicolumn{1}{c|}{  (\GeVns{})  }
      & \multicolumn{1}{c}{ (pb/GeV) } & (\%) & (\%) & (\%)
      & \multicolumn{1}{c}{ $(\GeVns^{-1})$ } & (\%) & (\%) & (\%) \\
    \hline
10,\,  60   & 2.6 , 10^{-4}  & 60  & 24  & 65 & 8.2 , 10^{-3} & 64 & 23 & 68\\
60,\,  100  & 1.7 , 10^{-4}  & 118  & 42  & 125 & 5.5 , 10^{-3} & 104 & 41 & 112\\
100,\,  170 & 5.0 , 10^{-5}  & 142  & 49  & 151 & 1.6 , 10^{-3} & 135 & 47 & 142 \\
170,\,  400 & 2.9 , 10^{-5}  & 64  & 44  & 77 & 9.4 , 10^{-4} & 66 & 45 & 80 \\
    \hline
    \multicolumn{9}{c}{} \\ [-10pt]
    \hline
    \multicolumn{1}{c|}{ \multirow{2}{*}{ $\DR_{{\PQb\PQb}}$ bin range } }
      & \multicolumn{1}{c}{ $ { \rd\sigma^{\text{vis}} } / { \rd \DR_{{\PQb\PQb}} } $ }
      & stat. & syst. & tot.
      & \multicolumn{1}{c}{ \multirow{2}{*}{ $( 1/{\sigma^{\text{vis}}} )( {\rd \sigma^{\text{vis}} }/{\rd \DR_{{\PQb\PQb}} } )$ } }
      & stat. & syst. & tot. \\ [-2pt]
      & \multicolumn{1}{c}{ (pb) } & (\%) & (\%) & (\%)
      & & (\%) & (\%) & (\%)   \\
    \hline	
0.5,\,  1.0 & 2.5 , 10^{-3}  & 327 & 99 & 342 & \multicolumn{1}{c}{0.1}  & 334  & 98   & 348 \\
1.0,\,  2.0 & 7.7 , 10^{-3}  & 75 & 39 & 84 & \multicolumn{1}{c}{0.2}  & 63  & 36   & 72 \\
2.0,\,  5.0 & 9.8 , 10^{-3}  & 29 & 14 & 32 & \multicolumn{1}{c}{0.3}  & 19  & 15   & 24 \\
\hline	
  \end{tabular}
  }
  \end{center}
\end{table*}

\begin{table*}[h]
  \renewcommand{\arraystretch}{1.2}
  \begin{center}
    \topcaption{Absolute (left) and normalized (right) differential \ttbar cross sections as a function of the \pt and the $|\eta|$ of the leading ($\pt^{\mathrm{b1}}$, $|\eta^{\mathrm{b1}}|$) and subleading ($\pt^{\mathrm{b2}}$, $|\eta^{\mathrm{b2}}|$) additional \PQb jet in the event (not coming from the top quark decay products), along with their statistical, systematic, and total uncertainties. The results are presented at particle level in the full phase space of the \ttbar system, corrected for acceptance and branching fractions.}
    \label{tab:dilepton:SummaryResultsBJetFullPS}
   \resizebox{\textwidth}{!}{
    \begin{tabular}{y|x{4}ccc|x{10}ccc}
    \hline
    \multicolumn{1}{c|}{   $ {\pt}^{\,\mathrm{b1} } $ bin range  }
      & \multicolumn{1}{c}{ $ { \rd\sigma^{\text{full}} } / { \rd {\pt}^{\,\mathrm{b1}} } $ }
      & stat. & syst. & tot.
      & \multicolumn{1}{c}{  $( 1/{\sigma^{\text{full}}} )( {\rd \sigma^{\text{full}} }/{\rd {\pt}^{\,\mathrm{b1}} } )$  }
      & stat. & syst. & tot. \\ [-2pt]
    \multicolumn{1}{c|}{  (\GeVns{})  }
      & \multicolumn{1}{c}{ (pb/GeV) } & (\%) & (\%) & (\%)
      & \multicolumn{1}{c}{ $(\GeVns^{-1})$ } & (\%) & (\%) & (\%) \\
\hline
20,\, 45   & 1.1 , 10^{-1} & 33 & 25 & 41 & 1.7 , 10^{-2} & 24 & 24 & 34\\
45,\, 80   & 6.3 , 10^{-2} & 17 & 19 & 25 & 9.5 , 10^{-3} & 19 & 19 & 27 \\
80,\, 200  & 1.2 , 10^{-2} & 22 & 20 & 29 & 1.8 , 10^{-3} & 26 & 20 & 33\\
200,\, 400 & 1.0 , 10^{-3} & 53 & 39 & 66 & 1.5 , 10^{-4} & 55 & 39 & 67 \\
    \hline
    \multicolumn{9}{c}{} \\ [-10pt]
    \hline
    \multicolumn{1}{c|}{ $ |\eta^{\,\mathrm{b1} } | $ bin range }
      & \multicolumn{1}{c}{ $ { \rd\sigma^{\text{full}} } / { \rd | {\eta^{\,\mathrm{b1}} } | } $ }
      & stat. & syst. & tot.
      & \multicolumn{1}{c}{ \multirow{2}{*}{ $( 1/{\sigma^{\text{full}}} )( {\rd \sigma^{\text{full}} }/{\rd | { \eta^{\,\mathrm{b1} } } | } )$ } }
      & stat. & syst. & tot. \\ [-2pt]
    \multicolumn{1}{c|}{ (\GeVns{}) }
      & \multicolumn{1}{c}{ (pb) } & (\%) & (\%) & (\%)
      & & (\%) & (\%) & (\%)   \\
\hline
0.0,\, 0.6 & \multicolumn{1}{c}{3.5} & 26 & 7 & 27 & \multicolumn{1}{c}{0.5}  & 26 & 7 & 27 \\
0.6,\, 1.2 & \multicolumn{1}{c}{2.9} & 24 & 6 & 25 & \multicolumn{1}{c}{0.4}  & 23 & 6 & 24 \\
1.2,\, 1.8 & \multicolumn{1}{c}{2.4} & 28 & 9 & 30 & \multicolumn{1}{c}{0.4}  & 26 & 9 & 27 \\
1.8,\, 2.4 & \multicolumn{1}{c}{2.7} & 29 & 10 & 31 & \multicolumn{1}{c}{0.4}  & 26 & 10 & 28 \\
\hline
    \multicolumn{9}{c}{} \\ [-10pt]
    \hline
    \multicolumn{1}{c|}{   $ {\pt}^{\,\mathrm{b2} } $ bin range  }
      & \multicolumn{1}{c}{ $ { \rd\sigma^{\text{full}} } / { \rd {\pt}^{\,\mathrm{b2}} } $ }
      & stat. & syst. & tot.
      & \multicolumn{1}{c}{  $( 1/{\sigma^{\text{full}}} )( {\rd \sigma^{\text{full}} }/{\rd {\pt}^{\,\mathrm{b2}} } )$  }
      & stat. & syst. & tot. \\ [-2pt]
    \multicolumn{1}{c|}{  (\GeVns{})  }
      & \multicolumn{1}{c}{ (pb/GeV) } & (\%) & (\%) & (\%)
      & \multicolumn{1}{c}{ $(\GeVns^{-1})$ } & (\%) & (\%) & (\%) \\
\hline
20,\, 45  & 4.2 , 10^{-2} & 40 & 10 & 42 & 3.0 , 10^{-2} & 18 & 7 & 20 \\
45,\, 80  & 7.3 , 10^{-3} & 50 & 25 & 56 & 5.3 , 10^{-3} & 57 & 24 & 62 \\
80,\, 200 & 6.8 , 10^{-4} & 108 & 35 & 113 & 4.9 , 10^{-4} & 114 & 35 & 120 \\
    \hline
    \multicolumn{9}{c}{} \\ [-10pt]
    \hline
    \multicolumn{1}{c|}{ $ |\eta^{\,\mathrm{b2} } | $ bin range }
      & \multicolumn{1}{c}{ $ { \rd\sigma^{\text{full}} } / { \rd | {\eta^{\,\mathrm{b2}} } | } $ }
      & stat. & syst. & tot.
      & \multicolumn{1}{c}{ \multirow{2}{*}{ $( 1/{\sigma^{\text{full}}} )( {\rd \sigma^{\text{full}} }/{\rd | { \eta^{\,\mathrm{b2} } } | } )$ } }
      & stat. & syst. & tot. \\ [-2pt]
    \multicolumn{1}{c|}{ (\GeVns{}) }
      & \multicolumn{1}{c}{ (pb) } & (\%) & (\%) & (\%)
      & & (\%) & (\%) & (\%)   \\
\hline
0.0,\, 0.6 & \multicolumn{1}{c}{1.0}  & 48 & 18 & 52 & \multicolumn{1}{c}{0.7}  & 46 & 18 & 50 \\
0.6,\, 1.2 & 5.8 , 10^{-1}  & 48 & 15 & 50 & \multicolumn{1}{c}{0.4}  & 41 & 12 & 43 \\
1.2,\, 2.4 & 3.4 , 10^{-1}  & 73 & 29 & 79 & \multicolumn{1}{c}{0.3}  & 66 & 29 & 72 \\
\hline
    \end{tabular}
   }
  \end{center}
\end{table*}

\begin{table*}[h]
  \renewcommand{\arraystretch}{1.2}
  \begin{center}
    \topcaption{Absolute (left) and normalized (right) differential \ttbar cross sections as a function of the invariant mass of the two leading additional \PQb jets in the event ($\mbb$) and the angle $\DR_{{\PQb\PQb}}$, along with their statistical, systematic, and total uncertainties. The results are presented at the particle level in the full phase space of the \ttbar system, corrected for acceptance and branching fractions.}
  \label{tab:dilepton:SummaryResultsBJet12FullPS}
    \resizebox{\textwidth}{!}{
    \begin{tabular}{y|x{4}ccc|x{-1}ccc}
    \hline
    \multicolumn{1}{c|}{   $\mbb$ bin range  }
     & \multicolumn{1}{c}{ $ { \rd\sigma^{\text{full}} } / { \rd \mbb } $ }
      & stat. & syst. & tot.
      & \multicolumn{1}{c}{ $( 1/{\sigma^{\text{full}}} )( {\rd \sigma^{\text{full}} }/{\rd \mbb } )$ }
      & stat. & syst. & tot. \\ [-2pt]
    \multicolumn{1}{c|}{  (\GeVns{})  }
      & \multicolumn{1}{c}{ (pb/GeV) } & (\%) & (\%) & (\%)
      & \multicolumn{1}{c}{ $(\GeVns^{-1})$ } & (\%) & (\%) & (\%) \\
    \hline
10,\, 60   & 1.1 , 10^{-2} & 83 & 23 & 86 & 8.4 , 10^{-3} & 69 & 23 & 73\\
60,\, 100  & 7.9 , 10^{-3} & 92 & 31 & 97 & 5.8 , 10^{-3} & 89 & 30 & 94 \\
100,\, 170 & 2.5 , 10^{-3} & 107 & 38 & 113 & 1.8 , 10^{-3} & 111 & 35 & 117 \\
170,\, 400 & 1.1 , 10^{-3} & 58 & 41 & 71 & 8.4 , 10^{-4} & 66 & 42 & 78 \\
    \hline
    \multicolumn{9}{c}{} \\ [-10pt]
    \hline
    \multicolumn{1}{c|}{ \multirow{2}{*}{ $\DR_{{\PQb\PQb}}$ bin range } }
      & \multicolumn{1}{c}{ $ { \rd\sigma^{\text{full}} } / { \rd \DR_{{\PQb\PQb}} } $ }
      & stat. & syst. & tot.
      & \multicolumn{1}{c}{ \multirow{2}{*}{ $( 1/{\sigma^{\text{full}}} )( {\rd \sigma^{\text{full}} }/{\rd \DR_{{\PQb\PQb}} } )$ } }
      & stat. & syst. & tot. \\ [-2pt]
      & \multicolumn{1}{c}{ (pb) } & (\%) & (\%) & (\%)
      & & (\%) & (\%) & (\%)   \\
    \hline	
0.5,\, 1.0 & 5.5 , 10^{-3} & 7508 & 2063 & 7786 & 4.1 , 10^{-3} & 7506 & 2063 & 7784 \\
1.0,\, 2.0 & 2.7 , 10^{-1} & 65 & 46 & 80 & 2.0 , 10^{-1} & 56 & 44 & 71 \\
2.0,\, 5.0 & 3.6 , 10^{-1} & 28 & 16 & 32 & 2.7 , 10^{-1} & 22 & 16 & 28 \\
\hline	
  \end{tabular}
  }
  \end{center}
\end{table*}

\section{Migration matrices}
\label{sec:migrationmatrix}
\begin{figure*}[htbp!]
 \begin{center}
     \includegraphics[width=0.385\textwidth]{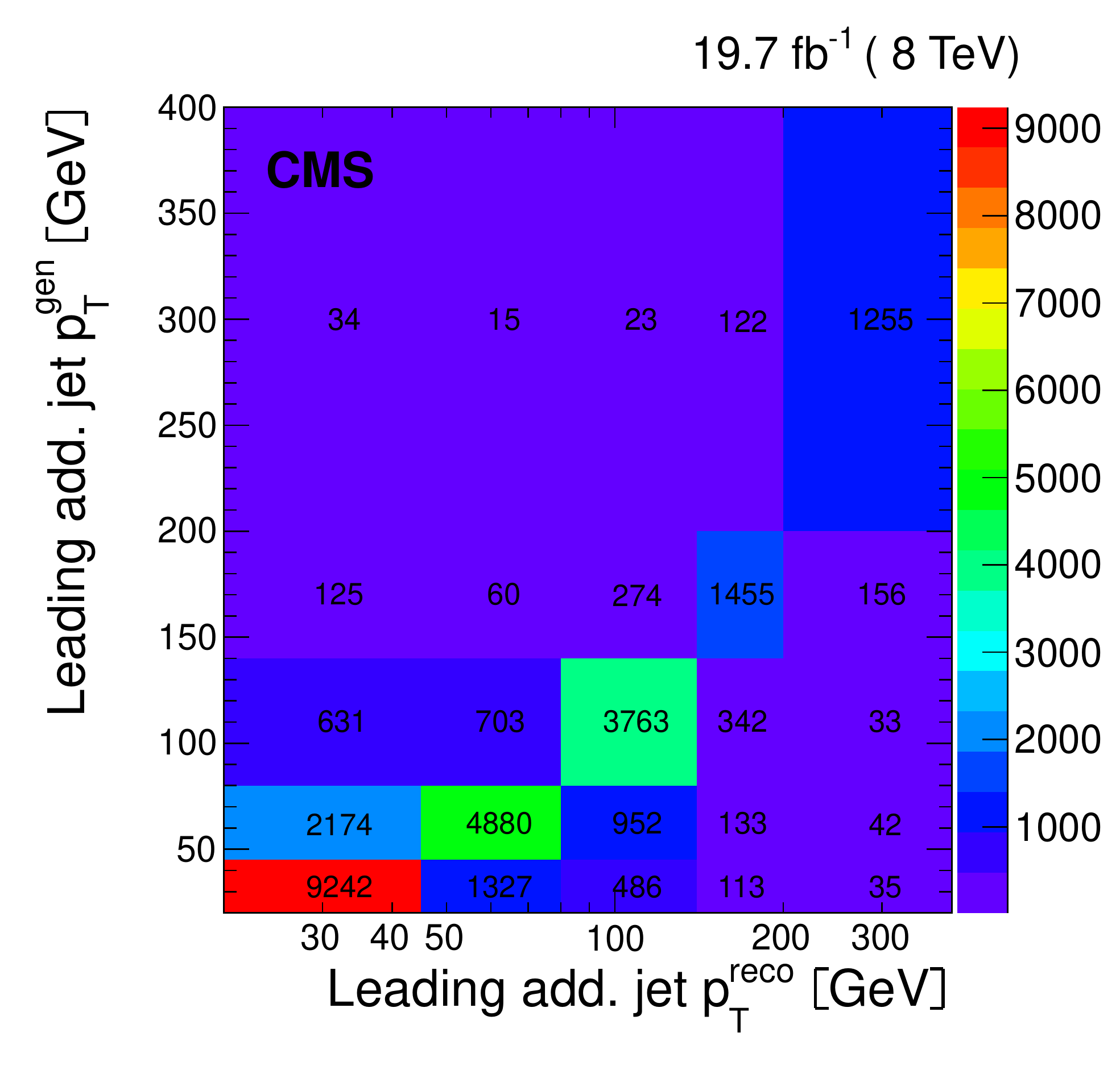}%
     \includegraphics[width=0.385\textwidth]{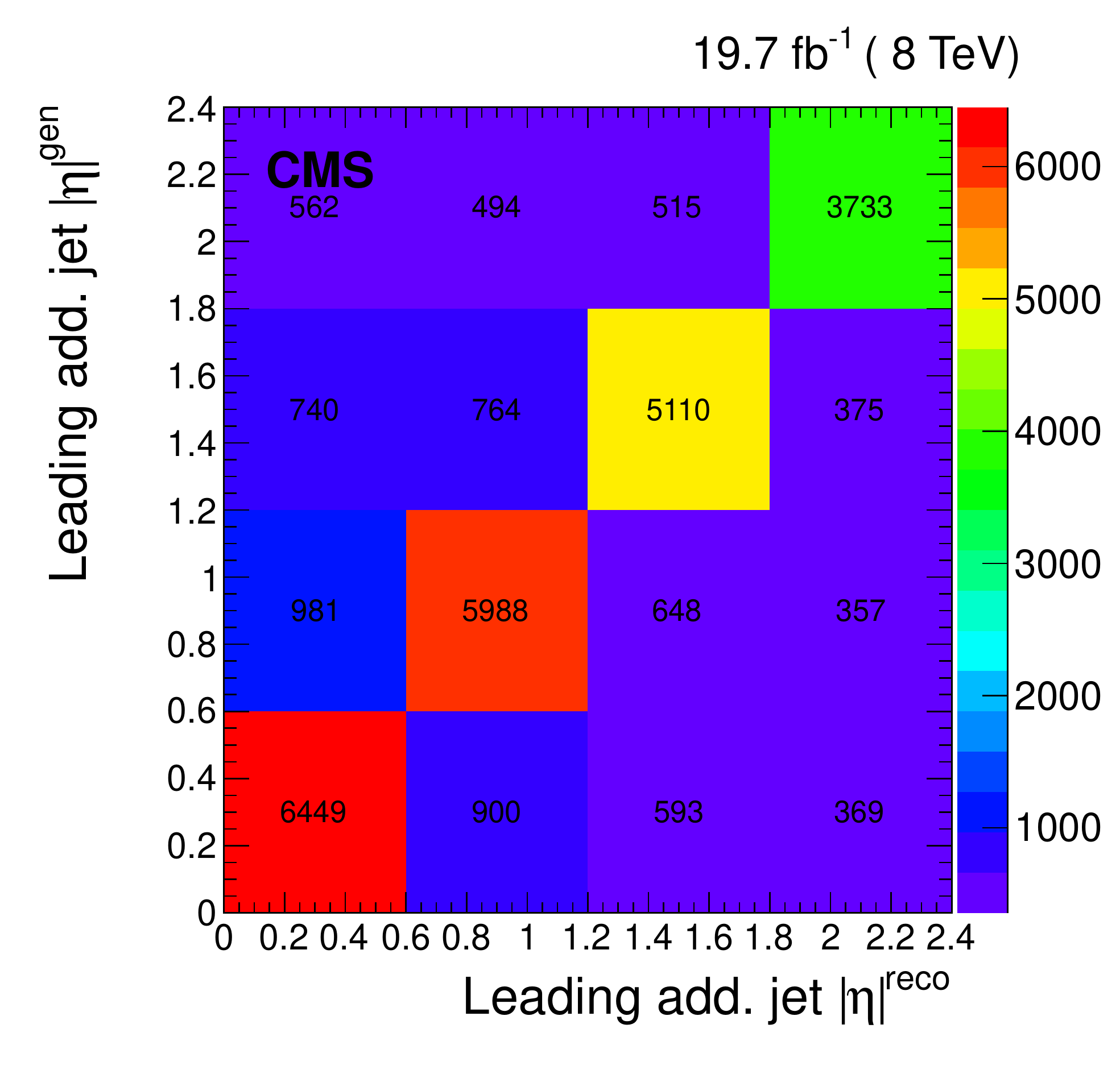}\\
     \includegraphics[width=0.385\textwidth]{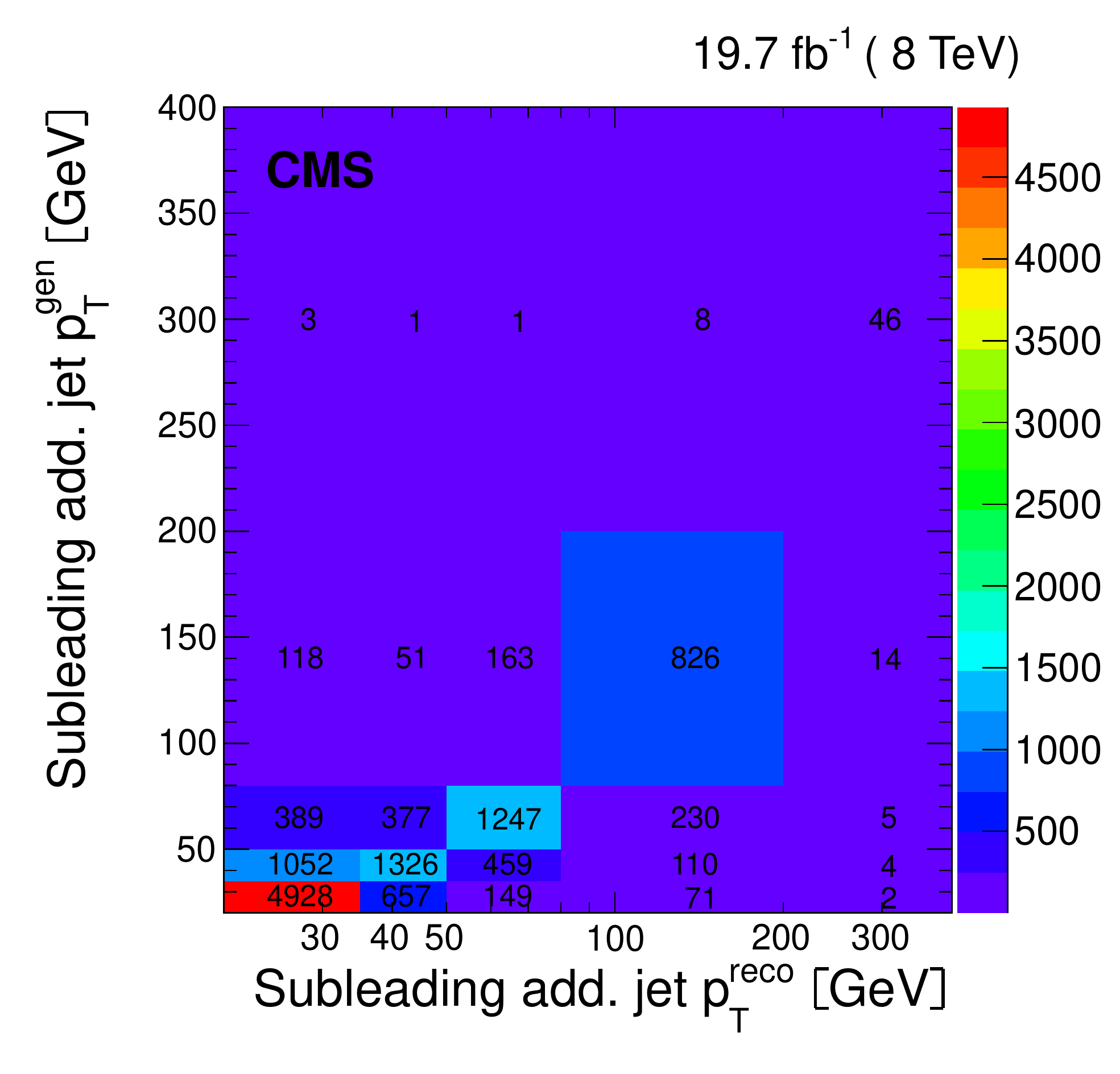}%
     \includegraphics[width=0.385\textwidth]{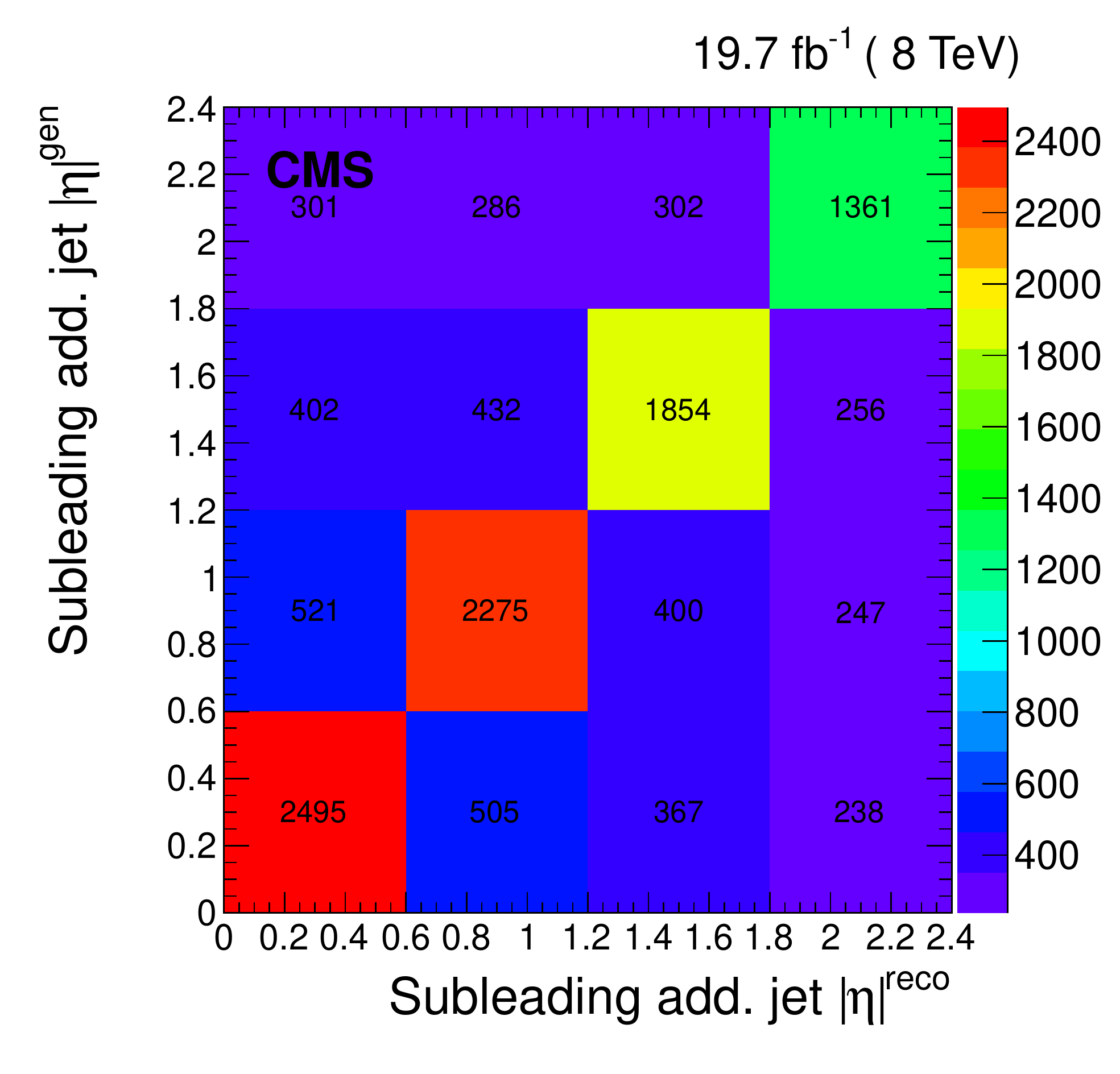}\\
     \includegraphics[width=0.385\textwidth]{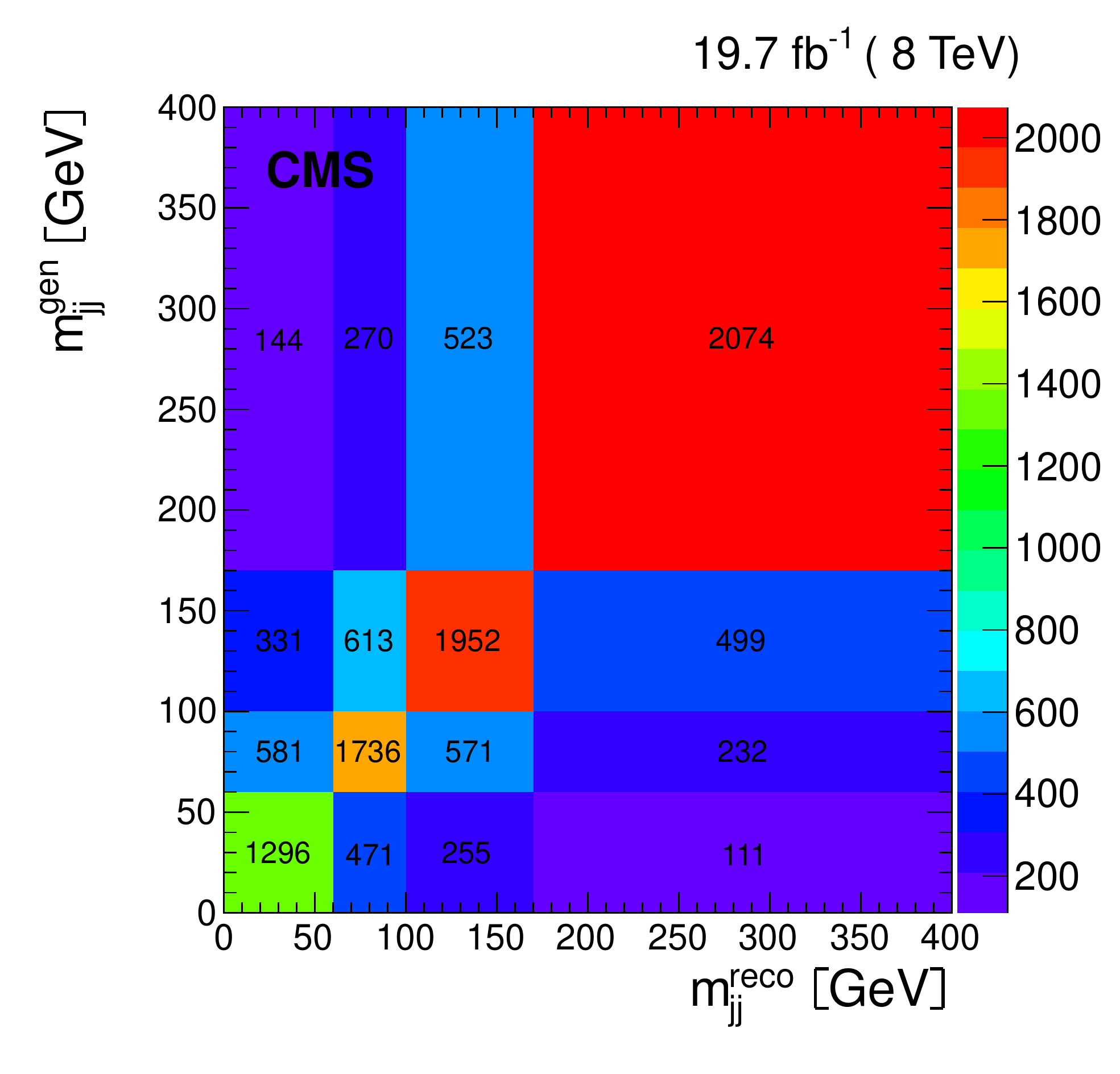}%
     \includegraphics[width=0.385\textwidth]{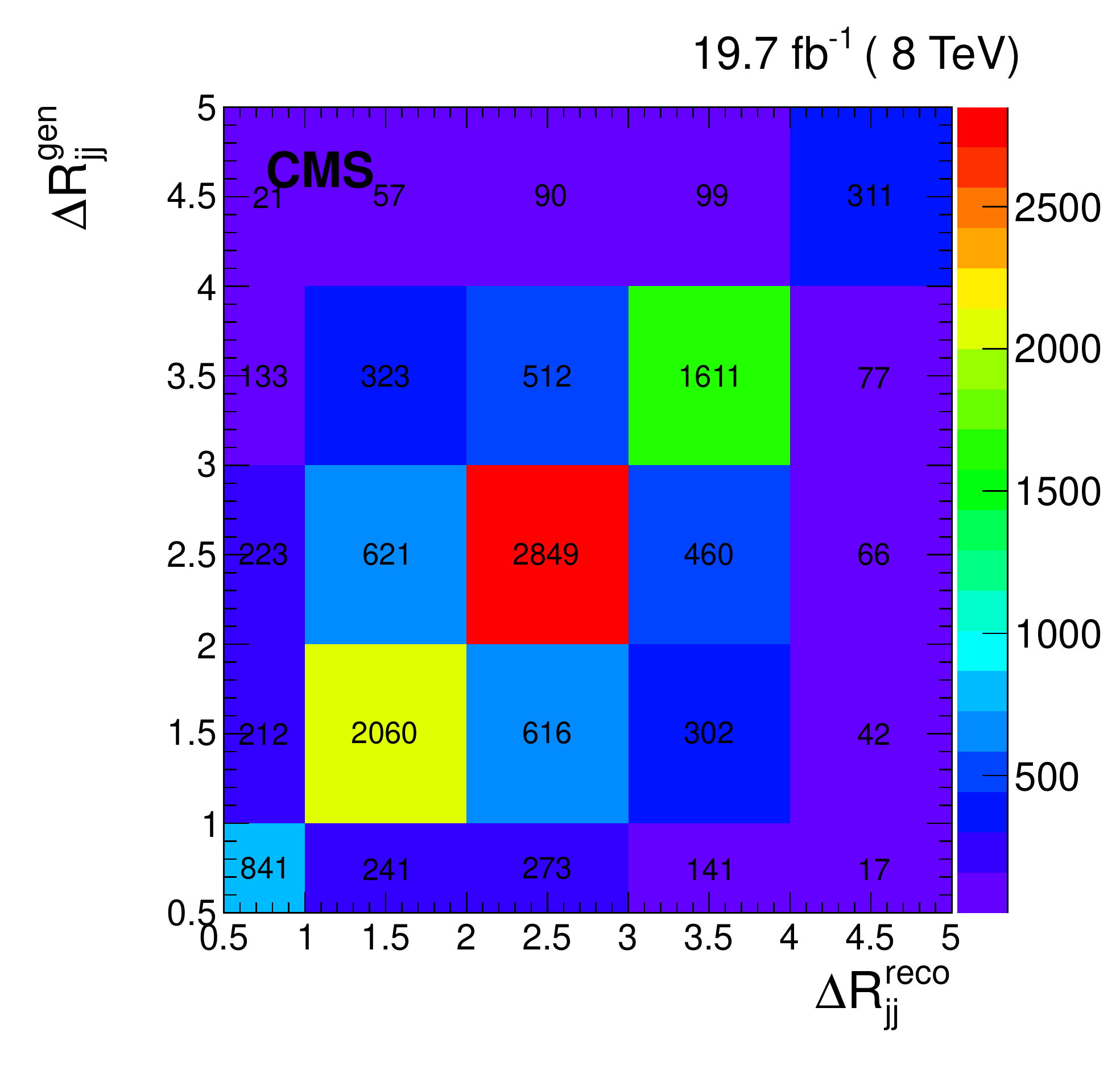}
\caption{The migration matrices relating the reconstructed level and the particle level in the visible phase space of the \ttbar decay products and the additional jets for the \pt (left) and \abseta (right) of the leading (top row) and subleading (middle row) additional jets in the event, \mjj (bottom left) and \Djj (bottom right). The matrices are obtained from simulated \ttbar events using \MADGRAPH{}+\PYTHIA{6}.}
\label{fig:migration}
 \end{center}
\end{figure*}

\begin{figure*}[htbp!]
 \begin{center}
      \includegraphics[width=0.385\textwidth]{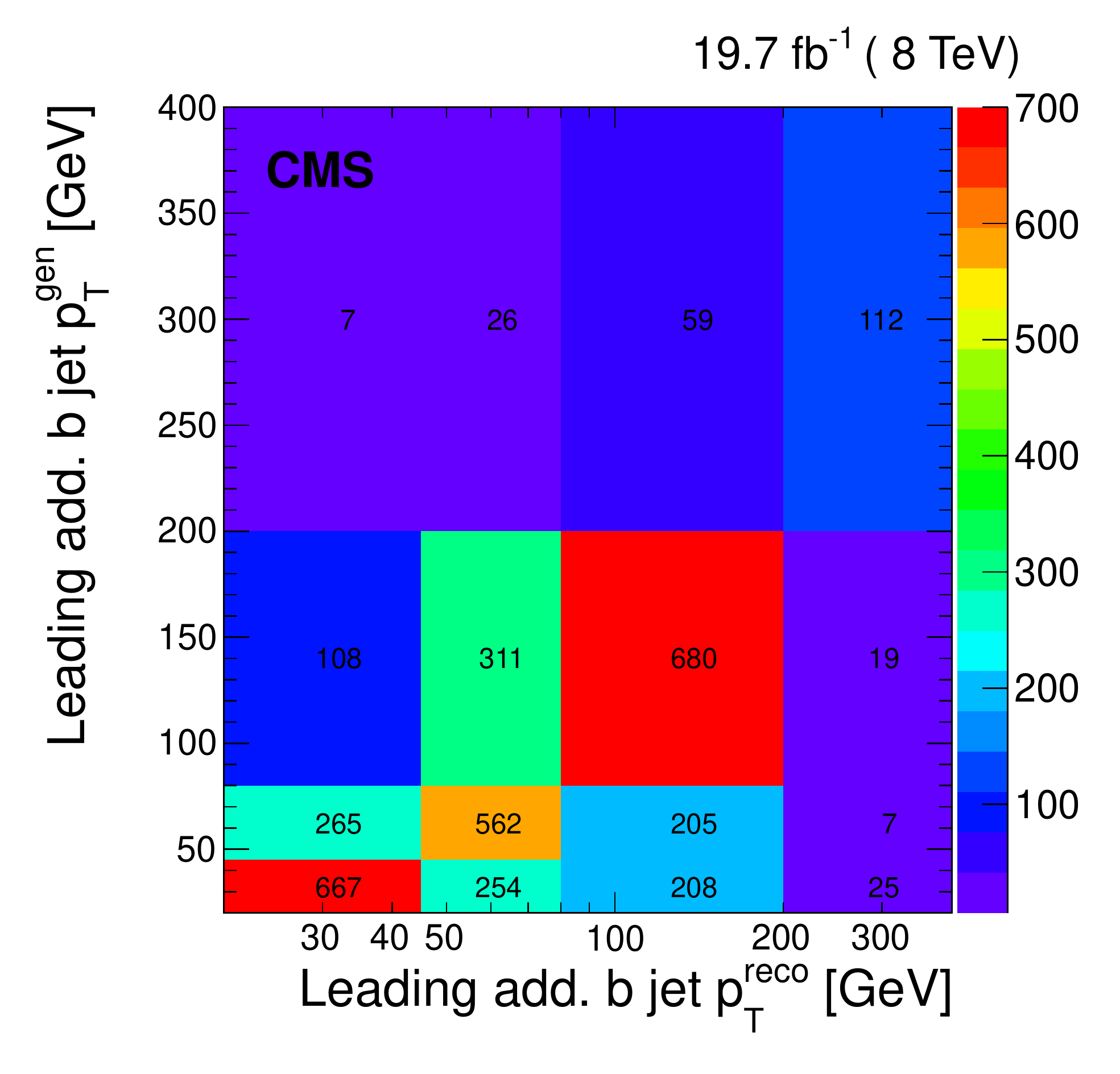}%
     \includegraphics[width=0.385\textwidth]{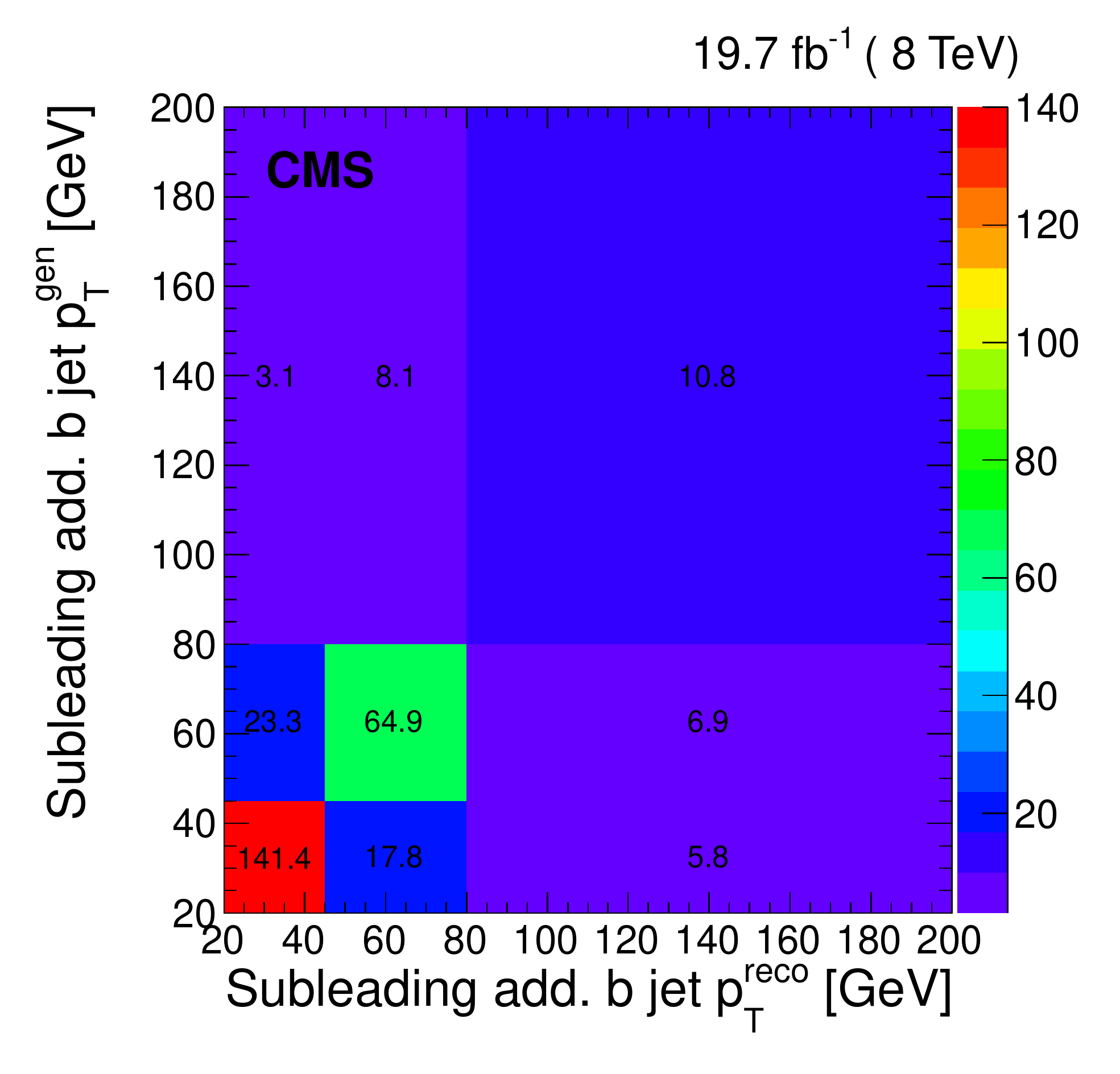}\\
      \includegraphics[width=0.385\textwidth]{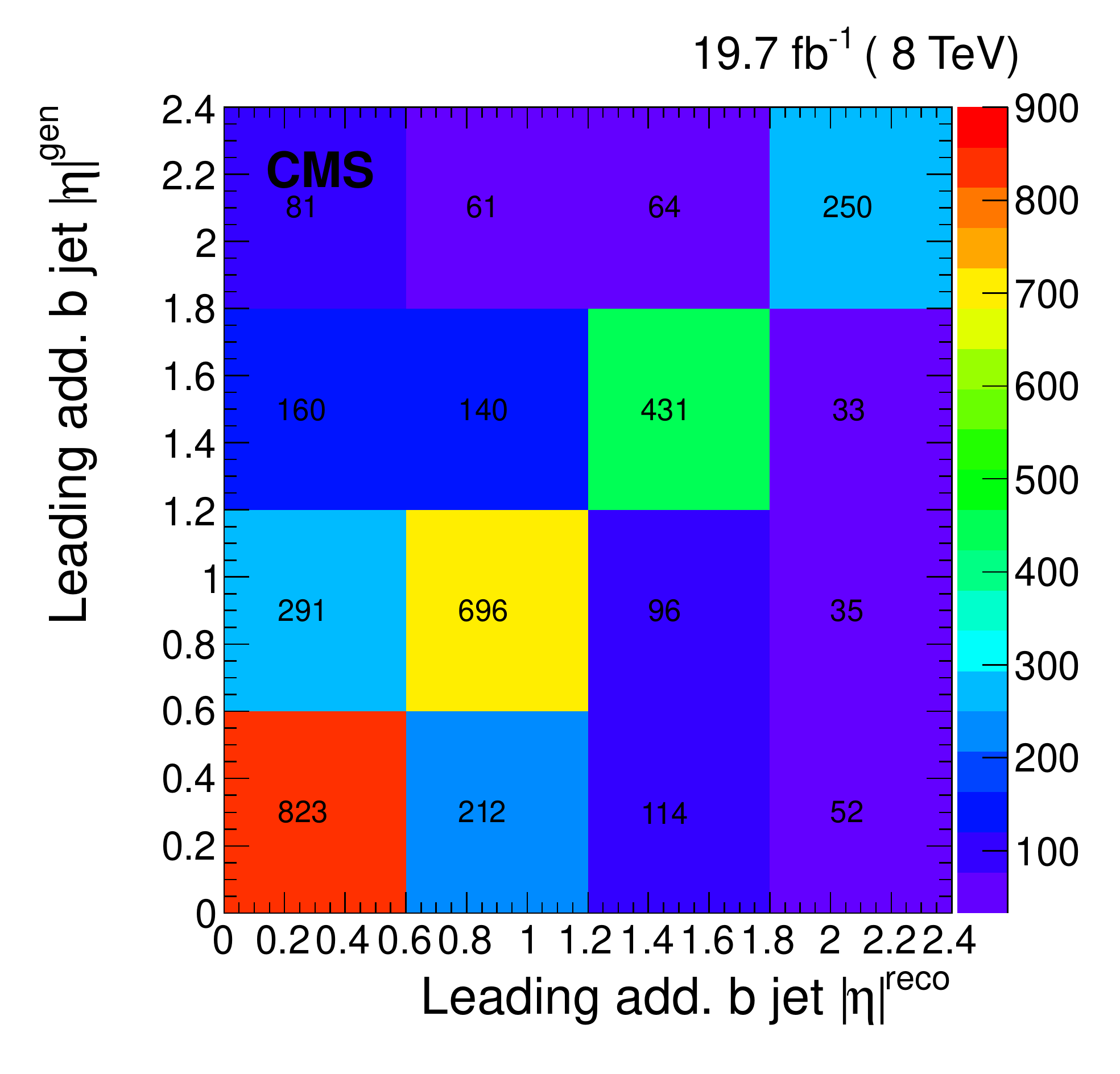}%
     \includegraphics[width=0.385\textwidth]{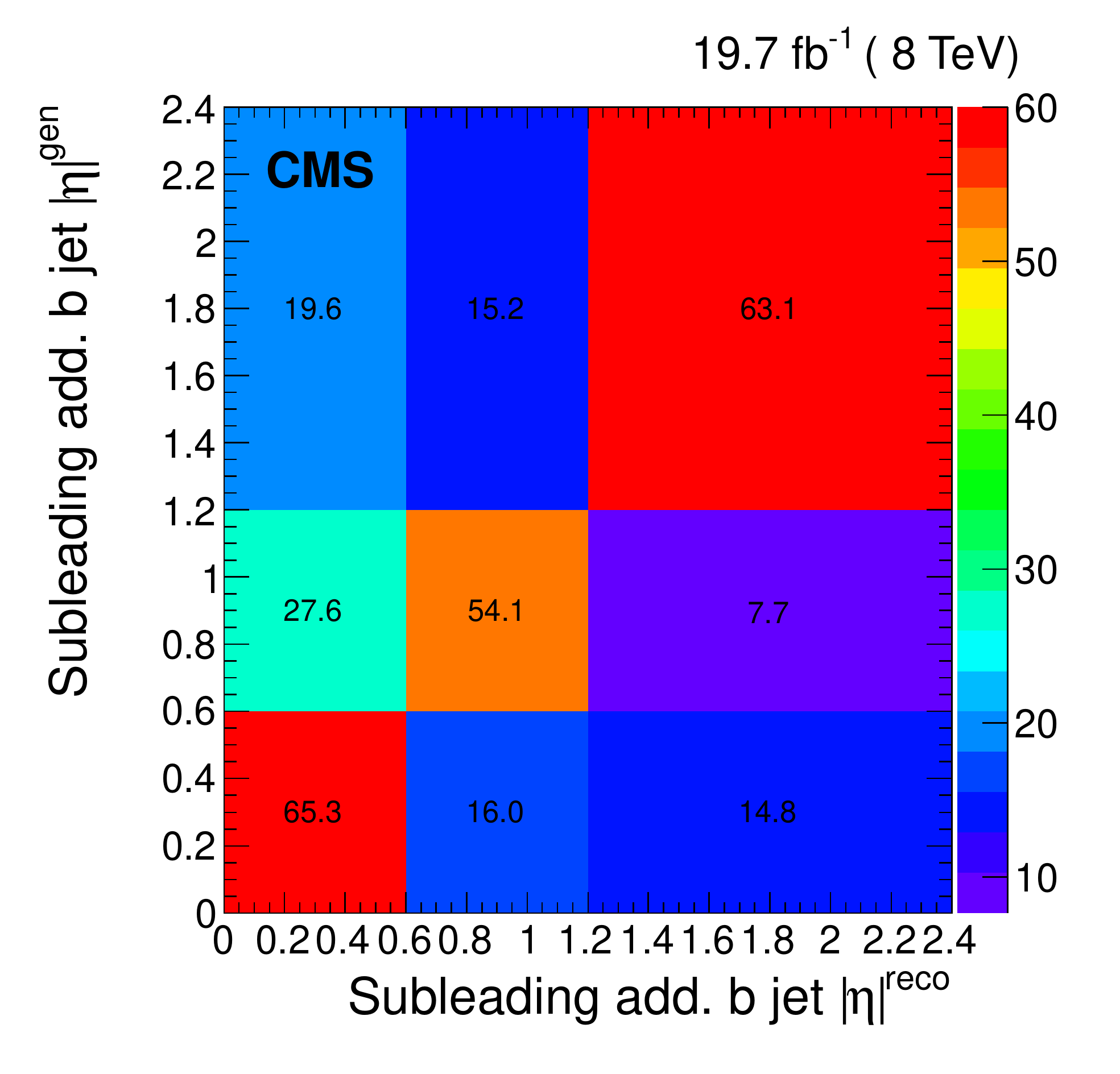}\\
     \includegraphics[width=0.385\textwidth]{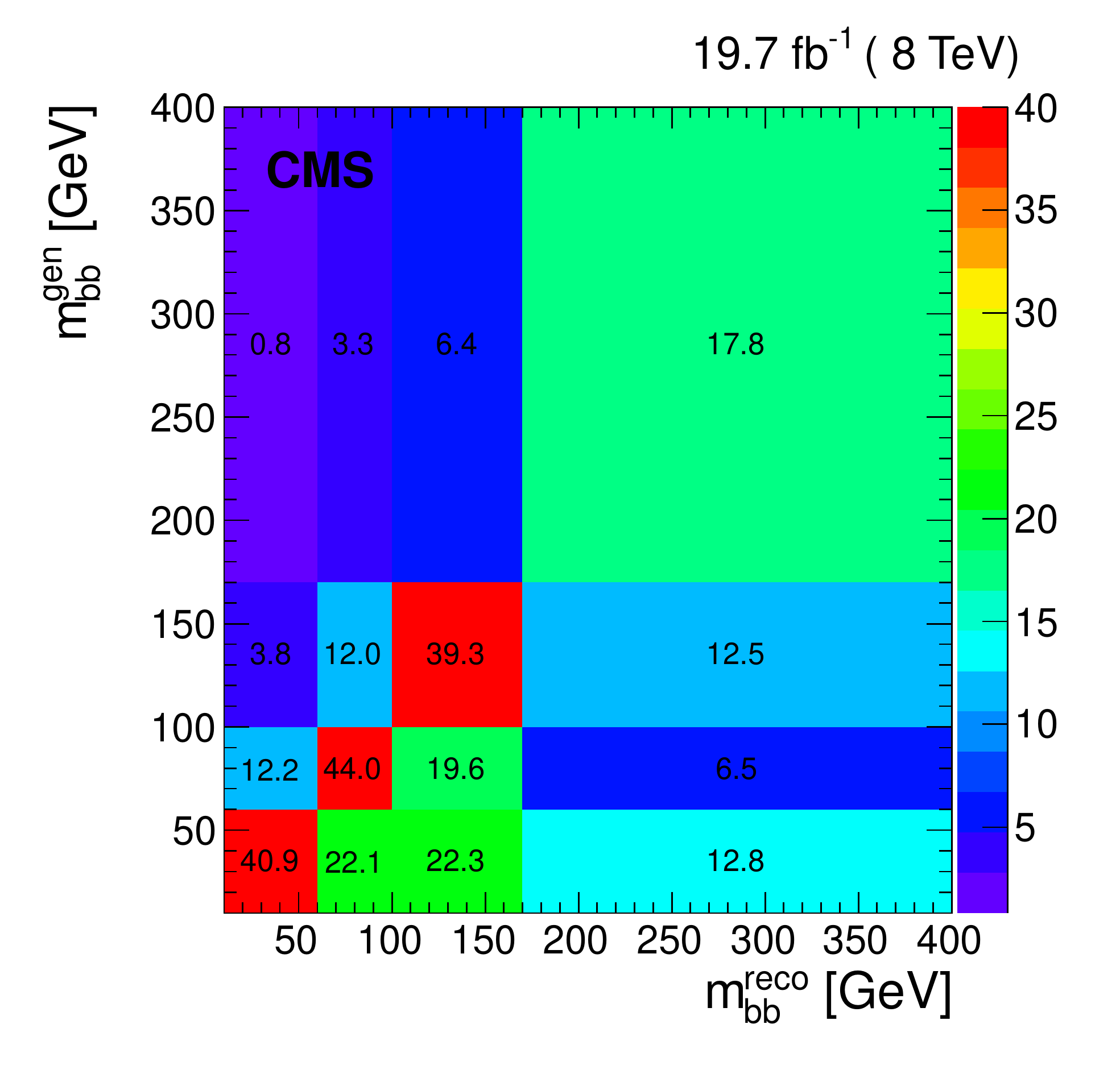}%
     \includegraphics[width=0.385\textwidth]{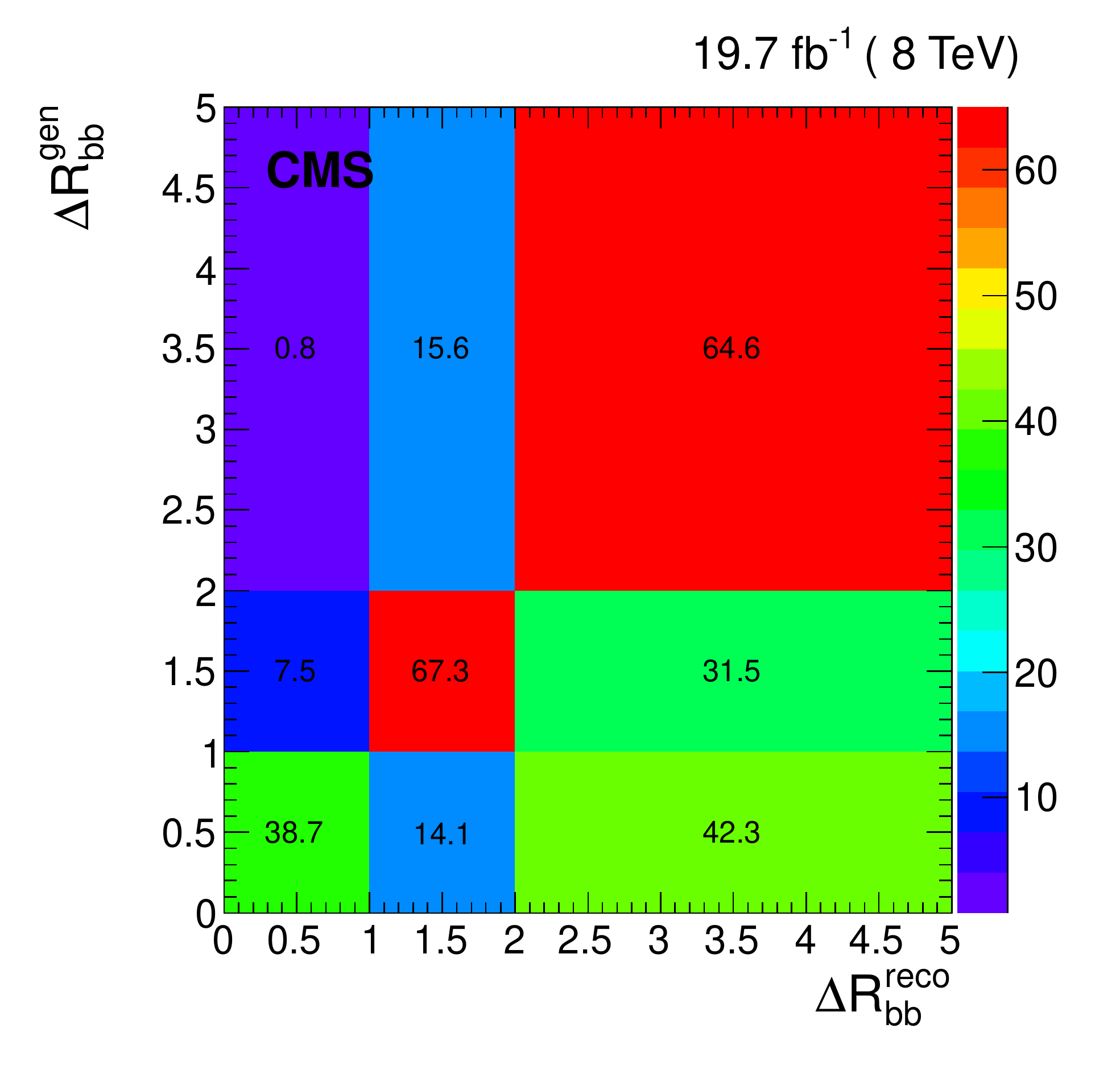}\\
\caption{The migration matrices relating the reconstructed level and the particle level in the visible phase space of the \ttbar decay products and the additional jets for the \pt (left) and \abseta (right) of the leading (top row) and subleading (middle row) additional \PQb jets in the event, $\mbb$ (bottom left), and $\DR_{{\PQb\PQb}}$ (bottom right). The matrices are obtained from simulated \ttbar events using \MADGRAPH{}+\PYTHIA{6}.}
\label{fig:migrationttbb}
 \end{center}
\end{figure*}

The migration matrices relating the kinematic properties of the additional jets and \PQb jets at the reconstruction level and particle level in the visible phase space of the \ttbar decay products and the additional jets are presented in Figs.~\ref{fig:migration} and ~\ref{fig:migrationttbb}, respectively.

\cleardoublepage \section{The CMS Collaboration \label{app:collab}}\begin{sloppypar}\hyphenpenalty=5000\widowpenalty=500\clubpenalty=5000\textbf{Yerevan Physics Institute,  Yerevan,  Armenia}\\*[0pt]
V.~Khachatryan, A.M.~Sirunyan, A.~Tumasyan
\vskip\cmsinstskip
\textbf{Institut f\"{u}r Hochenergiephysik der OeAW,  Wien,  Austria}\\*[0pt]
W.~Adam, E.~Asilar, T.~Bergauer, J.~Brandstetter, E.~Brondolin, M.~Dragicevic, J.~Er\"{o}, M.~Flechl, M.~Friedl, R.~Fr\"{u}hwirth\cmsAuthorMark{1}, V.M.~Ghete, C.~Hartl, N.~H\"{o}rmann, J.~Hrubec, M.~Jeitler\cmsAuthorMark{1}, V.~Kn\"{u}nz, A.~K\"{o}nig, M.~Krammer\cmsAuthorMark{1}, I.~Kr\"{a}tschmer, D.~Liko, T.~Matsushita, I.~Mikulec, D.~Rabady\cmsAuthorMark{2}, B.~Rahbaran, H.~Rohringer, J.~Schieck\cmsAuthorMark{1}, R.~Sch\"{o}fbeck, J.~Strauss, W.~Treberer-Treberspurg, W.~Waltenberger, C.-E.~Wulz\cmsAuthorMark{1}
% ...
\vskip\cmsinstskip
% ...
1:~~Also at Vienna University of Technology, Vienna, Austria\\
2:~~Also at CERN, European Organization for Nuclear Research, Geneva, Switzerland\\
% ...

\end{sloppypar}
\end{document}